\documentclass[a4paper,12pt,twoside]{report}
\usepackage{subfiles}
\usepackage[english]{babel}
\usepackage[utf8]{inputenc}
\usepackage[paperwidth=21cm,
            paperheight=29.7cm,
            lmargin=2.5cm,
            rmargin=2.5cm,
            tmargin=3.5cm,
            bmargin=3.5cm]{geometry}

\usepackage{commands}

\newacro{SM}{Standard Model}
\newacro{BSM}{Beyond the Standard Model}
\newacro{DM}{Dark Matter}
\newacro{DS}{Dark Sector}
\newacro{CMB}{Cosmic Microwave Background}
\newacro{LHC}{Large Hadron Collider}
\newacro{CE}{Chemical Equilibrium}
\newacro{KE}{Kinetic Equilibrium}
\newacro{FO}{Freeze-Out}
\newacro{FI}{Freeze-In}
\newacro{CDFO}{Conversion Driven Freeze-Out}
\newacro{WIMP}{Weakly Interacting Massive Particle}
\newacro{FIMP}{Feebly Interacting Massive Particle}
\newacro{SW}{SuperWIMP}
\newacro{QCD}{Quantum ChromoDynamics}
\newacro{EW}{ElectroWeak}
\newacro{BBN}{Big Bang Nucleosynthesis}
\newacro{MB}{Maxwell-Botlzmann}
\newacro{BE}{Bose-Einstein}
\newacro{FD}{Fermi-Dirac}
\newacro{MD}{Matter Dominated}
\newacro{EMD}{Early Matter Dominated}
\newacro{RD}{Radiation Dominated}
\newacro{LLP}{Long-Lived Particle}
\newacro{ECAL}{Electromagnetic CALorimeter}
\newacro{HCAL}{Hadronic CALorimeter}
\newacro{MS}{Muon Spectrometer}
\newacro{HSCP}{Heavy Stable Charged Particle}
\newacro{RH}{R-Hadron}
\newacro{WDM}{Warm Dark Matter}
\newacro{NLSP}{Next to Lightest Supersymmetric Particle}
\newacro{SUSY}{SUper SYmmetry}
\newacro{MSSM}{Minimal SuperSymetric Model}
\newacro{IR}{Infra-Red}
\newacro{UV}{Ultra-Violet}
\newacro{TOF}{Time-Of-Flight}
\newacro{MET}{Missing Transverse Energy}
\newacro{DV}{Displaced Vertex}
\newacro{DJ}{Delayed Jet}
\newacro{DLV}{Displaced Lepton Vertex}
\newacro{DT}{Disappearing Track}
\newacro{DL}{Displaced Lepton}
\newacro{KT}{Kinked Track}
\newacro{LEP}{Large Electron-Positron}
\newacro{SR}{Signal Region}
\newacro{NLO}{Next to Leading Order}
\newacro{LO}{Leading Order}
\newacro{CL}{Confidence Level}
\newacro{BR}{Branching Ratio}
\newacro{MACHO}{Massive Astrophysical Compact Halo Object}

\begin{document}

\includepdf[pages=-]{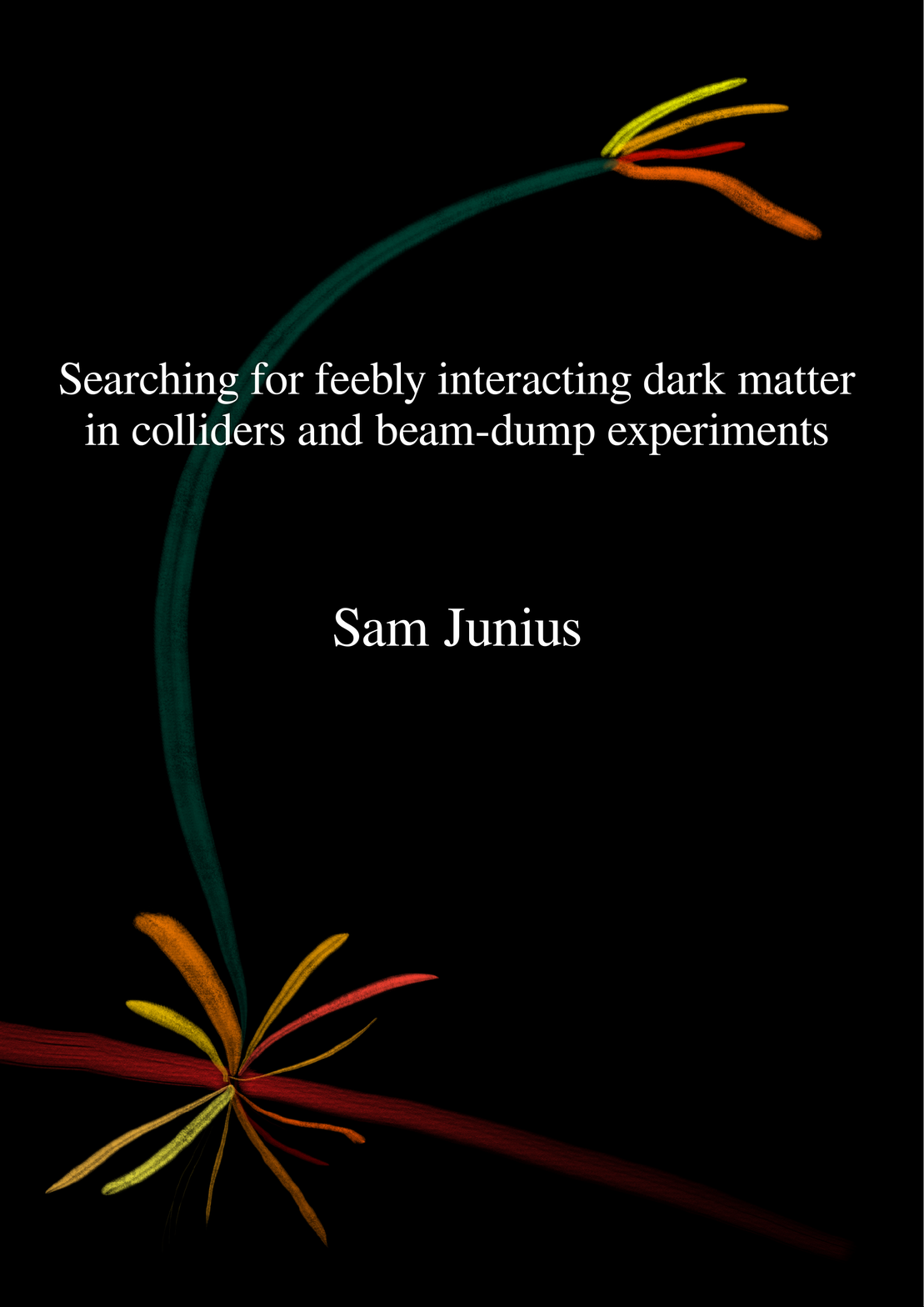}

\clearpage
\thispagestyle{empty}
\hfill
\clearpage

\begin{titlepage}

\includegraphics[height=2.8cm]{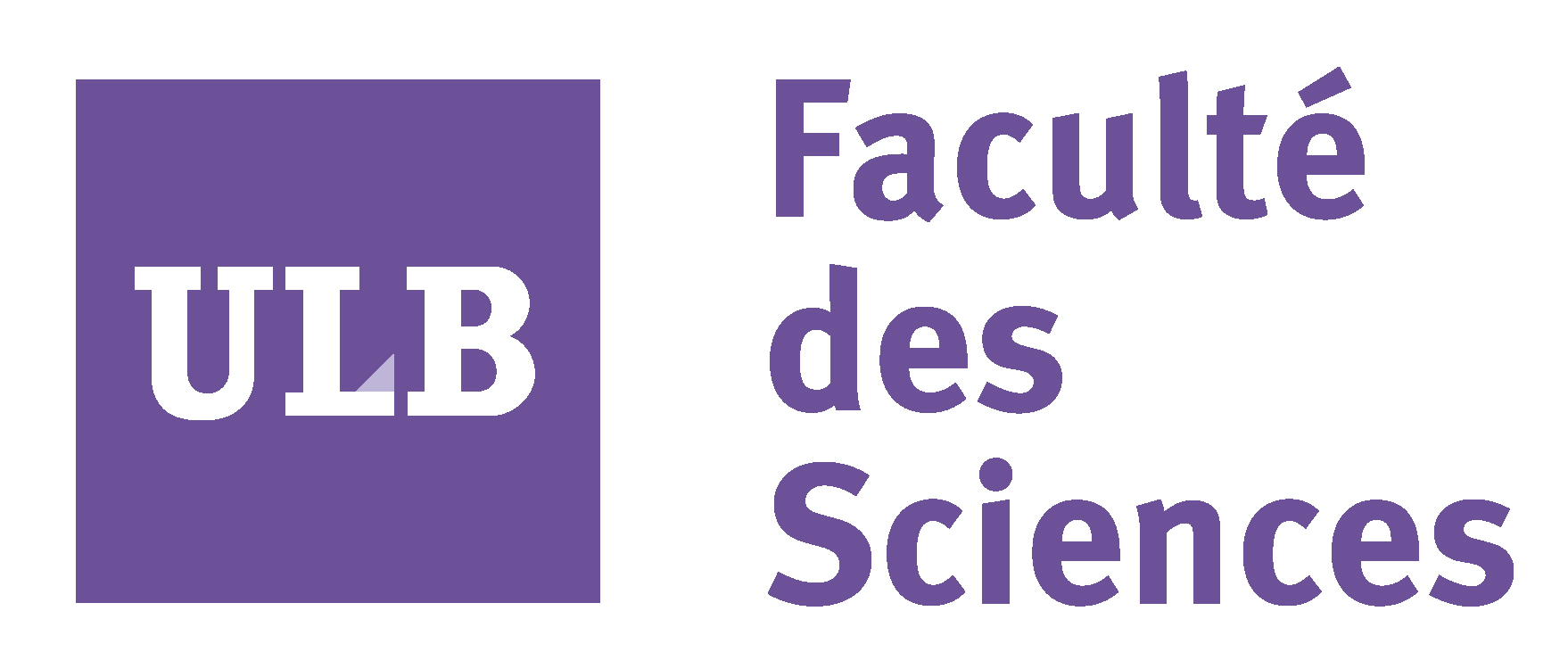} 
\hfill
\includegraphics[height=2.8cm]{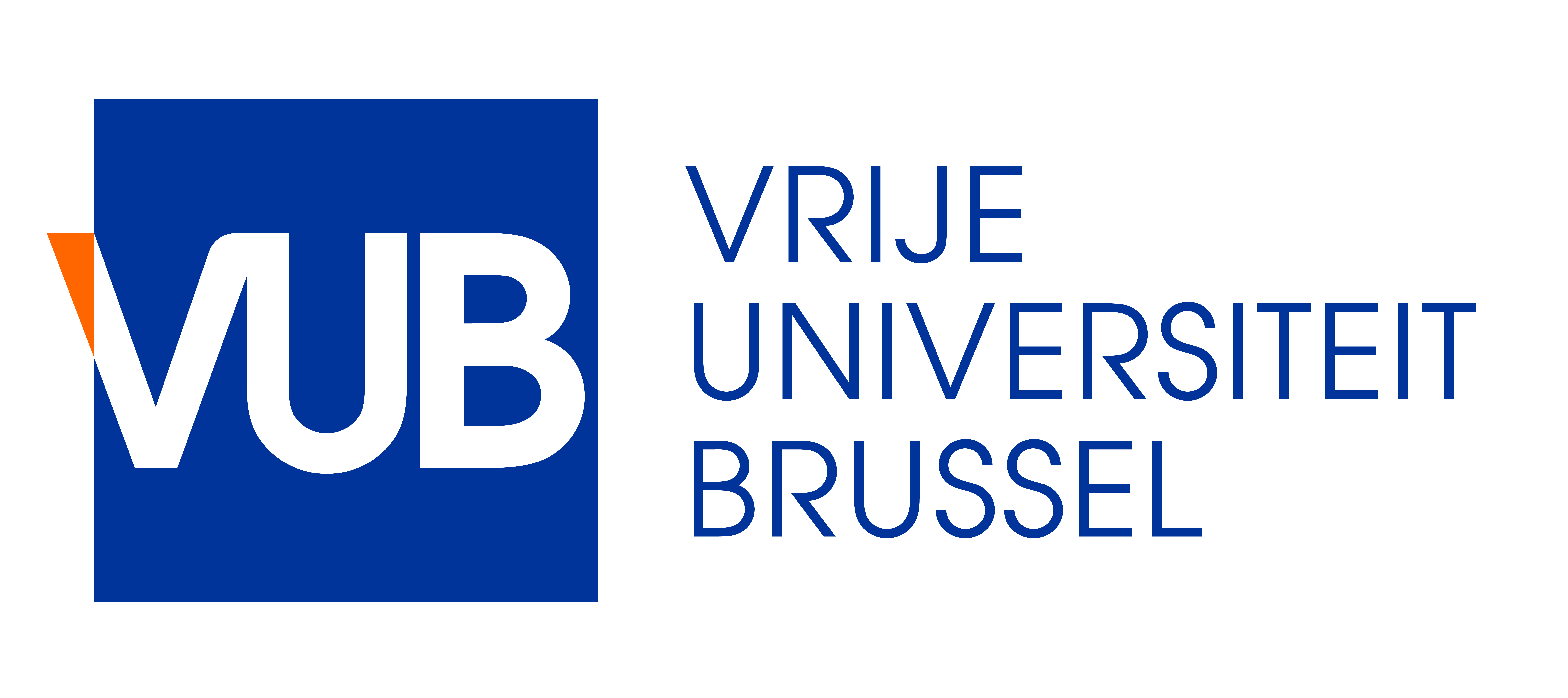}
     
\vspace{1cm}

\begin{titlebox}
    \Large{\textbf{Searching for feebly interacting dark matter}} \\
    \Large{\textbf{at colliders and beam-dump experiments}}
\end{titlebox} 

\vspace{1cm}

\begin{tcolorbox}[halign=left]
    \color{darkblue}
    \normalsize{\textbf{Thesis presented by Sam JUNIUS}} \\
    \color{black}
    \normalsize{in fulfilment of the requirements of the PhD Degree in Science (ULB - ``Docteur en Sciences Physiques'') (VUB - ``Doctor in de Wetenschappen'') \\
    Academic year 2022-2023}
\end{tcolorbox}
 
\vspace{1cm}

\begin{tcolorbox}[halign=right]
    \color{darkblue}
    \small{Supervisors: Professor Laura LOPEZ HONOREZ (Université libre de Bruxelles)}\\
    \color{black}
    \small{Service de Physique Théorique} \\
    \small{Theoretische Natuurkunde} \\
    \small{The International Solvay Institutes} \\
    \color{darkblue} 
    \small{and Professor Alberto MARIOTTI (Vrije universiteit Brussel)} \\
    \color{black}
    \small{Theoretische Natuurkunde} \\
    \small{The International Solvay Institutes} \\
    \small{Inter-University Institute for High Energies}
\end{tcolorbox}

\vfill

\begin{tcolorbox}
    \textbf{Thesis jury : }\\
    {\color{white} creating empty line} \\
    \small{Steven LOWETTE (Vrije universiteit Brussel, Chair)} \\
    \small{Thomas HAMBYE (Université libre de Bruxelles, Secretary)} \\
    \small{Felix KAHLHOEFER (Karlsruhe Institute of Technology)} \\
    \small{Suchita KULKARNI  (Universität Graz)}
\end{tcolorbox}

\end{titlepage}

\clearpage
\thispagestyle{empty}
\hfill
\clearpage

\chapter*{Summary in English}
Despite the compelling amount of evidence for the presence of dark matter in our universe through gravitational effects, the exact nature of dark matter is still one of the main open questions in modern physics. A great experimental effort has been performed in order to probe one of the most well studied dark matter candidates, the Weakly Interacting Massive Particle or WIMP. Despite many searches performed, there has not been any conclusive detection of dark matter yet. Hence, these searches place strong constraints on the WIMP paradigm, restricting the amount of models containing a viable WIMP candidate. The community has therefore started to consider different dark matter candidates who are able to evade the strong experimental constraints.

A novel dark matter candidate that has gained a lot of attention in the recent years is the Feebly Interacting Massive Particle or FIMP. Unlike the WIMP, which is produced through the freeze-out mechanism, for which it needs sizable interactions with the standard model particles, the FIMP has very feeble interactions. The FIMP is therefore unable to reach a state of equilibrium which is needed in order to be produced trough freeze-out. There is a variety of mechanisms able to produce the correct amount of FIMPs to explain all the dark matter in our universe, despite its feeble interactions. In this work, a selection of these mechanisms, such as freeze-in or conversion driven freeze-out, are discussed in detail.

Since these FIMPs are so feebly interacting with standard model particles, it is difficult to probe these particles through conventional direct and indirect detection searches. In contrast, the feeble couplings give rise to long-lived particles coupling to the DM, which can hence be smoking gun signatures for FIMPs. This thesis will therefore focus on searches for long-lived particles, both at hadron colliders and beam-dump experiments. In order to do so, a classification of FIMP models has been put forward within the context of this PhD, in which potential production mechanisms are identified in order to define a viable parameter space. Within this framework, both existing and potential future searches for long-lived particles are discussed and applied to a subset of the proposed classification.

\chapter*{Samenvatting in het Nederlands}
Ondanks de overweldigende hoeveelheid bewijs voor het bestaan van donkere materie in ons universum, blijft de exacte aard ervan één van de belangrijkste open vragen in de moderne fysica. Vele pogingen zijn gewaagd om één van de populairste kandidaten voor donkere materie, het ‘Weakly Interacting Massive Particle’ of WIMP deeltje, te detecteren. Ondanks de vele experimenten die zijn uitgevoerd, werd het WIMP deeltje nooit gedetecteerd. Hierdoor zijn vele WIMP modellen uitgesloten waardoor er slechts een gelimiteerd aantal WIMP kandidaten over blijven. De fysica gemeenschap is daarom op zoek gegaan naar potentiële alternatieven voor het WIMP paradigma.

Een nieuwe kandidaat voor donkere materie die de laatste jaren onder de aandacht kwam is het ‘Feebly Interacting Massive Particle’ of FIMP deeltje. In tegenstelling tot het WIMP deeltje, dat geproduceerd wordt door het `freeze-out' mechanisme waarvoor het WIMP deeltje sterke interacties nodig heeft met deeltjes uit het standaard model, heeft het FIMP deeltje zeer zwakke interacties met die deeltjes. Daarom kan het FIMP deeltje niet geproduceerd worden door het `freeze-out' mechanisme. Er zijn echter vele alternatieve mechanismes, zoals bijvoorbeeld het `freeze-in' of `conversion driven freeze-out' mechanisme, die FIMP deeltjes in de correcte hoeveelheden produceren om alle donkere materie dat we waarnemen in ons universum te verklaren. In dit werk zal een selectie van deze mechanismes in detail besproken worden.

Aangezien de FIMP deeltjes zo zwak interageren met deeltjes uit het standaard model, is het moeilijk om deze te detecteren via conventionele directe en indirecte donkere materie experimenten. De kleine koppelingen zorgen er echter voor dat in modellen met FIMP deeltjes als kandidaten voor donkere materie, deeltjes die enkel interageren met de donkere materie vaak een lange vervaltijd hebben. Het detecteren van zo een deeltjes kan dus ook wijzen op het bestaan van FIMP deeltjes. Deze thesis zal daarom focussen op onderzoeken naar deeltjes met een lange vervaltijd in zowel deeltjes versnellers als zogenoemde ‘beam-bump’ experimenten. Om dit te doen stellen we een een classificatie voor van FIMP modellen waarin we de verschillende productie mechanismes identificeren om de parameters te bepalen die de correcte hoeveelheid donkere materie voorspellen. Hierdoor kunnen we bestaande of potentieel nieuwe onderzoeken toepassen op bepaalde modellen uit onze classificatie.

\chapter*{Résumé en Fran\c{c}ais}

Malgré le nombre impressionnant de preuves de la présence de matière noire dans notre univers par le biais des effets gravitationnels, la nature exacte de la matière noire reste l'une des principales questions ouvertes de la physique moderne. De nombreux efforts expérimentaux ont été déployés afin de mettre en évidence l'un des candidats de matière noire les plus étudiés, la particule massive à interaction faible ou WIMP (Weakly Interacting, Massive Particle). Malgré les nombreuses recherches effectuées, il n'y a pas encore eu de détection conclusive de la matière noire. Ces résultats imposent donc de fortes contraintes au paradigme WIMP, ne laissant qu'un nombre limité de modèles contenant un candidat WIMP viable. La communauté a donc commencé à analyser d'autres candidats de matière noire capables d'échapper aux fortes contraintes expérimentales.

Un nouveau candidat de matière noire qui a suscité beaucoup d'attention ces dernières années est la particule massive à interaction faible (Feebly Interacting, Massive Particle ou FIMP). Contrairement au WIMP, qui est produit par le mécanisme de freeze-out, associé à des tous d'interaction relativement importants avec du modèle standard, le FIMP a des interactions très faibles. Le FIMP est donc incapable d'atteindre l'équilibre thermodynamique nécessaire pour pouvoir être produit par le mécanisme de freeze-out. Il existe cependant une variété de mécanismes capables de produire l'abondance correcte de FIMPs afin de rendre compte de toute la matière noire de notre univers, malgré sa faible interaction. Dans ce travail, une sélection de ces mécanismes est discutée en détail.

Puisque ces FIMPs interagissent si faiblement avec les particules du modèle standard, il est difficile de tester les propriétes de ces particules par des expériences de détection conventionnelles de détection directe et indirecte. En revanche, les couplages faibles donnent lieu à des particules à longue durée de vie, qui peuvent donc être les des FIMPs. Cette thèse se concentre donc sur l'étude des particules à longue durée de vie dans les accélérateurs de hadrons et dans les expériences à cible fixe. Pour ce faire, une classification des modèles de FIMPs a été proposée dans le cadre de cette thèse, dans laquelle les mécanismes de production potentiels de FIMPs sont identifiés afin de définir un espace de paramètres viable. Dans ce cadre, les expériences actuelles et futures de recherche de particules à longue demi-vie sont discutées et les contraintes correspondantes sont appliquées à un sous-ensemble de la classification proposée.

\clearpage
\thispagestyle{empty}
\hfill
\clearpage

\chapter*{Acknowledgments}

A long journey of more than nine years is coming to an end. When I decided to study physics at the age of 18, I never thought it would end up the way it eventually did. I was not even sure if I would enjoy it, but I decided to take the leap. It turned out completely different than I expected, but also way better. Until my final year of my master's program, I thought I would leave the world of physics after obtaining my master's degree. This was until I started working on my master thesis together with my supervisors Alberto Mariotti and Laura Lopez Honorez. During this time, I discovered the joy of doing research in dark matter physics, and this is what convinced me to start my PhD. I am very grateful that Alberto and Laura offered me this opportunity, as they have put a lot of effort in making this possible. Thanks a lot for the wonderful collaboration of the past five years, we had a lot of interesting discussions, and I could always knock on your doors any time I got a problem, whether it was physics related or not. We also had many nice moments completely unrelated to physics, which for me is also really important, so I am really happy I was able to work on my PhD together with you. Working with you shaped me as a physicist, but also as a person in general.

Further, I really want to thank everyone I collaborated with, namely Lorenzo Calibbi, Francesco D'Eramo, Susanne Westhoff and Anastasiia Filimonova. Thanks to the really nice physics input and the interesting discussions we had (unfortunately mainly online due to ...), we were able to publish qualitative papers of which I am really proud. Thanks to our collaborations, I was able to develop many new skills. The same goes out to Simon Knapen, Diego Redigolo, Andrei Messinger, Yuxiang Qin and Gaétan Facchinetti, with who I collaborated but was not able publish a paper (yet!). It also goes without saying that I had many interesting discussions with all the people I met during my PhD, both in the PhysTh group at ULB and the pheno group at VUB. This always boosted my enthusiasm and interest about physics. I especially noticed how important these informal discussions were to me during the multiple periods of obliged home office. For me, it was not easy to always move between the two groups, however, I always felt welcome and a part of the team. For that, I want to thank you all. I also want to thank the former and present sectaries of the VUB and ULB groups I was part of, for all the support with administration (which is clearly not my strong suit) and practical arrangements.

I also should not forget the members of my PhD jury, namely Steven Lowette, Thomas Hambye, Felix Kahlhoefer and Suchita Kulkarni. Thank you first of all for taking the time to read the manuscript. We also had very nice discussions during the private defense, so nice that I could feel the stress fading away as the discussions progressed. This gave me the confidence to answer all of your questions as good as I possibly could. The comments you provided to improve my manuscript were also very useful and I am really grateful for that.

Als laatste moet ik natuurlijk ook mijn familie bedanken voor alle steun die ik kreeg tijdens mijn negenjarig avontuur doorheen de wereld van de fysica. Dankzij alle steun en flexibiliteit die ik van jullie kreeg, kon ik mij op de juiste momenten focussen op mijn studies en doctoraat, maar had ik ook genoeg tijd om mijn gedachten vrij te maken. Ook tijdens de lockdown periodes zorgden jullie voor de nodige afleiding. Deze afleiding is zeker iets wat heeft geleid tot het succes van mijn doctoraat, en daarom wil ik ook zeker al mijn vrienden die ik leerde kennen tijdens mijn vele scoutsavonturen en mijn ploeggenoten en trainingsmakkers van mijn badmintonclub bedanken. Het lijkt misschien raar om te zeggen, maar ook jullie hebben een groot aandeel in het voltooien van mijn doctoraat. Telkens ik vast zat met een groot probleem, kon ik het dankzij jullie even naast mij neerleggen, en er de volgende dag met een frisse blik terug aan beginnen, wat meermaals tot een doorbraak heeft geleid. Bedankt hiervoor!

\clearpage
\thispagestyle{empty}
\hfill
\clearpage

\chapter*{Content of the manuscript}
This manuscript contains a very detailed review of the main subjects I have studied during the four years of my PhD and the final year of my master studies. I have mainly focused on feebly interacting dark matter and how this elusive particle can be probed at accelerator experiments. This work has led to the publication of three articles in peer-reviewed international journals:
\begin{itemize}
    \item S.~Junius, L.~Lopez~Honorez and A.~Mariotti, {\it A feeble window on leptophilic dark matter}, JHEP 07 (2019), 136, doi:10.1007/JHEP07(2019)136 [arXiv:1904.07513 [hep-ph]],
    \item L.~Calibbi, F.~D'Eramo, S.~Junius, L.~Lopez~Honorez and A.~Mariotti, {\it Displaced new physics at colliders and the early universe before its first second}, JHEP 05 (2021), 234, doi:10.1007/JHEP05(2021)234 [arXiv:2102.06221 [hep-ph]],
    \item A.~Filimonova, S.~Junius, L.~Lopez~Honorez and S.~Westhoff, {\it Inelastic Dirac dark matter}, JHEP 06 (2022), 048, doi:10.1007/JHEP06(2022)048 [arXiv:2201.08409 [hep-ph]].
\end{itemize}
For these articles, I was the main contributor of the numerical analyses (for the latter paper, this was in collaboration with Anastasiia Filimonova). I developed my own code to simulate the dark matter production in the early universe for all three papers. Further, I also did all the reinterpretation analysis of the LHC searches as described in detail in the articles and this manuscript. All the figures you can find in the articles mentioned above are produced by me, and I also contributed to writing the articles. 

Further, I have also worked on two projects who are not mentioned in this thesis and are not yet published. A first project I started with Laura Lopez Honorez, Gaétan Facchinetti, Andrei Mesinger and Yuxiang Qin. In this project, we are investigating how observations of the cosmological 21cm signal can constrain annihilating and decaying dark matter. More specifically, we are obtaining prospects of the potential bounds the HERA radio telescope could place on these types of DM models. In order to obtain these prospects, I developed together with Gaétan an extension of the code 21cmFAST that simulates the expected 21cm signal in the presence of DM annihilations and/or decays. The second project is a collaboration with Alberto Mariotti, Simon Knapen and Diego Redigolo. For this project, we try to investigate if we can propose alternative searches using the muon scouting data produced by the LHC experiments in order to better probe inelastic dark matter models in the 1 to 100~GeV mass regime. I contributed to this by setting up a pipeline to simulate and analyse events in the experiments. For both projects, no final results are yet obtained, and are hence not included in this manuscript.

\tableofcontents

\printacronyms[template=longtable]

\clearpage

\fancyhf{}
\fancyhead[RE]{\nouppercase{\leftmark}}
\fancyhead[LE]{\thepage}
\fancyhead[LO]{\nouppercase{\rightmark}}
\fancyhead[RO]{\thepage}
\pagestyle{fancy}
\pagenumbering{arabic}

\chapter{A brief overview of dark matter}
\label{chap:overview}

Uncovering the nature of \ac{DM}, an unknown form of matter responsible for most of the matter content in our universe while seemingly being invisible to us, is arguably one to the greatest challenges in modern day physics. The dark matter problem has a surprisingly long history. After the publication of Isaac Newtons {\it Philosophiae Naturalis Principia Mathematice}, in which Newton described his universal law of gravitation, scientist were provided with a tool to determine the gravitational mass of astronomical bodies by measuring their dynamical properties. Ever since, the question rose if there was a sizable amount of mass hidden from us in for instance dark stars or planets. It was not however until 1933 that the first real hint for dark matter was uncovered, when the Swiss-American astronomer Fritz Zwicky published his study on the Coma cluster~\cite{Zwicky:1933gu}. He observed the velocity dispersion of the member galaxies and by using the virial theorem, he deduced that in order to keep the system stable, the Coma cluster must have an average mass density much larger than that of the visible matter alone. Similar results were soon after found by Sinclair Smith, analyzing the Virgo cluster~\cite{Smith:1936mlg}, and Erik Holmberg, studying systems of galaxies~\cite{Holmberg:1937}. 

\begin{figure}
    \centering
    \includegraphics[width=0.5\textwidth]{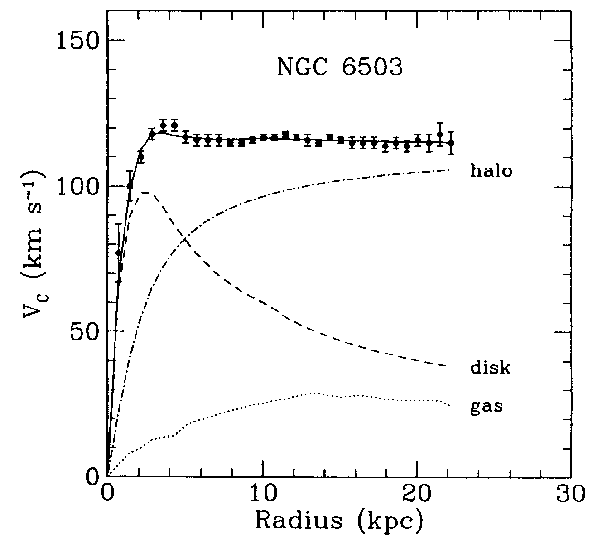}
    \caption{Galactic rotation curve for NGC 6503 showing disk and gas contribution plus the dark matter halo contribution needed to match the data. Picture taken from Ref.~\cite{Freese:2008cz}.}
    \label{fig:rot_curve}
\end{figure}

Despite this discrepancy, the belief in the existence of dark matter only started to grow in the 1970s, when observations started to show that rotation curves of galaxies were flat at large radii (see e.g. the work of Vera Rubin~\cite{Rubin:1970zza}). Rotation curves are diagrams representing the orbital velocity $v$ of gas and stars in galaxies as a function of their distance to the galactic center $r$. This velocity is expected to decline as $v\sim 1/\sqrt{r}$ at large radii when most of the mass is interior, as predicted by Keplers laws of planetary motion. If instead the rotational velocity stays constant, more mass should be present in the outskirts of galaxies to keep the system bound, mass which is not observable in the electromagnetic spectrum. This is illustrated in Fig.~\ref{fig:rot_curve}, where the velocity profile of galaxy NGC 6503 is displayed as a function of radial distance from the galactic center. The baryonic matter in the gas and disk cannot alone explain the galactic rotation curve. Adding a dark matter halo component resolves this discrepancy and allows for a good fit to data. 

The belief that there exists an invisible matter component only came when this missing mass problem was connected to cosmology. In the 1970s, the total energy density of our universe became an important quantity since it predicts whether we live in an open, flat or closed universe. For cosmologists at the time, the option of a closed universe seemed to be the most attractive. The only problem with a closed universe is that its energy density $\rho$ must be larger than the critical density $\rho_{c}$. The visible matter in our universe only equals a fraction of the critical density, so in order to have a closed universe, a dark matter component is needed to have $\Omega=\rho/\rho_{c}>1$.\footnote{Later, it was discovered that also DM was not able to close the universe, however, the existence of dark energy could.} With this new cosmological insight, scientist started to conclude that these hints for missing matter can be seen as a single anomaly that can be explained by a halo of invisible, i.e. dark, matter~\cite{EINASTO1974,Ostriker:1974lna}.\footnote{More details about the history of dark matter theory is provided in Ref.~\cite{Bertone:2016nfn,deSwart:2017heh}.}

Since then, the field has developed enormously. Many more evidence has been put forward for the existence of dark matter, of which we will discuss the most important ones in Sec.~\ref{sec:evidence}. This evidence has increased the belief that dark matter consists of a new particle forming halos in our universe. We will elaborate on the most studied candidates for dark matter in Sec.~\ref{sec:dm_candidates}. A large theoretical and experimental effort has been made to explain the nature of this particle, unfortunately without a conclusive result yet. In Sec.~\ref{sec:DM_detection}, we will elaborate on different ways on how to detect dark matter apart from its gravitational interactions.

\section{Evidence for (particle) dark matter}
\label{sec:evidence}
Throughout the further history, scientists gathered more and more evidence for the existence of DM. Here, we discuss some of the most convincing pieces of evidence coming from both cosmology and astrophysics. 

\subsubsection{Cosmic microwave background}
One of the most convincing pieces of evidence for the existence of DM comes from the \ac{CMB}. This background radiation originates from the propagating photons in the early universe, once they decoupled from matter. It is observed that the CMB is isotropic up to the $10^{-5}$ level and behaves like a black body with temperature $T=2.726K$ at an extraordinary level of precision. However, when COBE\cite{Boggess:1992xla} (COsmic Background Explorer) and later WMAP\cite{Komatsu:2014ioa} (Wilkinson Microwave Anisotropy Probe) started to take data, it became clear that there are small anisotropies ($\frac{\delta T}{T}<10^{-5}$), which was already predicted in order to explain the formation of large scale structures in our universe~\cite{Peebles:1982ff}. Even before the detection of these anisotropies, it was already pointed out that only baryonic matter is not able to seed such relatively large anisotropies necessary to explain the large scale structures~\cite{Peebles:1970ag}. Another matter component was needed, and a heavy DM particle could perfectly fulfill this role, as its density perturbations set by inflation could start to grow due to gravitational effects and seed the anisotropies observed in the CMB. Later experiments, such as the Planck satellite~\cite{Planck:2018vyg}, were able to measure more precisely the CMB anisotropies, so that scientists were able to precisely predict the abundance of DM in our universe, often denoted as $\Omega h^2=0.11933 \pm 0.00091$, where $h=H_0/(100$~km~s$^{-1}$~Mpc$^{-1}$) and $H_0$ is the Hubble constant today.

\subsubsection{Gravitational lensing}
One of the consequences of the general theory of relativity (GR) described by Einstein is that light rays do not always follow a straight line trajectory as they travel through curved spacetime. Thus, heavy celestial bodies can, just like optical lenses, serve as gravitational lenses to bend the trajectory of light. Celestial bodies with a large mass can therefore shift, distort or magnify the image of a background source. The strength of these effects depend heavily on the mass contained withing the lens, but does not depend on the nature of the matter since this is a purely gravitational effect. Hence, gravitational lensing can be a very important tool in the search for dark matter. The three main methods of lensing (micro-, weak- and strong lensing) can be used to map out the matter distribution in our universe, whether it is baryonic or dark. These maps have already uncovered dark matter halos extending way further outward compared to the visible matter. Galaxies and clusters are thus contained or ``hosted'' within a DM halo. More details about gravitational lensing surveys for DM can be found in Ref.~\cite{Massey:2010hh}.

\subsubsection{The bullet cluster}
\begin{figure}
	\centering
	\includegraphics[width=0.65\textwidth]{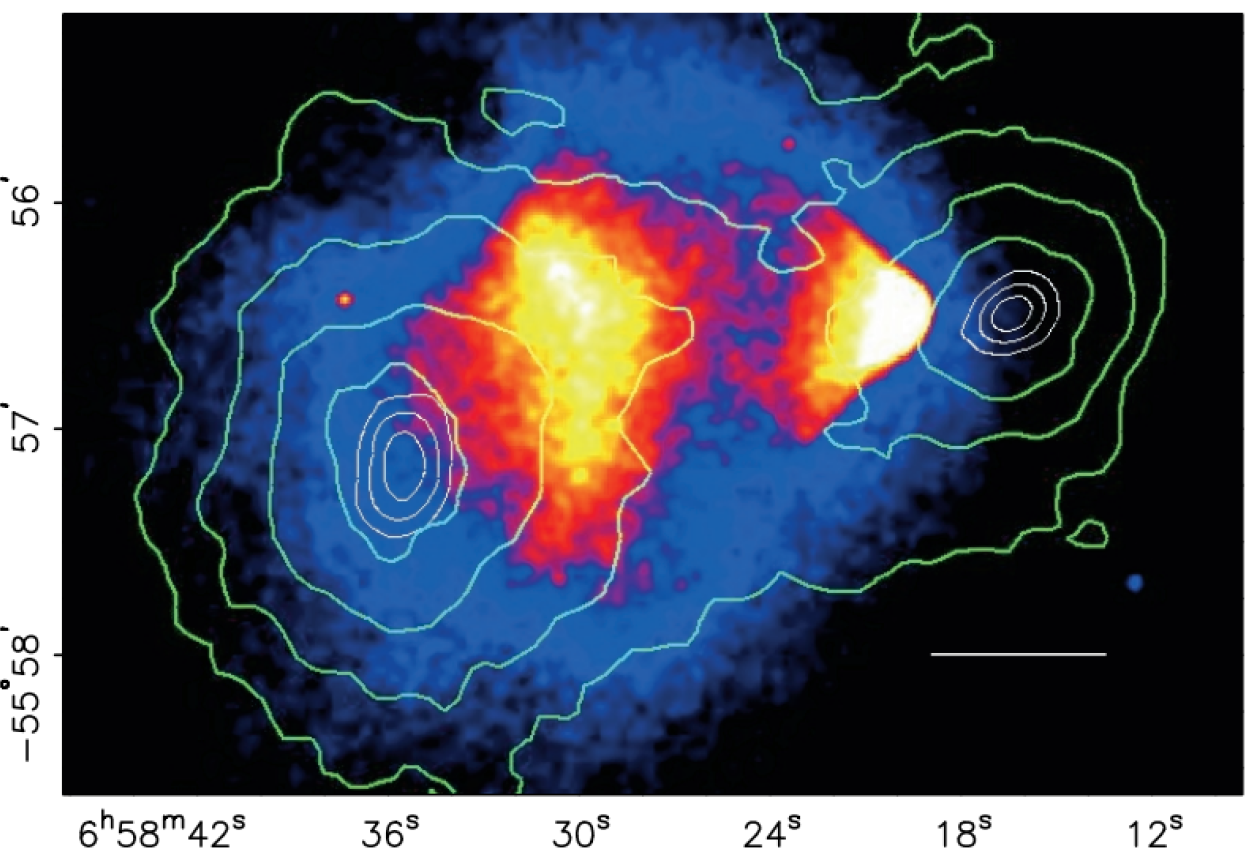}
	\caption{An X-ray image of the bullet cluster, obtained with a 500 second exposure with Chandra. The colors denote the X-ray temperature of the gas where blue indicates the coolest. The white bar represents a distance of 200~kpc and the green contours denote the reconstructed lensing signal, proportional to the projected mass in the system~\cite{Bertone:2016nfn}.}
	\label{fig:bullet}
\end{figure}

More recent evidence for DM comes from the properties of the bullet cluster. This cluster was created by the collision of two smaller clusters passing through each other disturbing their distribution of individual stars, galaxies and baryonic matter. The distribution of baryonic mass can be resolved from the hot X-ray emitting gas (which constitutes the majority of the baryonic mass in the system). One can compare this X-ray map with results from weak lensing where the total mass distribution is tracked. As can be seen in Fig.~\ref{fig:bullet}, the comparison clearly reveals that most of the mass does not trace the distribution of the baryonic mass in the system. This again can be explained by the presence of DM halos hosting the original two clusters. Indeed, if DM does not interact sufficiently with itself or baryonic matter, the DM halos would just have passed through each other without any disturbance, while the baryonic matter of both clusters got slowed down and its distribution got altered. The bullet cluster can therefore set at the same time an upper limit on DM self-interactions~\cite{Randall:2008ppe,Kahlhoefer:2013dca}.

\section{Dark matter candidates}
\label{sec:dm_candidates}
As the presence of dark matter in our universe became more and more established, the nature of dark matter remained unknown. Hence, many different candidates were proposed, ranging from massive object to new elementary particles. Below, we present the most popular candidates proposed (and some already excluded) by the physics community throughout the recent years.

\subsubsection{Compact objects}
The first object proposed to explain the missing mass anomalies were dubbed \acp{MACHO}, referring to objects that where less luminous, but otherwise qualitatively the same as ordinary astrophysical objects consisting out of baryons. Possibilities for such objects included planets, brown dwarfs, neutron stars and black holes. However, the hypothesis that \acp{MACHO} make up all the DM in the universe has been disproved by two lines of investigation. First, weak lensing searches have been performed to resolve \acp{MACHO}, however, these surveys did not observe enough \ac{MACHO} events, allowing them to place an upper limit of 8\% of the DM halo mass fraction to be in \acp{MACHO} within a certain \ac{MACHO} mass range~\cite{EROS-2:2006ryy}. Also, as mentioned in Sec.~\ref{sec:evidence}, the anisotropies of the CMB cannot be explained by baryonic matter alone. Further, predictions from \ac{BBN} place limits on the amount of baryonic matter in our universe. \ac{BBN} is the process where heavy nuclei such as hydrogen and helium are formed for the first time, and in order to match the observed abundances of these nuclei, theoretical analyses predict that the amount of baryonic matter at the time of \ac{BBN} must be roughly a factor five smaller than the DM abundance predicted by the \ac{CMB}~\cite{Burles:2000zk}. Hence, the consensus arose that DM cannot fully consist of baryonic matter. However, \acp{MACHO} are not yet completely ruled out, as dark matter might still consist of black holes formed before the epoch of \ac{BBN} from curvature perturbations. These Primordial Black Holes (PBHs) are also well constrained in certain mass regimes, by for instance the weak lensing surveys mentioned above, however, there are still regions in parameter space left for PBHs contributing to 100\% of the DM abundance~\cite{Green:2020jor}.

\subsubsection{Particle Dark Matter}

In modern days, the majority of DM research is devoted to particle DM making up the large halos observed in our universe. Throughout history, many particles have been proposed as DM candidates, starting with neutrinos, since this was the only known particle from the \ac{SM} that was neutral and stable, which are essential characteristics of any DM candidate. However, the first cosmological simulations in the 1980s investigated in dept the impact of DM particles on the formation of large scale structures. These simulations showed that relativistic (hot) DM particles can erase the density perturbations seeded by inflation, and hence no large scale structures would be able to form. Non-relativistic (cold) DM is needed to predict the large scale structures we observe in our universe today, and hence, neutrinos cannot fulfill the role of DM seeding structure formation~\cite{White:1983}. 

Since no candidate for DM was found in the \ac{SM}, new particles were proposed \acf{BSM}. One of the most popular candidates was the \acf{WIMP}. This particle interacts weakly (i.e. through the nuclear weak force or a new force with comparable strength) with the SM, and has a mass of roughly the weak scale, i.e. $m_{\rm WIMP} \sim \mathcal{O}(100~\GeV-1~\TeV)$. Such a particle can perfectly fulfill the role of cold DM throughout the cosmological evolution of our universe. The DM relic abundance measured by the Planck satellite can also be predicted, since \acp{WIMP} are produced through the freeze-out mechanism, where the \acp{WIMP} where initially in thermal contact with the SM thermal bath until they decoupled when the universe reached a temperature roughly equal to $m_{\rm WIMP}/25$, see Sec.~\ref{sec:std_fo}. The fact that the DM relic abundance was reproduced with weak scale interactions and weak scale masses, also referred to as the ``WIMP miracle'', increased the interest for \acp{WIMP} as candidates for cold DM. \acp{WIMP} are still one of the most popular candidates to explain DM, even though the extensive search for this elusive particle has not provided us with a clear detection yet, see Sec.~\ref{sec:DM_detection}.

The standard model of particle physics exhibits some theoretical problems, so that it is believed that a more complete theory must exist. One such a problem is the hierarchy problem. Within the SM, the masses of the particles can obtain contributions from quantum loop corrections which can be very large due to the large hierarchy between the weak and the Planck scale. Since the SM particles are observed with a mass at the weak scale or below, it can only be assumed that these corrections have to be small. The theory of supersymmetry (SUSY) tries to explain this by introducing a superpartner for every SM particle. These superpartners have the same charge under the SM gauge group, but have a different spin. For every boson, a new fermion is introduced and vice versa. The quantum corrections of these superpartners naturally cancel out the corrections arising from the SM particles themselves, alleviating the hierarchy problem (see Ref.~\cite{Martin:1997ns} for a more in dept overview). Within the spectrum of SUSY particles, new neutral and stable particles arise that can be natural candidates for DM. For instance, the superpartners of the neutral gauge and Higgs bosons, called neutralinos, or of the graviton, called gravitino, can fulfill this role. Although the belief in the existence of SUSY is slightly fading away due to the absence of a discovery, SUSY models still keep driving the theoretical motivation of many \ac{BSM} searches.

Another flaw in the SM can be found in the \ac{QCD} sector, namely the strong-CP problem. The \ac{QCD} Langrangian contains the following term,
\begin{align}
    \mathcal{L}_{\rm QCD} \in \Bar{\Theta} \frac{g_s^2}{32 \pi^2} G^{a\mu\nu} \Tilde{G}_{a\mu\nu},
\end{align}
where $G^{a\mu\nu}$ is the gluon field strength tensor, $g_s$ is the strong coupling constant and $\Bar{\Theta}$ is a quantity closely related to the phase of the QCD vacuum. If the latter would be of order unity, this term would induce charge-parity (CP) violating effects causing the neutron electric dipole moment to be many orders of magnitude larger than its experimental bound. Therefore, to be consistent with observations, $\Bar{\Theta}$ must be smaller than roughly $10^{-10}$, which is believed to be unnatural in the physics community. A more elegant solution has been proposed by Peccei and Quinn~\cite{Peccei:1977hh}, where the quantity $\Bar{\Theta}$ is promoted to a pseudoscalar field called the axion field and is dynamically driven to zero. These axions could have been produced in the early universe through the misalignment mechanism, see Refs.~\cite{Sikivie:2006ni,Marsh:2017hbv} for more details. For a certain range of parameters, this mechanism can reproduce the correct DM relic density and can hence solve two shortcomings of the SM at once. Axions have inspired a large class of light DM models. If the DM candidate in such models is not able to solve the strong CP problem, they are usually referred to as Axion-Like Particles (ALPs).  

Many other exotic \ac{BSM} scenarios have been considered in the literature to explain the existence of DM. In this thesis, we will focus on \acfp{FIMP}, which have been proposed as DM candidate in order to evade the strong experimental constraints placed on the \ac{WIMP} paradigm. In contrast to the \ac{WIMP}, a \ac{FIMP} interacts far more feebly with the standard model, so feeble that it is usually unable to reach thermal equilibrium in the early universe. Different production mechanisms have been proposed to produced the amount of \acp{FIMP} needed to explain the structure of the \ac{CMB}. Since \acp{FIMP} are in general way less constrained than \acp{WIMP}, the main focus of this thesis will be to discuss those production mechanisms (See Chapters~\ref{chap:DM_prod} and~\ref{chap:alt_cosmo}) and how to constrain them (see Chapters~\ref{chap:model_class},~\ref{chap:LHC_cons} and~\ref{chap:lightDM}).

\subsubsection{Modified Newtonian Dynamics}
To end this section, let us comment on an alternative to dark matter in order to explain the missing mass anomalies. In the 1980s, theorist presented adaptations to Newtons laws of motion to explain the anomalous velocities of stars and clusters. The first proposal, known as MOdified Newtonian Dynamics (MOND), was a seemingly simple one. If instead of obeying Newtons second law of motion, $F=ma$, the force due to gravity scaled as $F=ma^2/a_0$, so that for very low accelerations ($a\ll a_0$), it would be possible to explain the flatness of the rotation curves without postulating any new form of matter~\cite{Milgrom:1983ca,Milgrom:1983pn,Milgrom:1983zz}. This first proposal of a MOND theory was merely meant as a setup for research to pursue the route of altering the physical laws we know to explain the missing mass anomalies. In the following years, MOND theories gained some success by explaining the flat rotation curves of galaxies and giving an explanation for the empirical Tully-Fisher relation~\cite{McGaugh:2000sr}. However, after the discovery of the bullet cluster and during the era of precision cosmology, MOND theories were not able to explain these new pieces of evidence~\cite{Clowe:2006eq}, and the dark matter postulation became the favored explanation of the missing mass anomalies. 

\section{Detecting dark matter}
\label{sec:DM_detection}
\begin{figure}[t]
    \centering
    \includegraphics[width=0.65\textwidth]{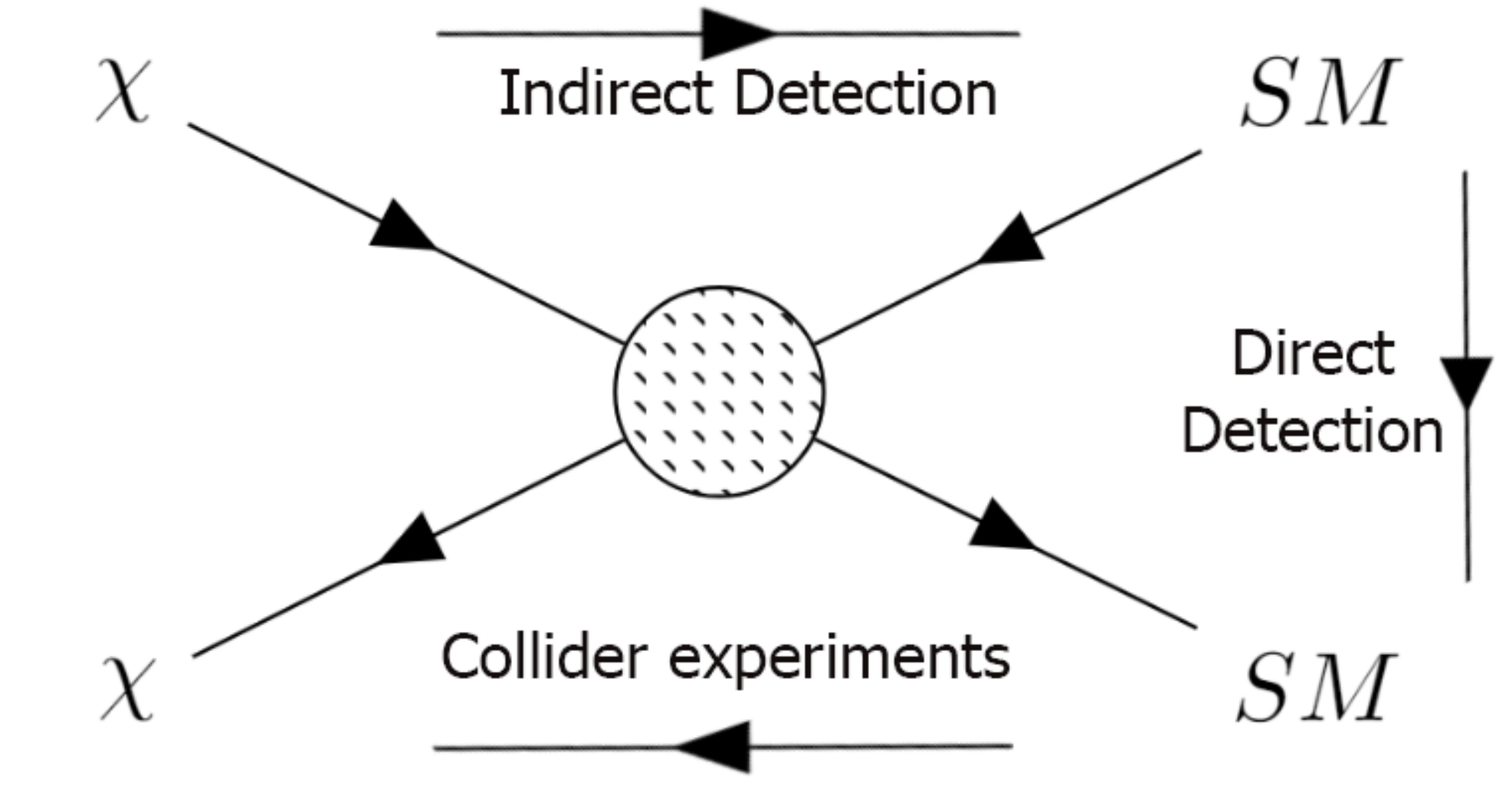}
    \caption{A schematic representation of the three main detection strategies for DM.}
    \label{fig:dm_detection}
\end{figure}

All the evidence leading to the postulation of DM presented before, was only based on the gravitational interactions of DM with baryonic matter. Hence, the exact nature of DM remains an open question, despite the large experimental effort of the past decades. Many DM search strategies are based on the hypothesis that DM is a WIMP like particle interacting with the SM as depicted in Fig.~\ref{fig:dm_detection}. Depending on the direction one looks at this diagram, three different search strategies can be identified, namely direct detection, indirect detection and detection through collider experiments. We will briefly comment on these strategies below. Further, we also discuss how cosmological probes can play an important role in detecting DM.

\subsubsection{Direct Detection}
If one reads the diagram in Fig.~\ref{fig:dm_detection} from top to bottom, DM can elastically scatter of SM particles. This reaction forms the basis of direct detection experiments. Since galaxies are submerged in a DM halo, also our planet Earth moves through the halo that hosts the Milky Way. Hence, DM particles continuously pass through the Earth and can potentially scatter with nuclei. Direct detection experiments host big detectors filled with a high density material to increase the probability of scattering to happen. If such a scattering event would happen, the recoil energy of the nuclei can be measured to identify the event since it is typically emitted as light or vibrations. A smoking gun signature for DM would be if an observed scattering rate would exhibit an annual modulation. Indeed, due to the Earth's rotation around the sun, the Earth can move both parallel and against the DM flux, depending on its relative direction of travel compared to the one of the sun. The DAMA/LIBRA experiment has observed such a modulation~\cite{DAMA:2008jlt}, however, since this was never confirmed by other direct detection experiments, it is believed that this is caused by some seasonal variations, although their is no real certainty yet about the origin of this modulation.

\begin{figure}
    \centering
    \includegraphics[width=0.65\textwidth]{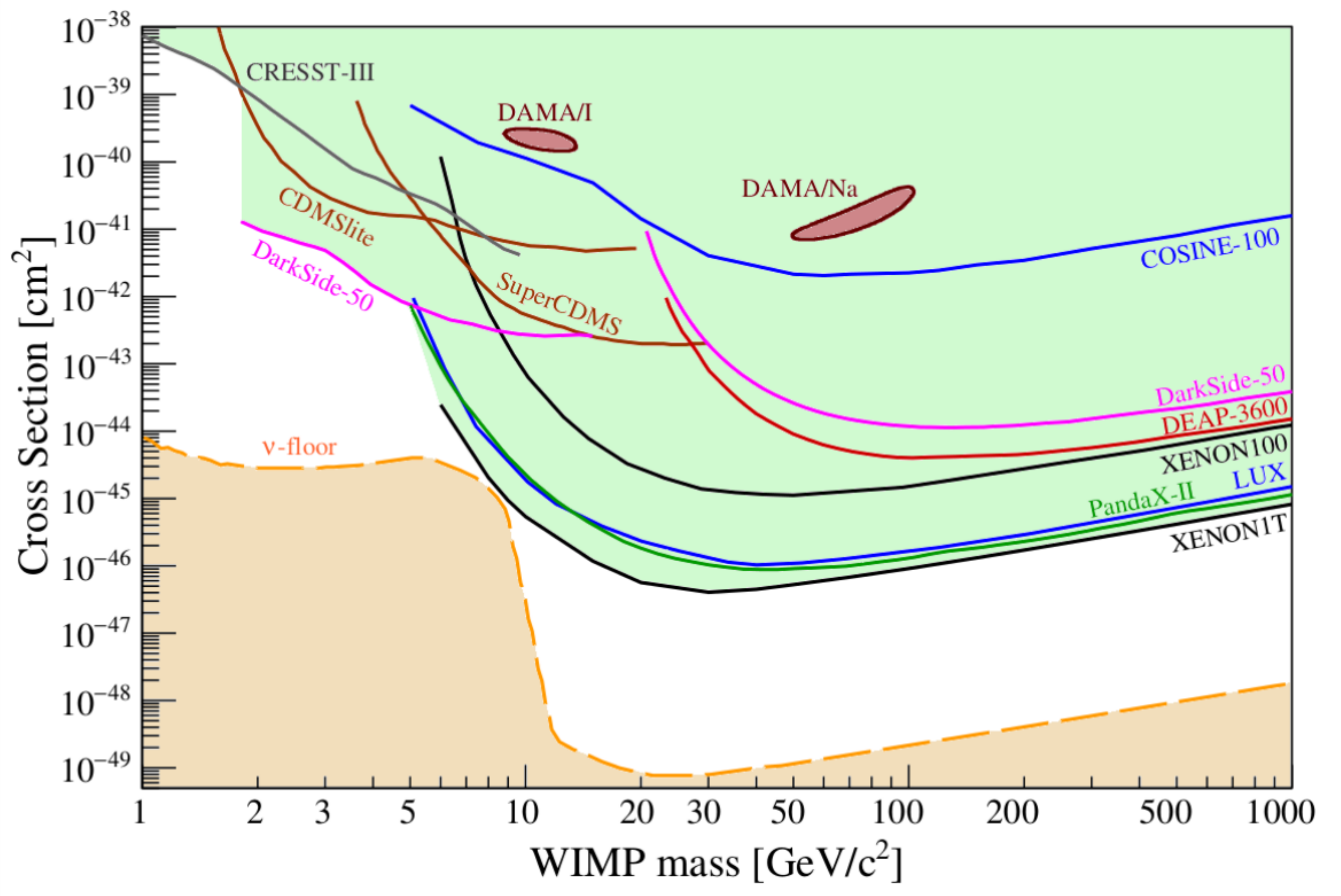}
    \caption{A selection of current experimental parameter space for spin-independent WIMP-nucleon scattering cross sections. Not all published results are shown. The space above the lines is excluded at a 90\% \ac{CL}. The two contours for DAMA interpret the observed annual modulation. The dashed line limiting the parameter space from below represents  the ``neutrino floor'', from the irreducible background from coherent neutrino-nucleus scattering. Figure taken from~\cite{Schumann:2019eaa}.}
    \label{fig:dd_limits}
\end{figure}

There are two main types of detectors, crystal and noble gas detectors. The latter are able to set the strongest constraints in the high mass regime ($m\gtrsim$~few~GeV) due to their low background rates and high scalability to obtain a large target mass to improve the event rate. The strongest limits in the $m\gtrsim$~10~GeV come from the Xenon~\cite{XENON:2018voc}, PandaX~\cite{PandaX-4T:2021bab} and LUX experiments~\cite{LUX:2017ree}, see Fig.~\ref{fig:dd_limits}. These experiments are however limited by their low recoil energy resolution and hence have almost no sensitivity at low, sub-GeV masses. This is where small crystal detectors such as CRESST~\cite{CRESST:2017cdd} can set limits. They do not provide such a high mass target and exhibit a high background rate and hence can not compete with the the strong limits from noble gas detectors. However, their high recoil energy resolution allows them to place limits on sub-GeV DM. In Fig.~\ref{fig:dd_limits}, we show some of the strongest constraints on the spin-independent WIMP-nucleon scattering cross section. An in depth discussion on the current status of direct detection can be found in~\cite{Schumann:2019eaa,Cooley:2022ufh}

\subsubsection{Indirect Detection}
The diagram in Fig.~\ref{fig:dm_detection} shows that DM can annihilate into SM particles. Since DM is abundantly present in the universe, we can expect it to annihilate constantly, especially in regions in space where the DM density is high. Hence, excesses of SM particles in the direction of those regions can hint to the existence of annihilating DM particles. The flux for those particles can be estimated as,\footnote{For charged particles, extra terms need to be added to model their interactions with cosmic rays.}
\begin{align}
    \frac{d\Phi}{dE d\Omega} = \frac{\sigmav}{m_\chi^2}\frac{dN}{dE} \int \rho^2(r) dl,
\end{align}
where $\sigmav$ is the thermally averaged cross section, $m_\chi$ is the DM mass and $dN/dE$ denotes the energy spectrum of the produced particles. The integral is often referred to as the $J$-factor, where we integrate the DM energy density profile $\rho$ over the line of sight. To obtain precise predictions, we have to know the DM density profile. However, there is still a debate about the exact form of this profile, since hydrodynamical simulations does not seem to agree perfectly with lensing observations~\cite{Weinberg:2013aya}.

Several distinctive signals have already been observed that could be accounted to DM annihilation. The bright 511-keV line emission for the center of our galaxy detected by the INTEGRAL satellite~\cite{Knodlseder:2003sv}, the galactic GeV excess observed by Fermi-LAT~\cite{DiMauro:2021raz} or the rise in the positron fraction above 10~GeV observed by PAMELA~\cite{PAMELA:2008gwm} and AMS-02~\cite{AMS:2013fma} are a couple examples of such signals that can potentially be attributed to DM physics. However, all these indirect detection signals suffer from a large astrophysical uncertainty, and hence, a DM interpretation is far from clear. More details on indirect detection searches can be found in~\cite{Buckley:2013bha,Leane:2020liq,Cooley:2022ufh}.

\subsubsection{Cosmic probes}

The two previously mentioned detection strategies rely on the fact that DM is abundantly present in the universe today. As mentioned earlier in this chapter, DM has played a crucial role in the evolution of our universe and must have already been present before the time of \ac{BBN}. DM annihilation processes into SM particles could have therefore already injected energy earlier on in our universe. Cosmic probes can hence be an extension of indirect DM searches and put bounds on the annihilation cross section. For instance, if DM injects energy after recombination, i.e. when the first neutral atoms are formed, this energy injection increases the left over ionized fraction of the Inter-Galactic Medium (IGM). Current CMB measurements are sensitive to changes in the ionization fraction of just a few times $10^{-4}$, and can hence constrain the energy injected into the IGM, see e.g. Ref.~\cite{Slatyer:2015jla}. 

Energy injection during the dark ages (the period between recombination and re-ionization) can also alter the temperature of the IGM. Lyman-$\alpha$ and 21cm probes are sensitive to this temperature at different redshifts. Lyman-$\alpha$ radiation originates from the transition of hydrogen atoms from the first excited state to the ground state. Observations of the Lyman-$\alpha$ flux power spectra and absorption features have been used to derive limits on the DM annihilation cross section, see e.g. Ref.~\cite{Liu:2020wqz}. The 21cm line originates from a spin-flip of the electron in the ground state of hydrogen. An overall signal has already been claimed by EDGES~\cite{Bowman:2018yin}, although it is not yet independently confirmed. Future measurements will be able to probe the power spectrum of the 21cm line, which can then be used to probe DM energy injection, see e.g. Refs.~\cite{Evoli:2014pva,Lopez-Honorez:2016sur,Liu:2018uzy}.

\subsubsection{Collider experiments}
If DM interacts with the SM, it can potentially be created at high enough rate in collider experiments. For instance, if the cross section of the diagram in Fig.~\ref{fig:dm_detection} when reading it from right to left is large enough, pairs of DM particles will be produced and travel undetected through the detector as they are neutral particles. Therefore, search strategies including missing energy are employed in order to look for such events. However, a reconstructed object is needed to identify the event. Therefore, searches for DM at colliders typically require a high momentum jet or vector boson in combination with a large missing energy component. They are referred to as ``mono-X'' searches, where the X stands for the hard final state balancing against the missing energy. The most energetic collider today is the \acf{LHC} at CERN, accelerating protons at a center of mass energy of 13~TeV. The CMS and ATLAS detectors, two multi-purpose detectors each placed around one of the collision points of the \ac{LHC}, have already performed mono-jet~\cite{CMS:2021far,ATLAS:2021kxv}, mono-Z~\cite{CMS:2020ulv,ATLAS:2021gcn}, mono-Higgs~\cite{CMS:2019ykj,ATLAS:2021shl} and mono-photon searches~\cite{ATLAS:2020uiq,CMS:2018ffd}. At electron-positron colliders such as BaBar, mono-photon searches can also have a very strong reach on DM models~\cite{BaBar:2017tiz}. Since these are performed at lower center of mass energies, they are typically sensitive to lower masses of the DM. 

\chapter{Producing Dark Matter in the early universe}
\label{chap:DM_prod}

As discussed in Chapter \ref{chap:overview}, there is a compelling amount of evidence for the presence of \ac{DM} in the our universe, both today but also early on in its evolution. Regardless the exact nature of \ac{DM}, it must have been produced early on in the evolution of the universe, in order to explain the observed effects on the \ac{CMB} and structure formation, see Sec.~\ref{sec:evidence}. More specifically, we know from \ac{CMB} data gathered by the Planck collaboration the energy density of \ac{DM} in our universe today, $\Omega h^2=0.11933 \pm 0.00091$~\cite{Planck:2018vyg}. If within a specific model, we know how the \ac{DM} is produced, the relic \ac{DM} abundance observation can lead us in detecting this model in experiments, since it restricts the parameter space of the model. Studying the production mechanism of \ac{DM} is therefore an essential part in unraveling the true nature of \ac{DM}.

From now on, we will restrict the discussion to the class of particle \ac{DM} models. Within this class, there are still a number of possible production mechanisms depending on the particle content of the model under consideration, as well as the interactions with the \ac{SM}. We will discuss the most popular ones, and focus more on mechanisms where the \ac{DM} candidate is feebly interacting with the \ac{SM} and/or other \ac{DS} particles. In general, we will assume our \ac{DM} candidate to be stable, which is often assured by introducing for instance a $\mathbb{Z}_2$-symmetry under which \ac{DM} and potential other \ac{DS} particles are odd and all \ac{SM} particles are even. Indeed, if \ac{DM} is then the lightest \ac{DS} particle, its decay to other \ac{DS} states is kinematically forbidden while its decay to a pure \ac{SM} final states is forbidden by the $\mathbb{Z}_2$-symmetry. 

\section{The Boltzmann equation}
In order to know the number density of \ac{DM} particles present at a specific moment in the universe, we need to study the evolution of the \ac{DM} phase space density $f(p^\mu,x^\mu)$, since from this, we can derive the \ac{DM} number density ($n$) and energy density ($\rho$),
\begin{align}
    n(x^\mu) &= g \int \frac{d^3p}{(2\pi)^3} f(p^\mu,x^\mu), \\
    \rho(x^\mu) &= g \int \frac{d^3p}{(2\pi)^3} E f(p^\mu,x^\mu),
\end{align}
where $E$ is the \ac{DM} energy and $g$ is the internal degrees of freedom of the particle species. In general, the phase space density can be altered by the cosmological history through e.g. the expansion of the universe, parameterized by the Hubble rate $H$, and by particle interactions redistributing the particles momenta. These interactions are parameterized by the interaction rate $\Gamma$. A general rule of thumb tells you that a specific process can influence the \ac{DM} phase space density if the interactions happen fast enough compared to the expansion of the universe, i.e. $\Gamma > H$. In this case, we say that the process is efficient and \ac{DM} is (kinetically and/or chemically) coupled to the other particles involved in the interaction. When $\Gamma < H$, \ac{DM} is decoupled. We will make these statements more precise in the remainder of this chapter.

In order to precisely track the evolution of the phase space density $f(p^\mu,x^\mu)$, we need to solve the Boltzmann equation which is given by~\cite{Kolb:1990vq}
\begin{equation}
    L[f] = C[f],
    \label{eq:Boltz_unint}
\end{equation}
where $L$ is the Liouville operator describing the influence of the cosmological model on the phase space density, while the collision operator $C$ describes the effects from particle physics interactions. Both operators can hence be studied separately.

\textbf{Liouville operator:} The Liouville operator describes the effects of the expansion of the universe on the phase space density. The operator is given by
\begin{align}
    L[f] = p^\alpha \frac{\partial f}{\partial x^\alpha} - \Gamma^\alpha_{\beta \gamma} p^\beta p^\gamma \frac{\partial f}{\partial p^\alpha},
\end{align}
where $\Gamma^\alpha_{\beta \gamma}$ are the Christoffel symbols which depend on the metric of the universe. Assuming a Friedman-Robertson-Walker (FRW) metric, the isotropy and homogeneity imply that the phase space density only depends on the magnitude of the three momentum ($p$), or equivalent the energy ($E$), and time ($t$), i.e. $f(p^\mu,x^\mu)=f(E,t)$. Using this and the appropriate Christoffel symbols for a FRW metric, the Liouville operator simplifies to
\begin{align}
    L[f] = E \left( \frac{\partial f}{\partial t} - H \frac{p^2}{E} \frac{\partial f}{\partial E} \right).
\end{align}
In cases where we only need to track the number density of the \ac{DM}, we can integrate the Liouville operator over the \ac{DM} phase space,
\begin{align}
 \label{eq:int_liouville}
    g \int \frac{d^3p}{(2\pi)^3} \frac{1}{E} L[f] = \frac{dn}{dt}+3Hn.
\end{align}

\textbf{Collision term:} The right hand side of the Boltzmann equation contains the collision term, which encapsulates all information concerning all the particle interactions the \ac{DM} particle can undergo. For every possible reaction, one need to add a term to the right hand side. The collision term for a general reaction $a,b,c, \dots \to i,j,\dots$ is~\cite{Kolb:1990vq}
\begin{align}
    C[f_a] = \frac{1}{2} \int & d\Pi_b d\Pi_c \dots d\Pi_i d\Pi_j \dots \nonumber \\
     \times & (2\pi)^4 \delta^4(p_a+p_b+p_c+\dots-p_i-p_j-\dots) \nonumber \\
     \times [ \, & |\mathcal{M}_{i,j, \dots \to a,b,c,\dots}|^2 f_i f_j \dots (1\pm f_a) (1\pm f_b) (1\pm f_c ) \dots \nonumber \\
     - & |\mathcal{M}_{a,b,c, \dots \to i,j,\dots}|^2 f_a f_b f_c \dots (1\pm f_i) (1\pm f_j) \dots \, ],
     \label{eq:Coll_term}
\end{align}
where $f_a,f_b,f_i,f_j,\dots$ are the phase space densities of the particles $a,b,i,j,\dots$ involved in the reaction, the plus or minus sign depends on the nature of the particle ($+$ for bosons undergoing bose-enhancement, $-$ for fermions undergoing Pauli-blocking) and 
\begin{align}
    d\Pi_k = \frac{g_k}{(2\pi)^3} \frac{d^3p_k}{2E_k},
\end{align}
for each species $k$ involved. Finally, $|\mathcal{M}|$ is the Feynman amplitude averaged over all initial and final state degrees of freedom. When the process under consideration is invariant under charge and parity transformations, the amplitude is the same for both directions of the process, i.e.
\begin{align}
    |\mathcal{M}_{a,b,c, \dots \to i,j,\dots}| = |\mathcal{M}_{i,j, \dots \to a,b,c,\dots}| = |\mathcal{M}|.
\end{align}

\section{Freeze-out}
\label{sec:FO+coann}

\subsection{Kinetic and chemical equilibrium}
\label{sec:fo_eq}

\begin{figure}
    \centering
    \subfloat[]{\includegraphics[width=0.38\textwidth]{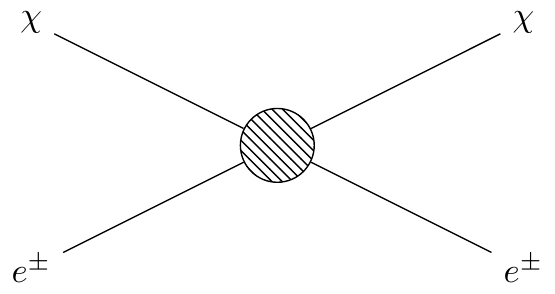}\label{feyn:el_scat}}
	\hspace{0.05\textwidth}
    \subfloat[]{\includegraphics[width=0.38\textwidth]{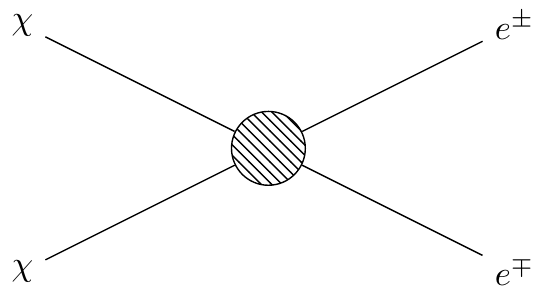}\label{feyn:ann}}
    \caption{Example of an elastic scattering (left) and annihilation (right) process for a DM candidate $\chi$ interacting with electrons.}
\end{figure}

In the vanilla \ac{WIMP} scenario,\footnote{The vanilla \ac{WIMP} scenario is a scenario where a $\mathcal{O}(\text{TeV})$ \ac{DM} particle is electroweakly interacting with the SM.} \ac{DM} is produced by mean of the \ac{FO} mechanism. In order for \ac{FO} to occur, the \ac{DM} must be in kinetic and \ac{CE} with the thermal bath of SM particles at early times, or equivalently, at high temperatures. \ac{KE} can typically be ensured by elastic scattering processes of SM particles that are in KE themselves, such as electrons or photons. An example is shown in Fig. \ref{feyn:el_scat}. If these reactions happen fast enough (i.e. faster than the expansion of the universe), the momenta of the DM particles can get redistributed such that the phase space density follows a generic \ac{BE} or \ac{FD} distribution (see App.~\ref{app:equilibrium} for more details),
\begin{align}
\label{eq:phase_space}
    f(E,T) = \frac{1}{\exp((E-\mu)/T) \pm 1},
\end{align}
where the $+$ ($-$) describes the \ac{BE} (\ac{FD}) case. Therefore, the evolution of the phase space density is completely determined by the evolution of the chemical potential $\mu$ and the temperature of the universe $T$, simplifying the analysis.

If besides \ac{KE}, \ac{DM} is also in \ac{CE}, it is straightforward to obtain the chemical potential. Typically, chemical equilibrium is ensured by annihilation processes of DM into thermal bath particles. If any such process happens efficiently, one can set the sum of the chemical potentials of the initial state particle species to the sum of the final state particle species (see App.~\ref{app:equilibrium}). For example, if the process in Fig.~\ref{feyn:ann} happens efficiently, we know that $2\mu_\chi = \mu_{e^-} + \mu_{e^+}$. For all \ac{SM} particles annihilating into photons, we know that the sum of the chemical potential of the particle and anti-particle equals to twice the one of the photon.\footnote{When no particle-antiparticle asymmetry exists, the chemical potential of both particles even equals to the one of the photon.} Since photon number is not conserved (2$\gamma \to 3\gamma$ processes are efficient in the early universe), $\mu_\gamma=0$, and hence, also $\mu_\chi=0$ when in \ac{CE}. Therefore, as long as DM is in kinetic and chemical equilibrium, $f(E,T)=f^{(0)}(E,T)$, where $f^{(0)}(E,T)$ is the equilibrium phase space density defined as in Eq.~\eqref{eq:phase_space} by setting the chemical potential to zero. The \ac{DM} number density hence also equals the equilibrium density $n^{(0)}_\chi$,
\begin{align}
    n^{(0)}_\chi = g_\chi \int \frac{d^3p}{(2\pi)^3} \frac{1}{\exp(E/T) \pm 1} \approx \frac{g_\chi}{2\pi^2} m^2 T K_2[m/T],
    \label{eq:n_eq}
\end{align}
where $K_2$ is the modified Bessel function of the second kind. In the final step, we assumed the quantum corrections to the \ac{FD}/\ac{BE} distributions to be negligible so that we could use the general equilibrium \ac{MB} distribution instead,
\begin{align}
    f^{(0)}_{\text{MB}}(E,T) = e^{-E/T}.
\end{align}
This is a good approximation when the temperature is well below the typical particle energy. As we will see later, the main \ac{FO} dynamics happen at temperatures below the DM mass, validating our approximation.

Typically, \ac{KE} holds longer than \ac{CE} if the \ac{DM} candidate interacts with light particles. The reason for this is that light particles are abundant at temperatures comparable to the \ac{DM} mass, so it is easy for \ac{DM} to scatter of them and stay in \ac{KE}. From Eq.~\eqref{eq:n_eq}, we can see that at these temperatures, the \ac{DM} number density gets suppressed (we refer to this as Boltzmann suppression), so it becomes harder for the \ac{DM} particles to annihilate. Therefore the annihilation process will not happen efficiently enough and DM chemically decouples before it kinetically does.\footnote{A more quantitative statement will be provided in Sec.~\ref{sec:std_fo}.} Since the scattering process does not influence the amount of DM in the universe, the DM number density will only redshift due to the expansion of the universe. This is what is referred to as the \ac{FO} mechanism, and we will discuss this in more detail in the next sections.






\subsection{Standard freeze-out equations}
\label{sec:std_fo}
We already established that in order for \ac{FO} to occur, the \ac{DM} must be in kinetic and chemical equilibrium at high temperatures (well above its mass), meaning that there are at least two processes between the DM and thermal bath particles (which from now on, we will refer to as $B$). We will assume that these bath particles are in full equilibrium throughout the whole \ac{FO} dynamics, which is in general true if they are lighter than the DM. In this section, we will restrict ourselves to the simplest case where DM only interacts with the thermal bath through these two processes: the elastic scattering process and the annihilation process. To study the evolution of the \ac{FO} mechanism, we need to solve the Boltzmann equation in Eq.~(\ref{eq:Boltz_unint}), meaning that for those two processes involved, we will have to add a collision term to the right and side. However, from Eq.~\eqref{eq:Coll_term}, we can see that for an elastic scattering process that is CP invariant, the collision term equals zero since both initial and final state are equivalent. Hence, we only have to take into account the annihilation process. 

As we mentioned in Sec.~\ref{sec:fo_eq}, neglecting quantum corrections to the phase space density is a good approximation since the main \ac{FO} dynamics happen at temperatures below the DM mass. Therefore, we will work with MB statistics and neglect the Bose-enhancement/Pauli-blocking terms in order to calculate the collision term for the annihilation process.\footnote{For results taking into account the quantum statistical effects, we refer the reader to \cite{Arcadi:2019oxh,Lebedev:2019ton}. The main effect is due to neglecting the quantum statistical effects for the SM final states, as their mass is typically below the FO temperature. However, annihilation processes of heavy DM particles produce SM states with high enough energies, so that our approximation is still valid.} These approximations allow us to analytically integrate the collision term over the DM phase space. The reason for this is that we can write the chemical potential in terms of the number density,
\begin{align}
    e^{\mu/T} = \frac{f_{\text{MB}}}{f^{(0)}_{\text{MB}}} = \frac{n}{n^{(0)}}.
\end{align}
Applying this relation to Eq.~\eqref{eq:Coll_term}, the collision term for a $2 \to 2$ process $a,b \to i,j$ (where $a$ plays the role of our \ac{DM} particle) reads
\begin{align}
    C[f_a] & = \frac{1}{2} \int d\Pi_b d\Pi_i d\Pi_j (2\pi)^4 \delta^4(p_a+p_b-p_i-p_j) |\mathcal{M}|^2 \nonumber \\
    & \hspace{3cm} \times \left[ f^{(0)}_i f^{(0)}_j \frac{n_i n_j}{n^{(0)}_i n^{(0)}_j} - f^{(0)}_a f^{(0)}_b \frac{n_a n_b}{n^{(0)}_a n^{(0)}_b} \right] \\
    & = \int g_b \frac{d^3p_b}{(2\pi)^3} \, E_a \, \sigma v \, f^{(0)}_a f^{(0)}_b \left[ \frac{n_i n_j}{n^{(0)}_i n^{(0)}_j} -  \frac{n_a n_b}{n^{(0)}_a n^{(0)}_b} \right],
\end{align}
where $v$ is the M{\o}ller velocity for the initial state particles
\begin{align}
\label{eq:moller_v}
  v_{a,b} = \frac{\sqrt{(p_a\cdot p_b)^2-(m_a m_b)^2}}{E_a E_b}
\end{align}
and in the last step, we have applied conservation of energy in order to relate the final state equilibrium densities to the initial state ones,\footnote{In general, this is the requirement of detailed balance, see App.~\ref{app:equilibrium}.}
\begin{align}
    f^{(0)}_a f^{(0)}_b = e^{-(E_a+E_b)/T} = e^{-(E_i+E_j)/T} = f^{(0)}_i f^{(0)}_j.
\end{align}
As mentioned before, by assuming \ac{KE}, we can trade the evolution of the full phase space density for the evolution of one single function, the chemical potential, which in turn defines fully the number density. Therefore, we can integrate the Boltzmann equation over the \ac{DM} phase space such that it becomes a differential equation we can solve to find the evolution of the number density,
\begin{align}
    \frac{dn_a}{dt}+3Hn_a = & \int d\Pi_a C[f_a]  \nonumber \\
     = & \gamma_{a,b \to i,j} \left[ \frac{n_i n_j}{n^{(0)}_i n^{(0)}_j} - \frac{n_a n_b}{n^{(0)}_a n^{(0)}_b} \right],
\end{align}
where $\gamma$ is the reaction density proportional to the thermally averaged cross section $\sigmav$,
\begin{align}
    \gamma_{a,b \to i,j} = \sigmav_{a,b \to i,j} n^{(0)}_a n^{(0)}_b = \int g_a g_b \frac{d^3p_a d^3p_b}{(2\pi)^6} \sigma v  f^{(0)}_a f^{(0)}_b
\end{align}
Often, this equation is written into dimensionless quantities. For instance, since after \ac{FO}, the number density only redshifts as $n_\chi \sim a^{-3} \sim T^3$, one can remove this redshift dependence by introducing the yield 
\begin{align}
    Y=\frac{n}{s},
\end{align}
where $s$ is the entropy density
\begin{align}
    s=\frac{2\pi^2}{45} h_{eff} T^3,
\end{align}
where $h_{eff}$ is the number of degrees of freedom contributing to the entropy. Assuming that the \ac{FO} process happens completely during the radiation dominated era, the time variable can be traded for 
\begin{align}
    x=\frac{m_{DM}}{T},
\end{align}
by making use of
\begin{align}
    \rho &= \rho_{R} = \frac{\pi^2}{30} g_{eff} T^4, \label{eq:rho_rad}\\
    H &= \frac{1}{M_{Pl}} \sqrt{\frac{\rho_{R}}{3}} = \frac{\pi}{M_{Pl}} \sqrt{\frac{g_{eff}}{90}} T^2, \label{eq:H_rad}
\end{align}
where $M_{Pl}=2.4\cdot10^{18}$~GeV is the reduced Planck mass and $g_{eff}$ is the number of degrees of freedom contributing to radiation. From the conservation of entropy, we can obtain the time-temperature relation,
\begin{align}
    \frac{dT}{dt} = -\Bar{H} T \quad \text{with} \quad \Bar{H} = H \left( 1+\frac{1}{3} \frac{d \ln h_{eff}}{d \ln T} \right)^{-1},
\end{align}
So that we can write the Boltzmann equation into its final form
\begin{align}
    \frac{dY_a}{dx} = \frac{\gamma_{a,b \to i,j}}{s\Bar{H}x}  \left[ \frac{Y_i Y_j}{Y^{(0)}_i Y^{(0)}_j} - \frac{Y_a Y_b}{Y^{(0)}_a Y^{(0)}_b} \right].
     \label{eq:int_boltz}
\end{align}
This expression can be generalized for any n-to-m process,
\begin{align}
    \frac{dY_a}{dx} = \frac{\gamma_{a,b,c,\dots \to i,j,\dots}}{s\Bar{H}x}  \left[ \frac{Y_i}{Y^{(0)}_i}\frac{Y_j}{Y^{(0)}_j} \dots - \frac{Y_a}{Y^{(0)}_a}\frac{Y_b}{Y^{(0)}_b}\frac{Y_c}{Y^{(0)}_c}\dots \right],
     \label{eq:int_boltz_gen}
\end{align}
where the reaction density in general is given by
\begin{align}
    \gamma_{a,b,c,\dots \to i,j,\dots} = \int &d\Pi_a d\Pi_b d\Pi_c d\Pi_i d\Pi_j \dots \nonumber \\
    &\times (2\pi)^4 \delta^4(p_a+p_b+p_c+\dots-p_i-p_j-\dots) \nonumber\\
    &\times |\mathcal{M}|^2 f^{(0)}_a f^{(0)}_b f^{(0)}_c \dots
    \label{eq:react_rate_gen}
\end{align}

\begin{figure}
    \centering
    \includegraphics[width=0.6\textwidth]{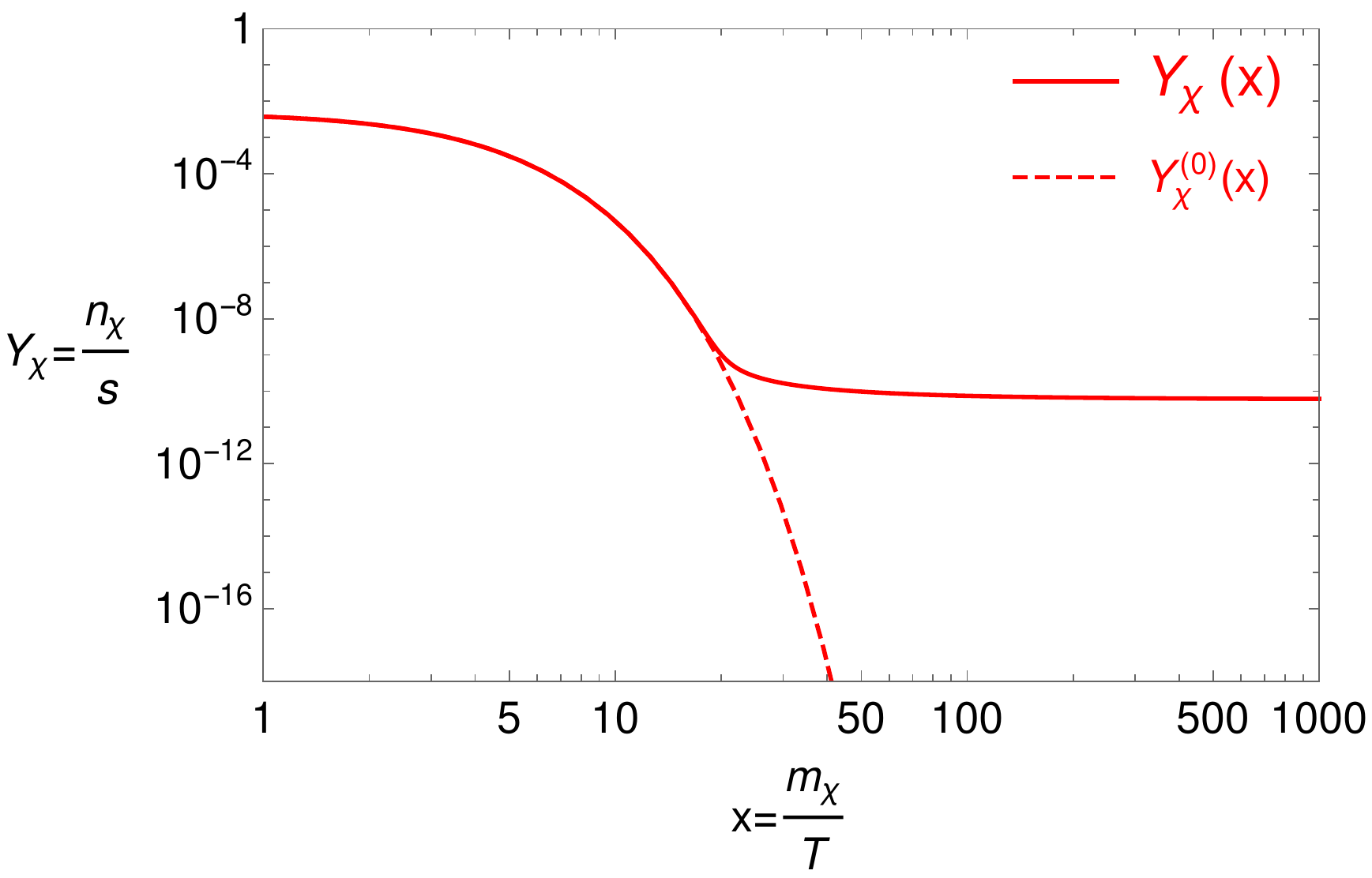}
    \caption{Evolution of the DM yield (red) compared to its equilibrium yield (red, dashed) for the freeze-out mechanism, assuming $m_\chi=1$~TeV, $g_\chi=2$, $\sigmav=3 \cdot 10^{-26}$~cm$^3$/s and $g_{eff}=h_{eff}=100$. }
    \label{fig:std_fo}
\end{figure}

Until now, we have not assumed anything about the specific particles involved in the process. As mentioned before, the only process we need to take into account in the simplest freeze-out case is the annihilation of \ac{DM} $\chi$ into thermal bath particles, $\chi \chi \to B B'$. Since the initial state particles are the same and the thermal bath particles can be assumed to be in equilibrium throughout the whole FO dynamics, the Boltzmann equation simplifies to
\begin{align}
    \frac{dY_\chi}{dx} &= \frac{\gamma_{\chi\chi \to BB'}}{s\Bar{H}x}  \left[ 1 - \left( \frac{Y_\chi}{Y^{(0)}_\chi } \right)^2 \right] \\
    &= \frac{s\sigmav_{\chi \chi \to BB'} }{\Bar{H}x} \left[ (Y^{(0)}_\chi)^2 - (Y_\chi)^2 \right].
\end{align}
This equation can not be solved analytically, hence, we show in Fig.~\ref{fig:std_fo} the numerical solution for a constant value of $\sigmav = 3 \cdot 10^{-26}$ cm$^3$/s, which is the value that reproduces the correct relic abundance~\cite{Steigman:2012nb}. As we expected, at high temperatures ($x\gg1$), the DM follows the equilibrium abundance, while around $x\approx25$, the annihilation rate $\Gamma_{ann} = \sigmav n^{(0)}_\chi$ drops below the Hubble rate such that the DM chemically decouples. Soon after this, there are no processes efficient enough to influence the DM number density, hence, the yield remains constant, i.e. it freezes out.

The \ac{FO} process happens so fast after chemical decoupling that the number density right after \ac{FO} can be approximated by the \ac{DM} equilibrium number density at the time of \ac{FO}, $x_{fo}$, after which it only redshifts. At that time, the annihilation rate roughly equals the Hubble rate, $\Gamma_{ann}(x_{fo}) = \sigmav n^{(0)}_\chi(x_{fo}) \approx H(x_{fo})$. Combining these two pieces of information leaves us with,
\begin{align}
    n_\chi(x_0) = n_\chi(x_{fo}) \frac{x_{fo}^3}{x_0^3} \approx \frac{H(x_{fo})}{\sigmav} \frac{x_{fo}^3}{x_0^3} \propto \frac{x_{fo}}{\sigmav M_{pl}},
\end{align}
where $x_0$ denotes the values of $x$ today. Increasing the cross section hence leaves us with a smaller \ac{DM} abundance, which is expected since a larger cross section keeps the DM longer in equilibrium, and the equilibrium number density decreases with time.

\subsection{Co-annihilating dark matter}
\label{sec:coann}
In the previous section, we discussed the freeze-out mechanism in the simplest setup possible, a single DM particle interacting with the thermal bath only through elastic scattering and annihilation processes. In the case that the DS consist of multiple particles (denoted by $\chi_i$, where $\chi_1$ is the lightest one and our DM candidate, $\chi_2$ the one but lightest on and so on), the \ac{FO} dynamics can drastically change. In this section, we will further assume that all \ac{DS} states are in full equilibrium at high temperatures (higher than the largest mass scale involved) and that all the heavy states $\chi_j$ can decay or down-scatter into at least one of the lighter states $\chi_i$, so that when all the heavy states are frozen-out, they eventually decay into $\chi_1$. This will give rise to co-annihilating DM.

\begin{figure}
    \centering
    \subfloat[]{\includegraphics[width=0.38\textwidth]{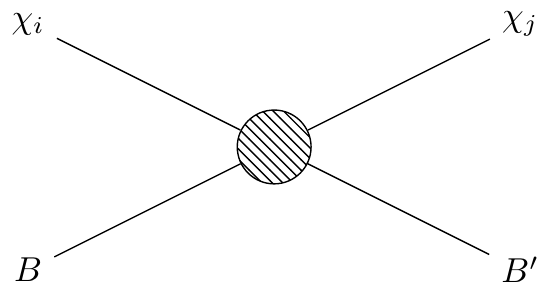}\label{feyn:coscat}}
	\hspace{0.05\textwidth}
    \subfloat[]{\includegraphics[width=0.38\textwidth]{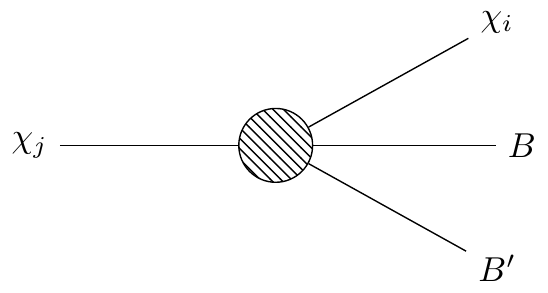}\label{feyn:decay}}
    \caption{Example of a coscattering (left) and decay (right) process between different \ac{DS} states and thermal bath particles.}
    \label{feyn:conv}
\end{figure}

Since we can assume \ac{KE} for all \ac{DS} states for the same reason as in the simple \ac{FO} case, we can use the integrated Boltzmann equation, Eq.~\eqref{eq:int_boltz} to study the evolution of the \ac{DM} abundance. Technically, we should solve a coupled system of these equations for every species where we add terms for annihilation and co-annihilation processes ($\chi_i \chi_j \to B B'$) and conversion processes, such as (inverse) decay ($\chi_j \to \chi_i B B'$) and coscattering ($\chi_j B \to \chi_i B'$), depicted in Fig.~\ref{feyn:conv},
\begin{align}
\label{eq:boltz_cdfo}
    \frac{dY_{\chi_i}}{dx} = \frac{1}{s\Bar{H}x}\sum_j &\gamma_{ij}  \left[ 1 - \frac{Y_{\chi_i} Y_{\chi_j}}{Y^{(0)}_{\chi_i} Y^{(0)}_{\chi_j}} \right] \nonumber \\
    & + \gamma_{i\to j}  \left[\frac{Y_{\chi_j}}{Y^{(0)}_{\chi_j}} - \frac{Y_{\chi_i}}{Y^{(0)}_{\chi_j}} \right],
\end{align}
where $\gamma_{ij}$ is the reaction density of the (co-)annihilation process $\chi_i \chi_j \to B B'$ and $\gamma_{i \to j}$ includes all the conversion processes,
\begin{align}
    \gamma_{i \to j} = \gamma_{\chi_i B B' \to \chi_j} + \gamma_{\chi_i B \to \chi_j B'}.
\end{align}






This complicated set of Boltzmann equations can be simplified by assuming the coscattering processes to be always efficient by using the same reasoning as for the elastic scattering processes. All the DS states are then in \ac{CE} and hence obtain the same chemical potential $\mu_\chi$ given by
\begin{align}
    e^{\mu_\chi/T} = \frac{Y_{\chi_i}}{Y_{\chi_i}^{(0)}} = \frac{Y_\chi}{Y_\chi^{(0)}},
\end{align}
where $Y_\chi=\sum_i Y_{\chi_i}$ is the total DS yield. Using this, we can trade the system of coupled equations for one single Boltzmann equation for $Y_\chi$ by summing over all the individual equations.\footnote{The heavy \ac{DM} partners will eventually decay to the \ac{DM} and enhance its relic abundance. By summing over all the number densities, we include this effect.} As one can see from Eq.~\eqref{eq:react_rate_gen}, the reaction density for the conversion and inverse conversion processes are the same. Hence, they will cancel out in the summation, leaving us with
\begin{align}
    \frac{dY_\chi}{dx} &= \frac{1}{s\Bar{H}x} \sum_{i,j} \gamma_{ij} \left[ 1 - \left( \frac{Y_{\chi}}{Y^{(0)}_{\chi} } \right)^2 \right] \nonumber \\
    &= \frac{s \sigmav_{eff} }{\Bar{H}x}  \left[ (Y^{(0)}_{\chi})^2 - (Y_{\chi})^2 \right],
    \label{eq:coan_boltz}
\end{align}
where the effective thermally averaged cross section can be written as
\begin{align}
    \sigmav_{eff} = \sum_{i,j} \sigmav_{ij} \frac{Y^{(0)}_{\chi_i}}{Y_\chi^{(0)}} \frac{Y^{(0)}_{\chi_j}}{Y_\chi^{(0)}} = \sum_{i,j} \sigmav_{ij} \frac{r_i}{r}\frac{r_j}{r}.
    \label{eq:sv_eff}
\end{align}
with
\begin{align}
    \label{eq:r_i}
    r_i = g_i \left( 1+\frac{m_i - m_1}{m_1} \right)^{3/2} e^{-x \frac{m_i - m_1}{m_1}}
\end{align}
and $r=\sum_i r_i$, where $m_i$ is the mass of $\chi_i$.

Comparing Eq.~\eqref{eq:coan_boltz} to~\eqref{eq:int_boltz}, we see that both equation are of the same form, only with the cross section swapped for an effective cross section. Therefore, the freeze-out dynamics will be exactly the same as in the simple case with one DS particle. The main difference is that it is not necessary for the DM particle to be efficiently annihilating into thermal bath particles, as long as it is efficiently coupled through conversion processes to its heavier partner(s) who can then efficiently annihilate in order to reproduce the correct relic abundance. Hence, this is the first and one of the most straightforward production mechanisms where the DM can be feebly interacting so that it is hard to observe in our present universe, since all the dark partners will already have been decayed away to the DM.  

\subsection{Caveats}
\label{sec:fo_caveats}
Throughout our discussion about freeze-out and co-annihilations, we made a number of assumptions which are usually true if the \ac{DS} interactions with the thermal bath are sizable. However, such sizable interactions can be very well constrained by DM experiments. When assuming smaller interactions to avoid such constraints, the underlying assumptions concerning the FO dynamics might break down. As a first example, we assumed that all DS particles were in full equilibrium with the thermal bath as an initial condition for our \ac{FO} computation. It might however be that the DM-SM interactions still allowed by experiments become too small to ensure this state of full equilibrium, meaning that the \ac{DM} will be produced in different ways than the ``vanilla'' \ac{FO} mechanism. There are a few different ways of producing the correct amount of very feebly interacting DM, of which we will discuss two, namely \ac{FI} in Sec.~\ref{sec:fi} and superWIMP in Sec.~\ref{sec:superWIMP}.

During the calculations of co-annihilating \ac{DM}, we made a second assumption that the conversion processes between the heavy \ac{DS} particles are efficient during the time of \ac{FO}. This is in general true if the conversion processes and (co-)annihilation processes are governed by the same coupling constant, or when all coupling constants are of the same order of magnitude. If there exists a hierarchy between the coupling constants dictating annihilation and conversion, for instance when the heavy partners can annihilate efficiently while the DM is only interacting feebly with the SM, this assumption can break down. In this scenario, one has to solve the full system of coupled Boltzmann equations (Eq.~(\ref{eq:boltz_cdfo})) to track the evolution of the DM abundance. We will discuss the physics behind this mechanism, which is referred to as \ac{CDFO}, in Sec.~\ref{sec:CDFO}.

In order for Eq.~(\ref{eq:boltz_cdfo}) to apply, \ac{KE} has been assumed. As mentioned before, heavy \ac{DS} particles decouple kinetically after \ac{FO} since elastic scattering processes maintaining \ac{KE} stay efficient well after DM chemically decouples when the annihilation processes become inefficient. This assumption might break down, for instance when the annihilation cross section gets resonantly enhanced, see e.g. Ref.~\cite{Binder:2017rgn}. In these scenarios, one should solve the Boltzmann equation at the level of the phase space density, see Eq.~(\ref{eq:Boltz_unint}). This is however a computationally very difficult task, but Ref.~\cite{vandenAarssen:2012ag} has pointed out that it can suffice to track not only the zeroth moment of Eq.~(\ref{eq:Boltz_unint}), but also its second moment in order to capture some of the effects of the DM not being in equilibrium. 

Finally, in the standard cosmological model, the universe would be dominated by SM radiation at high temperatures ($T \gtrsim 1$~eV). In the most simple FO mechanism as depicted here, \ac{DM} thermalizes with the SM bath particles so that it obtains the same temperature. Since the oldest probe of the universe that is available to us is \ac{BBN}, there is no certainty that at earlier times ($T\gtrsim1$~MeV), the universe is still dominated by SM radiation. We will explore the possibility of different cosmological histories in Chapter \ref{chap:alt_cosmo}. The possibility of DM thermalizing not with \ac{SM} radiation but within the \ac{DS} itself, often referred to as ``Secluded Freeze-out'', is something we will not discuss in this thesis, but is well studied in the literature, see e.g. \cite{Pospelov:2007mp,Chu:2011be,Evans:2017kti,Bringmann:2020mgx,Coy:2021ann}. 

\section{Freeze-in}
\label{sec:fi}

When the DM is not interacting strongly enough with the SM so that scattering and annihilation processes are not happening fast enough, the initial assumption of the FO mechanism, namely that DM should be in full equilibrium, breaks down meaning that DM is produced in a completely different way. The fact that all DM-SM interactions are inefficient does not mean that they cannot influence the DM relic abundance. DM can still be produced due to these interactions, for instance through the decay of a bath particle, as long as this bath particle is abundantly present in the thermal bath. This process will be a slow and gradual production, as if the DM was leaking from the thermal bath. This way of producing \ac{DM} is referred to as the \ac{FI} mechanism and it is a completely out-of-equilibrium production process as opposed to \ac{FO}, where annihilations happen very rapidly rendering DM to an equilibrium state.

To study the \ac{FI} dynamics, we also have to solve the Boltzmann equation shown in Eq.~(\ref{eq:Boltz_unint}). However, DM is now out of \ac{KE}, so that we cannot simply assume a \ac{FD}/\ac{BE} (or for simplicity, \ac{MB}) phase space density. Nevertheless, if we assume that there is no other production of \ac{DM} except for its \ac{FI} abundance,\footnote{\ac{DM} can however be produced through gravitational effects or inflation, see e.g. \cite{PhysRevD.35.2955,PhysRevD.37.3428,Fairbairn:2018bsw}} the initial abundance of \ac{DM} is negligible and we can safely assume $f_\chi(x^\mu,p^\mu) \approx 0$. 

In a simple example of the \ac{FI} mechanism, \ac{DM} is slowly produced via the out-of-equilibrium 2-body decay of a thermal bath particle $B \to \chi B'$. This mother particle can be both a \ac{SM} particle, like for instance the \ac{SM} Higgs boson, or a new \ac{DS} particle with sizable couplings to the \ac{SM}.\footnote{When this is not the case, $B$ can first undergo a FI production itself, before producing the DM. This is referred to as sequential FI, see e.g.~\cite{Hambye:2019dwd,Belanger:2020npe}.} The Boltzmann equation integrated over the \ac{DM} phase space for this process reads
\begin{align}
    s\Bar{H}x\frac{dY_\chi}{dx} &= \int d\Pi_\chi d\Pi_B d\Pi_{B'} (2\pi)^4 \delta^4(p_\chi+p_{B'}-p_B) |\mathcal{M}|^2 f_B^{(0)} \nonumber \\
    &= g_B \int \frac{d^3p_B}{(2\pi)^3} \frac{f^{(0)}_B \Gamma_{B\to \chi B'}}{\gamma_B} \nonumber \\
    &= n_B^{(0)} \Gamma_{B \to  \chi  B'} \frac{K_1[x]}{K_2[x]},
    \label{eq:Boltz_fi}
\end{align}
where we assumed that $B$ is always in equilibrium with the thermal bath, $\gamma_B=E_B/m_B$ is the Lorentz factor and we have neglected quantum corrections.\footnote{Just as we did for \ac{FO}, we will simplify this equation by neglecting Bose-enhancement and Pauli-blocking factors. In Ref.~\cite{Belanger:2018ccd}, they included quantum statistical corrections and concluded that the main error is due to the effects of the light final state particle $B'$. Especially when the mass splitting between $B$ and $\chi$ is small, $B'$ is be produced at low energy, so that neglecting quantum corrections might not be appropriate. Indeed, when $E \ll T$, $f\approx1-f\approx1/2$ for fermions and $f\approx1+f\approx\infty$ for bosons. Hence, producing fermionic $B'$ particles may lead to at most a factor of 2 difference, while the effects for bosonic $B'$ can be much more severe. For large mass splitting however, the effect is less severe as $B'$ can be produced at large energies.} We have also redefined the variable 
\begin{align}
    x=\frac{m_B}{T},
\end{align}
since the main freeze-in production happens around $T\approx m_B$. The FI production of DM shuts down at this point in time since the number density of the mother particle $B$ gets Boltzmann suppressed, also suppressing the production rate of DM as can be seen in Eq.~(\ref{eq:Boltz_fi}). However, for $x<1$, $\frac{K_1[x]}{K_2[x]} \simeq x$ and $n_B^{(0)} \simeq m_B^3 x^{-3}$ so that
\begin{align}
    \frac{dY_\chi}{dx} \simeq \frac{\Gamma_{B \to  \chi  B'} M_{Pl}}{m_B^2} x^2,
\end{align}
The amount of DM produced at any value of $x<1$ thus equals
\begin{align}
    Y^{\rm prod}_\chi(x) \simeq \frac{\Gamma_{B \to  \chi  B'} M_{Pl}}{m_B^2} x^3.
    \label{eq:fi_prod_ann_est}
\end{align}
$Y^{\rm prod}_\chi$ peaks at high values of $x$, or equivalent, low values of $T$. Hence, most of the \ac{DM} gets produced right before $B$ gets Boltzmann suppressed. 

\begin{figure}
    \centering
    \includegraphics[width=0.6\textwidth]{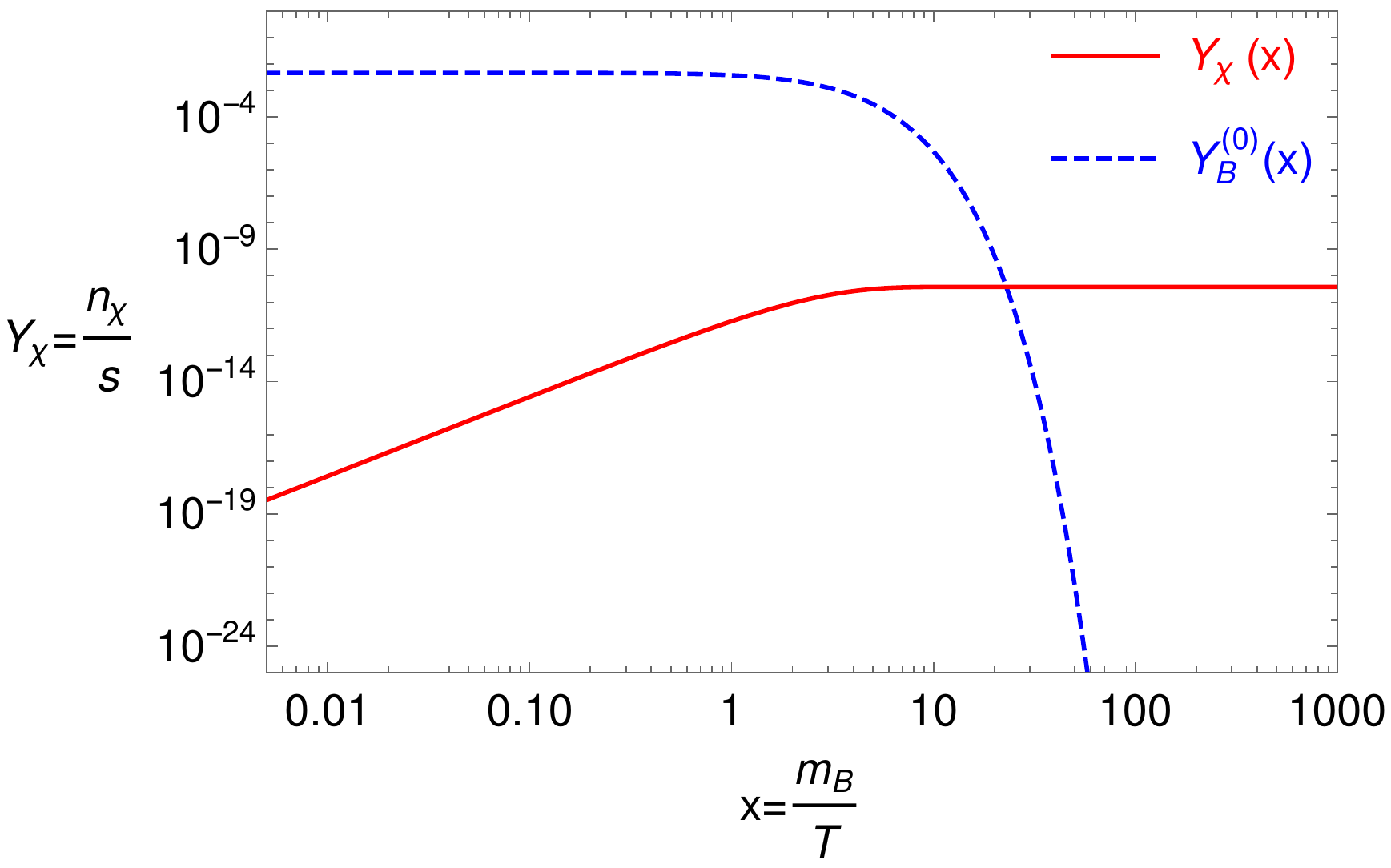}
    \caption{Evolution of the DM yield (red) compared to its equilibrium yield (blue, dashed) for the freeze-in mechanism, assuming $m_B=1$~TeV, $g_B=2$, $\Gamma=4.8 \cdot 10^{-21}$~cm$^3$/s and $g_{eff}=h_{eff}=100$. }
    \label{fig:std_fi}
\end{figure}

In Fig.~\ref{fig:std_fi}, we show the evolution of the DM yield which we obtained by solving Eq.~(\ref{eq:Boltz_fi}) numerically. Since the \ac{DM} production mainly happens at $T\approx 3 m_B$, the assumptions for the initial conditions at $T_{\rm init}$ do not have a great impact on the final results as long as $T_{\rm init} \gg m_B$ and $Y(T=T_{\rm init}) \ll Y(T=m_B)$. If there is no \ac{DM} production during the reheating phase of the universe, we can safely assume the initial DM abundance to be zero, and we start integrating at the reheating temperature $T_{\rm init}=T_R$, i.e. at the time when the universe was completely reheated and entered its radiation phase.\footnote{In Chapter \ref{chap:alt_cosmo}, we will see that this is a very good approximation when $T_R\gg m_B$.} We can even get a very good approximation of the relic abundance by integrating Eq.~(\ref{eq:Boltz_fi}) from $x=0$ to $\infty$, as the main FI production happens at intermediate times. Doing so, we obtain
\begin{align}
\label{eq:fi_ann_est}
    Y_\chi^\infty \simeq \int_0^\infty dx \frac{dY_\chi}{dx} = \frac{405 \sqrt{5}}{4 \sqrt{2} \, \pi^4} \frac{g_B}{h_{eff} g_{eff}^{1/2}} \frac{\Gamma_{B \rightarrow \chi B'} M_{\rm Pl}}{m_B^2} ,
\end{align}
or
\begin{align}
    \Omega_\chi h^2 = \frac{1.09 \cdot 10^{27} g_B}{h_{eff} g_{eff}^{1/2}} \frac{m_\chi \Gamma_{B \rightarrow \chi B'}}{m_B^2}.
    \label{eq:omega_fi}
\end{align}

In a complete model of \ac{DM}, there are typically more processes able to produce \ac{DM} besides the decay process. For instance, if the three level decay $B\to \chi B'$ is present, one can easily construct t-channel scattering processes $B A \to \chi A'$ with $B'$ in the t-channel, where $A$ and $A'$ are light thermal bath particles. These processes are typically suppressed compared to the decay, due to the extra phase space integration involved in the calculation of the cross section compared to the decay rate. However, in models where the mass splitting between the mother particle $B$ and the \ac{DM} is small, the decay is kinematically suppressed and scattering processes can start to play a non-negligible role, as we will discuss in Sec.~\ref{sec:fi_leptophilic}. In that case, one has to add an extra term to the Boltzmann equation,
\begin{align}
    s\Bar{H}x\frac{dY_\chi}{dx} &= n_B^{(0)} \Gamma_{B \to  \chi  B'} \frac{K_1[x]}{K_2[x]} + \sigmav_{B A \to \chi A'} n_B^{(0)} n_A^{(0)}.
    \label{eq:Boltz_fi_scat}
\end{align}

\section{SuperWIMP}
\label{sec:superWIMP}

While discussing the \ac{FI} mechanism in Sec.~\ref{sec:superWIMP}, we assumed that the mother particle $B$ was in equilibrium throughout the epoch where the \ac{FI} dynamics occur. However, also $B$ will eventually decouple from the thermal bath and freeze-out. Hence, there is a relic abundance of $B$ that can potentially decay fully to DM which gives an extra contribution to the DM relic abundance. This contribution is usually referred to as the superWIMP contribution, and can even dominate over the \ac{FI} contribution for sufficiently small values of the decay rate, see e.g. \cite{Garny:2018ali,Decant:2021mhj,No:2019gvl}.

If $B$ stays in \ac{CE} only by annihilating into other thermal bath particles, we can simply solve Eq.~(\ref{eq:int_boltz}) for $B$ to calculate its abundance $\Omega h^2_{B,FO}$ as if it would not decay, and assume all $B$ particles decay to DM so that
\begin{equation}
    \Omega h^2_{\chi,SW} = \frac{m_\chi}{m_B} \Omega h^2_{B,FO}.
    \label{eq:relic_ab_sw}
\end{equation}
This is a good approximation as long as the decay rate is sufficiently small so that we can completely decouple the \ac{FI} contribution coming from when $B$ was still in \ac{CE}, the \ac{FO} of $B$ and its subsequential decay. If not, one has to solve the following coupled system of equations in order to capture fully the combination of those effects
\begin{align}
    \frac{dY_\chi}{dx} &= \frac{Y_B \Gamma_{B \to  \chi  B'}}{\Bar{H}x}  \frac{K_1[x]}{K_2[x]}, \nonumber \\
    \frac{dY_B}{dx} &= \frac{s\sigmav_{B\Bar{B}} }{\Bar{H}x} \left[ (Y^{(0)}_B)^2 - (Y_B)^2 \right] - \frac{Y_B \Gamma_{B \to  \chi  B'}}{\Bar{H}x}  \frac{K_1[x]}{K_2[x]},
    \label{eq:Boltz_sw}
\end{align}
where $x=m_B/T$, as in the FI case.
\begin{figure}
    \centering
    \includegraphics[width=0.6\textwidth]{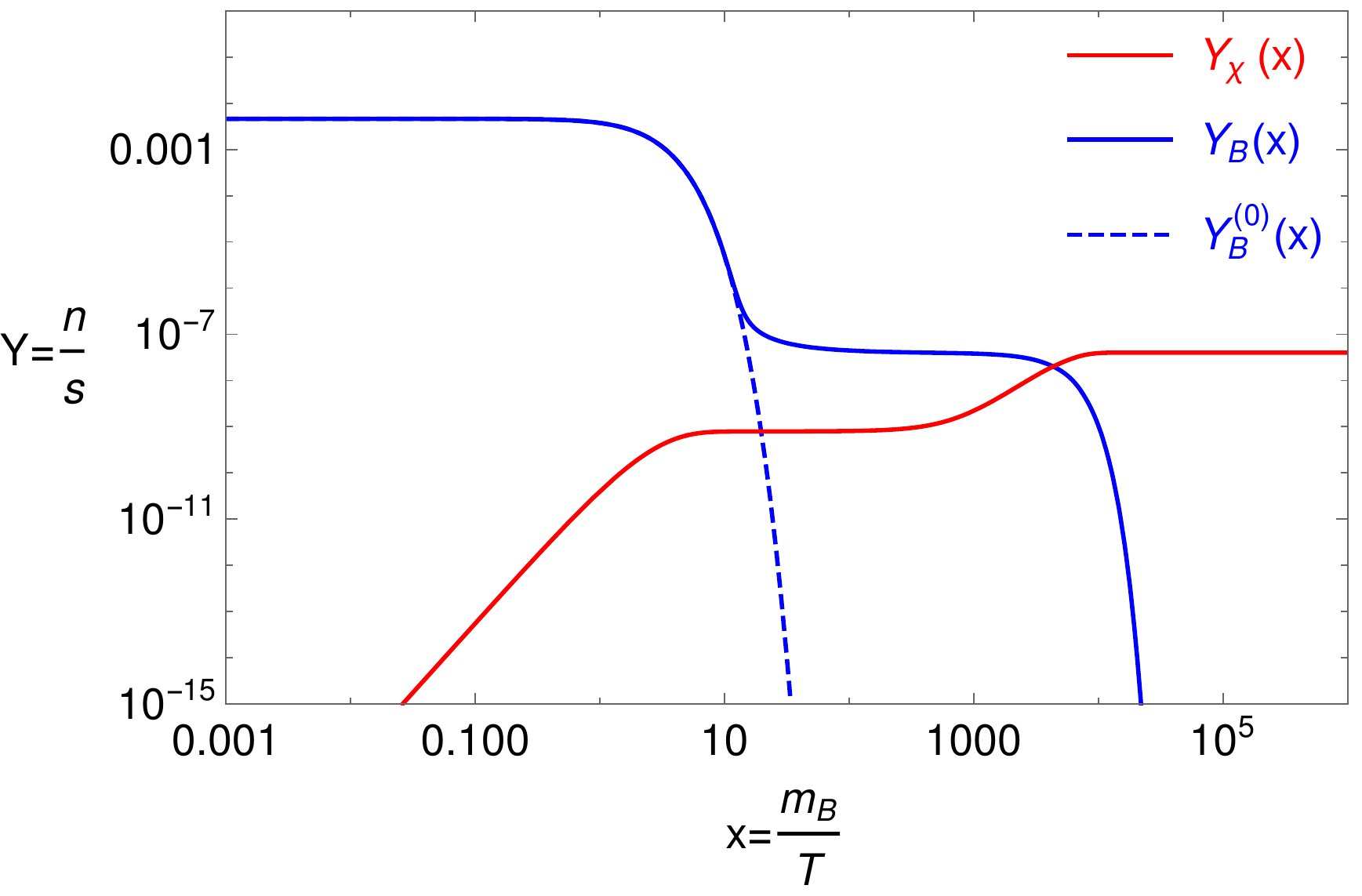}
    \caption{Evolution of the DM (red) and mediator (blue) yield compared to the mediator equilibrium yield (blue, dashed), assuming $m_{B}=1$~TeV, $m_{\chi}=0.9$~TeV, $g_{\chi}=g_{B}=2$, $\Gamma_{B \to \chi B'}= 10^{-19}$~GeV, $\sigmav_{B B}=10^{-14}$~GeV$^{-2}$, and $g_{eff}=h_{eff}=100$.}
    \label{fig:std_sw}
\end{figure}

In fig.~\ref{fig:std_sw}, we show the numerical evolution of Eq.~\eqref{eq:Boltz_sw}. Initially, the \ac{DM} yield grows as in the \ac{FI} regime which saturates around $x\approx3$. The mediator yield first tracks its equilibrium yield, until $x\approx25$, after which it freezes-out and finally decays away at $\Gamma_{B\to \chi B'}\sim H$, i.e. at $x\sim10^{7/2}$ for the Benchmark shown in Fig.~\ref{fig:std_sw}. At the time when the mediator starts decaying, we also see that the \ac{DM} yield starts rising until the mediator has decayed away completely. Since it it clear that the superWIMP contribution is much larger than the \ac{FI} contribution, Eq.~\eqref{eq:relic_ab_sw} will be a good approximation for the DM relic abundance for this benchmark.

\section{Conversion driven freeze-out}
\label{sec:CDFO}

In Sec.~\ref{sec:coann}, we mentioned that even if the \ac{DM} annihilation processes are completely inefficient during the evolution of the universe, DM can still undergo freeze-out due to its heavy partners. The only requirement is that \ac{DM} can easily convert to these partners via coscattering and/or (inverse-)decay processes so that DM achieves CE. Ref.~\cite{Garny:2017rxs} was the first to study what happens if this requirement of \ac{CE} breaks down. They identified a new regime of \ac{DM} production they dubbed \acf{CDFO}, a regime where conversion processes play a crucial role in setting the \ac{DM} relic abundance. 

The simplest scenario that exhibits \ac{CDFO} as a production mechanism is a \ac{DS} with two heavy particles, $\chi_2$ and $\chi_1$, where we assume the latter to be the lightest and our DM candidate. Studying the \ac{DM} production dynamics when we cannot simply assume \ac{CE} requires us to solve a system of coupled Boltzmann equations where we carefully track the conversion processes between $\chi_2$ and $\chi_1$, see Eq.~(\ref{eq:boltz_cdfo}).
\begin{figure}
    \centering
    \subfloat[]{
        \includegraphics[width=0.45\textwidth]{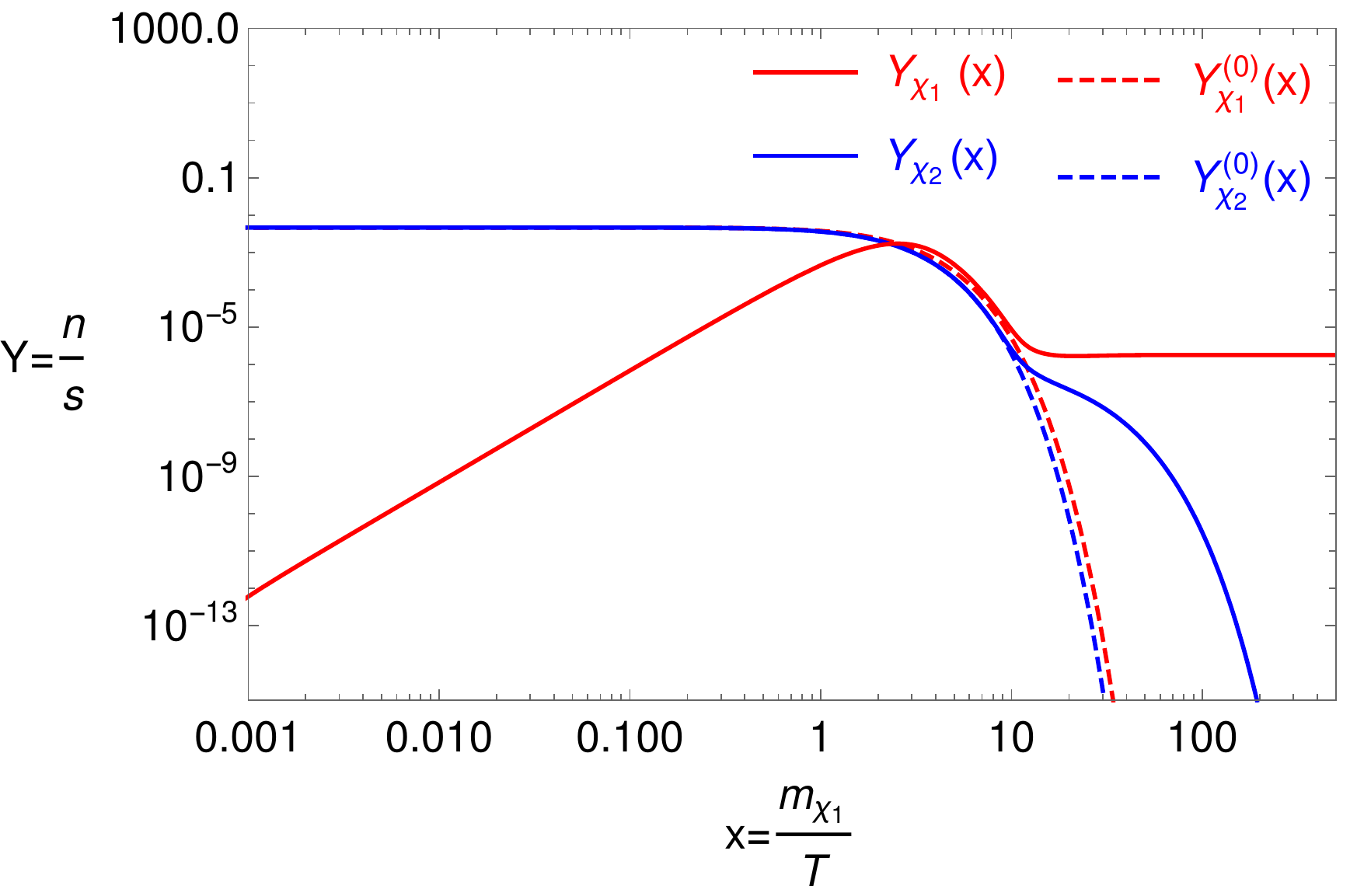}
        \label{fig:std_cdfo}}
	\hspace{0.05\textwidth}
    \subfloat[]{
        \includegraphics[width=0.45\textwidth]{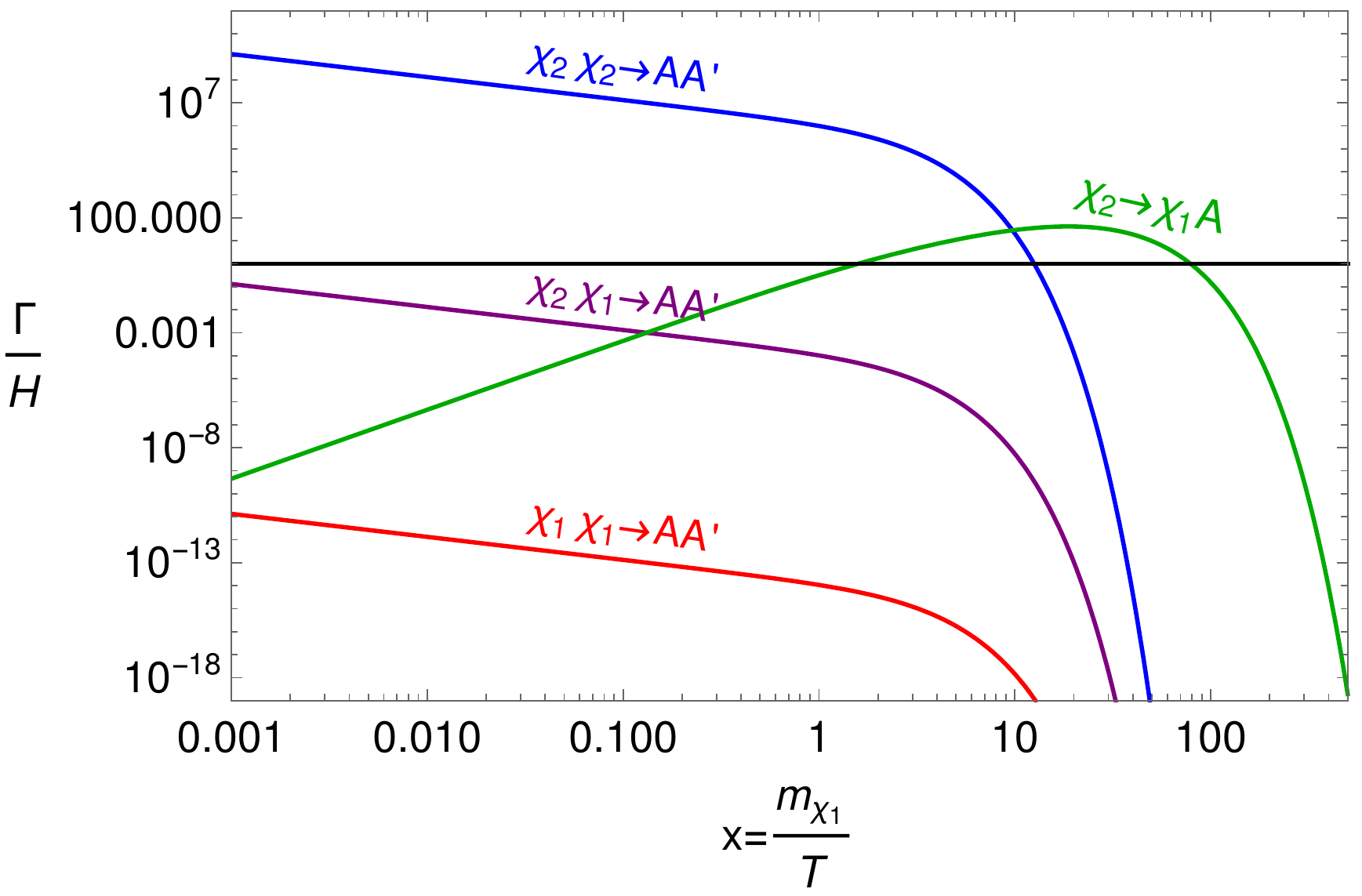}
        \label{fig:std_cdfo_rate}}
    \caption{LEFT: Evolution of the $\chi_1$ (red) and $\chi_2$ (blue) yield and their equilibrium yields (dashed) in the CDFO mechanism. RIGHT: Corresponding evolution of the interaction rates compared to the Hubble rate. For both figures, we assumed $m_{\chi_2}=1$~TeV, $m_{\chi_1}=0.9$~TeV, $g_{\chi_2}=g_{\chi_2}=2$, $\Gamma_{\chi_2 \to \chi_1 A}=10^{-12}$~GeV, $\sigmav_{\chi_1\chi_1 \to A A'}=10^{-34}$~GeV$^{-2}$, $\sigmav_{\chi_2\chi_1 \to A A'}=10^{-24}$~GeV$^{-2}$, $\sigmav_{\chi_2\chi_2 \to A A'}=10^{-14}$~GeV$^{-2}$, and $g_{eff}=h_{eff}=100$, where $A$ and $A'$ are SM thermal bath particles.}
\end{figure}
In Fig.~\ref{fig:std_cdfo}, we have solved numerically for the yield of both \ac{DS} particles. The associated rates are shown in Fig.~\ref{fig:std_cdfo_rate}. We readily see in Fig.~\ref{fig:std_cdfo} that the $\chi_1$ population does not follow its equilibrium number density evolution and that before the standard freeze-out temperature, it gives rise to $Y_{\chi_1} (x) > Y_{\chi_1}^{(0)} (x)$ due to the suppressed conversion processes that are unable to remove the overabundant $\chi_1$ efficiently.

Considering Fig.~\ref{fig:std_cdfo_rate}, we confirm that we are in a regime in which the most efficient process in reducing the dark sector abundance is the mediator annihilation process since we assumed that co-annihilation and especially \ac{DM} annihilation processes are always suppressed. On the other hand, this process is only efficient in depleting the DM when the inverse decays can efficiently convert $\chi_1$ into $\chi_2$, i.e. when the (inverse-)decay rate is faster than the Hubble rate. As we can see in Fig.~\ref{fig:std_cdfo_rate}, the (inverse-)decay rate is only barely faster than the Hubble rate so that we can confirm that we are indeed in the CDFO regime.\footnote{Requiring the inverse decay and Hubble rate to be of the same order requires some tuning of the other model parameters.} Therefore, both the decay and mediator annihilation processes will affect the \ac{DM} and the mediator abundance throughout their evolution. Given that we have considered $Y_{\chi_1}=0$ initially, there is first a large \ac{FI} contribution from the out-of-equilibrium decays, bringing the \ac{DM} yield close to its equilibrium yield. From the moment the decay becomes efficient, the \ac{DM} yield starts to converge to its equilibrium yield, but does not track it exactly because the decay process is only barely efficient. In Fig.~\ref{fig:std_cdfo}, we see that \ac{DM} yield always deviates from its equilibrium value. It is lower than the equilibrium yield at early times and larger when approaching the freeze-out time. It is also noticeable that when $Y_{\chi_1}(x)>Y_{\chi_1}^{(0)}(x)$, the \ac{DM} yield still stays close to its equilibrium yield until the mediator decouples from the thermal bath. This is due to barely but still efficient conversion processes.

\subsection{Coscattering}
\label{sec:std_coscat}

When discussing CDFO above, we considered a toy model where we only took into account the decay process as potential way of converting $\chi_1$ into $\chi_2$ and vice versa. This description is valid as long as no other conversion processes are efficient enough to equilibrate $\chi_1$ with $\chi_2$. For instance, for the model we will discuss in Sec.~\ref{sec:DM_prod_lepto}, the coscattering processes $\chi_2 A \to \chi_1 A'$, where $A$ and $A'$ are thermal bath particles, will only be barely efficient in a certain temperature range and inefficient anywhere else, so that the CDFO description is qualitatively correct. 

In models where the coscattering process is efficient at high temperatures, but becomes inefficient before the mediator freezes out, a slightly different behavior emerges. Around the time of FO, chemical equilibrium can still not be assumed so that we still need to solve the coupled system of Boltzmann equations depicted in Eq.~\eqref{eq:boltz_cdfo}. However, the coscattering process brings $\chi_1$ in equilibrium with $\chi_2$ at temperatures way above the FO temperature. Since the coscattering process becomes inefficient before $\chi_2$ chemically decouples from the thermal bath, it plays an important role in setting the DM relic abundance. So just as in the toy model above, the relic density is set by a conversion process, however, this coscattering mechanism is often distinguished from CDFO since there is state of initial chemical and kinetic equilibrium. This mechanism has first been studied by Refs.~\cite{DAgnolo:2017dbv,DAgnolo:2019zkf}.

\begin{figure}
    \centering
    \subfloat[]{
        \includegraphics[width=0.45\textwidth]{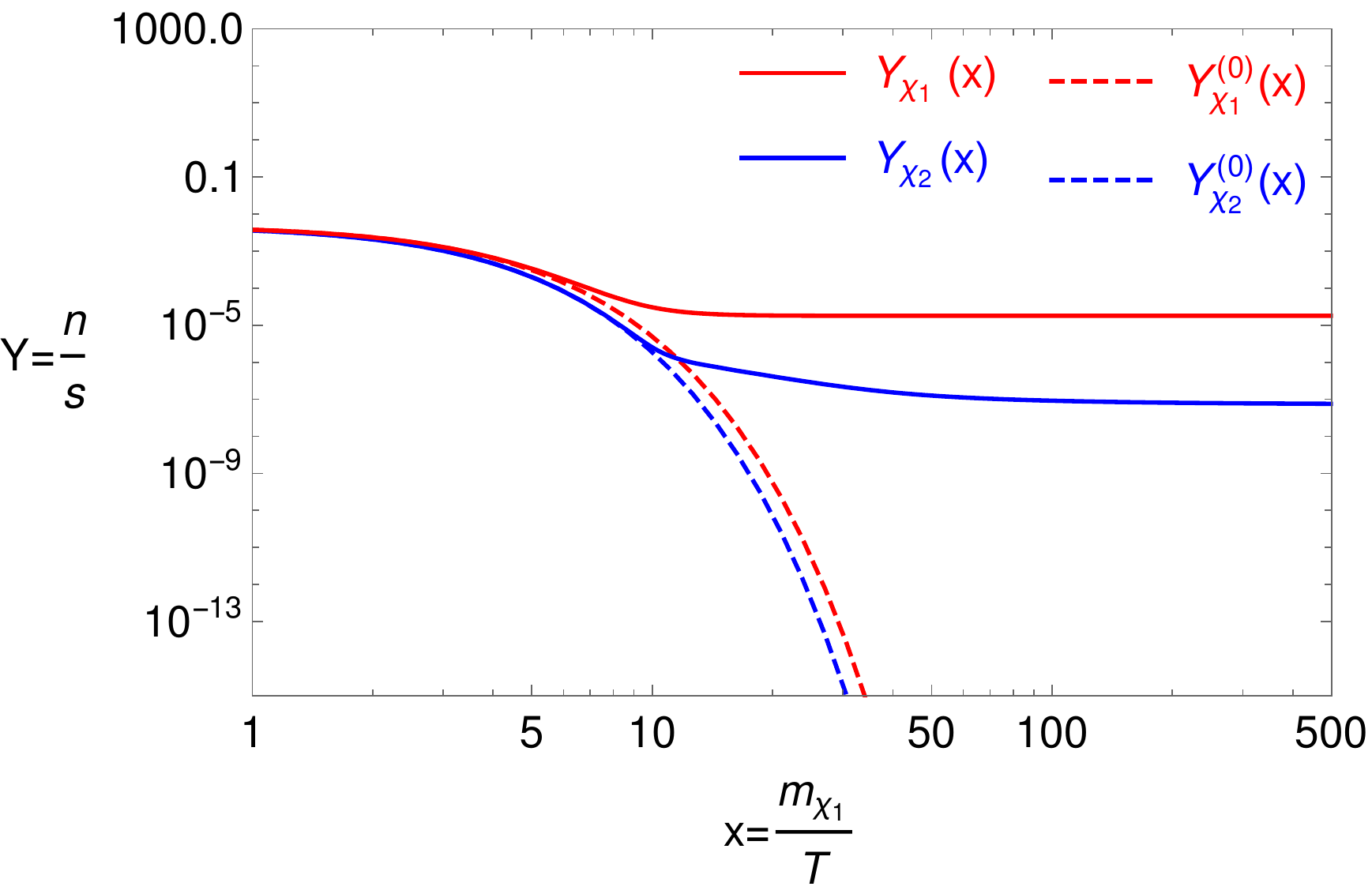}
        \label{fig:std_coscat}}
	\hspace{0.05\textwidth}
    \subfloat[]{
        \includegraphics[width=0.45\textwidth]{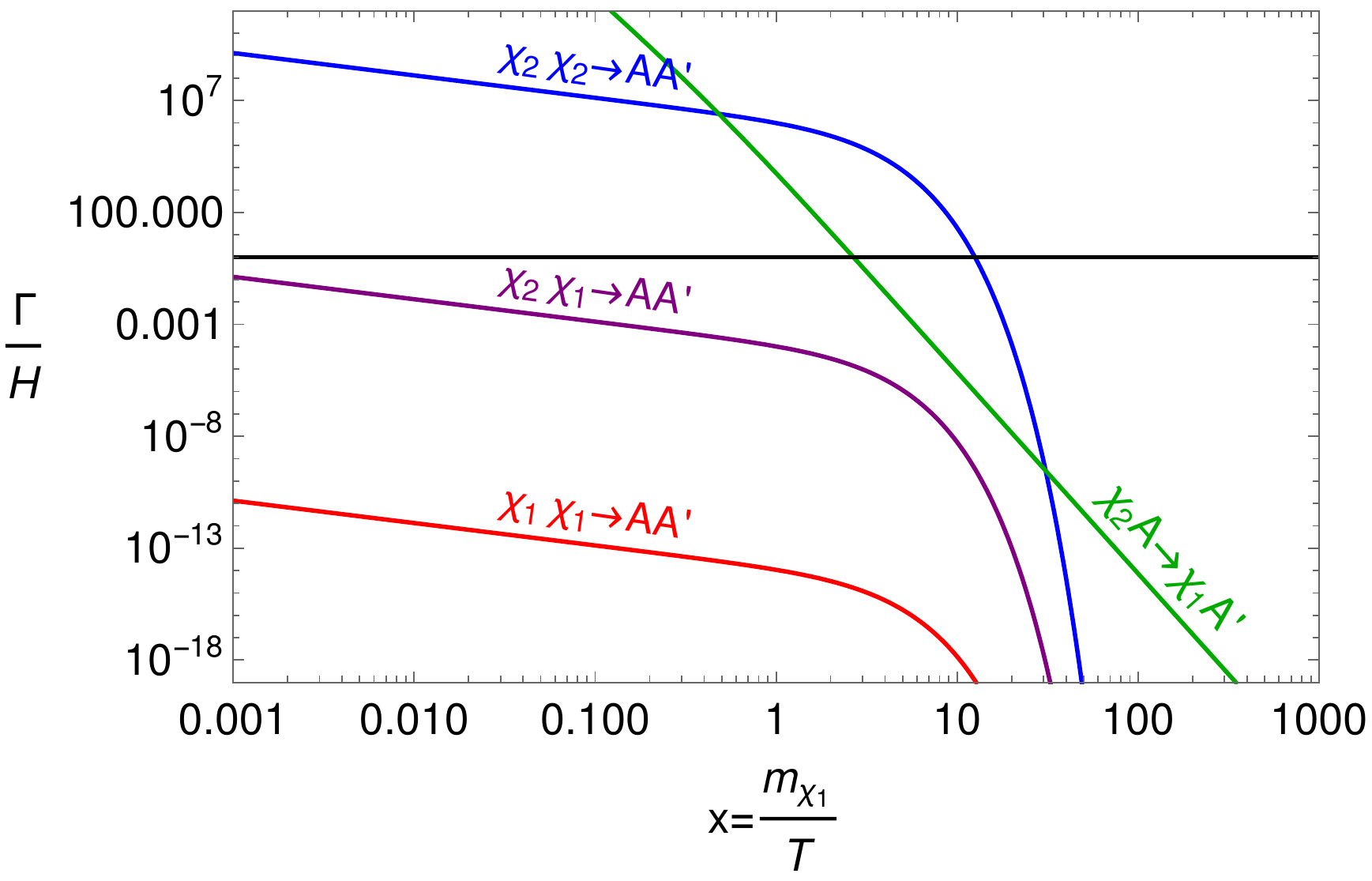}
        \label{fig:std_coscat_rate}}
    \caption{LEFT: Evolution of the $\chi_1$ (red) and $\chi_2$ (blue) yield and their equilibrium yields (dashed) in the coscattering mechanism. RIGHT: Corresponding evolution of the interaction rates compared to the Hubble rate. For both figures, we assumed $m_{\chi_2}=1$~TeV, $m_{\chi_1}=0.9$~TeV, $g_{\chi_1}=g_{\chi_2}=2$, $\sigmav_{\chi_1\chi_1 \to A A'}=10^{-34}$~GeV$^{-2}$, $\sigmav_{\chi_2\chi_1 \to A A'}=10^{-24}$~GeV$^{-2}$, $\sigmav_{\chi_2\chi_2 \to A A'}=10^{-14}$~GeV$^{-2}$, and $g_{eff}=h_{eff}=100$, where $A$ and $A'$ are SM thermal bath particles. For the coscattering process, we integrated Eq.~\ref{eq:coscat_ampl} with $m=10~\TeV$ and $\lambda=10^{-12}$ to obtain the thermally averaged cross section.}
\end{figure}

As an example, we assume the following Feynman amplitude for a t-channel coscattering process
\begin{align}
    |\mathcal{M}|^2=\lambda^2 \frac{s^2 + t^2}{(s + t + m^2)^2},
    \label{eq:coscat_ampl}
\end{align}
where $s$ and $t$ are the usual Mandelstam variables, $\lambda$ is a combination of coupling constants and we assumed the masses of the initial and final state to be much smaller than the mass $m$ of the particle in the t-channel and the temperature at which this process happens. We can integrate this to obtain the thermally averaged coscattering cross section. Further, we assume all thermally averaged annihilation cross sections to be velocity independent. We show the evolution of the relevant rates in Fig.~\ref{fig:std_coscat_rate} where we assumed $m=10~\TeV$ and $\lambda=10^{-12}$. 

Since equilibrium between the \ac{DS} states can not be simply assumed, we need to solve Eq.~\eqref{eq:boltz_cdfo} to track the evolution of the DM number density. We show in Fig.~\ref{fig:std_coscat} the numerical solution of this equation. As we can see, both $\chi_1$ and $\chi_2$ start out in chemical and kinetic equilibrium, since at high temperatures, both the $\chi_2$ annihilation and coscattering process are efficient. In contrast with the co-annihilation driven FO regime, we see that the DM yield deviates from its equilibrium yield before $\chi_2$ chemically decouples from the thermal bath. This happens when the coscattering process is becoming too inefficient to maintain a state of equilibrium between the DS states, as we can read from Fig.~\ref{fig:std_coscat_rate}. The coscattering rate is however not completely inefficient yet, so that the DM yield still gets depleted until $\chi_2$ chemically decouples. We also see that $\chi_2$ gets depleted after freeze-out, this is because it can still down-scatter to $\chi_1$.\\

One final word has to be devoted to a key assumption that goes into Eq.~\eqref{eq:boltz_cdfo}, which we need to solve in order to obtain the relic abundance. This assumption is the one of KE such that the DM phase space density obtains the predefined BE/FD (or MB) form. As mentioned before in Sec.~\ref{sec:fo_caveats}, if this assumption break down, the phase space density can get distorted and we have to solve the Boltzmann equation at the level of the phase space density, see Eq.~(\ref{eq:Boltz_unint}), which is a computationally difficult task. While in the co-annihilation regime, elastic scattering processes are typically efficient enough to maintain KE long after chemical decoupling, this is clearly not the case in the CDFO or coscattering regime as the processes dictating the FO dynamics also keep DM in KE, so that both CE and KE are lost at the same time. We will however see that in specific models (see Sec.~\ref{sec:DM_prod_lepto} and~\ref{sec:freeze-out}), it is not always necessary to solve the Boltzmann equation at the level of the phase space density as KE can be maintained roughly until the time of FO so that the error by assuming KE after FO is rather small. For completeness, we describe a procedure to solve Eq.~\eqref{eq:Boltz_unint} in the CDFO/coscattering regime in App.~\ref{app:BE_beyond_KE}.

\section{From freeze-out to freeze-in: a toy model}
\label{sec:DM_prod_lepto}
To conclude this chapter concerning DM production, we will go through all the previously discussed production mechanism within a specific model. As we will see, the phenomenology of all these mechanisms can be incorporated in one single model by varying just one coupling constant. This work is based on the analysis performed in Ref.~\cite{Junius:2019dci}. 

\subsection{A simplified leptophilic dark matter model}
\label{sec:simpl-lept-dark}

In this section, we work in a minimal extension of the \ac{SM} involving a Majorana fermion $\chi$ dark matter coupled to \ac{SM} leptons through the exchange of charged scalar mediator $\phi$. The Lagrangian encapsulating the \ac{BSM} physics reads
\begin{equation}
  {\cal L}\supset \frac{1}{2}  \bar{\chi} \gamma^{\mu} \partial_{\mu} \chi-\frac{m_{\chi}}{2} \bar{\chi} \chi+ (D_{\mu}\phi)^\dagger \ D^{\mu}\phi -m_\phi |\phi|^2 \ - \
  \lambda_{\chi} \phi  \bar \chi l_R \ + \ h.c.
\label{eq:lagr_lepto}
\end{equation}
where $m_{\chi} $ is the dark matter  mass, $m_\phi$ is the mediator mass and $\lambda_\chi$ denotes the Yukawa coupling between the dark matter, the right handed \ac{SM} lepton $l_R$ and the mediator. We have assumed that a $\mathbb{Z}_2$ symmetry prevents the dark matter to decay directly to \ac{SM} particles.  Both $\chi$ and $\phi$ are odd under the $\mathbb{Z}_2$ symmetry while the \ac{SM} fields are even and we assume $m_\phi>m_\chi$.

The standard \ac{WIMP} phenomenology of Eq.~\eqref{eq:lagr_lepto} has already been studied in length in Refs.~\cite{Garny:2011ii,Garny:2015wea,Garny:2013ama,Bringmann:2012vr,Kopp:2014tsa,Baker:2018uox,Khoze:2017ixx}, while the corresponding scalar \ac{DM} case has been studied in Refs.~\cite{Giacchino:2013bta,Giacchino:2014moa,Giacchino:2015hvk,Colucci:2018vxz,Toma:2013bka,Ibarra:2014qma}.\footnote{See also Ref.~\cite{Belanger:2018sti} for the \ac{FI} case.} Here, we will discuss how the phenomenology of the model in Eq.~\eqref{eq:lagr_lepto} non-trivially evolves between the two extremes of superWIMP and standard freeze-out by varying the coupling constant $\lambda_\chi$, focusing on small mass splittings between the dark matter and the mediator to incorporate co-annihilation and conversion driven freeze-out in our analysis. 

Considering the minimal SM extension of Eq.~\eqref{eq:lagr_lepto}, there  are three free parameters in our model, the dark matter mass $m_{\chi}$, the mediator mass $m_{\phi}$ and  the Yukawa coupling $\lambda_{\chi}$. Here in particular, we use
\begin{equation}
\label{eq:param}
	m_{\chi}, \Delta m, \lambda_{\chi}\, ,
\end{equation}
with $\Delta m=m_{\phi}-m_{\chi}$, as a minimal set of free parameters of the model.  No symmetries prevent the new scalar $\phi$ to couple to the \ac{SM} Higgs and this coupling will generically be radiatively generated. We will thus consider an extra interaction involving the quartic coupling $ \lambda_H$:
\begin{equation}
\label{eq:Lhiggs}
	{\cal L}_{H} = -\lambda_H H^{\dag} H \phi^{\dag} \phi.
\end{equation}
We can indeed always write a box diagram involving the $4$ scalars, through the exchanges of $Z$-boson or photon, giving rise to the Lagrangian in Eq.~\eqref{eq:Lhiggs}, with an effective coupling $\lambda_H^{eff}\sim g_{weak}^2/(16 \pi^2)\sim 10^{-2}$. This extra Higgs portal interaction shall thus definitively be taken into account. A non-negligible value of such a coupling will add to the gauge induced $\phi$ annihilation processes, potentially giving rise to a larger $\phi$ annihilation cross-section, and modify the freeze-out of the charged scalar.

We would like to mention that the simplified model involving the contributions of Eq.~\eqref{eq:lagr_lepto} and \eqref{eq:Lhiggs} to the Lagrangian serves as an illustrative case to discuss in details the early universe dark matter production for a larger class of feebly coupled dark matter scenarios with compressed mass spectrum. The leptophilic scenario considered here can easily emerge in non-minimal SUSY models (like extension of the \ac{MSSM}) where the \ac{NLSP} is one of the right handed sleptons and the \ac{DM} candidate is an extra neutralino (as for instance discussed in \cite{Aboubrahim:2019qpc}). The possibility that the \ac{NLSP} is one of right handed sleptons (not the stau) has been studied for instance in Refs.~\cite{Evans:2006sj,Calibbi:2014pza}.

\subsection{Relic abundance calculations}

\begin{table}[!t]
	\centering
	\renewcommand\arraystretch{1.5}
	\begin{tabular}{| c | c | c | c | c | }
		\hline 
		\multicolumn{2}{|c|}{initial state} & \multicolumn{2}{|c|}{final state} & scaling\\
		\hline	\hline
		$\chi$ & $\chi$ & $l^-$ & $l^+$ & $\lambda_{\chi}^4$ \\
		\hline	\hline
		\multirow{2}{*}{$\chi$} &  \multirow{2}{*}{$\phi$}
		& $l^-$ & $\gamma,Z,H$ & \multirow{2}{*}{$\lambda_{\chi}^2$} \\
		\cline{3-4}
		& & $W^-$ & $\nu_{l}$ & \\
		\hline \hline
		\multirow{4}{*}{$\phi$} &  \multirow{4}{*}{$\phi^{\dagger}$}
		& $\gamma,Z,W^+$ & $\gamma,Z,W^-$ & \multirow{4}{*}{$\lambda_{\chi}^0$} \\
		\cline{3-4}
		& & $q$ & $\bar{q}$ & \\
		\cline{3-4}
		& & $H$ & $Z$ & \\
		\cline{3-4}
		& & $l^-$ & $l^+$ & \\
		\hline
		$\phi$ & $\phi$ & $l^-$ & $l^-$ & $\lambda_{\chi}^4$\\
		\hline
	\end{tabular}
	\caption{List of all included co-annihilation processes where $l$ is one of the leptons ($e,\mu,\tau$), depending on the case we are studying. Also the dependence of the cross section on the coupling constant $\lambda_{\chi}$ is denoted in the last column. The $\phi \phi^{\dagger}$ annihilation into $l\bar{l}$ also has contributions scaling with $\lambda_{\chi}^2$ and $\lambda_{\chi}^4$.}
	\label{tab:coann}
\end{table}

\begin{table}[!t]
	\centering
	\renewcommand\arraystretch{1.5} 
	\begin{tabular}{| c | c | c | c | c | }
		\hline 
		\multicolumn{2}{|c|}{initial state} & \multicolumn{2}{|c|}{final state} & scaling \\
		\hline	\hline
		\multirow{4}{*}{$\chi$} & $l^-$ & \multirow{4}{*}{$\phi$} & $\gamma,Z,H$ & \multirow{4}{*}{$\lambda_{\chi}^2$} \\
		\cline{2-2} \cline{4-4}
		& $\gamma,Z,H$ & & $l^+$ & \\
		\cline{2-2} \cline{4-4}
		& $W^-$ & & $\bar{\nu_l}$ & \\
		\cline{2-2} \cline{4-4}
		& $\nu_l$ & & $W^+$ & \\
		\hline \hline
		\multicolumn{2}{|c|}{$\phi$} & $\chi$ & $l^-$ & $\lambda_{\chi}^2$\\
		\hline \hline
		$\chi$ & $\chi$ & $\phi$ & $\phi^{\dagger}$ & $\lambda_{\chi}^4$ \\
		\hline
	\end{tabular}
	\caption{List of all included conversion processes and their dependence of the cross section on $\lambda_{\chi}$. $l$ is one of the leptons ($e,\mu,\tau$), depending on the case we are studying. The dependence on $\lambda_\chi$ is denoted in the last column.}
	\label{tab:conv}
\end{table}

In the \ac{DM} model considered here, the mediator will always be in equilibrium with the thermal bath at early times, due to its gauge interactions leading to very efficient annihilation processes such as $\phi \phi^\dagger \to f \bar{f}$, where $f$ is a standard model fermion. These processes can even be enhanced by annihilations into Higgs bosons if the quartic coupling $\lambda_H$ is large enough. In contrast, the \ac{DM} annihilation cross section into SM leptons scales as $\lambda_\chi^4$, and since this is the only possible annihilation channel for \ac{DM}, the annihilation process quickly becomes too inefficient to keep DM in equilibrium when lowering the coupling constant. Similarly, the co-annihilation processes also depend on $\lambda_\chi$. All the (co)-annihilation processes and their dependence on the conversion parameter are summarized in Tab.~\ref{tab:coann}. 

As discussed in the Sec.~\ref{sec:CDFO}, conversion processes can play an important role in setting the \ac{DM} relic abundance in certain production regimes. For the leptophilic model studied here, we show all relevant conversion processes in Tab.~\ref{tab:conv}. When including all these processes, the Boltzmann equations for $\chi$ and $\phi$ read
\begin{align}
\label{eq:BEchi}
\frac{dY_{\chi}}{dx}=\frac{-2}{\bar{H}xs} & \left[ \gamma_{\chi \chi} \left (\left(\frac{Y_{\chi}}{Y^{(0)}_{\chi}}\right)^2-1\right) + \gamma_{\chi \phi} \left(\frac{Y_{\chi} Y_{\phi}}{Y^{(0)}_{\chi} Y^{(0)}_{\phi}}-1 \right) \right. \nonumber  \\
&+  \gamma_{\chi   \to \phi }\left( \frac{Y_{\chi}}{Y^{(0)}_{\chi}}- \frac{Y_{\phi}}{Y^{(0)}_{\phi}} \right) 
+ \left. \gamma_{\chi \chi \rightarrow \phi \phi^{\dagger}} \left(\left(\frac{Y_{\chi}}{Y^{(0)}_{\chi}}\right)^2- \left(\frac{Y_{\phi}}{Y^{(0)}_{\phi}}\right)^2\right) \right],
\end{align}
\begin{align}
\label{eq:BEsl}
\frac{dY_{\phi}}{dx}=\frac{-2}{ \bar{H}xs} & \left[\gamma_{\phi \phi^{\dagger}} \left( \left(\frac{Y_{\phi}}{Y^{(0)}_{\phi})}\right)^2- 1\right) + \gamma_{\chi \phi} \left(\frac{Y_{\chi} Y_{\phi}}{Y^{(0)}_{\chi} Y^{(0)}_{\phi}}-1 \right)  \right. \nonumber \\
  & -\gamma_{\chi   \to \phi }\left( \frac{Y_{\chi}}{Y^{(0)}_{\chi}}- \frac{Y_{\phi}}{Y^{(0)}_{\phi}} \right) 
- \left. \gamma_{\chi \chi \rightarrow \phi \phi^{\dagger}} \left(\left(\frac{Y_{\chi}}{Y^{(0)}_{\chi}}\right)^2- \left( \frac{Y_{\phi}}{Y^{(0)}_{\phi}}\right)^2\right) \right],
\end{align}
where $\gamma_{ij }=\gamma_{ij\to \alpha \beta }$, with $\alpha,\beta$ some SM particles in equilibrium with the bath, $\gamma_{\chi \to \phi}$ includes both decays and scatterings $\gamma_{\chi \to \phi }=\left(\gamma_{\chi \alpha \to \phi \beta}+\gamma_{\chi \alpha \to \phi }\right)$, $Y_{\phi}$ is the summed contribution of {\it both} the mediator and its antiparticle and we will use $x=m_\chi/T$.\footnote{In Eq.~\eqref{eq:BEchi}, in the terms for $\chi$ self-annihilations, the factor of 2 is due to two $\chi$'s disappearing in each process while, in the other terms, it is due to the contributions both from $\phi$ and from $\phi^{\dagger}$. In the Eq.~\eqref{eq:BEsl}, the factor of 2 is always due to the convention used here for $Y_\phi$.} The reaction densities have been obtained taking into account all processes given in Tab.~\ref{tab:coann} and \ref{tab:conv}. 

In our study of different production regimes within the above mentioned leptophilic DM model, we will numerically solve Eqs.~\eqref{eq:BEchi} and \eqref{eq:BEsl}. In general, we always assume zero initial dark matter abundance and begin to integrate our set of equations at $x=0.01$ i.e., in the small mass splitting limit, a time at which we expect the mediator is relativistic and in equilibrium with the SM bath. Also notice that the above set of equations assume both DM and the mediator to be in \ac{KE}. Elastic scattering processes of the mediator are driven by gauge interactions, and hence its equilibrium state is always assured. For DM, the elastic scattering and coscattering processes maintaining KE depend on the conversion parameter. This assures that in the (co)-annihilation driven freeze-out regime, DM is in kinetic equilibrium, while for other production mechanisms, it might be that DM does not reach \ac{KE}. We comment further on this assumption in the next section.

\subsection{Dependence on the conversion parameter}
\label{sec:relic}

\begin{figure}
  \begin{center}
  \includegraphics[width=0.85\textwidth]{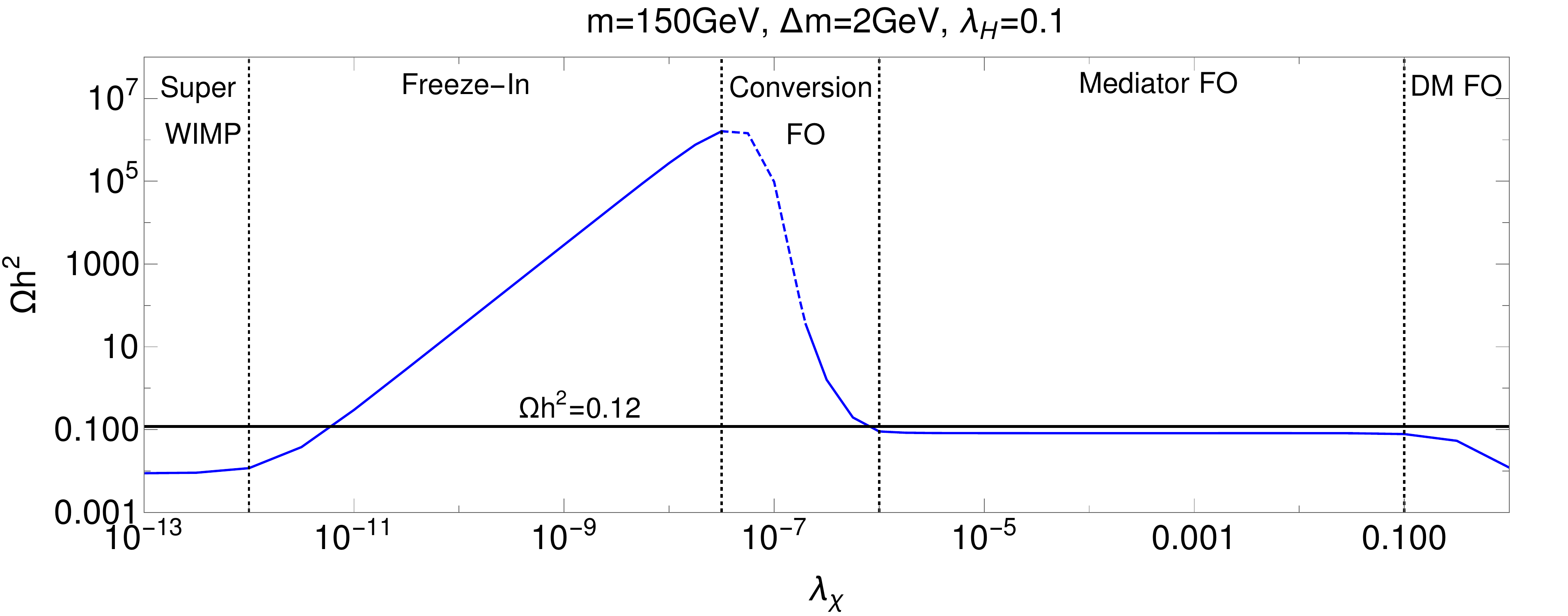}  
  \end{center}
  \caption{DM abundance as a function of the Yukawa coupling for $m_\chi=150$ GeV, $\Delta m=2$ GeV and $\lambda_H=0.1$ (i.e. for compressed DM-mediator mass spectrum).}
  \label{fig:ann-conv}
\end{figure}

Fig.~\ref{fig:ann-conv} shows the dependence of the \ac{DM} relic abundance with respect to the Yukawa coupling $\lambda_\chi$ in the case of a compressed mass spectrum. For illustration, we consider a coupling to $l_R=\mu_R$ with $m_{\chi} = 150$ GeV, $\Delta m = 2$ GeV and $\lambda_{H} = 0.1$, and we comment on the different regimes of \ac{DM} production that are realized by varying $\lambda_{\chi}$.
 
For the largest values of the coupling, one can assume chemical equilibrium between the two DS particles due to efficient conversion processes, and we can apply the procedure proposed in Ref.~\cite{Griest:1990kh} and elaborated on in Sec.~\ref{sec:coann}. When $\lambda_\chi \sim 1$, the effective annihilation cross section $\sigmav_{eff}$ defined in Eq.~\eqref{eq:sv_eff} is dominated by the dark matter annihilation cross section $\sigmav_{\chi\chi}$ into SM states with $\sigmav_{\chi\chi}\propto \lambda_\chi^4$. This is the well known ``standard \ac{WIMP} behavior'' (as outlined in Sec.~\ref{sec:std_fo}). As $\Omega h^2\propto 1/\sigmav_{eff}$, we see that the relic abundance line in Fig.~\ref{fig:ann-conv} rises for decreasing values of the coupling. Around $\lambda_\chi\sim 0.1$, the $\Omega h^2$ curve bends and we enter in the {\it co-annihilation driven} regime, where the co-annihilation processes $\chi \phi \to$~SM~SM dominate the effective cross section due to their milder dependency on the coupling constant, $\sigmav_{\phi \chi} \propto \lambda_\chi^2$. The effective cross section is also suppressed by the relative mass splitting $\Delta m/m_\chi$ ($\sigmav_{eff} \propto \sigmav_{\phi \chi} \exp(- x_f\Delta m/m_\chi) $, where $x_f\approx 25$), so that the exact value of $\lambda_\chi$ for which co-annihilation processes start playing an important role depends on the value of the mass splitting.

For even lower values of the coupling, $10^{-6}\lesssim\lambda_\chi\lesssim 0.1$, we see in Fig.~\ref{fig:ann-conv} that the relic abundance line reaches a plateau. In this regime, chemical equilibrium is still maintained, but DM annihilation and co-annihilation processes are rendered inefficient to reproduce the correct relic density due to their dependence on $\lambda_\chi$. Since the mediator annihilation cross section into SM states ($\sigmav_{\phi\phi^\dag}$) is purely dictated by gauge couplings and the quartic coupling $\lambda_H$, this mediator annihilation process starts dominating the effective cross section in Eq.~\eqref{eq:coan_boltz}, i.e. $\sigmav_{eff}\propto \sigmav_{\phi\phi^\dag} \exp(-2 x_f\Delta m/m_\chi)$. This explains why in Fig.~\ref{fig:ann-conv}, the relic density becomes independent on the exact value of the Yukawa coupling $\lambda_\chi$. We refer to this regime as the {\it mediator annihilation driven} FO regime. The fact that the mediator annihilation cross section is independent of $\lambda_\chi$ means that the correct relic abundance can be obtained for a broad range of values of $\lambda_\chi$. However, a small mass splitting is required due to the exponential suppression of the effective annihilation cross section, and hence, some tuning of the parameters is required.

If chemical equilibrium between the \ac{DM} and the other species could be maintained for even lower values of $\lambda_\chi$ we would expect the relic abundance to stay independent of this parameter. Ref.~\cite{Garny:2017rxs} however pointed out that a new window for dark matter production opens at sufficiently small values of $\lambda_\chi$. For a suppressed rate of conversion processes, here with $\lambda_\chi \lesssim 10^{-6}$, we are left with a dark matter abundance that is larger than in the equilibrium case due to inefficient $\chi\to\phi$ conversions. We enter in the {\it conversion driven freeze-out} regime, as already introduced in Sec.~\ref{sec:CDFO}, with a relic abundance that is larger than in the DM or mediator annihilation driven regime and that increases again with decreasing $\lambda_\chi$. This behavior is due to the rate of conversion processes becoming slower than the Hubble rate.  In contrast, in the ``standard WIMP'' freeze-out, it is the rate of the relevant (co-)annihilation processes that becomes inefficient. For even lower values of the coupling, $\lambda_\chi < 10^{-8}$, the dark matter relic abundance is set by a combination of the FI and superWIMP mechanism, i.e. the dark matter is produced through decays or scatterings of the mediator, either before its chemical decoupling or after, see Secs.~\ref{sec:fi} and \ref{sec:superWIMP}. For $10^{-12} < \lambda_\chi < 10^{-8}$, the main contribution comes from \ac{FI} such that $\Omega h^2\propto \Gamma_{\phi\to\chi l_R} \propto \lambda^2_\chi$. This is confirmed in Fig.~\ref{fig:ann-conv} where we see that the relic abundance decreases with decreasing coupling. When $\lambda_\chi$ drops below $10^{-12}$, the \ac{FI} contributions has decreased so much that the superWIMP contributions starts dominating. This contribution is completely set by the \ac{FO} of $\phi$, and hence does not depend on the conversion parameter $\lambda_\chi$.

With the dashed line in Fig.~\ref{fig:ann-conv} we show the ``naive'' transition between the \ac{CDFO} and \ac{FI} regime that one would obtain using Eqs.~(\ref{eq:BEchi}) and~(\ref{eq:BEsl}).  In this picture, one can compute $\phi$ and $\chi$ abundances assuming that they stay in \ac{KE} all the way from standard \ac{FO} to \ac{FI}. This is difficult to argue. In Ref.~\cite{Garny:2017rxs}, the authors compared the results of the un-integrated Boltzmann equations with the one of Eqs.~(\ref{eq:BEchi}) and~(\ref{eq:BEsl}). It was shown that even though \ac{KE} can not be maintained all along the process of \ac{CDFO}, the resulting error on the estimation of the relic dark matter was small ($\sim 10\%$). The authors argue that this is due to the thermally coupled mediator eventually decaying to the DM at the time of \ac{DM} \ac{FO} actually allowing for the dark matter to inherit back a thermal distribution (or at least close enough to it). For this effect to be relevant, the DM abundance should definitively not be too far higher than the mediator abundance around freeze-out.  In Fig.~\ref{fig:ann-conv}, the dashed line starts at $\lambda_\chi=2 \cdot 10^{-7}$ and denotes the cases when $Y_\chi- Y_\phi>0.1 \times Y_\chi$ around $\chi$ freeze-out. In the latter case, we assume that we are in the same situation as in~\cite{Garny:2017rxs} and we neglect departure from \ac{KE}. We have checked that all viable models considered in the following for \ac{CDFO} satisfy to this 10\% condition.

As we just showed, there are many different production regimes for \ac{DM} within the leptophilic DM model under study. Since the standard freeze-out regimes have already been studied in detail, we will focus only on the phenomenology of \ac{CDFO} and the \ac{FI} regime.

\subsection{Phenomenology of conversion driven freeze-out}

\begin{figure}[!t]
  \centering
  \subfloat[]{\label{fig:Rateplot3}{\includegraphics[width=0.53\textwidth]{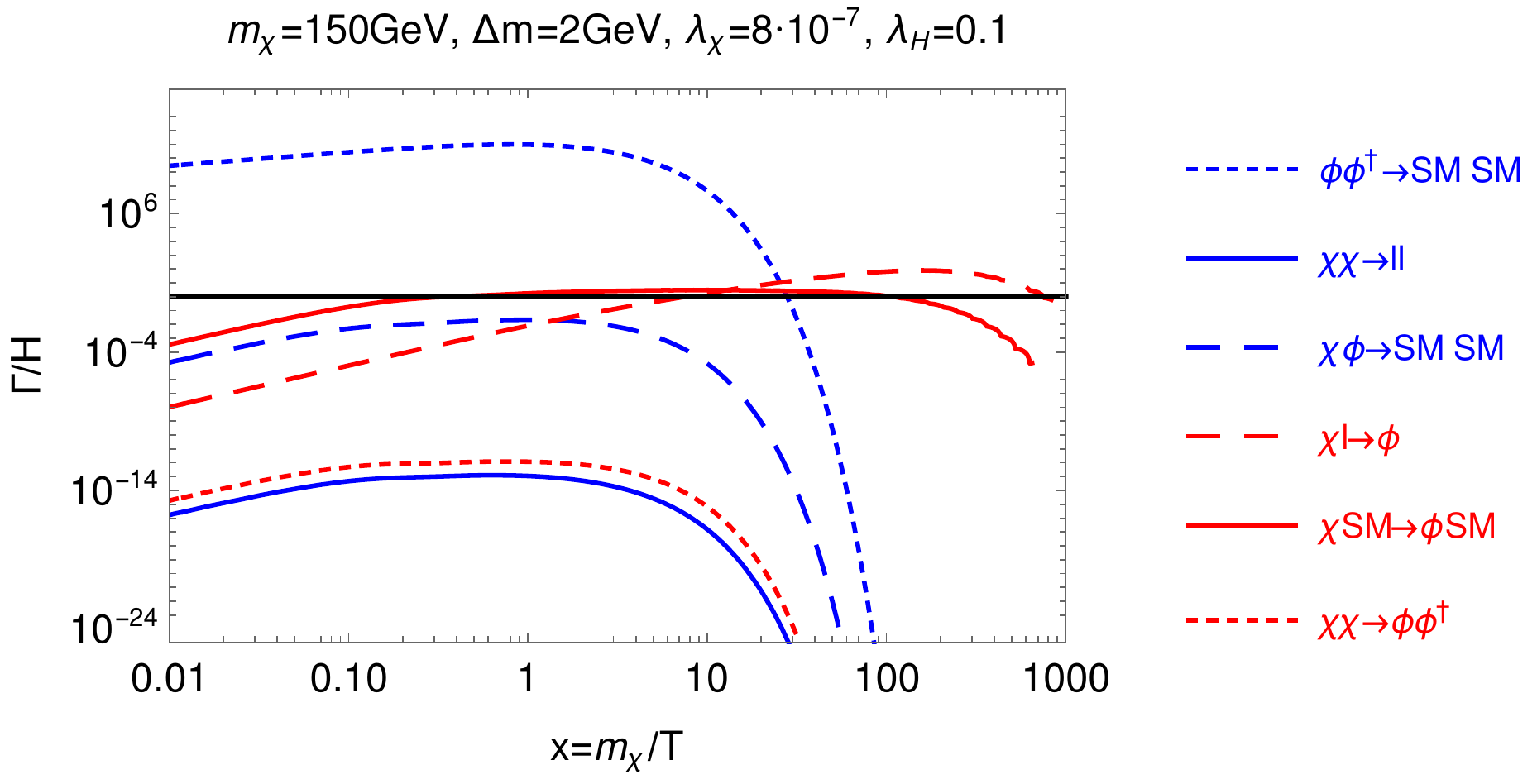}}}
  \hspace{0.03\textwidth}
  \subfloat[]{\label{fig:Yield3}{\includegraphics[width=0.4\textwidth]{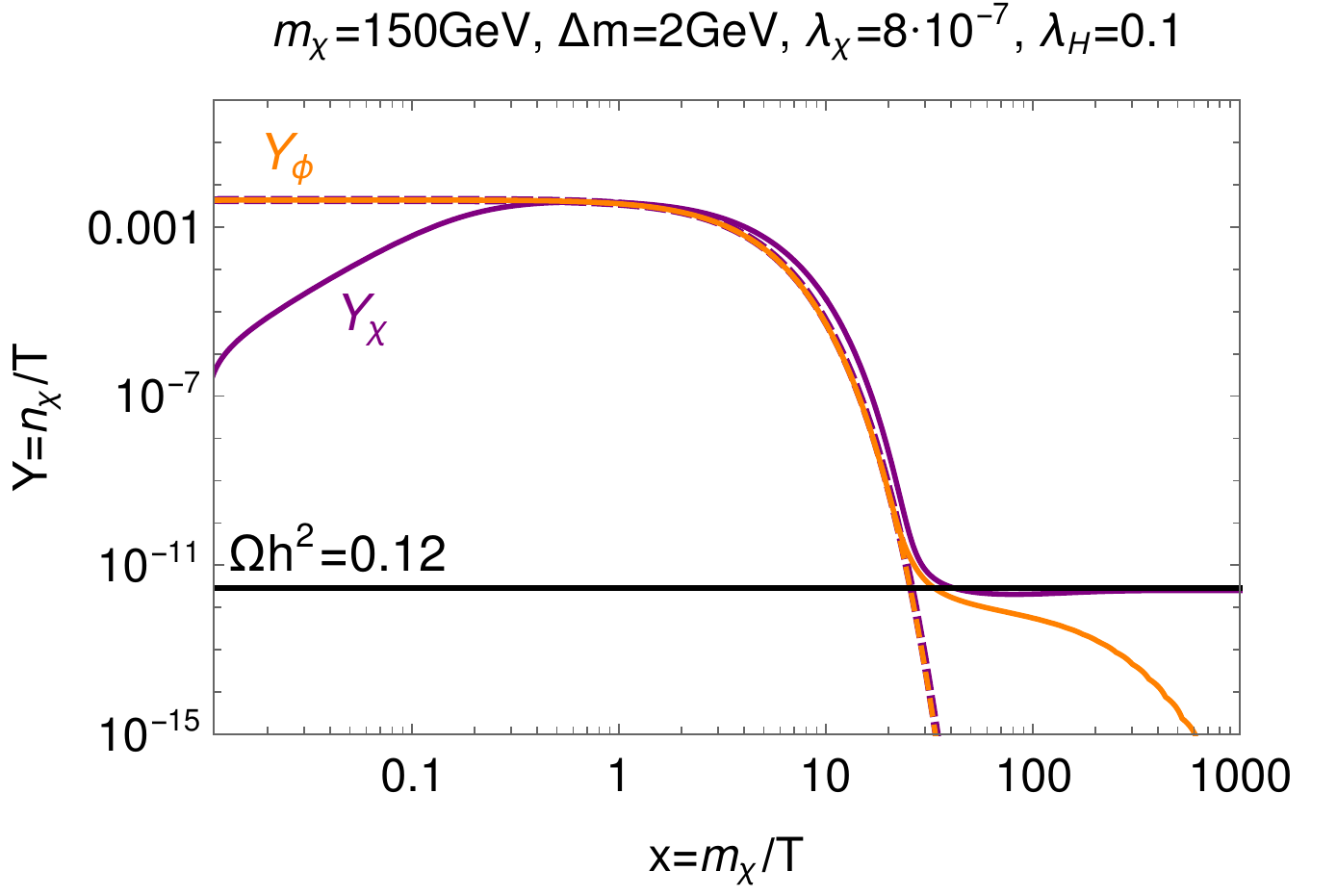}}}\\
  \vspace{0.002\textheight}
  \subfloat[]{\label{fig:Rateplot1}{\includegraphics[width=0.53\textwidth]{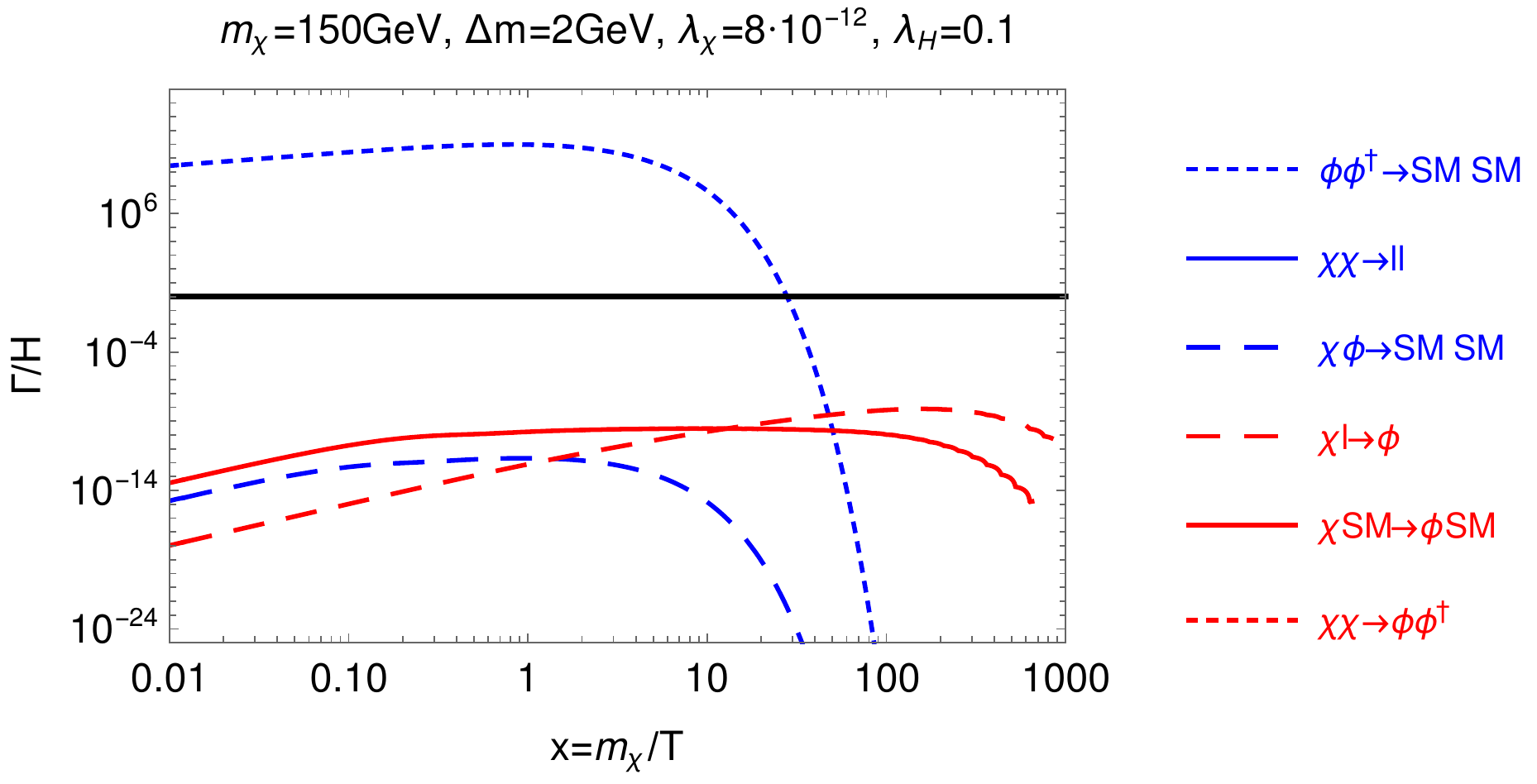}}}
	\hspace{0.03\textwidth}
   \subfloat[]{\label{fig:Yield1}{\includegraphics[width=0.4\textwidth]{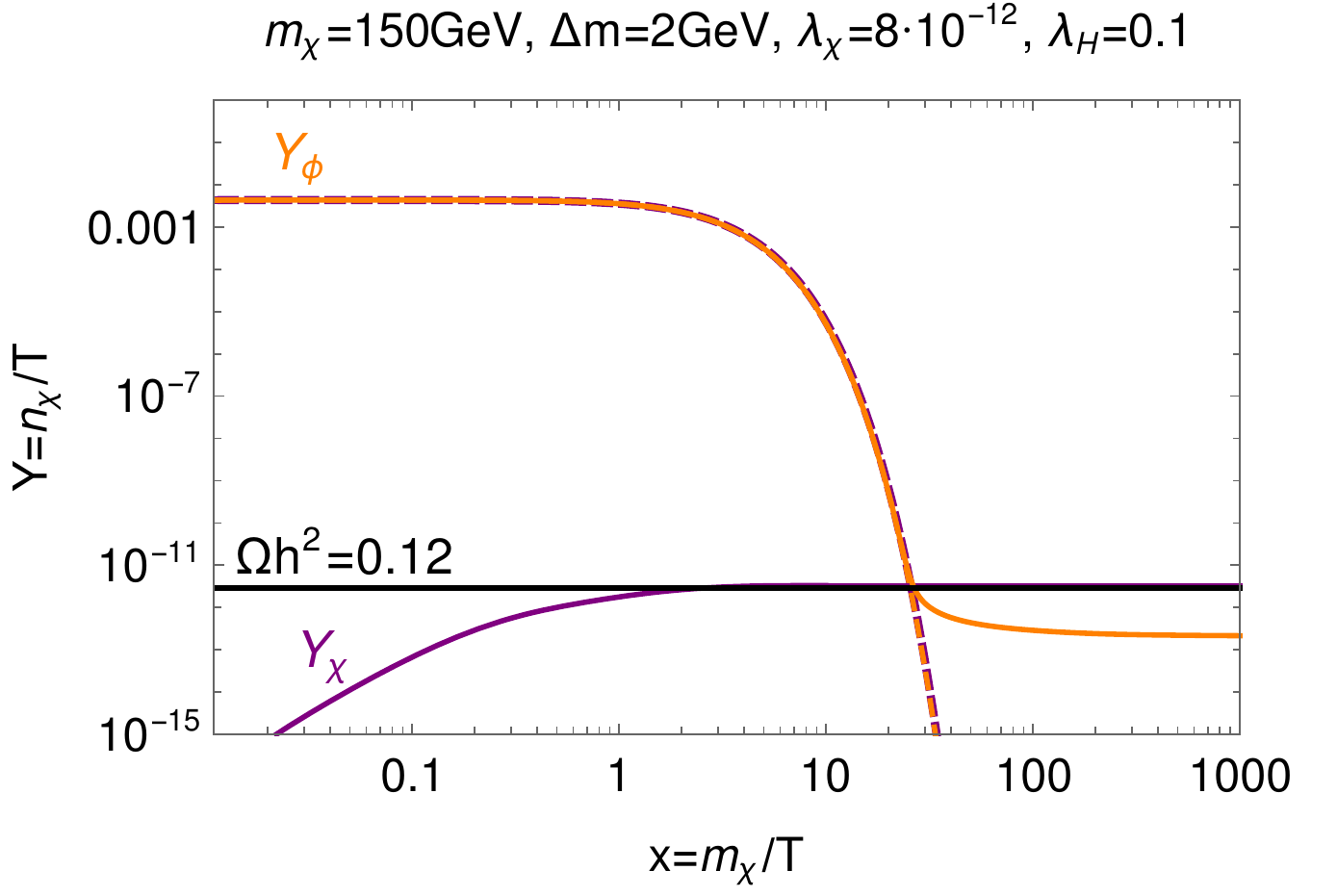}}}
  \caption{Benchmark points of Fig.~\ref{fig:ann-conv} giving rise to $\Omega h^{2}=0.12$. Left: Ratio of the rates of interactions depicted in the legend and of the Hubble rate as a function of $x=\frac{m_{\chi}}{T}$. The conversion processes are depicted in red, the (co-)annihilation ones in blue. The black line depicts $\Gamma/H=1$. Right: Evolution of the yield of $\chi$ (solid purple line) and $\phi$ (solid orange line) and their equilibrium yield (dashed lines) as a function of $x$. }
    \label{fig:EvolutionComp}
\end{figure}

We focus now on the new region of the parameter space with feeble couplings giving rise to all the \ac{DM} through the conversion driven freeze-out. This \ac{DM} production regime opens up for compressed mediator-DM mass spectrum, and suppressed conversion couplings such that the \ac{DM} is out of \ac{CE}, as described in Sec.~\ref{sec:CDFO}. In Fig.~\ref{fig:Yield3}, the evolution of the \ac{DM} and mediator yield is shown for a representative benchmark and the corresponding rates are shown in Fig.~\ref{fig:Rateplot3}. From the latter, we can see that the conversion processes are indeed only barely efficient, giving rise to \ac{CDFO}.

\begin{figure}[!t]
  \centering
  \subfloat[$\lambda_H = 0.01$]{\label{fig:ParamPlot1e-2}
    \includegraphics[width=0.4\textwidth]{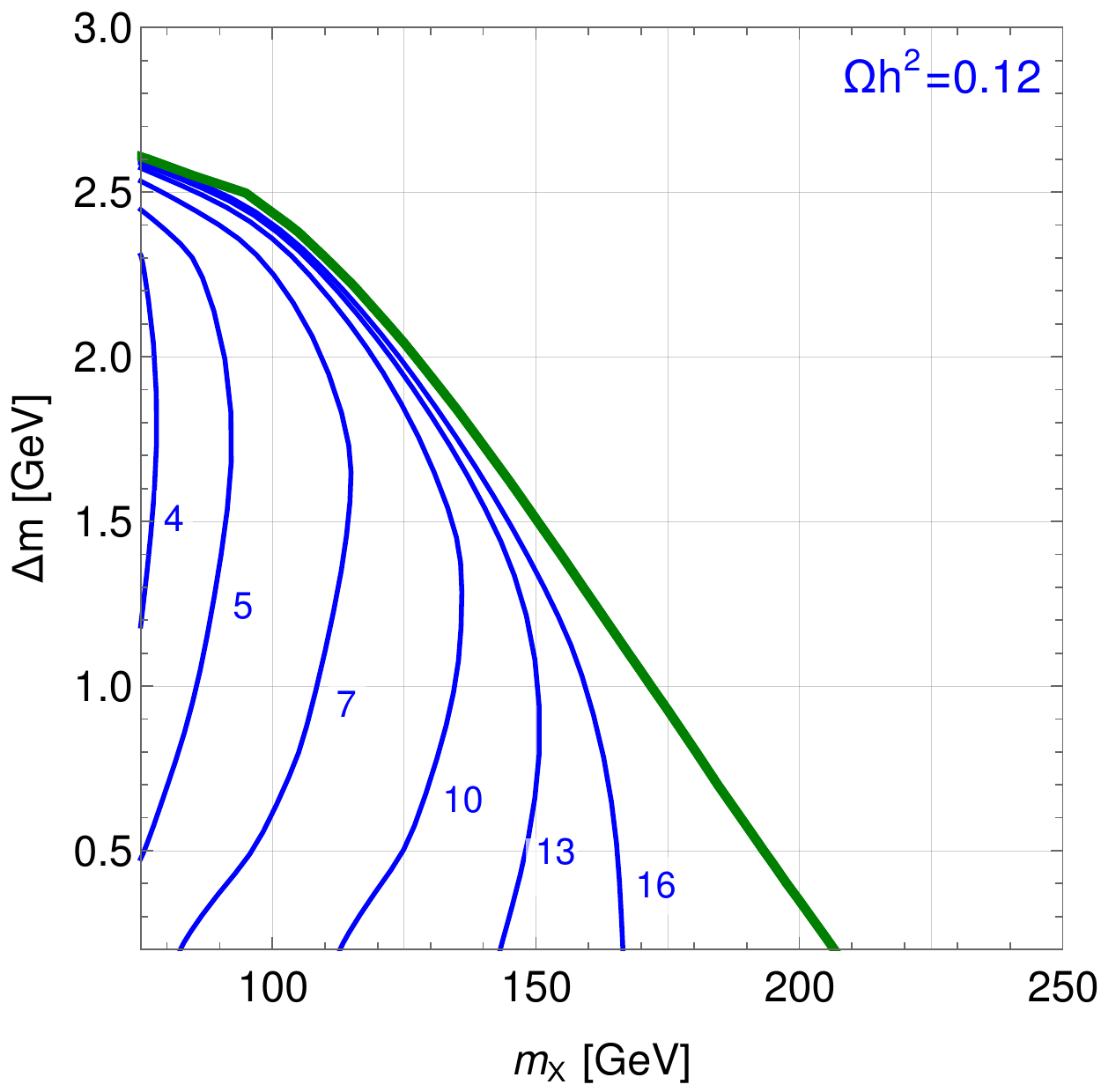}}
	\hspace{0.05\textwidth}
    \subfloat[$\lambda_H = 0.1$]{\label{fig:ParamPlot1e-1}
    \includegraphics[width=0.4\textwidth]{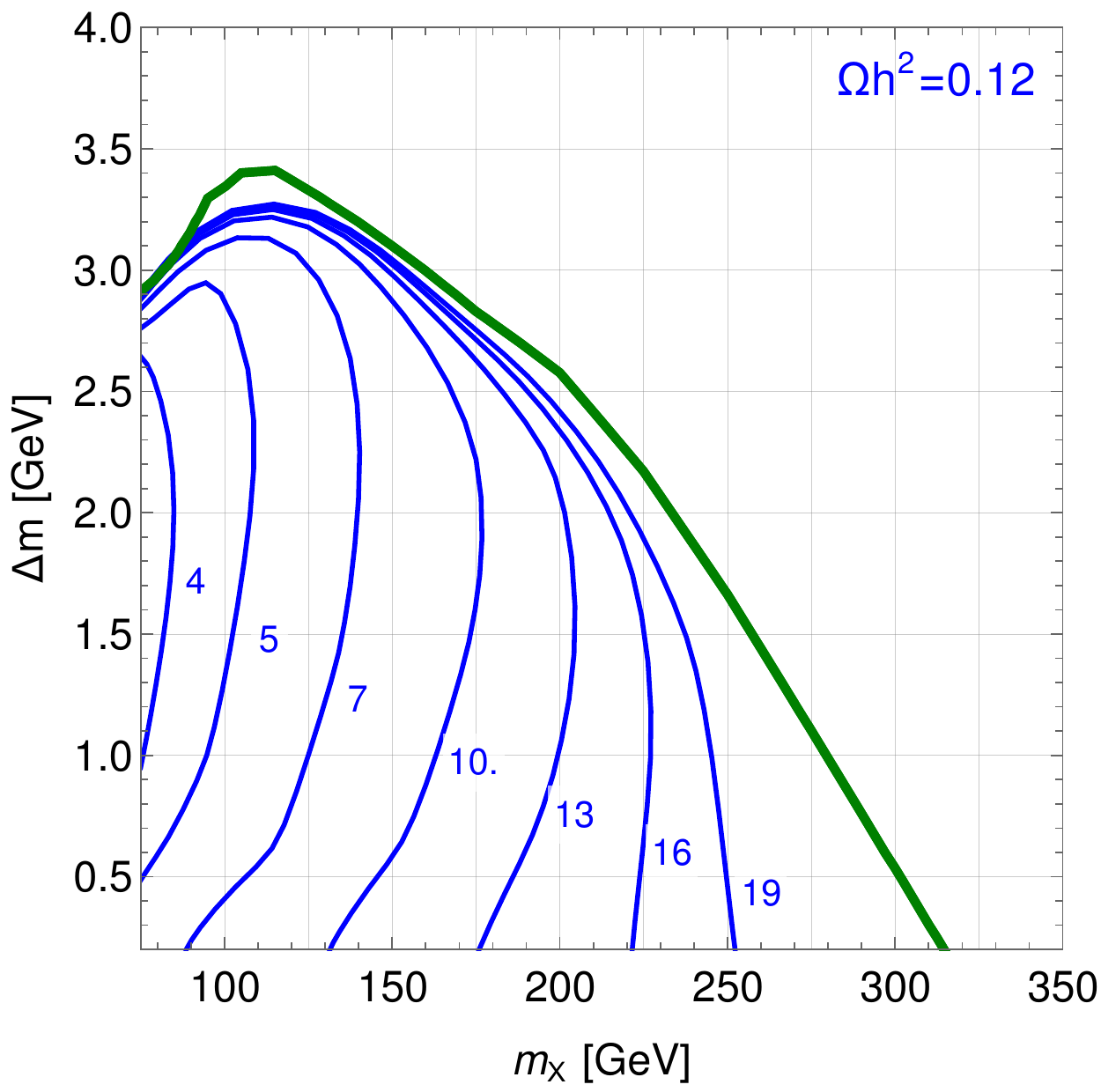}}
  \\
	\vspace{0.005\textheight}
    \subfloat[$\lambda_H = 0.5$]{\label{fig:ParamPlot5e-1}\includegraphics[width=0.4\textwidth]{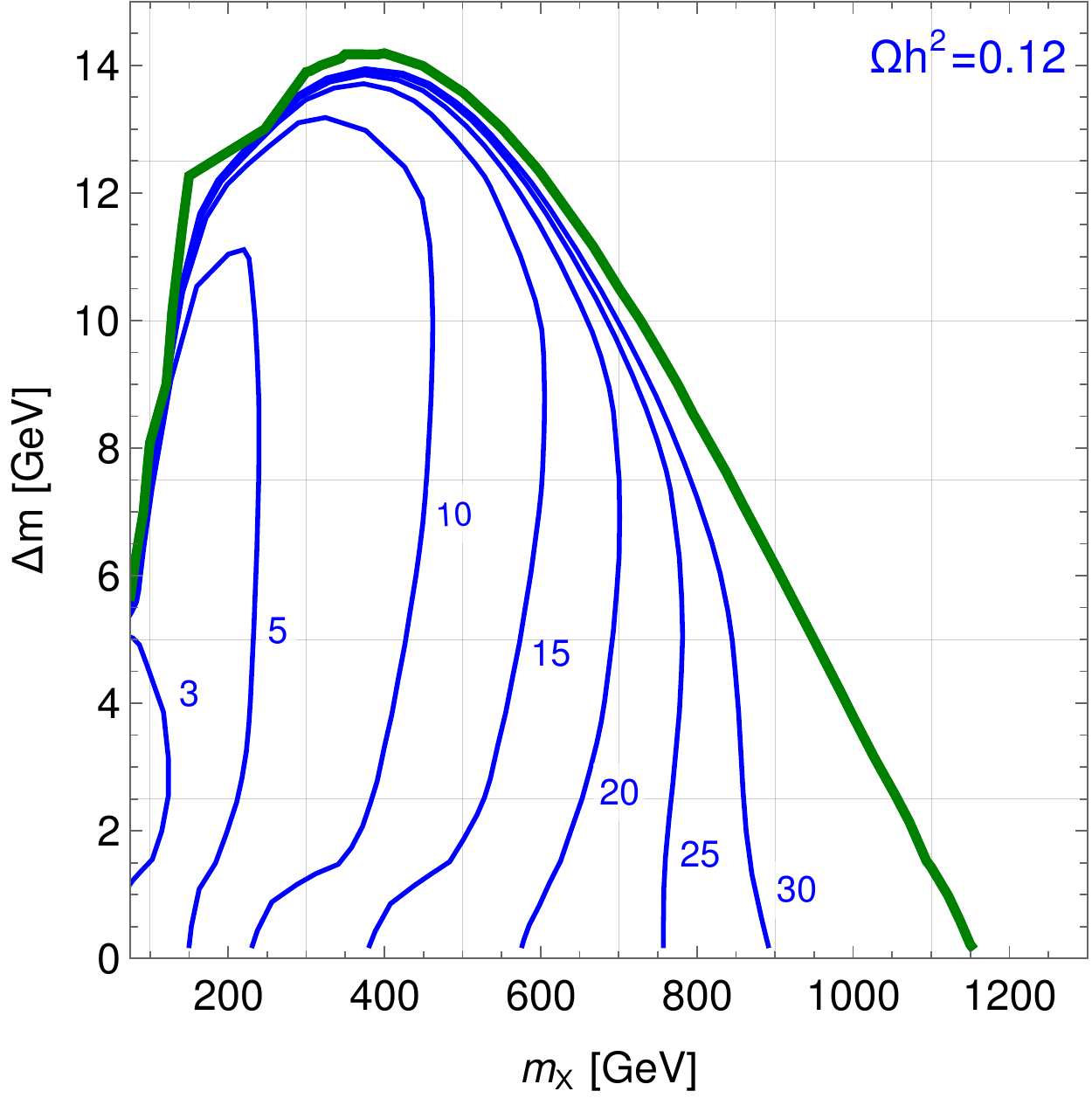}}
    \caption{Viable parameter space for DM abundance through conversion driven freeze-out for several value of the $H-\phi$ coupling $\lambda_H$. Contours denoting $\Omega h^2 = 0.12$ for fixed value of the Yukawa coupling $\lambda_\chi/10^{-7}$ are shown with blue lines. The border of the parameter space is delimited by a green contour corresponding the combinations of $\Delta m, m_\chi$ giving rise to the right dark matter abundance through mediator annihilation driven freeze-out. }
  \label{fig:ParamPlot}
\end{figure}

In the 3 plots of Fig.~\ref{fig:ParamPlot}, corresponding to 3 values of the Higgs-mediator coupling $\lambda_H$, the viable parameter space for \ac{CDFO} is enclosed by the green line in the plane $(m_\chi,\Delta m)$. We readily see that larger values of $\lambda_H$ give rise to a larger viable parameter space.  Let us first comment on the area above the green contour. In the latter region, the standard freeze-out mechanism (with \ac{DM} in \ac{CE}) can give rise to all the \ac{DM} for a specific choice of the coupling $\lambda_{\chi}$. For large $\Delta m$ or $m_\chi$ (above the green line), the freeze-out is {\it \ac{DM} annihilation driven}, i.e. the $\sigmav_{eff}$ of Eq.~(\ref{eq:sv_eff}) is directly given by the dark matter annihilation cross-section $\sigmav_{\chi\chi}$. Approaching the green line at fixed value of $m_\chi$ implies smaller mass splitting $\Delta m$ and thus a larger contribution of co-annihilation processes to $\sigmav_{eff}$. The green line itself delimiting the viable parameter space for \ac{CDFO} is obtained by requiring that {\it mediator annihilation driven} freeze-out gives rise to all the \ac{DM} (\ac{DM} still in \ac{CE}). In the latter case, $\sigmav_{eff}$ is directly proportional to the mediator annihilation cross section $\sigmav_{\phi\phi^\dag}$ and to the Boltzmann suppression factor $ \exp(-2 x_f\Delta m/m_\chi)$ with $x_f \approx 25$. Let us emphasize that the green border, in the 2-dimensional $(m_\chi,\Delta m)$ plane, can be realized for a wide range of suppressed conversion couplings. In for instance the specific benchmark case of Fig.~\ref{fig:ann-conv}, the mediator driven freeze-out region extends from $\lambda_{\chi} \sim 10^{-1}$ to $\lambda_{\chi} \sim 10^{-6}$ giving rise to a fixed value of $\Omega h^2$. Notice that the entire region of the parameter space above the green line has already been analyzed in details in previous studies, see e.g. Ref.~\cite{Garny:2015wea}, and we do not further comment on this region here.

Below the green line, the standard freeze-out computation (assuming \ac{DM} in \ac{CE}) would predict an underabundant \ac{DM} population. The conversion coupling gets however so suppressed that the \ac{DM} can not anymore be considered in \ac{CE} and the standard computation involving Eq.~(\ref{eq:sv_eff}) breaks down. In contrast, the treatment of the Boltzmann equations described in Sec.~\ref{sec:CDFO} can properly follow the \ac{DM} yield evolution in such a region. As a result, the blue contours in Fig.~\ref{fig:ann-conv} can give rise to $\Omega h^2=0.12$ through {\it conversion driven} freeze-out (\ac{DM} out of \ac{CE}) for fixed value of $\lambda_\chi \in$ few $\times [10^{-7},  10^{-6}]$.

We also notice that the maximum value of the allowed mass splitting and of the dark matter mass in the conversion driven region increases with the Higgs portal coupling, going from $\Delta m_{max}\simeq 2.6 $ GeV and $ m_\chi^{max}=180$ GeV for the minimal value of $\lambda_H= 0.01$ to $\Delta m_{max}\simeq 14 $ GeV and $ m_\chi^{max}=1$ TeV for e.g.~$\lambda_H= 0.5$. Increasing $\lambda_H$ effectively increases the resulting $\sigmav_{eff}$ and, as a consequence, decreases the DM relic abundance through mediator annihilation driven freeze-out which is relevant for the extraction of the green contour. In order to understand this behavior let us comment on the overall shape of the viable parameter space delimited by the green line.  This line corresponds to the parameter space giving $\Omega h^2=0.12$ considering mediator annihilation driven freeze-out, i.e. when $\sigmav_{eff}\propto \sigmav_{\phi\phi^\dag} \exp(-2 x_f\Delta m/m_\chi)$ where $\sigmav_{\phi\phi^\dag}$ mainly depends on $\alpha_{\rm EM}, \lambda_H$, but not on $\lambda_\chi$. The general dependence on the masses leading to the shape of the contour (hill shape) results from the competition between the two factors $\sigmav_{\phi\phi^\dag}$ and $\exp(-2x_f\Delta m/m_\chi)$. 

Let us first focus on the large mass range, where the green line displays a negative slope in the $(m_{\chi}, \Delta m)$ plane. At fixed values of $m_\chi$, decreasing $\Delta m$, one increases $\sigmav_{eff}$ due its to the exponential dependence on $\exp(-2x\Delta m/m_\chi)$. One way to compensate for this effect and anyway obtain the same effective cross section (and hence correct relic abundance) is to consider larger values of the mediator mass (or equivalently the \ac{DM} mass) as one can expect $\sigmav_{\phi\phi^\dag} \propto m_\phi^{-2}$ for large enough $m_\phi$ which dominates over the mass dependence in the exponential factor. As a result, for fixed $\Omega h^2$, lower $\Delta m$ implies larger $m_\phi$. This is indeed the shape of the green line that we recover for large $m_\phi$ in the plots of Fig.~\ref{fig:ParamPlot}. 

At some point, the mass splitting $\Delta m$ becomes smaller than the mass of the SM lepton involved in the Yukawa interaction in Eq.~\eqref{eq:lagr_lepto} and we get to the maximum allowed value of $m_\phi$ (focusing on 2 body decay $\phi \to \chi l$ only\footnote{For an analysis involving 3 body decays, see~\cite{Khoze:2017ixx}.}). One simple way to enlarge the parameter space allowing for larger $m_\chi$ at fixed $\Delta m$ is to increase the annihilation rate of the mediator. Indeed, since the effective cross section dominantly scales as $\sigmav_{eff} \propto m_\phi^{-2}$, this allows for larger values of $m_\phi$, or equivalent $m_\chi$, to account for the right DM abundance. Increasing the overall effective cross section can be done by increasing $\lambda_H$. Going back to the benchmark of Fig.~\ref{fig:ann-conv}, this would imply the horizontal part of the $\Omega h^2$ curve goes to lower $\Omega h^2$ value when increasing $\lambda_H$.

Finally, we comment on the shape of the green line in the small mass region, where it has a positive slope in the $(m_{\chi}, \Delta m)$ plane. This time, it is mainly the exponential factor $\exp(-2 x_f\Delta m/m_\chi)$ dominating the behavior of $\langle \sigma v\rangle_{eff}$ and hence determining the shape of the green contour. Larger $\Delta m$ thus requires larger $m_{\chi}$ in order to keep the dark matter abundance at the correct value. Note that this region is present also in the small $\lambda_{H}$ case, but it is realized for dark matter masses lower than the ones showed in the plots (and not phenomenologically interesting, see Sec.~\ref{sec:leptophilic@lhc}).

\subsection{Freeze-in production through mediator decay and scattering}
\label{sec:fi_leptophilic}
One can also account for all the DM through the \ac{FI} mechanism for the lowest values of the conversion coupling in Fig.~\ref{fig:ann-conv}, that is $\lambda_\chi= 8\cdot 10^{-12}$. The associated rates and abundance evolution are shown in Fig.~\ref{fig:Rateplot1} and~\ref{fig:Yield1}. In Fig.~\ref{fig:Yield1} we mainly recover the standard picture of the dark matter \ac{FI} which relic abundance is due to scatterings ($\phi SM \to \chi SM$) and decays ($\phi \to \chi SM$) of a mediator in thermodynamic equilibrium with the \ac{SM} bath (see Tab.~\ref{tab:conv} for all processes that can convert $\phi$ into $\chi$ and hence contribute to the freeze-in abundance). The relic dark matter abundance freezes-in around the time at which the rate of mediator decay/scatterings gets strongly suppressed ($x\sim 3$).  Notice that after the mediator freezes-out, its relic population eventually decay to \ac{DM} and leptons at a time characterized by its life-time $\tau_\phi$. This would correspond to the ``superWIMP'' contribution to the relic abundance which is around one order of magnitude lower than the freeze-in one for this specific benchmark. For a detailed study on the freeze-in and superWIMP interplay see e.g.~\cite{Garny:2018ali,Decant:2021mhj}.

As mentioned in Sec.~\ref{sec:fi}, in the simplest \ac{FI} models, the largest contribution to the abundance comes from a mediator decaying to \ac{DM}, and we can estimate the DM relic abundance with Eq.~\eqref{eq:Boltz_fi}. In the case considered here, this equation underestimates the relic dark matter abundance in the \ac{FI} regime. One reason for this is the small mass splitting between the mediator and the DM implying that the decay of the mediator is kinematically suppressed. The contribution to the DM abundance from scattering processes hence becomes more relevant.

\begin{figure}
  \begin{center}
  \includegraphics[width=0.55\textwidth]{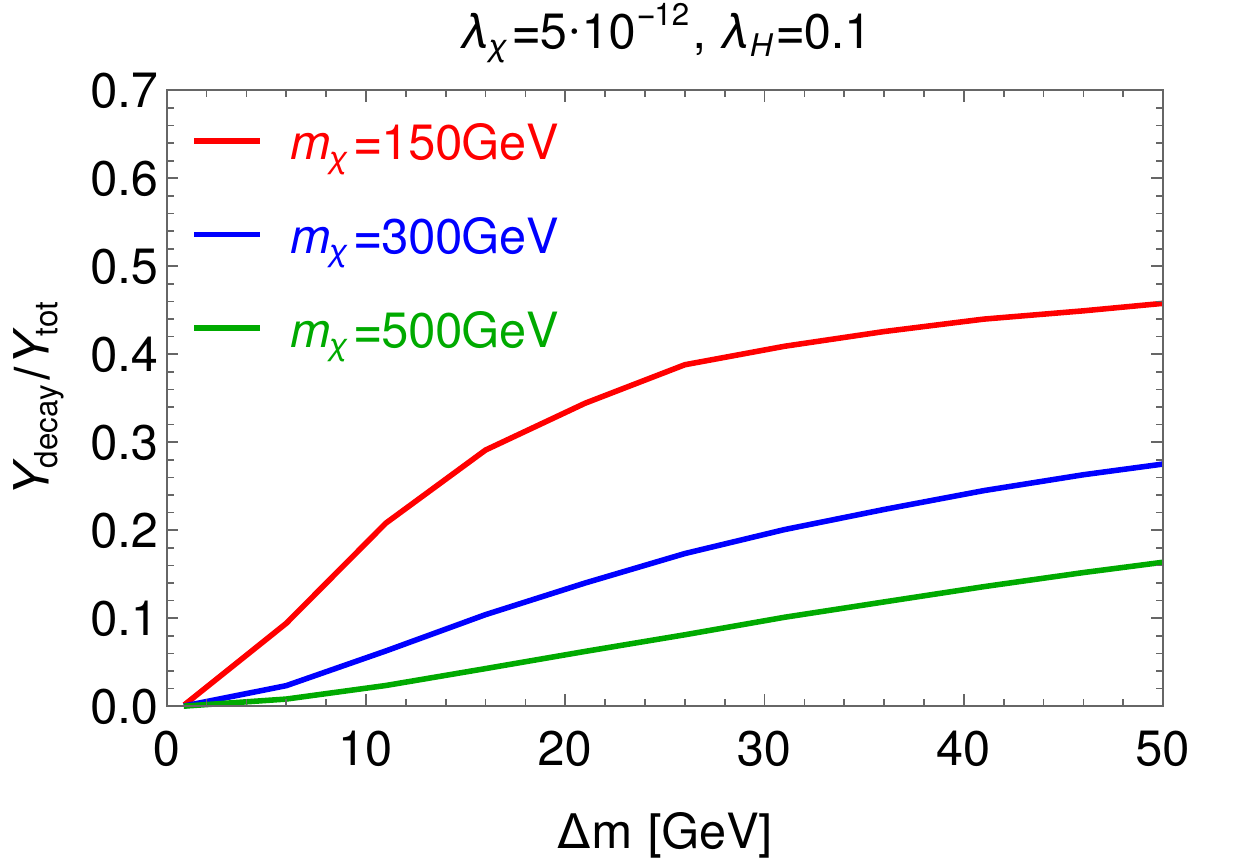}  
  \end{center}
  \caption{The ratio of the freeze-in yields $Y^{FI}_{\rm decay}/Y^{FI}_{\rm tot}$ as a function of $\Delta m$ when taking only decay processes into account for  $Y^{FI}_{\rm decay}$ and including all the relevant scattering and decay processes for $Y_{\rm tot}^{FI}$. We show the results for three different values of the DM mass $m_\chi \in \{150, 300, 500 \}$ GeV while keeping $\lambda_{\chi} = 5 \cdot 10^{-12}$ and $\lambda_H=0.1$ fixed.}
  \label{fig:FI_ratio}
\end{figure}

The relative importance of the decay contribution to the \ac{FI} is displayed as a function of the mass splitting in Fig.~\ref{fig:FI_ratio}. The red curve corresponds to an illustrative benchmark with $m_\chi= 150$ GeV, $\lambda_\chi= 5 \cdot 10^{-12}$ and $\lambda_H=0.1$.  We show the ratio between the DM abundances obtained through freeze-in $Y^{FI}_{\rm decay}/Y^{FI}_{\rm tot}$ considering the decay processes only for $Y^{FI}_{\rm decay}$ (setting the scattering processes to zero by hand) and including all the relevant scattering and decay processes for $Y_{\rm tot}^{FI}$. For small mass splittings, the mediator decay contribution is subleading.  For increasing mass splitting, instead, the decay process becomes the main player.  Let us emphasize though that even for the largest values of the mass splitting considered in Fig.~\ref{fig:FI_ratio}, the contribution from scattering is non-negligible.  This is due to the fact that, in the model under study, a large number of scattering processes can contribute to the DM production. There are indeed $\sim $ 10 possible scattering processes with a rate of DM production $\propto \lambda_\chi^2\alpha_{\rm EM}$ to be compared to one single decay process with a rate $\propto \lambda_\chi^2$, where $\alpha_{\rm EM}$ is the electromagnetic fine structure constant (Tab.~\ref{tab:conv} lists all the relevant processes). The multiplicity factor of the scattering process partially compensates for the extra SM gauge coupling suppression $\alpha_{\rm EM}$. As a result, scattering contribution to the dark matter relic abundance through freeze-in can still be $\sim {\cal O} (1)$ compared to the decays for sizable mass splitting. The relative contribution of the decay is also a function of the mediator mass. The larger the mediator mass, the smaller is the decay contribution at fixed value of the couplings and of the mass spitting (see the blue and green curves in Fig.~\ref{fig:FI_ratio}). This is because the decay rate scales like $\Gamma_{\phi} \sim m_{\phi}^{-1}$ when $\Delta m \ll m_\phi$. We conclude that both scatterings and decay contribution must be taken into account here to provide a correct estimate of the relic dark matter abundance through freeze-in.

\chapter{Freeze-in production in alternative cosmologies}
\label{chap:alt_cosmo}

Producing \ac{DM} can be done in a large amount of different ways, depending on the underlying models and parameter region of interest. In Chapter~\ref{chap:DM_prod}, we discussed in details a variety of production mechanisms together with the underlying assumptions and caveats. However, we never changed one the major underlying assumptions, which is that \ac{DM} production happens in the standard model of cosmology, where for $T\gtrsim$~eV, the universe is dominated by radiation such that $T\sim a^{-1}$, where $a$ is the scale factor. This era of radiation domination started at the end of inflation, where the universe got reheated and the radiation bath got created. In Chapter~\ref{chap:DM_prod}, we always assumed that the reheating temperature $T_R$, denoting the end of reheating and the start of the radiation dominated era, was much larger than the energy scale of interest for the corresponding production mechanism such that the main production dynamics always happened in a radiation dominated universe.

The standard model of cosmology, often referred to as $\Lambda$CDM, is able to explain a fair amount of phenomena such as the \ac{CMB} and \ac{BBN}. However, the latter is (for now) the earliest probe available to us, and hence, we have no way of testing the $\Lambda$CDM model at earlier times, i.e. at $T\gtrsim1$~MeV. We have therefore no clear indication of the exact value of the reheating temperature, or that the universe is \ac{RD} for temperature between $T_{R}$ and $T_{BBN}$. In this chapter, we will study the effects of relaxing the assumption that \ac{DM} has been produced in a \ac{RD} universe and explore what happens if we allow $T_R$ to be smaller than the relevant energy scales for \ac{DM} production. As we are mainly interested in FIMPs, the main focus will lie on the freeze-in production mechanism. 

\section{Alternative cosmological histories}
\label{sec:alt_cosmo}
The expansion of the universe is governed by the Hubble rate, which in general is given by the Friedmann equation,
\begin{align}
\label{eq:friedman}
    H = \frac{\Dot{a}(t)}{a(t)} = \frac{\sqrt{\rho_{\text{tot}}}}{\sqrt{3}M_{Pl}}, 
\end{align}
where $\rho_{\text{tot}}$ is the total energy density of the universe. In the \ac{RD} era, $\rho_{\text{tot}}$ is of course dominated by the radiation energy density $\rho_R$ such that $H\sim T^2$, see Eq.~\eqref{eq:H_rad}. This relation can be altered when a new species starts dominating the total energy density. We will denote this new species by $\Phi$, with equation of state $\omega=\mathrm{p}_\Phi/\rho_\Phi$, where $\rho_\Phi$ and $\mathrm{p}_\Phi$ are the energy and pressure density of $\Phi$. This new species must eventually decay away to radiation, since we know that at the time of \ac{BBN}, the universe was dominated by radiation\cite{deSalas:2015glj}. The evolution of the energy densities are governed by the following set of Boltzmann equations \cite{Chung:1998rq},
\begin{align}
\label{eq:BE_rhoPhi}
    \frac{d\rho_\Phi}{dt} + 3 (1+\omega)H\rho_\Phi &= - \Gamma_\Phi \rho_\Phi, \\
\label{eq:BE_rhoR}
    \frac{d\rho_R}{dt} + 4H\rho_R &= \Gamma_\Phi \rho_\Phi, 
\end{align}
where $\Gamma_\Phi$ is the decay rate of $\Phi$ into radiation. The solution to these Boltzmann equations depend on the initial densities $\rho_{R,i}$ and $\rho_{\Phi,i}$. At times earlier than $\tau_\Phi = 1/\Gamma_\Phi$, the effects of decays on $\rho_\Phi$ can be neglected, giving
\begin{align}
    \rho_\Phi &= \rho_{\Phi,i} \left( \frac{a_i}{a} \right)^{3(1+\omega)}, \label{eq:rho_phi} \\
    \rho_R &= \rho_{R,i} \left( \frac{a_i}{a} \right)^{4} + \frac{2}{5-3\omega}\rho_{\Phi,i}\Gamma_\Phi \left[ \frac{1}{H(a)} \left( \frac{a_i}{a} \right)^{3(1+\omega)} -  \frac{1}{H(a_{i})} \left( \frac{a_i}{a} \right)^4 \right],
\end{align}
where $a_i$ is the scale factor at some arbitrary initial point. In an era when the species $\Phi$ starts dominating the total energy budget of the universe, we can write $H(a) \sim \sqrt{\rho_\Phi} \sim a^{-3(1+\omega)/2}$. In this era, the radiation density can be rewritten as
\begin{align}
    \rho_R = \rho_{R,i} \left( \frac{a_i}{a} \right)^{4} + \frac{2\sqrt{3}}{5-3\omega}M_{Pl}\sqrt{\rho_{\Phi,i}}\Gamma_\Phi \left[  \left( \frac{a_i}{a} \right)^{\frac{3}{2}(1+\omega)} - \left( \frac{a_i}{a} \right)^4 \right].
    \label{eq:rho_r}
\end{align}

\subsection{Early matter domination during inflationary reheating}
\label{sec:emd_universe}

A motivated example for an alternative cosmology comes from inflation. It is believed that after inflation, the radiation bath was created from the decay of coherent oscillations of a scalar field, the inflaton, whose energy density dominated the universe before it started decaying. 
This scenario is captured by Eqs.~\eqref{eq:rho_phi} and \eqref{eq:rho_r} by assuming an equation of state $\omega=0$ (corresponding to an \ac{EMD} universe) and $\rho_{R,i}=0$. As initial condition for the inflaton density, we take $\rho_{\Phi,i}=E_I^4$, where $E_I$ is the scale of inflation. During the matter dominated era driven by the inflaton, the inflaton and radiation energy densities are given by~\cite{Co:2015pka}
\begin{align}
    \rho_\Phi &= \rho_{\Phi,i} \left( \frac{a_i}{a} \right)^{3}, \label{eq:rho_phi_MD} \\
    \rho_R &= \frac{2\sqrt{3}}{5}M_{Pl}\sqrt{\rho_{\Phi,i}}\Gamma_\Phi \left[  \left( \frac{a_i}{a} \right)^{\frac{3}{2}} - \left( \frac{a_i}{a} \right)^4 \right].
    \label{eq:rho_r_MD}
\end{align}

\begin{figure}
    \centering
    \includegraphics[width=0.6\textwidth]{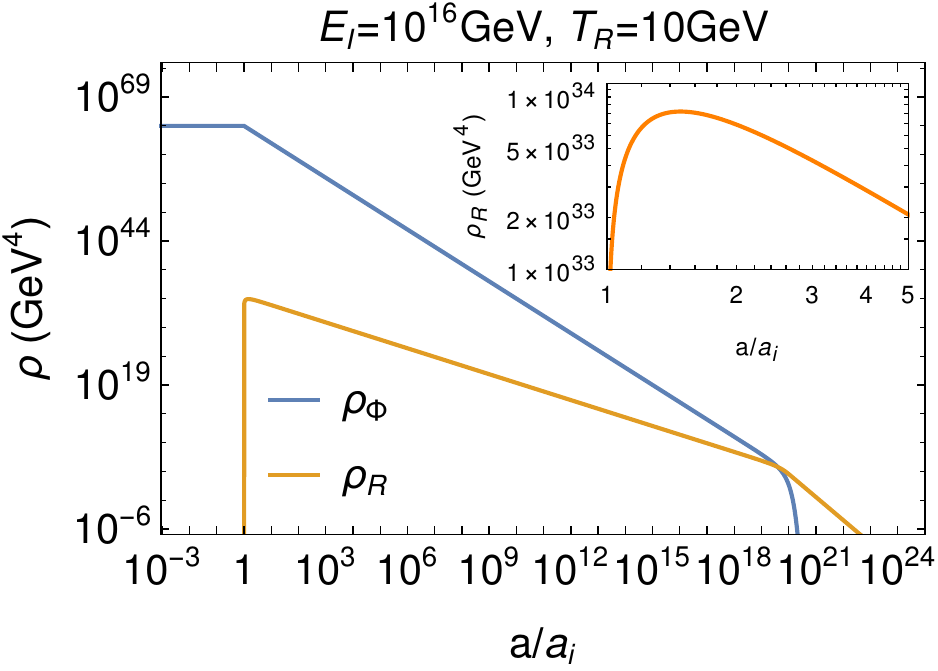}
    \caption{Inflaton (blue line) and radiation bath (orange line) energy densities as a function of the scale factor for inflationary reheating. In the top-right part of the plot we zoom in at early times and we show how the radiation bath energy density arises from inflaton decays.}
    \label{fig:reheating}
\end{figure}

\begin{figure}
\centering
\subfloat[]{\label{fig:reheating_T}
    \includegraphics[width=0.45\textwidth]{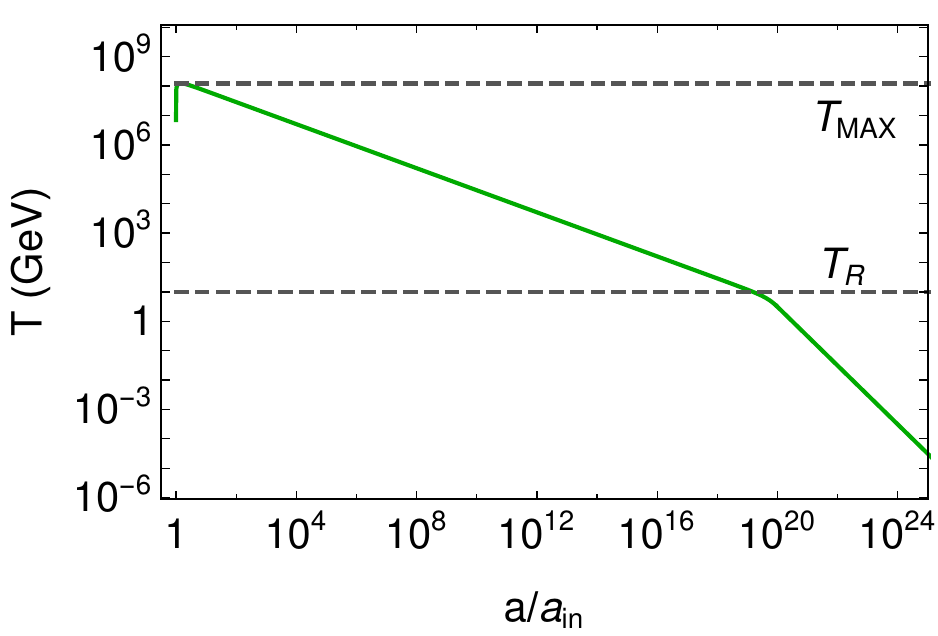}} 
	\hspace{0.05\textwidth}
\subfloat[]{\label{fig:reheating_s}
    \includegraphics[width=0.45\textwidth]{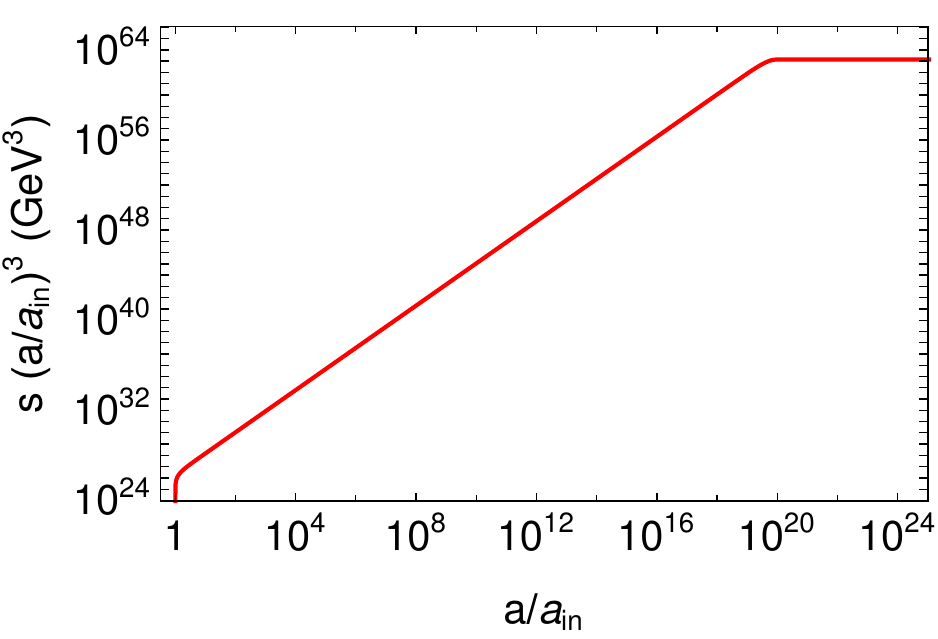} }
\caption{Radiation bath temperature (left) and the entropy in a comoving volume (right) as a function of the scale factor. We set $E_I$ and $T_R$ to the same values chosen in Fig.~\ref{fig:reheating}.}
\label{fig:EarlyMD2}
\end{figure}

In Fig.~\ref{fig:reheating}, we show the numerical solution of Eqs.~\eqref{eq:BE_rhoPhi} and~\eqref{eq:BE_rhoR}, which are well approximated by Eqs.~\eqref{eq:rho_phi_MD} and~\eqref{eq:rho_r_MD} during the EMD phase. Initially, for $a<a_i$, there is no radiation bath and the inflaton dominates the energy budget with a constant density $ E_I^4$. At some point, which we take corresponding to a value of the scale factor $a_{i}$, inflation ends and the inflaton begins damped oscillations around its minimum and starts decaying, populating a newly formed radiation bath. In Fig.~\ref{fig:reheating}, we see that the initial growth is very sudden. Indeed as can be seen from Eq.~\eqref{eq:BE_rhoR}, $\rho_R\approx0$ at early times so that the derivative $d\rho_R/dt \approx \Gamma_\Phi \rho_\Phi$ is very large, as $\rho_\Phi\approx E_I \gg \Gamma_\Phi$,\footnote{In the case where $E_I\approx\Gamma_\Phi$, reheating is roughly instantaneous.} explaining the steep increase of $\rho_R$. The radiation energy density and its temperature reach a maximum for a value of the scale factor $a_\textsc{max} = ( 8/3)^{2/5} \, a_{\rm in}$. Neglecting the $g_{eff}$ dependence on the temperature, the associated maximum temperature results in
\begin{equation}
T_\textsc{max} \simeq \frac{\left(M_{ Pl} \Gamma_\Phi \right)^{1/4} \rho_{\Phi,i}^{1/8}}{g_{eff}^{1/4}} \simeq \frac{\left(M_{Pl} \Gamma_\Phi E_I^2 \right)^{1/4}}{g_{eff}^{1/4}}  \ .
\end{equation}
Afterwards, the bath temperature decreases as follows
\begin{equation}
T(a) = T_\textsc{max} \left(\frac{a_\textsc{max}}{a}\right)^{3/8},
\label{eq:TaMD}
\end{equation}
since the decrease of the radiation energy density is controlled by the first term in the square brackets of Eq.~\eqref{eq:rho_r_MD}, i.e. $\rho_R \sim a^{-3/2}$ as opposed to the adiabatic expansion for a \ac{RD} universe where the energy density decrease is faster, $\rho_R \sim a^{-4}$. We show the evolution of the temperature as a function of $a$ in Fig.~\ref{fig:reheating_T}.

The approximation that leads us to Eq.~\eqref{eq:rho_r_MD} breaks down when most inflatons decay and the energy budget starts being dominated by the radiation bath: we recover the \ac{RD} universe that serves as a background for \ac{BBN}. The transition to a \ac{RD} universe happens when the temperature of the thermal bath has a value $T_R$ satisfying the condition 
\begin{equation}
H(T_R) \simeq \Gamma_\Phi \simeq \frac{\rho_R(T_R)^{1/2}}{\sqrt{3} M_{\rm Pl}} \ .
\end{equation}
The so called reheating temperature reads
\begin{equation}
T_ R = \left(\frac{90}{\pi^2 g_*}\right)^{1/4}\sqrt{\Gamma_\Phi M_{\rm Pl}} \ .
\label{eq:TRH}
\end{equation}
The radiation temperature spans a potentially large range from $T_\textsc{max}$ down to $T_R$, and they are connected via the relation $T_\textsc{max} \simeq \left(T_R E_I \right)^{1/2}$.

It is also instructive to investigate the evolution of one other important bath property during reheating, the entropy in a comoving volume $S(T)= s(T) (a(T)/a_{\rm in})^3$, which we show in Fig.~\ref{fig:reheating_s}. We compute this quantity for $T_\textsc{max}<T<T_R$ with respect to its value at the maximum temperature
\begin{equation}
\frac{S(T)}{S(T_\textsc{max})} = \frac{s(T) a^3}{s(T_\textsc{max}) a_\textsc{max}^3} =  \left(\frac{T \, a}{T_\textsc{max} \, a_\textsc{max}} \right)^3 = \left(\frac{T_\textsc{max}}{T} \right)^5 \ .
\label{eq:SvsT}
\end{equation} 
This entropy dump provides a dilution to the \ac{DM} yield produced during this era as we will discuss further in Sec.~\ref{sec:fi_emd}. In the \ac{RD} era, the entropy per comoving volume is conserved as expected. Finally, it is useful to find an approximate expression for the Hubble parameter during reheating
\begin{equation}
H(T) = \frac{E_I^2}{\sqrt{3} M_{\rm Pl}} \left( \frac{a_{\rm in}}{a}\right)^{3/2} \simeq \frac{E_I^2}{M_{\rm Pl}} \left( \frac{T}{T_\textsc{max}}\right)^{4} \simeq 
\frac{1}{M_{\rm Pl}} \frac{T^4}{T_R^2} \ .
\label{eq:HearlyMD}
\end{equation}
From this expression, we see that during reheating, the Hubble rate is larger than if the universe would be in a RD era at the same temperature. Hence, the universe expands faster and interaction rates are rendered inefficient at higher temperatures.

\subsection{Particle production during a non-radiation dominated universe}
\label{sec:part_prod_emd}
The optimal numerical technique to solve the Boltzmann equation is different if the particle production happens during an early non-radiation dominated era. The starting point is still Eq.~\eqref{eq:Boltz_unint}, but using the comoving number density or yield is rather inconvenient as the comoving entropy is not conserved. Indeed, after \ac{DM} production ends, which happens at $T=T_{\rm prod}$, the number density dilutes with the scale factor as $n_\chi \propto a^{-3}$. At the same time, $\Phi$ decays dump entropy in the primordial bath and the entropy in a comoving volume $s a^3$ grows with time. As a consequence, the comoving number density $Y_\chi$ decreases with time when the bath temperature spans between $T_{\rm prod}$ and $T_R$. 

This suggests a new variable accounting for the number density of \ac{DM} particles
\begin{equation}
\mathcal{X} \equiv n_\chi a^3 \ .
\end{equation}
Once \ac{DM} production has run its course, the quantity $\mathcal{X}$ remains constant until the present time. Integrating the Liouville operator, see Eq.~\eqref{eq:int_liouville}, and replacing $n_\chi$ by $\mathcal{X}$, the Boltzmann equation tracking the DM number density becomes
\begin{equation}
\frac{d \mathcal{X}}{d t} = a^3 \sum_\alpha \mathcal{C}_\alpha \ ,
\label{eq:BEMD}
\end{equation}
where $\mathcal{C}_\alpha$ is the integrated collision term of a process $\alpha$. The cosmic time $t$ is not an ideal evolution variable, the scale factor is quite convenient instead. We trade time derivative with derivative with respect to $a$ by using the Friedmann equation in Eq.~\eqref{eq:friedman}, and we find
\begin{equation}
\frac{d \mathcal{X}}{d \ln a} = \frac{\sqrt{3} M_{\rm Pl} \, a^3}{\sqrt{\rho_M(a) + \rho_R(a)}} \sum_\alpha \mathcal{C}_\alpha(a) \ .
\end{equation}

At late enough times, once the thermal bath temperature drops below $T_R$ and the universe is dominated by radiation, we can trust entropy conservation again. In order to compute the FIMP contribution to the current energy density, we identify a late temperature $T_* < T_R$ and we evaluate $n_X(T_*)$ and the entropy density $s(T_*)$. The corresponding comoving number density:
\begin{equation}
Y_X(T_*) = \frac{n_X(T_*)}{s(T_*)} =   \frac{\mathcal{X}(T_*)}{S(T_*)} = \frac{\mathcal{X}^\infty}{S(T_*)}   \quad [T_*< T_R]
\end{equation}
which is constant in the subsequent evolution of the universe,
i.e.~$Y_X^\infty= Y_X(T_*)$. The energy density today thus 
reads $\rho_X(t_0) = m_X  Y_X^\infty\, s_0$ where $s_0$ is the current entropy
density.

\section{Freeze-in during an early matter dominated era}
\label{sec:fi_emd}
We revisit freeze-in via $B$ decays when \ac{FIMP} production peaks during an early inflationary \ac{MD} era, i.e.~when $m_B >T_R$.\footnote{See e.g.~\cite{Giudice:2000ex,Arias:2019uol,Bernal:2022wck} for freeze-out during an early inflationary period.} Our working assumption throughout this analysis is that the maximum temperature of the radiation bath attained during reheating, $T_\textsc{max}$, is larger than $m_B$. Thus the thermal bath generated during reheating began its existence with a full relativistic abundance of $B$ particles. Once we ensure that $T_\textsc{max} > m_B$, the resulting \ac{FIMP} relic density depends only on $T_R$. The methodology to solve numerically the Boltzmann equation for FI during an early \ac{MD} era has been discussed in Sec.~\ref{sec:part_prod_emd}. Here, we present numerical solutions and we understand the underlying physics thanks to analytical estimates.

\subsection{Freeze-in through decays}
\label{sec:fi_emd_IR}

\begin{figure}
\centering
\includegraphics[width=0.55\textwidth]{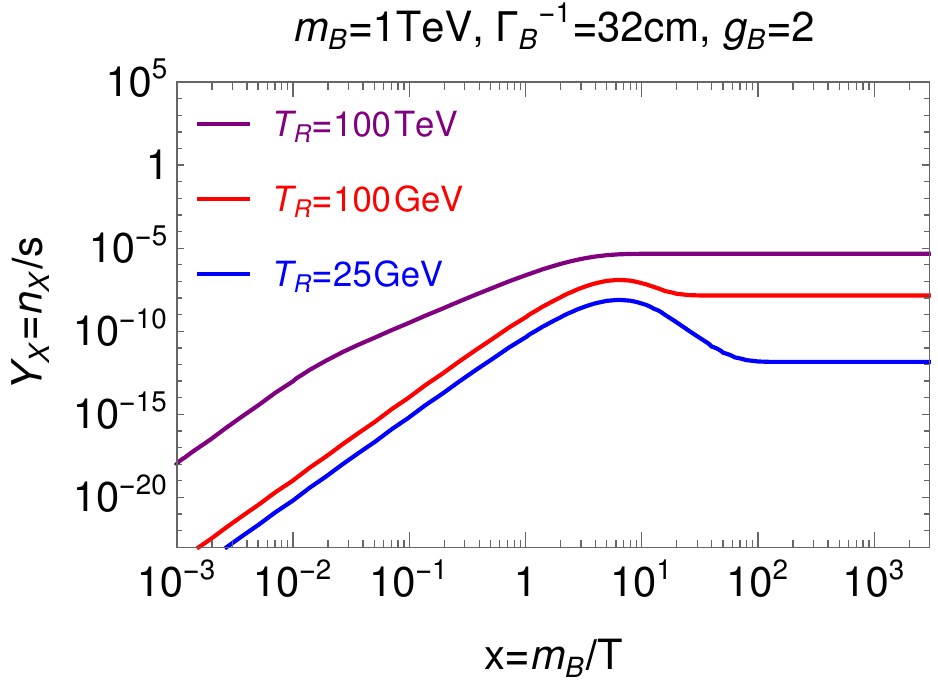}
\caption{Freeze-in comoving number density from $B$ decays for a Weyl fermion ($g_B=2$) during an early \ac{MD} era. We fix $m_B = 1 \, {\rm TeV}$ and $\Gamma_{B \to \chi B'}=32$~cm, and we consider three different values for the reheating temperature.  Relic density requires: $m_X = 100 \, {\rm keV}$ (purple line), $m_X = 30 \, {\rm MeV}$ (red line), $m_X = 300 \, {\rm GeV}$ (blue line).}
\label{fig:FIearlyMD}
\end{figure}

Fig.~\ref{fig:FIearlyMD} shows the evolution of the \ac{DM} comoving number density for fixed values of the $B$ mass and decay length, but different values of the reheating temperature. We can potentially reproduce the \ac{DM} relic density upon choosing an appropriate value for $m_X$. The purple line does not present any substantial difference with respect to the solutions for standard \ac{FI} as depicted for instance in Fig.~\ref{fig:std_fi}. Indeed, the reheating temperature in this case is much larger than the $B$ mass so \ac{FI} happens long after inflationary reheating is over and the cosmological background is a RD universe. An estimate for the relic yield $Y_\chi^{\infty}$ is still provided by Eq.~\eqref{eq:fi_ann_est}. 

The red and the blue lines feature a quite different behavior. The \ac{DM} comoving number density reaches a peak and decreases afterwards before settling down to its asymptotic value. The \ac{FI} production happens during an \ac{EMD} epoch for both cases since $T_R < m_B$. The amount of $\chi$ particles produced when the thermal bath had a temperature $T$ during a Hubble doubling time, namely the time it takes for the universe to double its size, can be estimated by
\begin{equation}
Y^{\rm prod}_\chi(T) \simeq \mathcal{R}_{B \rightarrow B' X}(T) \, t(T) \ .
\label{eq:Yapprox}
\end{equation}
Here, $\mathcal{R}_{B \rightarrow \chi B'}(T)$ is the (temperature dependent) rate of particle production and $t(T)$ is the age of the universe when the temperature is $T$. We estimate the former by multiplying the $B$ rest frame decay width by an additional factor of $m_B / T$ to account for time dilatation due to the kinetic energy of the thermal bath. The latter is approximately the inverse Hubble parameter, $t(T) \simeq 1 / H(T)$, and for an \ac{EMD} universe we have $H(T) \sim T^4/(T_R^2 M_{Pl})$, see Eq.~\eqref{eq:HearlyMD}. Thus the comoving number density scales with the temperature $T$ as follows 
\begin{equation} 
Y^{\rm prod}_\chi(T)  \simeq \frac{\Gamma_{B \rightarrow A_\textsc{sm} X} \, M_{\rm Pl} m_B T_R^2}{T^5} \quad {\rm for} \quad [T_\textsc{max}> T\gtrsim {\rm Max}(m_B,T_R)]\,.
\label{eq:YprodMD}
\end{equation}
The above expression approximates well the blue and red curves of Figure~\ref{fig:FIearlyMD} for temperatures larger than the $B$ mass ($x=m_B/T \lesssim 1$). The initial growth scales as $x^5$ instead of the $x^3$ found for a standard cosmological history in Eq.~\eqref{eq:fi_prod_ann_est}. The $x^5$ and $x^3$ dependencies are also well visible in the purple curve slope when going from \ac{FIMP} production in an \ac{EMD} era to production in a standard RD era, namely in the transition between $x < 10^{-2}$ and $10^{-2}< x <1$. Besides the different power-law behavior, the message about \ac{IR} domination is confirmed and actually reinforced. 

The expression in Eq.~\eqref{eq:YprodMD} can only account for $Y_X(T)$ at $x\lesssim 1$ because at later times, $B$ hits the Boltzmann suppression. In the case of \ac{FI} during a \ac{RD} universe, we could estimate the final yield $Y_X^\infty$ by evaluating $Y_X^{\rm prod}(T)$ at a temperature $T_{FI}$ of the order of the $B$ mass, since this is the temperature where the main DM production occurs, see Sec.~\ref{sec:fi}. In contrast, for \ac{FI} during an \ac{EMD} universe, we still have to account for the subsequent dilution of $Y_X(T)$ until the reheating time. The dilution, due to entropy released by inflaton decays, is well visible in the red and blue curves of Fig.~\ref{fig:FIearlyMD} for $x> 1$ up to $x\sim m_B/T_{R}$. We quantify the dilution factor by the ratio of the entropy in a comoving volume between  a production time at a given temperature $T$ and the time when we can trust entropy conservation again (namely temperatures below $T_R$)
\begin{equation}
  D(T) = \frac{S(T_R)}{S(T)} = \left(\frac{T}{T_R} \right)^5  \quad {\rm for} \quad [T_\textsc{max}> T> T_R]\,,
  \label{eq:D(T)}
\end{equation}
where we use the analytical estimate of Eq.~\eqref{eq:SvsT}. We define the ratio of the produced comoving \ac{FIMP} density $Y^{\rm   prod}_X (T)$  of
Eq.~(\ref{eq:YprodMD}) and of the dilution factor $D(T)$  as 
\begin{equation}
  \tilde Y_X(T) = \frac{Y^{\rm prod}_X(T) }{D(T)} \simeq \frac{\Gamma_{B \rightarrow A_\textsc{sm} X} \, M_{\rm Pl} m_B T_R^7}{T^{10}}
  \label{eq:Ytoday}
\end{equation}
 which results from the combination of \ac{FIMP} production in an \ac{EMD} era at temperature $T$ and subsequent entropy dilution between $T$ and $T_R$. We notice in Eq.~(\ref{eq:Ytoday}) an extreme \ac{IR} domination with a power-law scaling as $1 / T^{10}$, and such an \ac{IR} domination is the reason why the resulting relic density does not depend on $T_{\rm MAX}$. 
   
As for the \ac{RD} case, \ac{FIMP} production is more efficient at the lowest possible temperature compatible with the presence of $B$ particles in the thermal bath. Thus the same argument used above would suggest that the relic \ac{FIMP} comoving density is approximately $\tilde Y_X(T)$ evaluated at a temperature $T_{FI}$ that is of the order of the $B$ mass. However, Ref.~\cite{Co:2015pka} observed that \ac{FIMP} production is still in action when we are in the tail of the Maxwell-Boltzmann distribution for $B$; this is just a consequence of the extreme \ac{IR} domination observed in Eq.~\eqref{eq:Ytoday}. The precise calculation of the relic density would require numerically solving the Boltzmann equation provided in Sec.~\ref{sec:fi_emd}. Here, we report the analytical estimate for the \ac{DM} comoving density obtained by evaluating Eq.~\eqref{eq:Ytoday} at $T=m_B$,
 \begin{equation} Y_X^{\infty} \simeq 10^2 \times \left(
 \frac{T_R}{m_B} \right)^7 \times \frac{\Gamma_{B \rightarrow A_{\rm
       SM} X} \, M_{\rm Pl}}{m_B^2} \ ,
\label{eq:YXMDfinal}
\end{equation}
where the enhancement factor of $10^2$ is obtained by matching our analytic solution to the numerical ones shown in Fig.~\ref{fig:FIearlyMD}. This factor is quite large due to the strong temperature dependence of Eq.~\eqref{eq:Ytoday}. From this analytic estimate, we see that \ac{FI} during an \ac{EMD} era leads to a $(T_R / m_B)^7$ suppression to the \ac{FIMP} relic density that we observe today compared to the \ac{RD} case of Eq.~(\ref{eq:fi_ann_est}). Consistently with Eq.~\eqref{eq:YXMDfinal}, we notice how the red and the blue lines lead to an asymptotic relic density that is suppressed with respect to the purple line by approximately three and seven orders of magnitude, respectively.

\subsection{UV freeze-in through scatterings}
\label{sec:fi_emd_UV}

As we have illustrated in Secs.~\ref{sec:fi} and \ref{sec:fi_leptophilic}, scatterings can be an irreducible contribution to \ac{FI} production. We complete our analysis by discussing this additional channel in a \ac{EMD} era, and we show how its relative importance depends on whether the interaction between $\chi$ and the thermal bath is renormalizable or not. Let $d$ be the mass dimension of the operator in the Lagrangian responsible for the scattering processes. At high enough temperatures, larger than any mass of the particles involved in the collision, the scaling of the interaction rate with the temperature follows from dimensional analysis
\begin{equation}
\mathcal{R}_{B A_\textsc{sm} \rightarrow A^{'}_\textsc{sm} X}(T) \propto \frac{T^{2d - 7}}{\Lambda^{2d - 8}} \ ,
\label{eq:ratescattering}
\end{equation}
where $\Lambda$ is the mass scale appearing in the \ac{FIMP} interaction. For non-renormalizable interactions, $d > 4$, this is the scale suppressing the operator. The amount of \ac{FIMP} produced is still given roughly by Eq.~\eqref{eq:Yapprox}, so that for \ac{DM} production through scatterings in a \ac{RD} universe, the resulting comoving number density scales with the temperature as
\begin{equation}
Y_X(T) \propto \frac{M_{\rm Pl}}{\Lambda^{2d - 8}} \, T^{2d - 9} \quad \qquad \text{[RD epoch]}  \,.
\label{eq:YvsTscatteringRD}
\end{equation}

The value $d = 4.5$ divides two distinct regimes. While for $d<4.5$, \ac{DM} production peaks at low temperatures (see Sec.~\ref{sec:fi}), production is more efficient at high temperatures for $d > 4.5$ so that scattering production can start dominating over decays if production at large enough temperatures is allowed. The latter case, where the interaction is non-renormalizable, is often referred to as \ac{UV} freeze-in~\cite{Chen:2017kvz,Bernal:2019mhf,McDonald:2015ljz,Barman:2020plp}. On the contrary, as we have discussed in Sec.~\ref{sec:fi_leptophilic} for renormalizable operators, scatterings dominate over decays only if a larger number of scattering processes can contribute to DM production or when the mass splitting between $B$ and $\chi$ is small enough to suppress the decay rate (see also Refs.~\cite{Belanger:2018ccd,Garny:2018ali}). 

The abundance of \ac{DM} produced through scatterings via non-renormalizable interactions cannot be computed only in a \ac{RD} universe. As it is manifest from Eq.~\eqref{eq:YvsTscatteringRD}, the production diverges as we go back in time (i.e.,~higher $T$). However, the estimate in Eq.~\eqref{eq:Yapprox} is only valid as long as FIMPs are out of equilibrium, and eventually at high enough temperatures the rate would become so large that FIMPs thermalize. We need to know to what extent we can extrapolate the \ac{RD} universe back in time in order to provide a precise calculation. In other words, we need a temperature \ac{UV} cut-off corresponding to the highest temperature for the RD phase. In earlier work~\cite{Elahi:2014fsa,Barman:2020plp}, it has often been assumed that reheating is an instantaneous process such that there is no \ac{FI} production during the reheating era, and the standard \ac{FI} Boltzmann equation (Eq.~\eqref{eq:Boltz_fi_scat}) can be used where we simply start integrating at the reheating temperature. Here, we will use the full description of the reheating phase as elaborated in Sec.~\ref{sec:alt_cosmo} and Refs.~\cite{Co:2015pka,Calibbi:2021fld}.\footnote{See also~\cite{Bernal:2019mhf,Barman:2022tzk} for UV freeze-in scenarios assuming alternative cosmologies other than an EMD era coming from inflationary reheating.}

During the reheating era, the expression in Eq.~\eqref{eq:Yapprox} still accounts for the amount of FIMP produced in a Hubble doubling time when the temperature was $T$, and the scattering rate in Eq.~\eqref{eq:ratescattering} is still valid. However, as we have already seen in Sec.~\ref{sec:fi_emd_IR} for decays, there are two new ingredients: we need to use the Hubble parameter during reheating given in Eq.~\eqref{eq:HearlyMD}, and we need to account for dilution due to the entropy released by the inflaton decays. Taking this all into account, we find
\begin{equation}
Y_X(T) \propto \frac{M_{\rm Pl} T_R^2}{\Lambda^{2d - 8}} \, \frac{T^{2d - 11}}{D(T)} = \frac{M_{\rm Pl} T_R^7}{\Lambda^{2d - 8}} \, T^{2d - 16} \qquad \text{[MD epoch]}  \ .
\label{eq:YvsTscatteringMD}
\end{equation}
From this, we see that \ac{DM} production via scattering for renormalizable operators (i.e.~$d \leq 4$) is \ac{IR} dominated during both a \ac{RD} and \ac{EMD} era. On the contrary, in the range $5 \leq d \leq 8$, \ac{DM} production is dominated at small temperatures during reheating and at large temperatures during \ac{RD}. In other words, when fully accounting for the reheating phase, \ac{DM} production is most efficient either around $T=T_{R}$ or $T\sim m_B$, depending on the hierarchy between these two scales. For $d > 8$, \ac{DM} production is \ac{UV} dominated for both \ac{RD} and \ac{EMD}. In the latter case, the resulting relic density is sensitive to $T_\textsc{max}$ and to the details of reheating. Interestingly, for $d < 8$ \ac{DM} production via scattering depends only on $T_R$ and not on $T_\textsc{max}$~\cite{Co:2015pka,Garcia:2017tuj,Chen:2017kvz}.

One concrete example of non-renormalizable interactions is for a Weyl fermion singlet \ac{FIMP} $\chi$. We consider the dimension 5 operator coupling $\chi$ to gluons and to a new fermion $\lambda^a$ in the adjoint representation of the $SU(3)$ color gauge group 
\begin{equation}
\mathcal{L} \supset \frac{1}{\Lambda} G_{\mu \nu }^a \lambda^a \sigma^{\mu} \bar \sigma^{\nu} \chi \ ,
\label{eq:Ldim5sec2}
\end{equation}
where $G_{\mu\nu}^a$ the gluon field strength. This is for instance the well known case of gluino-gluon-goldstino coupling in supersymmetric theories~\cite{Martin:1997ns}. Such an interaction induces \ac{DM} production both via decays $\lambda \rightarrow G \, \chi$ and scatterings. The rate for the former scales proportionally to the decay width of the mother particle, $\Gamma_\lambda \simeq m_\lambda^3/\Lambda^2$, and therefore production via decays is \ac{IR} dominated even for non-renormalizable operators. The operator in Eq.~\eqref{eq:Ldim5sec2} induces also \ac{DM} production through t-channel scattering processes. Considering for example the t-channel exchange of a gluon, we have the matrix element
\begin{equation}
	|\mathcal{M}|^2 \sim \frac{g_s^2}{\Lambda^2} \frac{st(s+t)}{(t-\Pi(T)^2)^2} \, ,
\label{eq:Mt}
\end{equation}
where $g_s$ is the strong coupling, $s$ and $t$ are Mandelstam variables, and $\Pi(T)$ is a thermal mass that we insert to regulate the t-channel \ac{IR} divergence.

\begin{figure}[!t]
\centering
\includegraphics[width=0.44\textwidth]{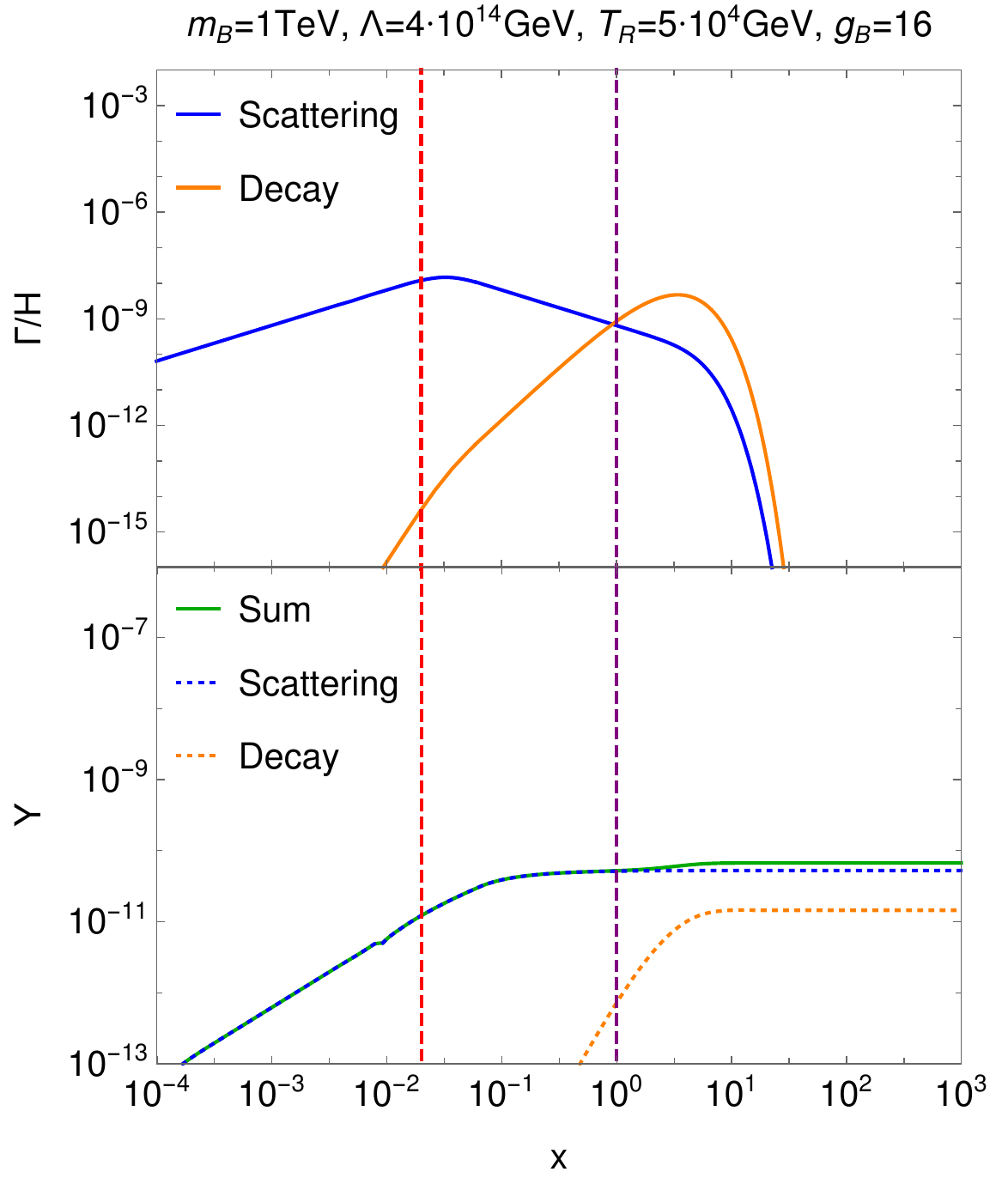} 
\hspace{0.05\textwidth}
\includegraphics[width=0.44\textwidth]{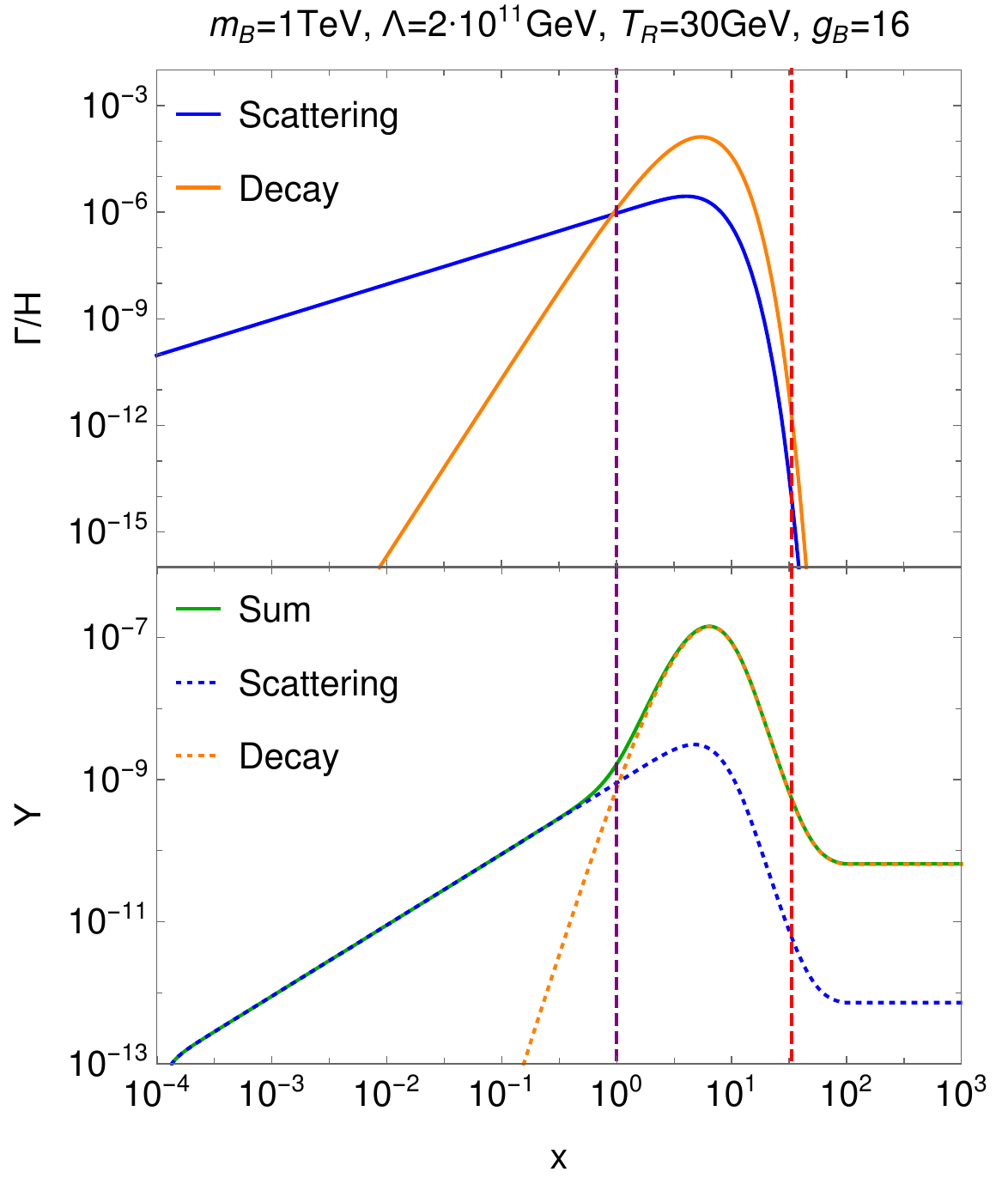} 
\caption{\ac{FIMP} production via decays (orange) and scatterings (blue). The top panel depicts the ratio between interaction and Hubble rates while the bottom panel shows the total \ac{DM} yield. We show two benchmarks with $T_{R}>m_B$ (left) and $T_{R}<m_B$ (right). The dashed vertical lines denote $T=T_{R}$ (red) and $T = m_B$ (purple). For these parameters, the observed \ac{DM} abundance is reproduced for $m_{\rm DM} = 10$~GeV.}
\label{fig:rates}
\end{figure}

Taking $\Pi (T)= g_s T$ for illustration, Figure~\ref{fig:rates} presents numerical results for the operator in Eq.~\eqref{eq:Ldim5sec2}. Fixing the value of $m_B$, we consider two different choices for the combination $(\Lambda, T_R)$ in the two panels. For both cases, we take $\Lambda$ to be much larger than the weak scale\footnote{This is appealing since it could explain the tiny \ac{FIMP} coupling from a hierarchy of mass scales.} and we choose numerical values such that we reproduce similar asymptotic \ac{DM} comoving densities. The reheating temperature is always below the new physics scale, $T_R < \Lambda$, in order to provide consistency with the contact interaction. As argued above, we do not need to specify the value of $T_\textsc{max}$ for $d = 5$ but it is implicit that we assume the hierarchy $T_\textsc{max} < \Lambda$ for the same reason. The left and right panels have reheating temperature larger and smaller than the $B$ mass, respectively. We show the evolution of the production rates with respect to the Hubble parameter (upper part) and the \ac{DM} comoving density (lower part) as a function of $x$. We keep the two contributions, decays (orange lines) and scatterings (blue lines). Vertical dashed lines identify the two key temperatures when \ac{FI} production is mostly efficient around $x_{FI}\sim {\cal O}( 1)$ (the purple lines are shown for $x=1$ as a guide for the eye) and when inflationary reheating is over at $x=m_B/T_R$ (red lines). As expected, the scattering rate for the case $T_R > m_B$ is mostly efficient around $T_R$ whereas the decay rate is maximized when the temperature is around $m_B$. The relative height of the peaks is set by the reheating temperature: the higher $T_{R}$, the higher the contributions from the scattering processes. Consequently, the comoving density features two plateaus corresponding to the two production channels. In contrast, both scattering and decay rates peak around $T \approx m_B$ for the case $T_R < m_B$. For the specific interaction of Eq.~\eqref{eq:Ldim5sec2}, the scattering rate is smaller than the decay rate around the peak and therefore scatterings provide a negligible contribution to the total \ac{DM} abundance. Thus it does matter whether the operator mediating the decay is renormalizable or not because the resulting scattering production may have a relatively large importance and gets \ac{UV} dominated~\cite{Moroi:1993mb,Rychkov:2007uq,Strumia:2010aa,Elahi:2014fsa}.

\chapter{Displaced signatures as a probe for feebly interacting DM}
\label{chap:model_class}

As we have illustrated in Chapters \ref{chap:DM_prod} and \ref{chap:alt_cosmo}, FIMPs have the potential to be viable \ac{DM} candidates as there is a variety of production mechanisms able to explain the total amount of \ac{DM} observed in our universe today. If FIMPs were to be the main component of DM, there is one major drawback, as they are very difficult to detect since their interactions with SM particles are very much suppressed by construction. The typical detection techniques for \ac{DM} as illustrated in Sec.~\ref{sec:DM_detection} are therefore insufficient to probe FIMPs. It is hence very timely to go beyond these techniques and explore alternative ways to probe FIMPs as \ac{DM} candidates.

For all the \ac{FIMP} production mechanisms discussed before, the \ac{DS} does not consist solely of the \ac{DM} candidate. There is at least one other \ac{DS} particle present we often refer to as the mediator. This mediator can interact more strongly with the SM through for instance annihilation processes, so that it is in thermal contact with the SM bath in the early universe. We can also benefit from these sizable interactions when trying to probe \ac{DS} containing a \ac{FIMP} as DM candidate. However, since the mediator has decayed away in the early universe, there is no relic mediator abundance of it today and direct or indirect searches remain insensitive to \ac{FIMP} models. Nonetheless, the mediator can be produced copiously at collider experiments making them very important in probing a \ac{DS} including a \ac{FIMP} as \ac{DM} candidate.

\begin{figure}
    \centering
    \includegraphics[width=0.45\textwidth]{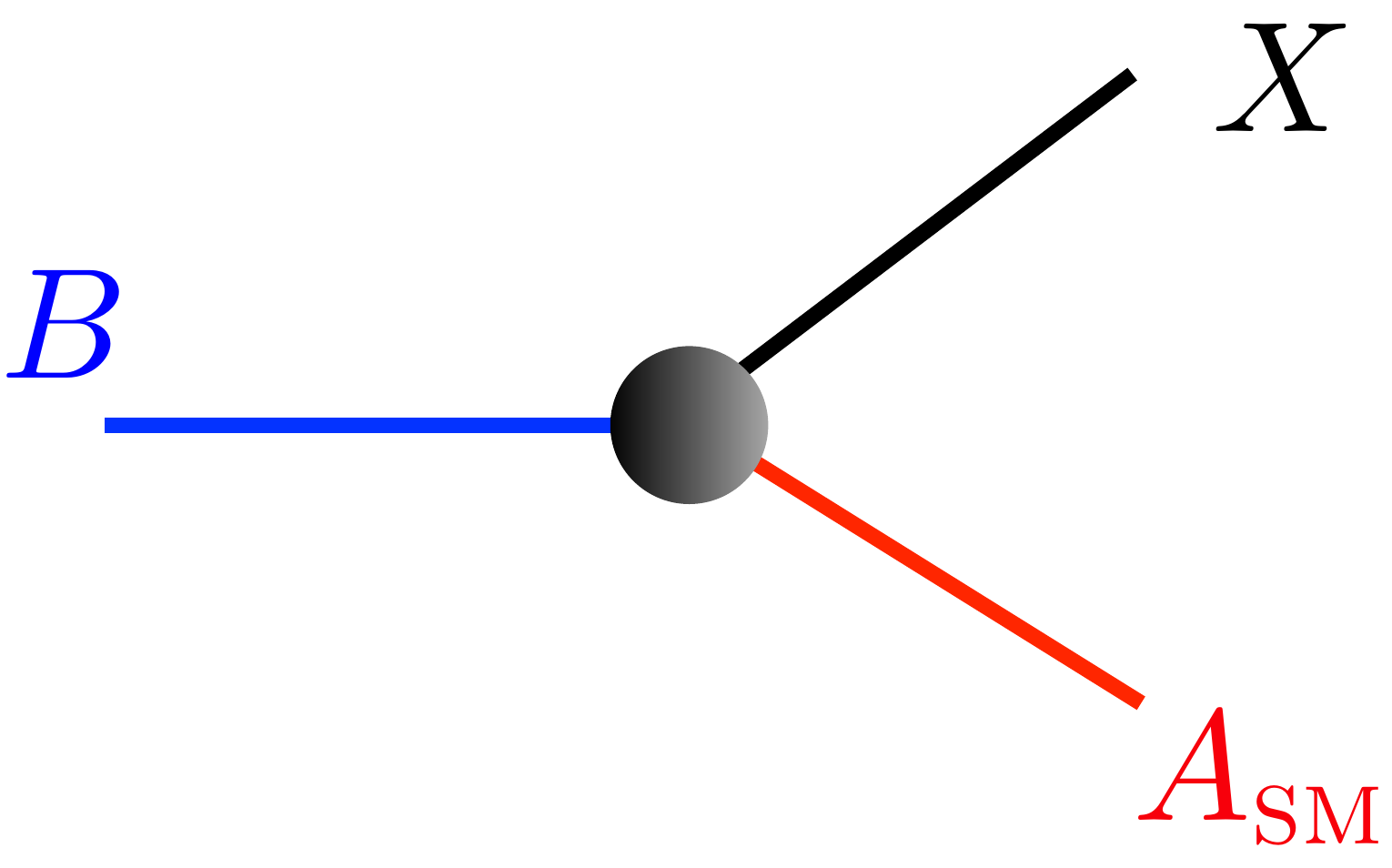}
    \caption{Diagram of a \ac{FIMP} $X$ coupling to the mediator $B$ and a SM particle $A_\textsc{SM}$. The coupling governing this interaction is assumed to be feeble.}
    \label{fig:fi_model_diagram}
\end{figure}

When the mediator is produced at a colliders such as the \ac{LHC}, there is a whole variety of signatures that can be looked for at multi-purpose detectors such as CMS and ATLAS, which are each placed around one of the collision points of the \ac{LHC}. The signatures that can arise depend on the model under study and its parameter values. As an instructive example, we will focus on \ac{FIMP} models where the \ac{DM} interacts with the SM through the three body interaction depicted in Fig.~\ref{fig:fi_model_diagram}. The \ac{FIMP} is a SM gauge singlet and we take the cubic interaction to be very suppressed in order to prevent its thermalization. In contrast, the bath particle $B$ carries \ac{SM} gauge quantum numbers and it is in thermal equilibrium with the primordial plasma at early times. Known examples for $B$ can be supersymmetric partners (gluino, wino, squarks, etc.), but we take a bottom-up approach and we do not commit to any specific realization.

If the decay of the mediator is only governed by the suppressed cubic interaction of Fig.~\ref{fig:fi_model_diagram}, $B$ can obtain macroscopic lifetimes such that after production in a collision at LHC, it can travel partly or completely through the detector before decaying. This gives rise to so-called ``displaced'' or ``long-lived signatures''. Since such signatures happen only rarely in the SM, they are a very good probe for new physics. They are especially important for probing \ac{FIMP} models as it is often one of the few ways in which these models can be probed. In this chapter, we will outline a systematic way of probing \ac{FIMP} models based on the interaction depicted in Fig.~\ref{fig:fi_model_diagram} using searches for displaced signatures caused by a \ac{LLP}.

\section{FIMP model classification}
\label{sec:models}

The main goal of this chapter is to emphasize the importance of displaced physics searches in detecting feebly interacting DM. We will do this by considering a set of well defined scenarios giving rise to the cubic interaction shown in Figure~\ref{fig:fi_model_diagram}. Restricting ourselves to \ac{BSM} particles of spin $0$ and $1/2$,\footnote{For a FI model with instead spin-1 \ac{DM} produced by the decay of a long-lived vectorlike lepton, see~\cite{Delaunay:2020vdb}.} and assuming \ac{DM} to be a singlet under the SM gauge group, the resulting list is compactly reported in Tab.~\ref{tab:classification}.  These ``simplified models'' are minimal in the sense that they feature the lowest dimensional operators (with $d\leq 5$), containing one SM field and two extra fields of the dark sector, giving rise to the three-body interaction. Hence for each model only two fields need to be added to the SM field content. Furthermore, we impose an unbroken $\mathbb{Z}_2$-symmetry under which the SM fields are even and the new fields are odd; this new parity, together with the fact that we always assume the \ac{DM} field to be the lightest of the dark sector fields, ensures \ac{DM} stability and restricts the three-field interactions to the form shown in Fig.~\ref{fig:fi_model_diagram}.  In the first column of Tab.~\ref{tab:classification}, we consider the finite set of options for the SM particle $A_\textsc{sm}$: (i) a SM fermion field $\psi_\textsc{sm}$, that is, a left-handed or right-handed lepton or quark; (ii) a SM gauge boson (photon, $W$, $Z$, gluon), following from an interaction involving the field strength $F^{\mu\nu}$ of $U(1)_Y$, $SU(2)_L$ or $SU(3)_c$; (iii) the Higgs boson $H$. The second and third columns contain all the corresponding possible choices for the spins of $X$ and $B$, respectively. The gauge quantum numbers of $B$ under the SM gauge group should be equal to the ones of $A_\textsc{sm}$ since the \ac{DM} field is a gauge singlet. Notice that the case when $A_\textsc{sm}$ is the $U(1)_Y$ hypercharge gauge boson is of no phenomenological relevance at colliders because $B$ would be a complete SM singlet for gauge invariance. In the fourth column, we display the Lagrangian term relevant for freeze-in \ac{DM} production and for the decay of $B$ into \ac{DM} at colliders.  We write these interactions denoting a fermionic \ac{DM} particle (bath particle) as $\chi$ ($\Psi_B$) and a scalar \ac{DM} particle (bath particle) as $\phi$ ($\Phi_B$). In the last column, we give a label to each minimal model: the first letter indicates whether the bath particle $B$ is a fermion ($\cal F$) or a scalar ($\cal S$), the second letter indicate the SM field involved, while the third letter indicate the nature of DM.

\renewcommand{\arraystretch}{1.0}
\begin{table}[!t]
	\centering
	\begin{tabular}{ | c | c | c | c | c | }
		\hline
		$\boldsymbol{A_\textsc{sm}}$ & \textbf{Spin DM} & \textbf{Spin B} & \textbf{Interaction} & \textbf{Label} \\ 
		\hline \hline
		\multirow{2}{1cm}{\centering$\psi_\textsc{sm}$} & 0 & 1/2 & $\bar{\psi}_\textsc{sm} \Psi_B \phi $ & $\mathcal{F}_{\psi_\textsc{sm}\phi}$ \\
		& 1/2 & 0 &  $\bar{\psi}_\textsc{sm} \chi \Phi_B $ &$\mathcal{S}_{\psi_\textsc{sm}\chi}$ \\  
		\hline
		$F^{\mu\nu}$ & 1/2 & 1/2 & $\bar{\Psi}_B \sigma_{\mu\nu} \chi F^{\mu\nu} $ & $\mathcal{F}_{F\chi}$ \\
		\hline
		\multirow{2}{1cm}{\centering$H$} & 0 & 0 & $H^\dagger \Phi_B \phi$ & $\mathcal{S}_{H\phi}$ \\
		& 1/2 & 1/2 & $\bar{\Psi}_B \chi H$ & $\mathcal{F}_{H\chi}$ \\
		\hline
	\end{tabular}
	\caption{Classification of the simplest possible operators featuring a cubic interaction of the type of Fig.~\ref{fig:fi_model_diagram}. The \ac{DM} particle is a SM singlet denoted by $\phi$ if it is a scalar, $\chi$ if it is a fermion. The bath field ($\Psi_B$ for fermions, $\Phi_B$ for scalars) has therefore the same quantum numbers as the corresponding SM field $A_\textsc{SM}$. See the text for further details.}	\label{tab:classification}
\end{table}

It is worth noticing that all the interactions of Tab.~\ref{tab:classification} are renormalizable except for the ones of the models $\mathcal{F}_{F\chi}$ which are of dimension $d=5$. The interaction term $\mathcal{F}_{F\chi}$ resembles the gluino-gluon-goldstino coupling or Wino-$W$-goldstino interactions in supersymmetry, see e.g. Ref.~\cite{Martin:1997ns}.\footnote{One could in principle add analogous models featuring scalar \ac{DM} and a scalar bath particle $B$ in an adjoint representation of $SU(2)_L$ or $SU(3)_c$ with couplings to the field strengths given by dimension-6 operators of the kind $\partial_\mu\Phi_B \partial_\nu\phi F^{\mu\nu}$. However, as one can show by integrating by parts, such an interaction is   proportional to the gauge boson masses, thus vanishing in the $SU(3)_c$ case and for the unbroken phase of $SU(2)_L$ (up to  thermal corrections). We defer to future work a detailed study of this peculiar scenario.}  Among the renormalizable interactions, three classes have $d=4$.  Some instances of models involving interactions of this kind have already been studied within the context of freeze-in in e.g.~\cite{Garny:2018ali} for $\mathcal{F}_{\psi_{\rm SM}\phi}$, in~\cite{Ibarra:2008kn,Belanger:2018sti} for $\mathcal{S}_{\psi_{\rm SM}\chi}$ and in~\cite{Calibbi:2018fqf,No:2019gvl} for $\mathcal{F}_{H\chi}$. Finally, $\mathcal{S}_{H\phi}$ is a dimension-three operator involving scalars only.

This simple scheme encodes all possible cubic interactions involving one FIMP, another \ac{BSM} field and one SM particle (restricting to spins 0 and 1/2 for the \ac{BSM} sector). They amount to a total of $35$ simplified models (counting all possible combinations of chirality and flavor of the SM fermions, and the three different choices for the gauge boson field strength $F^{\mu\nu}$). This classification can serve as a guide in the discussion of probing \ac{FIMP} models. Since the \ac{LEP} collider has already provided strong bounds on charged particles with a mass below roughly 100~GeV, we only assume the mediator mass in our FIMP models in Tab.~\ref{tab:classification} to be at least 100~GeV, so that they can serve as physics targets for the \ac{LHC}. Based on the classification presented here, we can hence make a selection of predicted signatures arising from these models, as we will do in Sec.~\ref{sec:LHC}.

\section{Displaced new physics at the LHC}
\label{sec:LHC}
Many searches for displaced new physics have already been performed at the CMS and ATLAS experiment. These searches are based on a variety of signatures arising in different parts of the detector. We will discuss all of these signatures after we have first reviewed the main components of the CMS and ATLAS detector, which is necessary to understand how the different signatures arise. 

\subsection{The CMS and ATLAS detector}
\label{sec:detector_description}
\begin{figure}
    \centering
    \includegraphics[height=0.25\textheight]{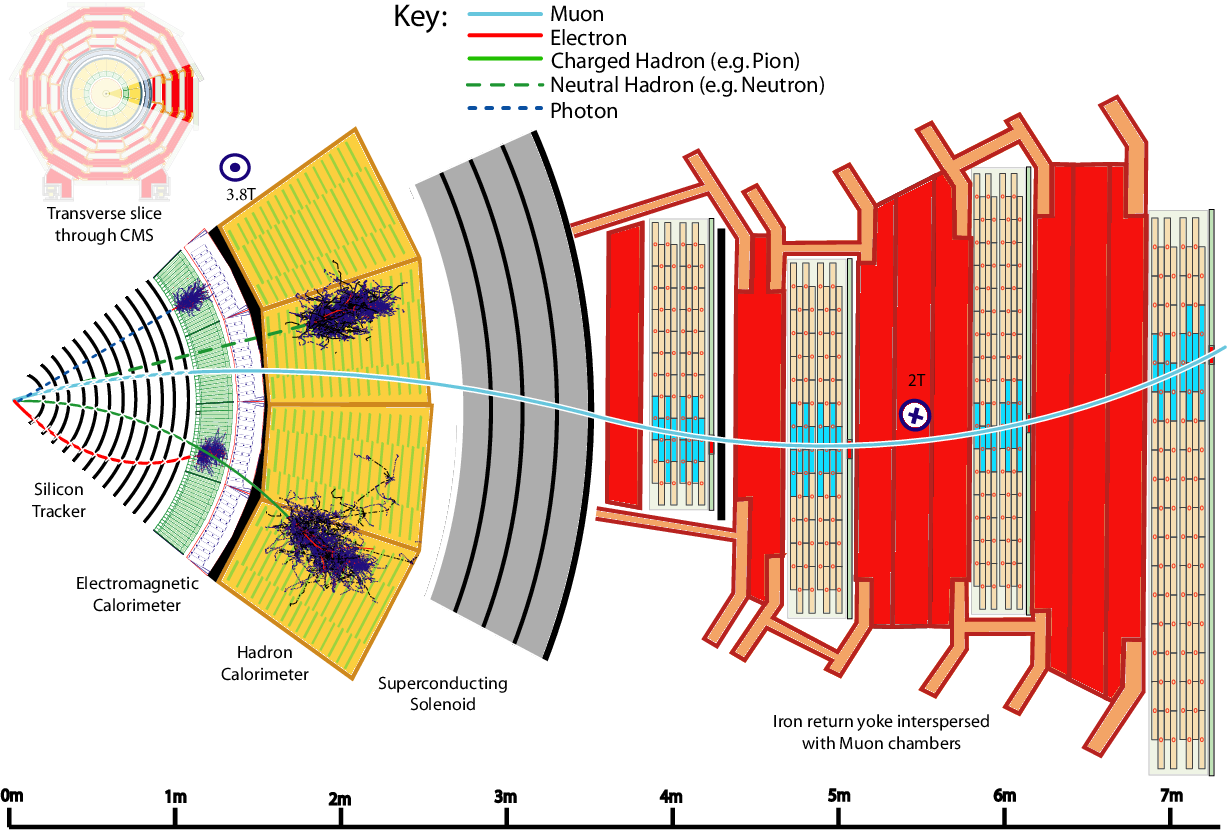}
	\hspace{0.005\textwidth}
    \includegraphics[height=0.25\textheight]{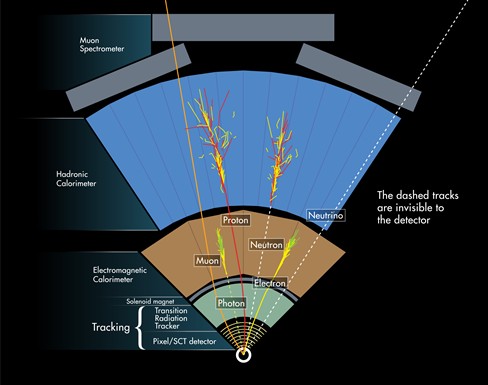}
    \caption{A cross sectional slice of the CMS (left) and ATLAS (right) experiment depicting the main components of the detector. Pictures are taken from \cite{CMS:2017yfk,Pequenao:1505342}.}
    \label{fig:cms_atlas}
\end{figure}

The ATLAS and CMS experiments are two multi-purpose detectors located at the \ac{LHC}. They are both cylindrical detectors completely surrounding there respective collision points. The basic design of both detectors is outlined in Fig.~\ref{fig:cms_atlas}, however, as one can see, both detectors are composed of the main concentric sub-systems, namely the tracker, the \ac{ECAL} and \ac{HCAL} and the \ac{MS}. We will briefly discuss them below. 
\\

\textbf{Tracker} The innermost part of the detector surrounding the beampipe is called the tracker. It consist of multiple layers so that when charged particles pass through the different layers, they leave behind a tiny electrical signal. This is often referred to as a hit, and by connecting hits in the different tracking layers, one is able to reconstruct the path the particle has travelled. Since the tracker is submerged into a strong magnetic field, these paths will be curved, depending on the charge and the momentum of the particle. Hence, reconstructing tracks also helps us inferring basic properties of the particles we observe.

\textbf{Calorimeters} The \ac{ECAL} and \ac{HCAL} are specifically designed to stop and measure the energy of the particles passing through them. This is done by a combination of absorbing the energy and measuring the energy loss throughout the calorimeter, hence, calorimeters usually consist of two types of material: a high-density material such as lead to absorb the energy, and an active medium to measure the energy loss. The \ac{ECAL} does this for electromagnetic particles (photons, electrons and positrons) while the \ac{HCAL} stops and measures hadrons (e.g. pions). The calorimeter can stop most known particles, except for muons and neutrinos.

\textbf{Muon spectrometer} Since muons interact only very little with detector material, they do not get stopped by the calorimeters and travel almost freely to the outer layers of the detector. Here, the muon spectrometer is installed in order to better track the muons, since they do not always leave a track in the inner tracking system. 
\\

In order to identify particles, one has to measure the particles velocity, as this, together with the momentum measured by the tracker, reveals the mass, and hence also the identity of the particle. There are different ways of measuring the velocity, one is for instance by using a \ac{TOF} detector to measure the time it took for the particle to cross a certain distance. More details on this and the whole detector design can be found in Refs.~\cite{Chatrchyan:1129810} and~\cite{Aad:1129811} for the CMS and ATLAS detector respectively.

The ATLAS and CMS detector fully encapsulate the collision so that all observable particles produced in the collision can be detected. This is vital since it can be used to infer information about the unobserved particles. Indeed, since the collision happens along one axis, the total momentum in the transverse plane before collision is zero. Conservation of energy and momentum dictates that the vectorial sum of the transverse momenta of all particles created in the collision must add up to zero as well. If for instance \ac{DM} is produced in the collision, it will not be observed and the vectorial sum of the momenta of all observed particles will not add up to zero. The momenta necessary to bring the vectorial sum to zero is then referred to as missing transverse momentum or \ac{MET}. Hence, events with a large amount of \ac{MET} can hint to the presence of DM. 

A final important part of the detector is the triggering system. Since at the \ac{LHC}, a collision occurs every 25~ns, a lot of data is produced because at each collision, many particles are produced. Saving this amount of data at such high rates is not possible. Trigger systems are designed to select the potentially interesting events in order to save them, and discard the others, so that the rate at which data has to be stored can be contained to a manageable amount. Both ALTAS~\cite{ATLAS:2016wtr} and CMS~\cite{Tosi:2017hjj} have a list of different triggers they use during their runs. These triggers are based on reconstructed objects, for instance jets or leptons with high transverse momentum that typically arise in events with new heavy particles. For DM models, a very interesting trigger strategy is the one requiring a high amount of MET, which is an event property often met in events where DM particles are produced. 

\subsection{Searching for long-lived particles}
\label{sec:searches}
\begin{figure}[t]
	\centering
	\includegraphics[width=0.8\textwidth]{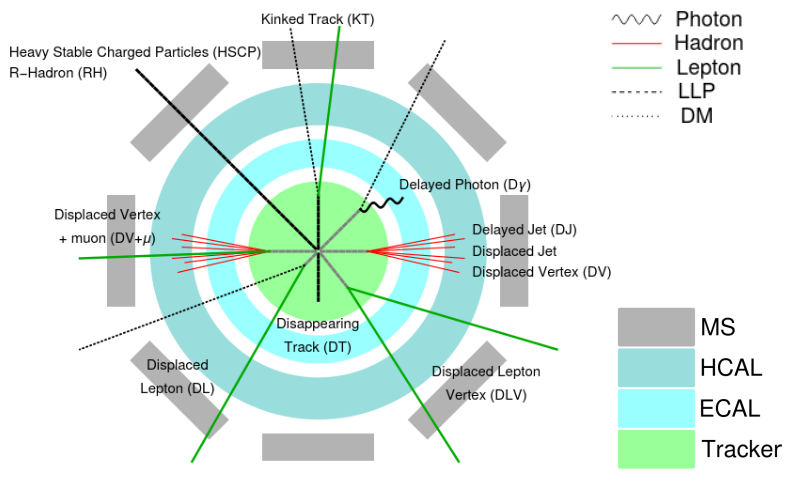}
	\caption{Schematic illustration of a subset of potential \ac{LLP} searches. The green area represents the tracker, the blue (cyan) region denoted by \ac{HCAL} (\ac{ECAL}) refers to the hadronic (electromagnetic) calorimeter, the gray boxes represent the muon spectrometer (MS). Dotted lines refer to DM, dashed lines refer to the \ac{LLP} and they are darker for searches that require  a charged track associated to it.}
    \label{fig:searches}
\end{figure}

In recent years there has been an important effort of the \ac{LHC} community to explore the sensitivity of the \ac{LHC} to \ac{BSM} signatures involving displaced decays and/or long-lived particles (together denoted as \ac{LLP} searches)~\cite{Alimena:2019zri}. As argued in the introduction of this chapter, \ac{LLP} searches can play a crucial role in testing the nature of feebly interacting DM. Both ATLAS and CMS have published a variety of searches for LLPs, exploring many different signatures. Below, we give an overview of the potential displaced signatures and point the reader to the respective CMS and ATLAS analyses (focusing on the analyses performed at a center of mass energy $\sqrt{s}=13$~TeV). As pictures can sometimes summarize information better than a long text, we also illustrate some of these displaced signatures in Fig.~\ref{fig:searches}. In Chapter~\ref{chap:LHC_cons}, we will apply a subset of these searches to an example set of the simplified models of Tab.~\ref{tab:classification} to show the potential reach of \ac{LLP} searches to feebly interacting \ac{DM} models.

\subsubsection{Heavy stable charged particle and R-hadrons}

When a heavy charged mediator has a long enough lifetime ($c\tau > \mathcal{O}(10)$~m), it can cross the detector completely, leaving a track in the detector. This type of signature is usually referred to as a \acf{HSCP}. When the charged particle also carries \ac{QCD} color, the long-lived particle is expected to hadronize, hence one refers to \ac{RH} searches. When a neutral mediator decays outside the detector, it is invisible and such models can hence only be probed by mono-X searches, see Sec.~\ref{sec:DM_detection}.

There are two ways of identifying \acp{HSCP}. Firstly, one can look for the tracks they leave in the tracker of the detector. When a particle traverses the tracker, it loses ionization energy which can by measured in the different layers of the tracker. The amount of ionization energy loss $dE/dx$ is described by the Bethe-Bloch relation, stating that the slower a particle traverses the detector, the larger the ionization loss. Since \acp{HSCP} are in general heavy ($m>\mathcal{O}(100)$~GeV), they travel slower compared to SM particles who have a velocity $\beta \sim 1$. Thus, \acp{HSCP} leave an anomalously large amount of $dE/dx$ in the detector. Secondly, \acp{HSCP} require a longer time to reach the outermost parts of the detector so that \ac{TOF} measurements can also help distinguishing SM background from \acp{HSCP}. Searches for \acp{HSCP} and \acp{RH} have been performed by both CMS~\cite{CMS:2016ybj} and ATLAS~\cite{ATLAS:2018lob,ATLAS:2019gqq} at $\sqrt{s}=13$~TeV, both using $dE/dx$ and \ac{TOF} measurements to infer the mass of potential \ac{HSCP} candidates, and often a combination of the two. \acp{RH} are typically stopped in the \ac{HCAL}, but \acp{HSCP} can potentially survive until the \ac{MS}. This is why often multiple signal regions are constructed, with and without the requirement of a track in the \ac{MS} connected to the track in the inner tracking system.

Recently, the ATLAS collaboration has observed an excess (with global significance of 3.3$\sigma$)~\cite{ATLAS:2022pib} that could correspond to an \ac{HSCP} of mass $m\approx1.4$~TeV. However, when reconstructing $\beta$ using both $dE/dx$ and \ac{TOF} measurements, the ATLAS collaboration found two incompatible values. The $dE/dx$ measurement reconstructed low velocities as one would have expected for an \ac{HSCP}, while the \ac{TOF} measurements are compatible with $\beta\approx1$. Hence, there is still debate about this measurement and it has to be confirmed by other experiments. 

\subsubsection{Disappearing and kinked tracks}
\acp{DT} are tracks that seem to disappear halfway through the inner tracking system of ATLAS and CMS, i.e. they exhibit hits in the innermost layers of the tracker but not in the outermost ones. This signature can arise in models with a charged mediator that has a proper decay length $c\tau_B \sim \mathcal{O}(10-100)$~cm, i.e.~smaller than the size of the tracker, and its decay products cannot be reconstructed, either because they are neutral or they carry low momentum due to e.g. a small mass splitting. Near degenerate mass states occur naturally as a result of a symmetry. Indeed, mass splittings between different components of an electroweak multiplet are often $\mathcal{O}(100)$~MeV due to gauge boson loops~\cite{Cirelli:2009uv}. Hence, \ac{DT} tracks are often a good probe for such models, see e.g.~\cite{Fukuda:2017jmk,Saito:2019rtg}.

Searches for disappearing tracks have been performed at both ATLAS~\cite{ATLAS:2017oal,ATLAS:2022rme} and CMS~\cite{CMS:2018rea,CMS:2020atg}. There is however one major difference between the CMS and ATLAS detector impacting the \ac{DT} search. Both searches require a certain amount of hits in the innermost layers of the tracking system. Before run 2 of the \ac{LHC}, ATLAS installed a new innermost pixel detector having a larger density of layers so that the track can be shorter and still reach the required amount of hits, making it more sensitive to smaller lifetimes compared to the CMS detector. Hence, both detectors probe slightly different lifetimes making them complementary to each other, as we will see in Sec.~\ref{sec:leptophilic@lhc}.

One has to note that if one of the decay products of the \ac{LLP} is reconstructed, we do not have a \ac{DT} topology, but more a \ac{KT} topology where the track of the mother and daughter particle seemingly form one track with a kink occurring at the decay position. No searches on such topology exist yet, however, it might be captured by searches for \ac{DT}. For this to happen, the track of the mother and daughter particle can not be reconstructed as one single track, because then the track would display outer hits in the tracker caused by the daughter particle and hence it will not be disappearing. One can imagine that when the angle between the mother and daughter particle is larger than a certain value $\Delta R_{\rm min}$,\footnote{$\Delta R = \sqrt{\Delta \phi^2 + \Delta \eta^2}$, where $\Delta \phi$ and $\Delta \eta$ are the differences between the azimuthal angle and pseudorapidity respectively.} both tracks will not be reconstructed as one so that the \ac{DT} searches might be sensitive to the \ac{DT} topology. A first theorist study of the potential sensitivity has already been done, see~\cite{Jung:2015boa,Evans:2016zau,Belyaev:2020wok}. In Sec.~\ref{sec:leptophilic@lhc}, we will apply our own methodology on which we elaborate in App.~\ref{app:recast}.

\subsubsection{Displaced leptonic decays}
In cases where the long-lived mediator decays to a lepton within the tracker, one would observe a lepton whose track does not extrapolate back to the primary collision vertex. We refer to this as a \ac{DL}. Searches have been performed for events containing leptons with large impact parameters, i.e.~displaced leptons, by both CMS~\cite{CMS:2016isf,CMS:2021kdm} and ATLAS~\cite{ATLAS:2020wjh}. These searches target topologies with a pair of long-lived mediators each decaying to a lepton, so that three \acp{SR} can be defined, one where a displaced $e^\pm \mu^\mp$ pair is observed and two others where same flavor leptons are observed ($e^\pm e^\mp$ or $\mu^\pm \mu^\mp$). Hence, they are also straightforwardly applicable to cases where the \ac{DM} couples to one lepton flavor only, compared to the earlier CMS searches~\cite{CMS:2014xnn,CMS:2016isf} that were only probing the different flavor lepton pairs. One important difference between the CMS and ATLAS searches for \acp{DL} is their reach in lifetime. The CMS search accepts \acp{DL} with a transverse impact parameter $|d_0|$ as small as 0.1~mm, while the minimal required transverse impact parameter in the ATLAS search is 3~mm. However, the maximally allowed $|d_0|$ in the CMS search is only 100~mm, compared to 300~mm in the ATLAS analysis. Hence the CMS search is more sensitive to smaller lifetimes, and both searches are complementary to each other.

LLPs can potentially also decay to two or more leptons. In this scenario, both leptons can be reconstructed and their tracks will form a vertex coinciding with the decay position of the \ac{LLP}. Special algorithms are developed to reconstruct these displaced vertices in the tracker. The ATLAS collaboration has performed a search~\cite{ATLAS:2019fwx} for such a \ac{DLV} where the invariant mass of the diplepton pair is used to discriminate signal from background.\footnote{The invariant mass of a dilepton pair equals $m_{ll}=\sqrt{(E_1+E_2)^2 - (p_1+p_2)^2}$, where $E_i$ and $p_i$ are the energies and momenta of both leptons. When the \ac{LLP} decays only to this lepton pair, the invariant mass of the dilepton system equals the \ac{LLP} mass.} This has been done in the $\mu^+\mu^-$, $e^+e^-$ and $\mu^\pm e^\mp$ channel. Searches only for displaced dimuon pairs have also been performed at CMS~\cite{CMS:2022qej} and ATLAS~\cite{ATLAS:2018rjc}.

\subsubsection{Displaced all-hadronic decays}
Displaced hadronic decays can be searched for in a variety of ways. Firstly, when a \ac{LLP} decays to quarks, they shower and create a cluster of tracks within the tracking system. This showering process happens very fast, and hence all these tracks will seemingly originate from one point and form a vertex. A search for a \ac{DV} with multiple tracks is therefore a good probe for displaced hadronic decays. ATLAS has performed such a search~\cite{ATLAS:2017tny}, where they require a large amount of MET in order to trigger and select the events used in the analysis, making this search optimal for our \ac{DM} models discussed in Sec.~\ref{sec:models} since they naturally exhibiting a large amount of MET due to the presence of the \ac{DM} candidate. Furthermore, searches for DVs in multijet events have been performed by both CMS~\cite{CMS:2018tuo,CMS:2021tkn} and ATLAS~\cite{ATLAS:2022fag}, where events are selected by a multijet trigger. In the CMS analysis, two DVs are required, while both ATLAS searches require only one.

Searches for \acp{DV} suffer from a large background coming from interactions with the detector material. To tackle this background, both ATLAS searches use a map of the material dense regions of the detector to veto any \ac{DV} coming from such a region. In contrast, the CMS searches require the decay of the \ac{LLP} to happen inside the beam pipe to make sure none of the accepted \acp{DV} originates from a material dense region. Therefore, these CMS searches are less sensitive to hadronically decaying \acp{LLP} with longer lifetimes compared to the ATLAS searches. To remain sensitive to these longer lifetimes, CMS has performed searches for displaced jets where they require a jet reconstructed in the \ac{HCAL} and check if the tracks associated to this jet form a secondary vertex~\cite{CMS:2018qxv,CMS:2020iwv}. For these searches, CMS also used a material veto map to suppress the background from material interactions, and hence they can better probe longer lifetimes than the CMS \ac{DV} analysis. 

Another way to discriminate a prompt from a displaced jet is by making use of the timing capabilities of the \ac{HCAL}, since a jet originating from the decay of a heavy \ac{LLP} arrives later to the \ac{HCAL}. This is often referred to as a Delayed Jet (DJ) signature and has been searched for by the CMS collaboration~\cite{CMS:2019qjk}, using a \ac{MET} trigger to select the events. There is no need to reconstruct the \ac{DV}, making this search strategy more sensitive to even longer lifetimes. In contrast, the time delay for small displacements will not be enough to distinguish between a displaced or a prompt jet. This makes all previously discussed search strategies very complementary as they probe very similar topologies in different lifetime ranges.

CalRatio searches are searches sensitive to models where a neutral \ac{LLP} decays to jets inside the \ac{HCAL} and hence deposits little to none energy in the \ac{ECAL}. These jets can hence be distinguished from QCD jets by looking at the ratio of the energy deposited in the \ac{HCAL} versus the \ac{ECAL}. Another discriminator can be that QCD jets leave many tracks in the tracking system, while our target model does not as the tracks are too displaced to be reconstructed. The latter is often referred to as trackless jets. Searches for such signature are performed by ATLAS~\cite{ATLAS:2019qrr,ATLAS:2022izj,ATLAS:2022zhj}.

Finally, a \ac{LLP} can travel through the detector and decay hadronically inside the \ac{MS}. In this scenario, a \ac{DV} can be reconstructed within the MS only. This topology has already been considered by both ATLAS~\cite{ATLAS:2018tup,ATLAS:2022gbw} and CMS~\cite{CMS:2021juv}, significantly extending the lifetime range in which hadronically decaying LLPs can be probed.  

\subsubsection{Displaced semi-leptonic decays}
Another type of displaced vertex search is performed by ATLAS looking for events containing displaced vertices together with a displaced muon track~\cite{ATLAS:2020xyo}. The search defines two orthogonal signal regions, one containing events triggered by missing energy, the other one containing events triggered by a high $p_T$ muon. Since we are interested in probing the simplified \ac{DM} models outlined in Sec.~\ref{sec:models}, the \ac{MET} signal region is expected to be the most efficient.

\subsubsection{Displaced photonic decays}
If the long-lived mediator decays to a photon plus \ac{DM}, the expected LLP-type signature is a displaced photon plus missing energy. When a photon is emitted from a secondary vertex, it reaches the \ac{ECAL} at a later time compared to a photon produced at the primary vertex. The reason for the delay is twofold. Firstly, there is a difference in trajectory, and secondly, the long-lived mediator traverses the detector at a lower speed. Another distinctive feature of a displaced photon is that it enters the \ac{ECAL} at non-normal impact angles. Precision measurements by the \ac{ECAL} of these impact angles can be used to deduce the direction the photon originates from, and a displaced photon would not point back to the primary vertex, hence, this signature is referred to as non-pointing photons. Both delayed and non-pointing photons have been covered by the CMS~\cite{CMS:2019zxa} and ATLAS~\cite{ATLAS:2022vhr} collaborations.

\section{Collider meets cosmology}
\label{sec:BoundTR}
The large variety of displaced signatures at the \ac{LHC} discussed in Sec.~\ref{sec:searches} can cover a wide range of decay lengths of the mediator in our FIMP models presented in Sec.~\ref{sec:models}. To understand what range of parameters we can probe, it is useful to have a prediction for the lifetime dictated by the relic density requirement. For instance, if the DM candidate $X$ of our FIMP models is produced through the freeze-in mechanism via the decay of its mediator $B$ in a \acf{RD} universe, the expected lifetime can be estimated by making use of Eq.~\eqref{eq:omega_fi},
\begin{align}
    c\tau_B \simeq 3.58 \times 10^3 \, {\rm m} \, \left( \frac{g_B}{2} \right) \, \left(\frac{m_X}{1\, \GeV}\right) \,  \left(\frac{1 \TeV}{m_B}\right)^{2}, \qquad \text{[FI during RD era].}
\label{eq:ctauRD}
\end{align}
For most DM masses ($m_X\gtrsim1$~MeV), the lifetime is so long so that we only expect signatures from the mediator traversing the whole detector. For masses below roughly 1~MeV, assuming $m_B \sim 1~\TeV$, the decay of the mediator happens mainly inside the detector so that displaced signatures can arise. FIMPs can however not be as light as we want, otherwise they would behave similarly to thermal \ac{WDM} and erase small scales structures in the universe, which is in contradiction with observations. Data from the Lyman-$\alpha$ forest put bounds on how warm these relics can be~\cite{Viel:2005qj,Viel:2013fqw,Yeche:2017upn,Irsic:2017ixq,Decant:2021mhj}. In particular, Ref.~\cite{Irsic:2017ixq} put bounds on the \ac{WDM} mass by implementing the velocity distribution into Boltzmann codes to get the linear transfer function and input the latter into dedicated hydrodynamical simulations. As a result, we have the bounds $m_\textsc{wdm}\gtrsim 3.5 \, {\rm keV}$ (conservative) and $m_\textsc{wdm}\gtrsim 5.3 \, {\rm keV}$ (stringent), corresponding to different assumptions on the thermal history of the universe. Refs.~\cite{Boulebnane:2017fxw,Heeck:2017xbu,Kamada:2019kpe,McDonald:2015ljz,Ballesteros:2020adh,DEramo:2020gpr,Decant:2021mhj} used the results of this analysis for thermal \ac{WDM} to obtain mass bounds on non-thermal FIMPs, produced by decays or scatterings in a \ac{RD} era, by comparing the associated linear transfer functions.\footnote{More accurate tools in the form of Lyman-$\alpha$ likelihoods~\cite{Archidiacono:2019wdp} have not yet been applied to freeze-in.} For the case of decays, assuming that the final state particles masses are negligible compared to initial state particle masses, Refs.~\cite{Boulebnane:2017fxw,Ballesteros:2020adh,DEramo:2020gpr,Decant:2021mhj} have obtained bounds in the range $m_{X}\gtrsim 7-16 \, {\rm keV}$. As a guide for the eye, we will make use of the bound $m_{X} > 10 \, {\rm keV}$ as a Lyman-$\alpha$ bound on FIMPs. This bounds narrows the window in mass where we can probe our FIMP models by searching for displaced signatures at colliders other than a highly ionized tracks coming from the mediator.

Searches for displaced new physics can nevertheless still be relevant for FIMP models. Indeed, in Chapter~\ref{chap:alt_cosmo}, we have established that when FI production happens during an \acf{EMD} era, larger decay widths, i.e. smaller lifetimes, are needed to reproduce the same amount of DM compared to \ac{FI} production in a \ac{RD} universe. From Eq.~\eqref{eq:YXMDfinal}, we find an expected decay length of
\begin{align}
    c\tau_B \simeq 1.1 \times 10^8 \, {\rm m} \, \left( \frac{T_R}{m_B} \right)^7 \, \left(\frac{m_X}{1\, \GeV}\right) \,  \left(\frac{1 \TeV}{m_B}\right)^{2}, \qquad \text{[FI during EMD era]}.
\label{eq:ctauEMD}
\end{align}
Hence, due to the strong $(T_R/m_B)^7$ dependence, larger DM masses can be reconciled with lifetimes of the order of a few centimeter while still reproducing the correct relic density, as long as $T_R<m_B$.

Searches for \acp{LLP} have unfortunately not yet observed new displaced physics. However, in the case a discovery occurs, the measurement of kinematic observables can reveal properties of the \ac{DM} model and cosmological history, through the connection provided by the \ac{DM} production mechanism. Assuming that we can measure the masses from reconstructing the collider event in combination with information from measurements of the $B$ production cross section,\footnote{see e.g.~\cite{Bae:2020dwf} for new methods to read off the mass spectrum involved in displaced events.} the observed decay length for $B \to A_\textsc{sm} X$ will directly pinpoint the reheating temperature $T_R$ if we require that $X$ makes up all the observed DM. If we are less demanding and we require that $X$ is just a subdominant component, it would at least provide an upper bound on $T_R$. 

Reconstructing the complete spectrum at collider is rather challenging, so we should explore the less optimistic scenario where we cannot reconstruct the \ac{DM} mass. As mentioned above, we know from Lyman-$\alpha$ measurements that the DM mass cannot be below roughly 10~keV.\footnote{It is worth keeping in mind that the freeze-in production via scatterings or decays during an EMD epoch might also affect these bound, see e.g.~\cite{McDonald:2015ljz}.} So even if we cannot reconstruct the \ac{DM} mass, for a known $B$ mass and decay length we can still determine a specific value of the reheating temperature $T_R$ once we set $m_X$ to the lowest value allowed by Lyman-$\alpha$ forest data. The \ac{DM} mass cannot get any lower than that, and for such a smallest allowed value we need the smallest amount of dilution after freeze-in because $\rho_X \propto m_X$. And the smallest amount of dilution corresponds to the largest value of $T_R$, hence this procedure provides an upper bound on $T_R$.

Finally, if we consider a non-renormalizable interaction, \ac{DM} production can be controlled by scatterings. If this is the case, the amount of \ac{DM} grows with $T_R$. If we fix the \ac{DM} mass to the smallest value compatible with the Lyman-$\alpha$ bound, for a given $m_B$ and $\tau_B$ we can derive an upper bound on $T_{R}$, which is valid for any other (higher) \ac{DM} mass.  Thus, even when the dominant  production channel would be freeze-in through scatterings, instead of decays, colliders could indirectly provide hints about cosmology.

To summarize, displaced events at colliders could indirectly provide constraints on the cosmological history of the early universe under the assumption that we have observed a DM signal and that this DM candidate constitutes all the DM in the universe.

\chapter{Feebly interacting dark matter at the LHC}
\label{chap:LHC_cons}

Due to the small DM-SM interactions arising in FIMP DM models, direct and indirect detection are basically rendered useless in probing this type of models.\footnote{There are some exceptions, see e.g.~\cite{Hambye:2018dpi}.} Hence, searches for \acfp{LLP} can play a key role in the probing DM if the latter is feebly interacting with the SM, and they can even provide vital information about the history of our universe, as we have established in Sec.~\ref{sec:BoundTR}. We have also seen in Sec.~\ref{sec:searches} that a whole variety of displaced signatures can arise from the decay of an \ac{LLP}. We collect in Tab.~\ref{tab:searches} a selection of existing LLP searches performed by ATLAS and CMS employing data from collisions with center of mass (c.o.m.)~energy of $13$ TeV, and we assign a label to each of them. This chapter is based on the analysis performed in Ref.~\cite{Calibbi:2021fld} (except for Sec.~\ref{sec:lepto@lhc_cdfo}, which is based on Ref.~\cite{Junius:2019dci}), and therefore only searches available at that time are included in Tab.~\ref{tab:searches} (with exception of the displaced lepton analysis by CMS). The selection is dictated by the attempt to have a full coverage over the possible SM final states with LLP signatures, and, when available, searches including \acf{MET} are preferred. Tab.~\ref{tab:searches} shows how ATLAS and CMS are spanning already a quite wide landscape of possible LLP-type signatures, involving leptons, gauge bosons, charged tracks, and also MET.

\renewcommand{\arraystretch}{1.2}
\begin{table}[t]
	\centering
	\resizebox{\textwidth}{!}{
	\begin{tabular}{ | c | c | c | c | c | }
		\hline
		\multirow{2}{2cm}{\textbf{Signature}} &  \multirow{2}{2.1 cm}{\centering \textbf{Exp.~\&~Ref.}} & \multirow{2}{1.5cm}{\centering 
		$\boldsymbol{\mathcal{L}}$} &   \textbf{Maximal} & \multirow{2}{1.5cm}{\centering \textbf{Label}} \\ 
		& & &  \textbf{sensitivity} & \\
		\hline
		\hline
		R-hadrons  & CMS~\cite{CMS:2016ybj} &  $12.9~\text{fb}^{-1}$ & \multirow{2}{2cm}{\centering $c\tau\gtrsim 10$~m} & RH\\
		Heavy stable charged particle & ATLAS~\cite{ATLAS:2019gqq} & $36.1~\text{fb}^{-1} $ &  & HSCP\\
                \hline
		\multirow{2}{4cm}{\centering Disappearing tracks}  &ATLAS~\cite{ATLAS:2017oal}&  $36.1~\text{fb}^{-1}$ &  $c\tau \approx 30$~cm & \multirow{2}{2cm}{\centering DT} \\
		& CMS~\cite{CMS:2018rea,CMS:2020atg} &  $140~\text{fb}^{-1} $& $c\tau \approx 60$~cm & \\
                \hline
		\multirow{2}{4cm}{\centering Displaced leptons} & \multirow{1}{1.5cm}{\centering CMS~\cite{CMS:2021kdm} }  &  $118~\text{fb}^{-1}$ & \multirow{2}{1.8cm}{\centering $ c\tau\approx 3$~cm}&\multirow{2}{2cm}{\centering DL} \\
        & \centering ATLAS~\cite{ATLAS:2020wjh}&  $139~\text{fb}^{-1}$ & & \\
        \hline
		Displaced vertices + MET & ATLAS~\cite{ATLAS:2017tny} &  $32.8~\text{fb}^{-1}$ & $c\tau \approx 3$~cm & DV+MET \\
		\hline
		Delayed jets + MET  &CMS~\cite{CMS:2019qjk} &  $137~\text{fb}^{-1}$ & $c\tau \approx 1-3$~m & DJ+MET \\
		\hline
		Displaced vertices + $\mu$ &ATLAS~\cite{ATLAS:2020xyo} &  $136~\text{fb}^{-1}$ & $c\tau \approx 3$~cm& DV+$\mu$ \\
			\hline
		Displaced dilepton vertices &ATLAS~\cite{ATLAS:2019fwx} & $32.8~\text{fb}^{-1}$ & $c\tau \approx 1-3$~cm& DLV \\
		\hline
		Delayed photons & CMS~\cite{CMS:2019zxa} &  $77.4~\text{fb}^{-1}$ &  $c\tau \approx 1$~m & D$\gamma$ \\
		\hline
	\end{tabular}}
	\caption{Searches for \acf{LLP} at the LHC experiments performed on $\sqrt{s} = $ 13 TeV run data. See text and App.~\ref{app:recast} for details. The maximal sensitivities in the proper decay length refers to the models employed in the original Refs.~of the second column to interpret the searches.} 
	\label{tab:searches}
\end{table}

A large MET component is clearly a smoking gun signature of DM. Only two of these experimental analyses explicitly require missing energy in the final state. Nevertheless, the other searches are often sensitive to our simplified DM models presented in Sec.~\ref{sec:models} due to a rather inclusive event selection. We remark that missing energy is not typically required in LLP searches since the SM background is already very much suppressed by the displacement selection. However, the presence of a large MET in the targeted BSM signature could also be exploited for triggering the relevant events. Hence for the purpose of detecting DM models with LLP signatures, the MET requirement could be beneficial.

We can confront the minimal simplified model setup presented in Tab.~\ref{tab:classification} to the series of LHC searches listed in Tab.~\ref{tab:searches}. The result of this comparison is illustrated in Tab.~\ref{tab:models-searches}. We distinguish models involving the first and second generation fermions ($\ell$ and $q$) from models involving the third generation ($\tau$ and $t$), because of their different collider signatures. This table can serve as a tool in the search for feebly interacting DM models, as it gives a quick and visual overview of which signatures can arise in which models. It illustrates which signatures can probe the most amount of models, but also that it is of great importance to keep probing a large variety of signatures as to stay sensitive to as much \ac{FIMP} models as possible, since no search will be able to cover all these models. In case of a discovery of new displaced physics, it can also serve as a tool to help pinpointing down the exact model responsible for the observed signal. 

We also include the so-called ``kinked track'' (KT) signature in the last column of Tab.~\ref{tab:models-searches}. No dedicated and fully experimental analysis has been performed addressing this type of signature. We nevertheless include it to highlight the range of physics targets that such a search would have. Moreover, constraints on KT signatures can be obtained by re-interpreting the disappearing track searches as we already mentioned in Sec.~\ref{sec:searches}. 

For our phenomenological discussion, we select within the model classes of Table~\ref{tab:classification} the following three that, we argue, represent an exhaustive set of simplified FIMP models:  
\begin{itemize}
 \item $\mathcal{S}_{\ell_R\chi}$, that features a scalar mediator with the quantum numbers of a right-handed lepton, coupling to DM through a renormalizable operator;
 \item  $\mathcal{F}_{t_R\phi}$, another example of renormalizable cubic interaction but involving strongly-interacting fields, and the third generation (namely, a right-handed top);
 \item $\mathcal{F}_{W\chi}$, an explicit case of freeze-in controlled by a non-renormalizable operator, involving the $SU(2)_L$ field strength, hence a fermion mediator in a triplet representation.
\end{itemize}
Our selection includes renormalizable interactions (with first/second and third generation) and a non-renormalizable interaction. We perform a detailed analysis of the collider implication for early-universe cosmology for these models. As shown in Sec.~\ref{sec:fi_emd}, DM freeze-in production for renormalizable interactions will be IR dominated. In addition, for simplicity, we will assume  that the DM only couples to one of the lepton/quark flavors at a time. Notice that even for a mediator coupling to multiple families, we do not expect any relevant constraint from flavor violation because of the smallness of the couplings considered, see e.g.~\cite{Kopp:2014tsa} for a leptophilic scenario of such kind. The elected models are highlighted in green in Tab.~\ref{tab:models-searches}. From this, we can see that we can exploit all the available signatures presented in Tab.~\ref{tab:searches}.

As the main focus of this chapter is the interplay between LLP searches at the LHC and DM production in the early universe, we focus on production mechanism for feebly interacting dark matter that reproduce the observed amount of DM and give rise to decay lengths of the typical size of the detectors at the LHC, i.e. $c\tau \sim$ 1mm to 1m. As discussed already in Sec.~\ref{sec:BoundTR}, models in Tab.~\ref{tab:models-searches} where the DM candidate is produced through FI in a RD universe do not give rise to such lifetimes. In contrast, when the FI production happens during an EMD epoch, such lifetimes can be obtained by varying the reheating temperature. Macroscopic lifetimes can also be obtained when DM is produced through CDFO. Hence, we mainly focus in this chapter on the latter two production mechanisms. 

\renewcommand{\arraystretch}{1.1}
\begin{table}[t]
	\centering
	\begin{tabular}{|c||c|c|c|c|c|c|c|c|c||c|c|c|}
		\hline
		& DV & DJ & DV & & & & & & &  \\
		\textbf{Label} & + & + & + & DL & DLV & D$\gamma$ & DT & RH & HSCP &  KT \\
		& MET & MET & $\mu$ & & & & & & &  \\
		\hline
		\hline
		\rowcolor{Green}
		$\mathcal{F}_{\ell\phi} \ \& \ \mathcal{S}_{\ell\chi}$ &  &  & & \cmark & &  &  &  & \cmark  & \cmark \\
		\hline
		$\mathcal{F}_{\tau\phi} \ \& \ \mathcal{S}_{\tau\chi}$ & \cmark & \cmark &  & \cmark & & & & & \cmark  & \cmark \\
		\hline
		$\mathcal{F}_{q\phi} \ \& \ \mathcal{S}_{q\chi}$ & \cmark & \cmark & & & & & & \cmark &  &  \\
		\hline
		\rowcolor{Green}
		$\mathcal{F}_{t\phi} \ \& \ \mathcal{S}_{t\chi}$ & \cmark & \cmark & \cmark & \cmark & & & & \cmark &  & \\
		\hline
		$\mathcal{F}_{G\chi}$ & \cmark & \cmark & & & & & & \cmark &  & \\
		\hline
		\rowcolor{Green}
		$\mathcal{F}_{W\chi}$ & \cmark & \cmark & \cmark & \cmark & \cmark & \cmark & \cmark & &  & \cmark \\
		\hline
		$\mathcal{S}_{H\phi} \ \& \ \mathcal{F}_{H\chi}$ & \cmark & \cmark & \cmark & \cmark & \cmark & & \cmark & &  & \cmark \\
		\hline
	\end{tabular}
	\caption{Sensitivity of the LHC searches, shown in Tab.~\ref{tab:searches} and discussed in App.~\ref{app:recast}, to our simplified models as labeled in Tab.~\ref{tab:classification}. The fermion $\psi_\textsc{sm}$ of Tab.~\ref{tab:classification} can be either a light lepton $\ell\in \{ e,\mu \}$, a tau lepton $\tau$, a light quark $q\in \{ u,d,s,c,b \}$ or a top quark $t$. Fermions can be either left-handed ($L$) or right-handed ($R$), a choice that has little impact on the final state but it can affect LHC production of the mediator $\Phi_B/\Psi_B$. $\mathcal{F}_{W\chi}$ and $\mathcal{F}_{G\chi}$ refer to models of the class $\mathcal{F}_{F\chi}$ where we consider the $SU(2)_L$ (coupling to electroweak gauge bosons) and the $SU(3)_c$ (coupling to gluons) field strength, respectively. The models we study in this work belong to the classes highlighted in light green.}
        \label{tab:models-searches}
\end{table} 
 
The interplay between LHC and cosmology has already been considered by Ref.~\cite{Belanger:2018sti} for models similar to our category $\mathcal{F}_{\ell\phi}$ (thus with scalar instead of fermion DM) and $\mathcal{F}_{q\phi}$ (hence with first and second generation quarks involved, instead of the top).\footnote{Another difference is that we do not make the instantaneous reheating approximation. This would not change dramatically the overall picture presented in Ref.~\cite{Belanger:2018sti} for this class of models.} Also, the model $\mathcal{S}_{t_R\chi}$ was already studied in the context of freeze-in in~\cite{Garny:2018ali}, but without the special emphasis on the role of $T_{R}$ and on displaced events that we put here. In addition, R-hadron searches applied to a model $\mathcal{F}_{G\chi}$ have been considered in~\cite{Davoli:2018mau} but in the context of freeze-out through co-annihilations with feeble conversion processes (see also~\cite{Brooijmans:2020yij} in the case of $\mathcal{S}_{b\chi}$). Finally, the \ac{CDFO} regime for $\mathcal{S}_{b_R\chi}$ and $\mathcal{S}_{t_R\chi}$ have been studied in~\cite{Garny:2017rxs,Garny:2021qsr} and~\cite{Garny:2018icg} respectively.

\section{Leptophilic scenario}
\label{sec:leptophilic@lhc}
For the first simplified model we will discuss, we augment the \ac{SM} field content by a new charged scalar $\Phi_B$ with the quantum numbers of a right-handed lepton and a Majorana fermion singlet $\chi$: this is the $\mathcal{S}_{\ell_R\chi}$ scenario from Tab.~\ref{tab:classification}. Both extra fields are odd under an unbroken $\mathbb{Z}_2$ symmetry. For simplicity, we choose to couple the new fields to right-handed muons only, i.e.~$\ell_R=\mu_R$. The Lagrangian reads
\begin{equation}
{\cal L}_{\mathcal{S}_{\ell_R\chi}} \supset \frac{1}{2} \bar{\chi} \gamma^{\mu} \partial_{\mu}
\chi-\frac{m_{\chi}}{2} \bar{\chi} \chi+ (D_{\mu}\Phi_B)^\dagger
\ D^{\mu}\Phi_B -m_\phi^2 |\Phi_B|^2 \ - \ \lambda_{\chi} \Phi_B \bar \chi
\mu_R \ + \ \text{h.c.}\,,
\label{eq:lagr}
\end{equation}
where $\lambda_\chi$ is a dimensionless Yukawa coupling. We have already explored various production mechanisms of this model in Sec.~\ref{sec:DM_prod_lepto}. Here, we discuss the collider phenomenology of two production regimes. First, we consider the case when $m_X \lesssim m_B$, so that displaced signatures arise in a regime of couplings $\lambda_\chi \sim \mathcal{O}(10^{-6}-10^{-7}$) where \ac{DM} is produced through conversion driven freeze-out. In the scenario where $m_X \ll m_B$, \ac{DM} is overproduced by the freeze-in mechanism for similar values of $\lambda_\chi$ giving rise to macroscopic lifetimes of the mediator $c\tau_\phi$. However, by assuming \ac{DM} production happens in an \ac{EMD} era, the \ac{DM} abundance will get diluted due to entropy production so that one can reproduce the observed relic abundance. 

Besides the terms in Eq.~\eqref{eq:lagr}, the mediator $\Phi_B$ can also couple to the Higgs field via $\lambda_H |\Phi_B|^2 |H|^2$. As we have shown in Sec.~\ref{sec:DM_prod_lepto}, the exact value of $\lambda_H$ greatly influences the DM relic abundance computation in the conversion driven freeze-out regime, since it partly defines when the $\Phi_B$ chemically decouples. Since freeze-in happens before this time, the freeze-in calculations do not depend on $\lambda_H$ and hence, we will not consider it in this regime.

The mediator $\Phi_B$ can be pair produced at colliders due to its electroweak gauge interactions. The production cross section at \ac{NLO} can be obtained by interpolating the tables available at~\cite{slepton-xsec}. We have checked that the NLO cross section is well reproduced by the \ac{LO} cross section obtained with {\tt MadGraph5\_aMC@NLO}~\cite{Alwall:2014hca} times a K-factor of approximately~1.3. Constraints from \ac{LLP} searches at the \ac{LHC} will be discussed below for both scenarios. 

\subsection{Conversion-driven freeze-out}
\label{sec:lepto@lhc_cdfo}
\begin{figure}
	\centering
	\subfloat[$\lambda_H = 0.01$]{\includegraphics[width=0.45\textwidth]{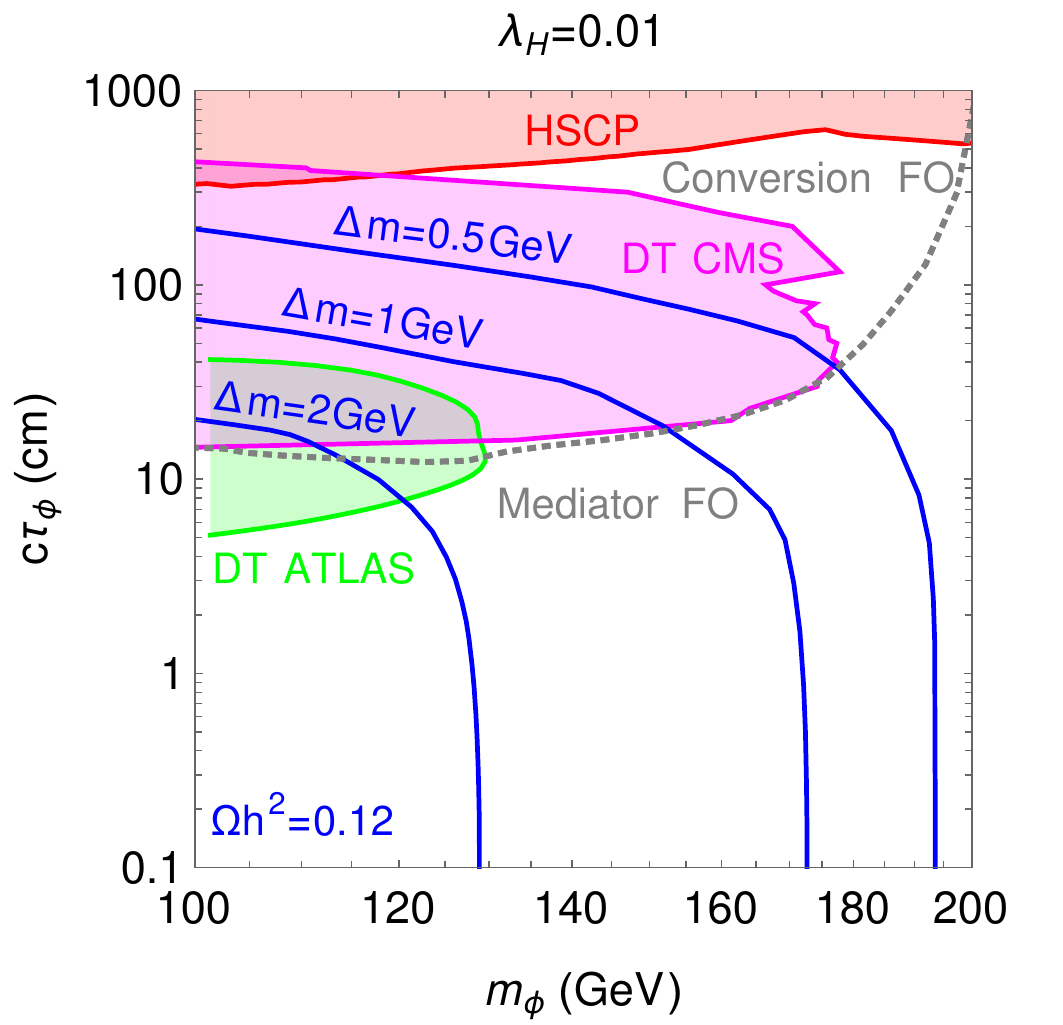}}
	\hspace{0.05\textwidth}
	\subfloat[$\lambda_H = 0.1$]{\includegraphics[width=0.45\textwidth]{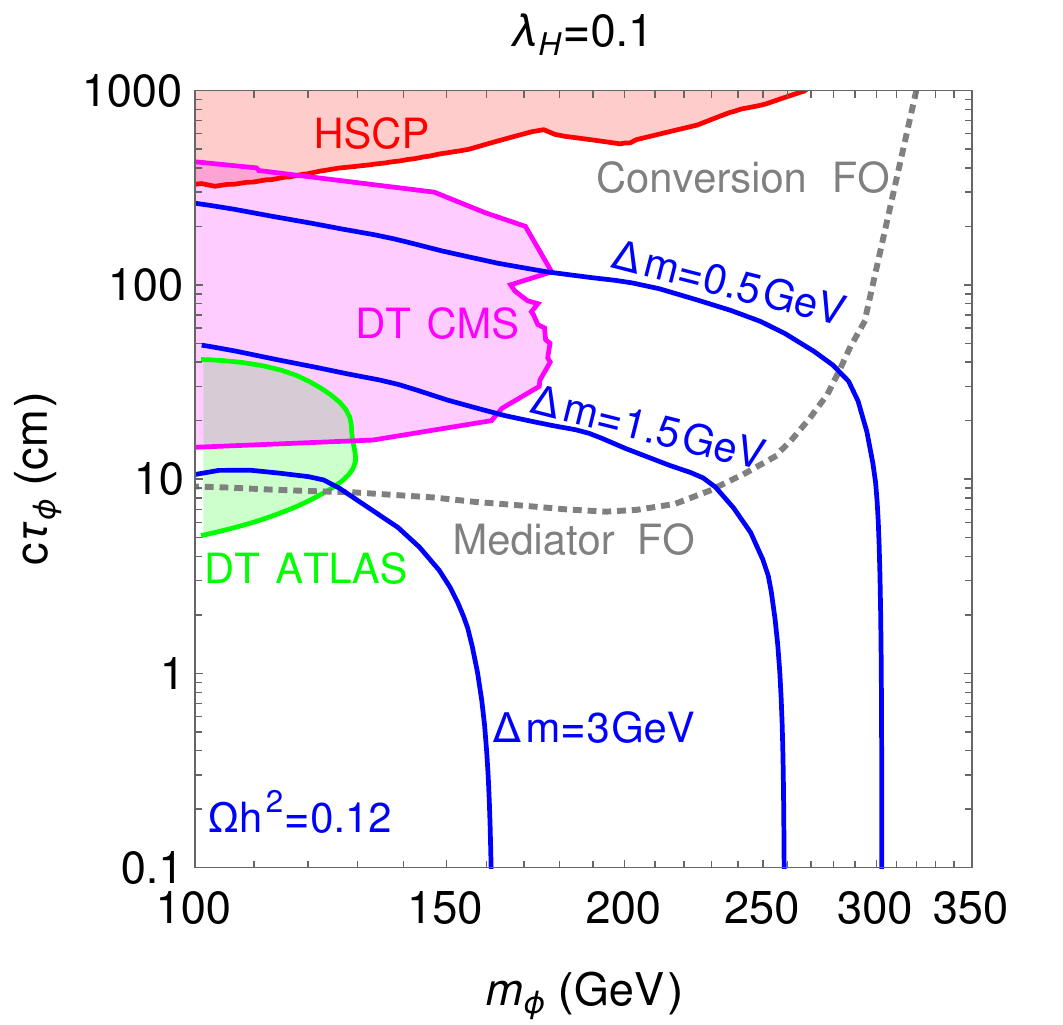}}
	\vspace{0.005\textheight}
	\subfloat[$\lambda_H = 0.5$]{\includegraphics[width=0.45\textwidth]{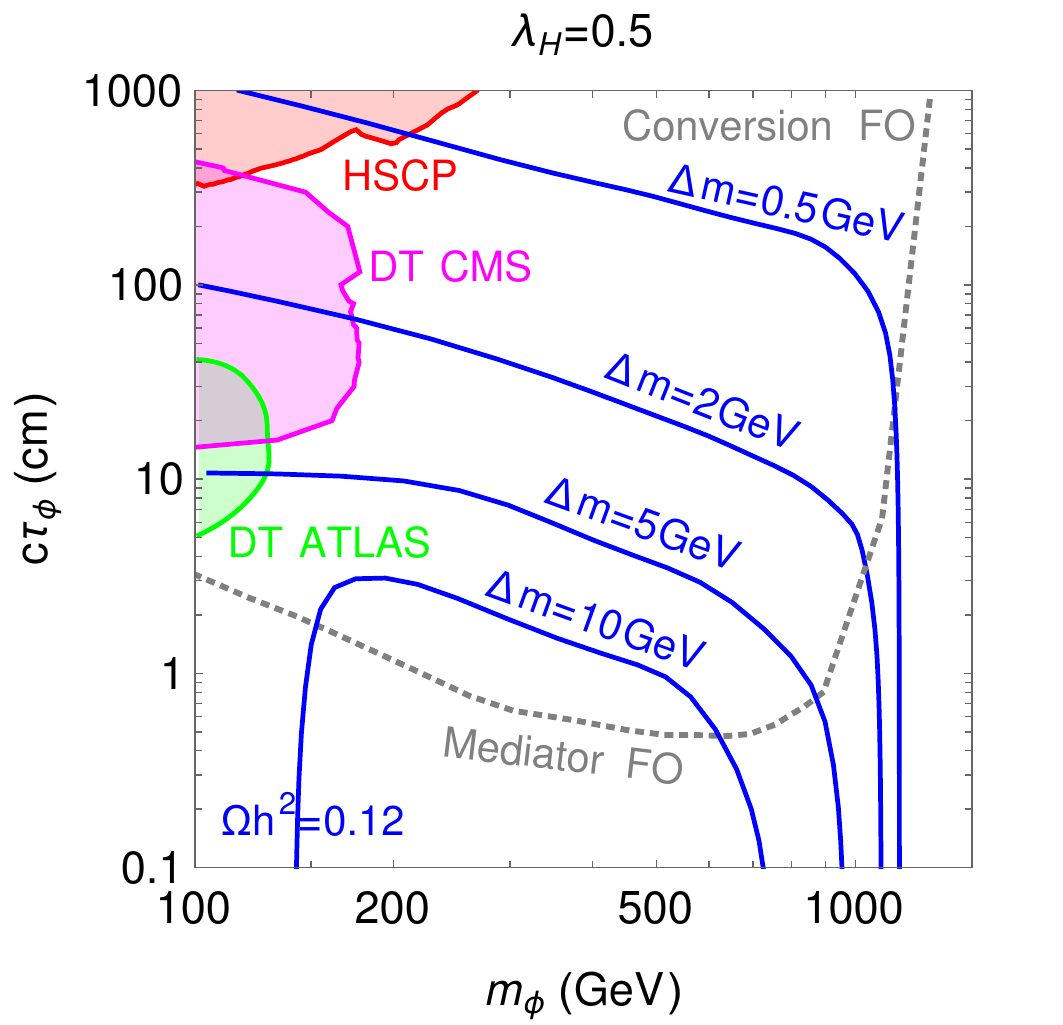}}
	\caption{ Proper life time of the mediator as a function of
          the dark matter mass. Each panel corresponds to a different
          value of $\lambda_H$ contributing to the mediator
          annihilation. The blue contours reproduce the correct relic
          abundance in the $(m_\chi, c\tau_\phi)$ plane for different
          values of the mass-splitting $\Delta m$.  The gray dotted
          line separates the conversion driven (top) from the mediator
          annihilation driven (bottom) freeze-out regime. The excluded
          regions resulting from heavy stable particle (HSCP, red
          region) and disappearing tracks (DT, green region)
          searches at LHC are also shown.}
	\label{fig:ctauLimits}
\end{figure}

In this section we discuss the collider constraints on the conversion driven regime. The small Yukawa couplings necessary to reproduce the correct DM relic abundance shown in Fig.~\ref{fig:ParamPlot} (blue contours), together with the mass compression, imply a small decay width of the charged mediator through the process $\Phi_B \rightarrow \chi l$. For $\Delta m\ll m_\chi$, the decay rate for $\Phi_B \rightarrow \chi l$ reduces at first orders in $\Delta m$ to
\begin{equation}
  \Gamma_{\phi} \approx \frac{\lambda_{\chi}^2 \Delta m^2}{4\pi m_{\chi}} \left[ 1- \frac{2 \Delta m}{m_{\chi}}\right]
  \sim \frac{1}{25 \text{ cm}} \left(\frac{\lambda_{\chi}}{10^{-6}}\right)^2 \left(\frac{\Delta m}{1 \text{ GeV}} \right)^2 \left( \frac{100 \text{ GeV}}{m_{\chi}} \right)
\label{eq:decaysmalldm}
\end{equation}
when neglecting the lepton mass ($m_l \ll m_{\chi}, m_{\phi}$). The testable signature at colliders for this class of model is hence the pair production of charged mediators through gauge interaction and possibly their subsequent macroscopic decay into DM plus leptons.\footnote{The mediator pair production through s-channel Higgs adds up to the Drell Yan production considered here. We checked that this extra contribution does not qualitatively affect the constraints shown in Fig.~\ref{fig:ctauLimits} and hence, we conservatively neglect it.}

The overall result of our investigations is shown in Fig.~\ref{fig:ctauLimits} in the mediator proper decay length $c \tau_\phi= 1/\Gamma_\phi$ versus mediator mass plane. The blue contours give rise to $\Omega h^2=0.12$ for fixed values of the mass splitting.  We consider mediator masses with $m_\phi>100$ GeV as constraints from the \ac{LEP} collider typically exclude such charged particle mass~\cite{OPAL:2005pwy}. The gray dotted line indicates the region where we go beyond CE regime. Above the gray line, the DM abundance can be accounted for through conversion driven freeze-out while below the gray line, mediator annihilation driven freeze-out is at work. In the latter case, the relic abundance becomes essentially independent of the coupling $\lambda_\chi$ and is determined by a combination of the parameters $\lambda_H, m_\phi$ and $\Delta m$. For fixed values of the latter parameters, several values of $\lambda_\chi$ (or equivalently $c \tau_\phi$ given Eq.~(\ref{eq:decaysmalldm})) can thus account for the same DM abundance. As a result the blue contours become vertical in the bottom part the plots.  Notice that for e.g. the $\lambda_H=0.5$ case the inverse-$U$-shape of the blue contour at $\Delta m= 10$ GeV is to be expected given the form of the blue and green contours of the Fig.~\ref{fig:ParamPlot}. In particular the green contours tell us that the right abundance can be obtained at $\Delta m= 10$ GeV in the mediator annihilation freeze-out regime for two dark matter masses: $m_\chi\approx 150$ and 700 GeV. This is indeed what can be inferred from the lower region of the $\lambda_H=0.5$ plot of Fig.~\ref{fig:ctauLimits}. 

Further constraints can be derived exploiting the long lifetime of the mediator. The charged track limit derived from CMS search~\cite{CMS:2016ybj} for \acp{HSCP} is displayed in red in Fig.~\ref{fig:ctauLimits} and constrains long lifetimes ($c \tau > 1$m), as we expect since the mediator has to travel through the full detector to be captured by the \ac{HSCP} search. Details of the recasting method are discussed in App.~\ref{app:HSCP}. As can be seen in Tab.~\ref{tab:searches}, besides CMS, ATLAS also provides a search for a \ac{HSCP} signature performed at a larger data set ($\mathcal{L}=36.1$fb$^{-1}$ compared to $\mathcal{L}=12.9$fb$^{-1}$). However, at small \ac{HSCP} masses ($\mathcal{O}$(100-200GeV)), the CMS search is more constraining due to the fact that it defines its \acp{SR} so that every \ac{HSCP} fits in at least one, while the \acp{SR} for the ATLAS search are defined so that only \acp{HSCP} with a reconstructed mass above a certain threshold (depending on the type of particle) are accepted. Hence, the ATLAS search looses sensitivity for \acp{HSCP} with masses around 200~GeV, and we choose not to show it in Fig.~\ref{fig:ctauLimits}.

For a moderate decay length of the mediator, the relevant signature is covered by the \ac{DT} searches since the final state particles are not visible. Indeed, the dark matter is neutral and hence invisible, and due to the small mass splitting, the lepton will be too soft to reconstruct. The purple region shows the region excluded by the CMS analysis~\cite{CMS:2018rea}, while the bound obtained from the ATLAS search~\cite{ATLAS:2017oal} is displayed in green. The reinterpretation of these analyses, originally performed for long-lived charginos in supersymmetric models, for our DM model is discussed in App.~\ref{app:DT}. Note that, as already mentioned in Sec.~\ref{sec:searches}, ATLAS and CMS provide a complementary constraints on the lifetime of the mediator due to a different performance of the inner layers of the tracking system. Indeed, the CMS search is most efficient at lifetimes $c \tau_\phi \approx 1$~m, while the ATLAS search performs best around $\ctau_\phi\approx20$~cm.

\begin{figure}
	\centering
	\includegraphics[width=0.62\textwidth]{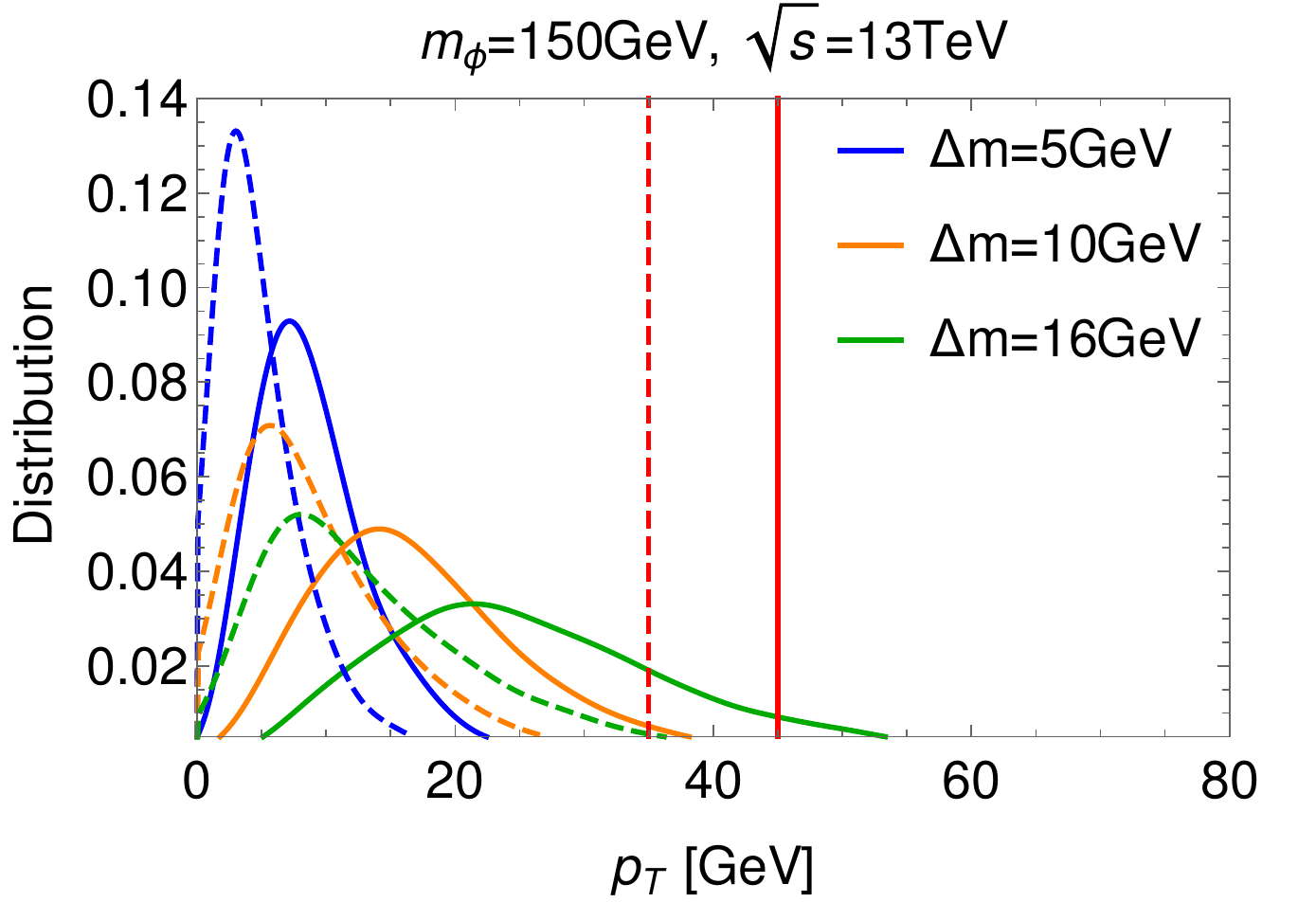}
	\caption{The $p_{T}$-distribution of the leading (subleading) muons denoted by the solid (dashed) lines for the case when $m_{\phi}=$ 150 GeV and different values of the mass splitting. The red vertical solid (dashed) line denotes the cut on the $p_T$ of the leading (subleading) muon that is made in the search \cite{CMS:2021kdm} at 35~GeV (45~GeV) for $\sqrt{s}=$ 13 TeV .}
	\label{fig:pTDistr}
\end{figure}

There is actually a significant portion of the parameter space where the decay of the charged mediators has a proper lifetime of a few centimeters or less, see Fig.~\ref{fig:ctauLimits}.  In the latter case, the final topology of the signal includes a pair of displaced leptons and a pair of dark matter particles. If the leptons can be reconstructed, this final state could be constrained by displaced lepton searches~\cite{ATLAS:2020wjh,CMS:2021kdm}. However, the small mass splitting between the mediator and \ac{DM} will give rise to a very soft spectrum of these displaced leptons making them difficult to reconstruct. In Fig.~\ref{fig:pTDistr}, we show the transverse momentum ($p_T$) spectrum (obtained with {\tt Madgraph}~\cite{Alwall:2014hca,Alwall:2011uj}) and see that only a small fraction of the leptons pass the $p_T$ requirements of~\cite{CMS:2021kdm} denoted by the vertical line. In this perspective, it would be interesting to explore how much the minimal $p_T$ cut on the leptons could be reduced in displaced lepton searches, as also suggested in~\cite{Filimonova:2018qdc,Blekman:2020hwr}.

\subsection{Freeze-in during EMD era}
\label{sec:lepto@lhc_emd}
\begin{figure}[!t]
	\centering
	\subfloat[]{\includegraphics[width=0.47\textwidth]{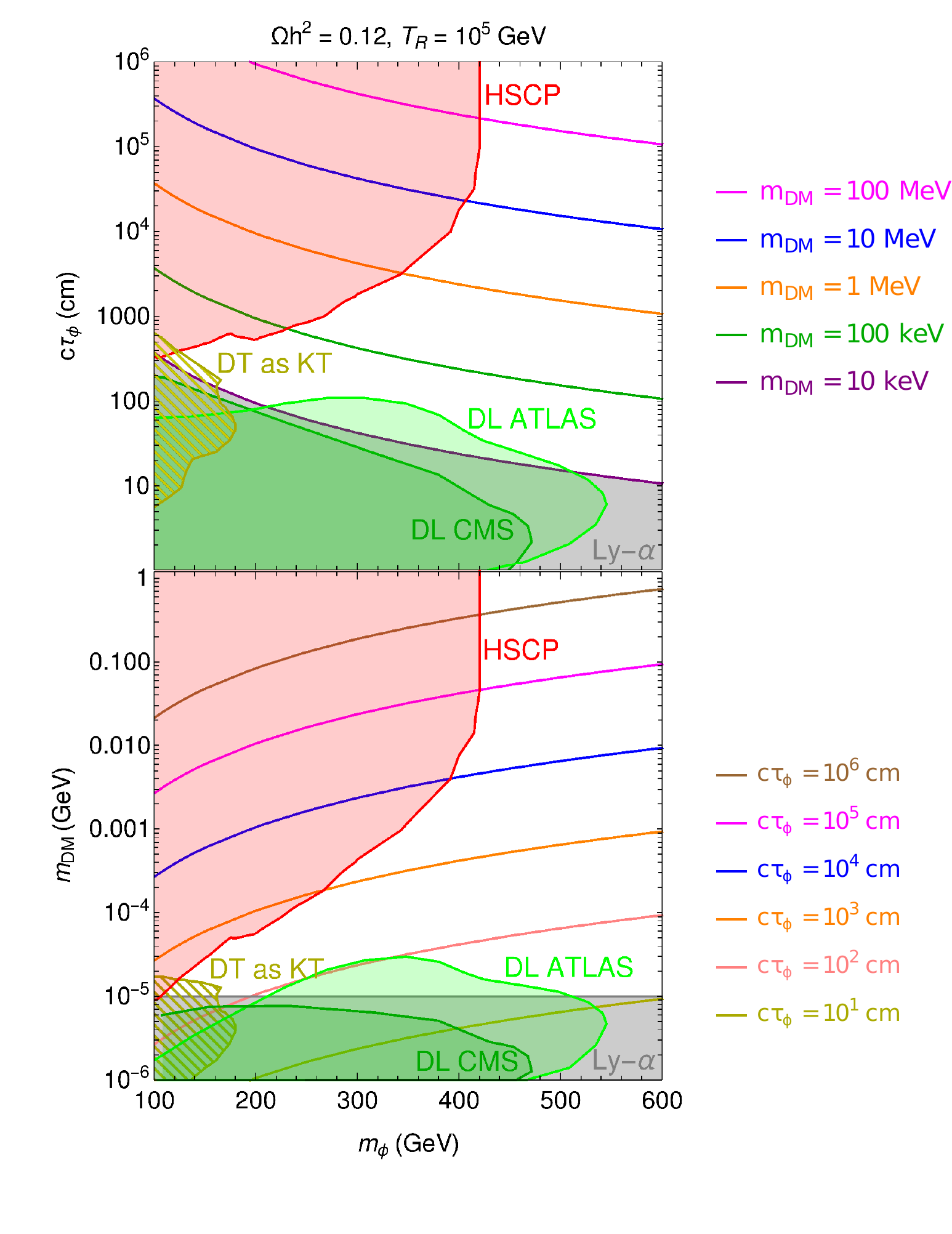}}
	\hspace{0.03\textwidth}
	\subfloat[]{\includegraphics[width=0.47\textwidth]{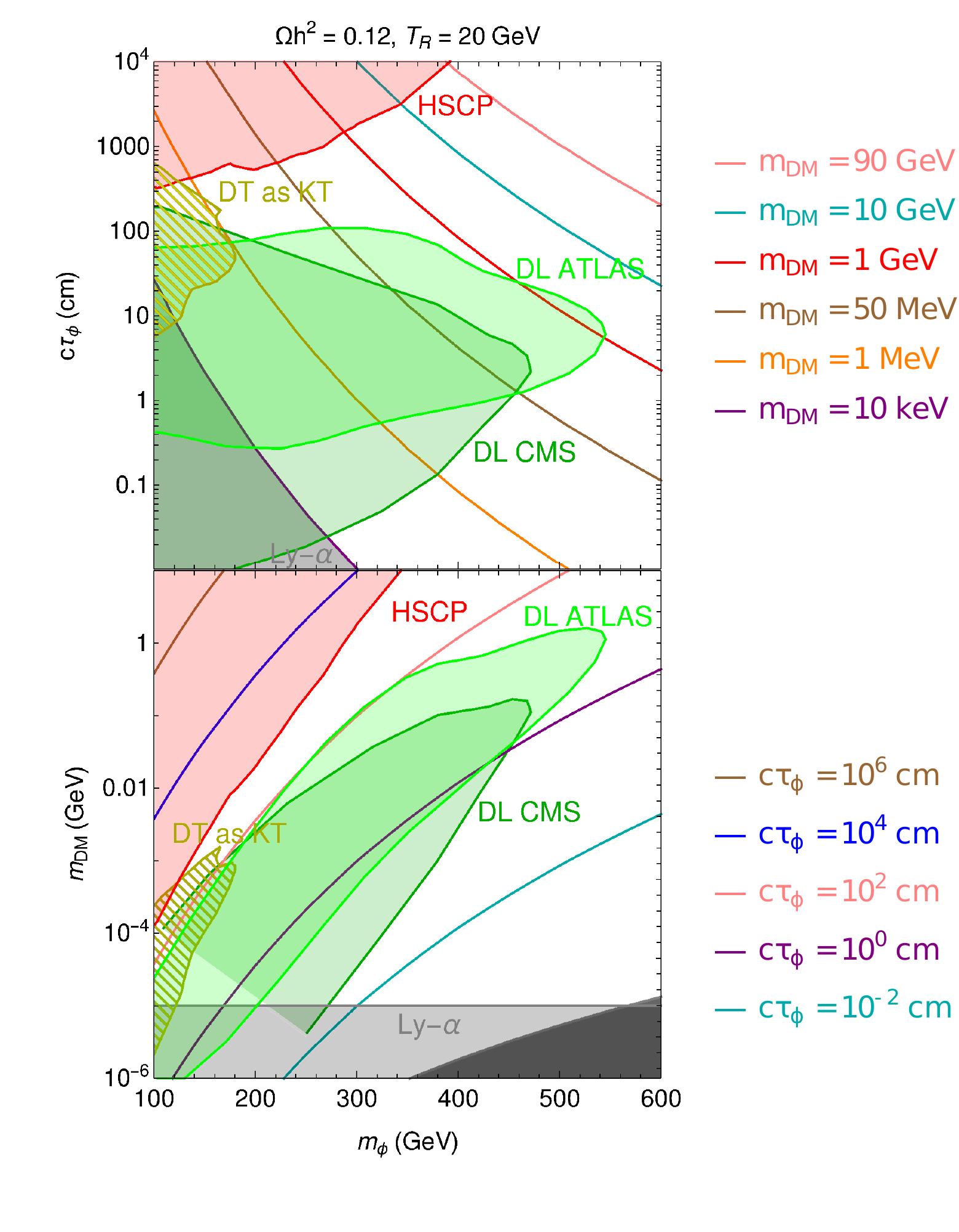}}
	\caption{Leptophilic DM with $T_{R} = 10^5$~GeV (left) and $T_{R} = 20$~GeV (right). In the top panels, solid lines in the plane $(m_\phi,\,c\tau_\phi)$ reproduce the observed DM relic density for the associated $m_{\rm DM} = m_\chi$. Similarly, in the bottom panel we show relic density contours for the proper mediator decay length in the plane $(m_\phi,\,m_{\rm DM})$. We identify regions excluded by HSCP searches from CMS and ATLAS (HSCP - red) and by displaced same-flavor lepton searches by ATLAS (DL ATLAS - light green) and CMS (DL CMS - dark green). The possible reach from our recasting of disappearing track searches in the case of a kinked track signature (DT as KT) is shown as a dark yellow densely hatched region. We denote with a light gray area regions excluded by the Lyman-$\alpha$ bound $m_{\rm DM}\gtrsim 10$~keV, see text for details. Finally, in the dark gray region DM thermalize in the early universe and therefore cannot be produced via FI.}
	\label{fig:lepto2}
\end{figure}

\begin{figure}[!t]
	\centering
	\subfloat[]{\includegraphics[width=0.47\textwidth]{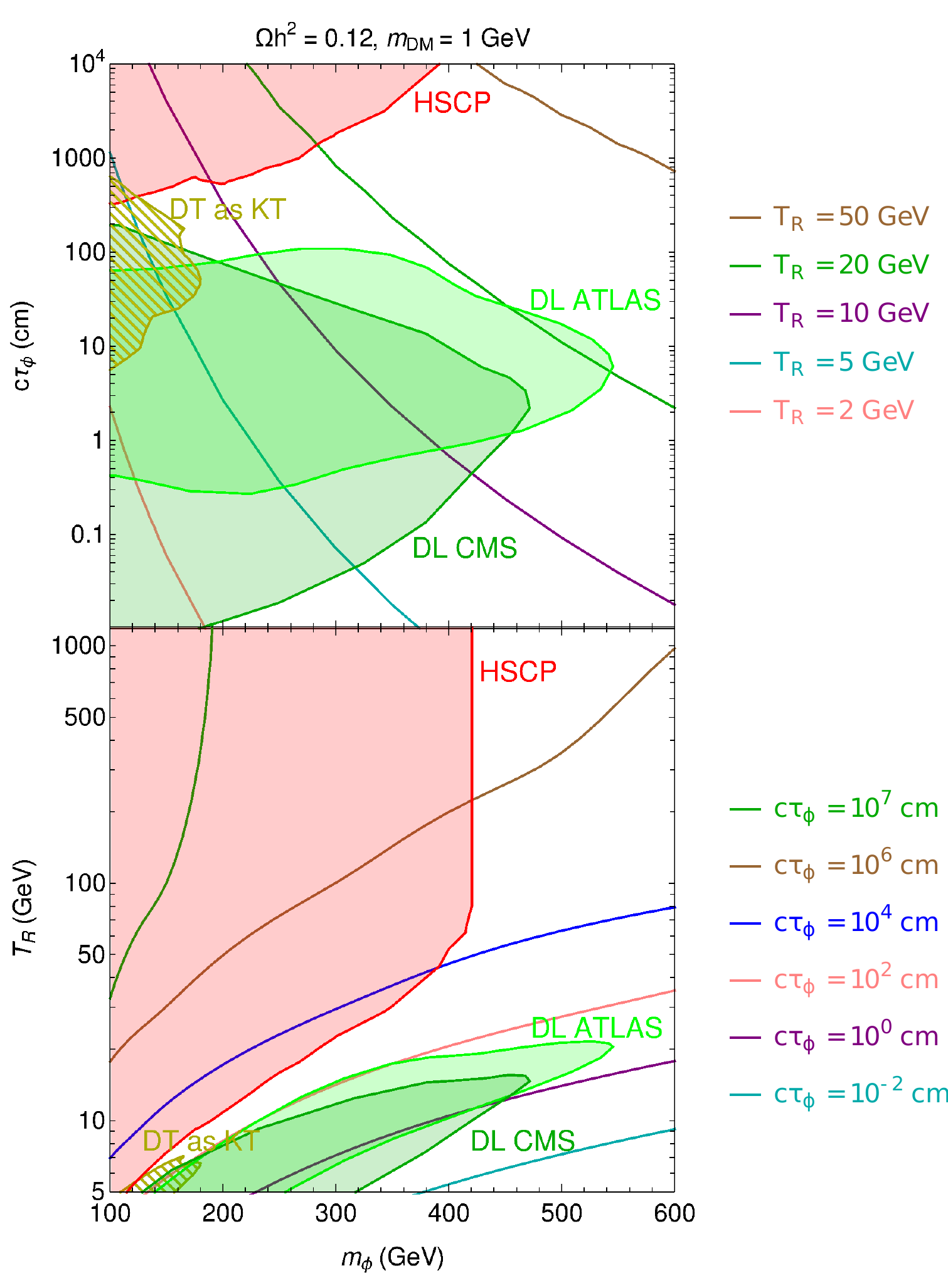}}
	\hspace{0.03\textwidth}
	\subfloat[]{\includegraphics[width=0.47\textwidth]{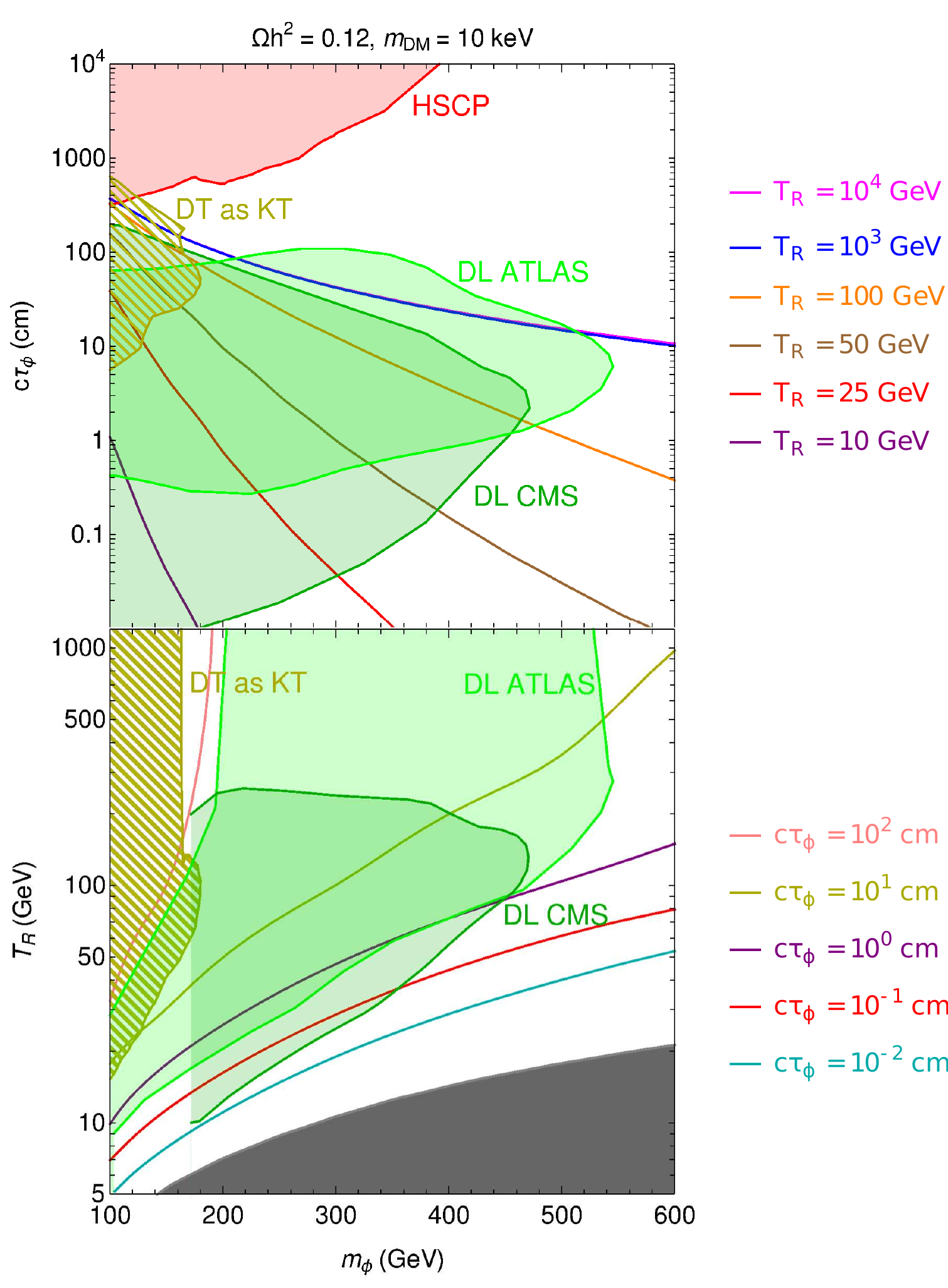}}
	\caption{Leptophilic DM with $m_\chi = 1$~GeV (left) and $m_\chi = 10$~keV (right). The upper panel shows constraints in the $(m_\phi,c\tau_\phi)$ plane while the lower panel identifies the viable parameter space in the $(m_\phi,T_{R})$ plane. Solid lines reproduce the observed relic density, and the color code is the same as in Fig.~\ref{fig:lepto2}. The value of $T_R$  associated to a given $(m_\phi,c\tau_\phi)$ in the right panel essentially provides an upper bound on $T_R$ for any dark matter mass as $m_\chi \gtrsim 10$~keV due to small scale structures constraints, see the text for details.}
	\label{fig:lepto}
\end{figure}

Let us now turn to the scenario where $m_\chi \ll m_\phi$. Conversion driven freeze-out only predicts the correct amount of \ac{DM} when both co-annihilation partners are compressed, hence, we need to rely on the freeze-in mechanism to reproduce the observed amount of \ac{DM}. When \ac{FI} happens during a \ac{RD} era (i.e. when $m_\phi\ll T_R$), the decay width needed to match the correct \ac{DM} relic density corresponds to lifetimes way longer than the detectors at the \ac{LHC}. In contrast, when $m_\phi > T_R$, larger decay widths are allowed as we have seen in Sec.~\ref{sec:fi_emd}, so that searches for \acp{LLP} are able to probe this scenario. Within this model, there are two types of processes contributing to the \ac{FI} production of \ac{DM}: (i) the decay of the mediator and (ii) the scattering processes converting the mediator into DM.\footnote{For t-channel processes involving massive bosons and a lepton in the propagator, t-channel divergences should be regulated by thermal corrections. We expect that the regulated process contribute similarly to other scatterings, with a few percent correction in the abundance that we neglect  in the leptophilic case.} In Figs.~\ref{fig:lepto2} and~\ref{fig:lepto}, we take both contributions into account and show how different searches for \acp{LLP} can probe this scenario.

\paragraph{Collider and cosmological constraints.} Just as in the \ac{CDFO} regime, lifetimes corresponding to a proper decay length exceeding a meter can be probed by the \ac{HSCP} searches performed by CMS and ATLAS displayed in Tab.~\ref{tab:searches}. In App.~\ref{app:HSCP}, we review the details of the recasting method. The corresponding regions excluded at 95\% \ac{CL} appear in red in Figs.~\ref{fig:lepto2} and~\ref{fig:lepto}. Compared to the CDFO regime, the correct DM relic abundance can be reproduced for long lifetimes and larger masses. As mentioned before, the ATLAS search is more sensitive in this region of parameters space due to a larger integrated luminosity. Hence, we show one line in Figs.~\ref{fig:lepto2} and~\ref{fig:lepto} corresponding to the strongest limits coming from both the CMS and ATLAS analysis. 

For intermediate lifetimes, when the decay of the charged mediators occurs inside the tracker, one expects kinked tracks since the lepton from the mediator decay will be non-soft due to the large mass splitting. \ac{DT} searches can have some sensitivity to this type of signatures, that we estimate in the following. First, the charged track of the mother particle (the mediator) and the one of the daughter particle (the muon) should have a sufficient angular separation such that the track of the mother particle is interpreted as a disappearing track. In our analysis we conservatively require that the angular separation of the daughter track with respect to the mother track should be larger than $\Delta R > 0.1$. Second, the DT searches include a lepton veto. However, by the same requirement above on $\Delta R$, the emitted muon does not generically point towards the primary vertex and hence we assume that it is not properly reconstructed and the lepton veto does not apply. We refer to App.~\ref{app:DT} for more details. In the figures of this section we will denote as ``DT as KT'' the sensitivity lines obtained by reinterpreting the DT search for a KT signature through the procedure explained above. For this purpose we have used the CMS search~\cite{CMS:2020atg} since it is the one with the largest luminosity. The area that would be excluded by KT following our reinterpretation appears as a dark-yellow densely-hatched region in Figs.~\ref{fig:lepto2} and~\ref{fig:lepto}. As can be seen, our reinterpretation demonstrates that a dedicated KT would nicely complement the other LLP searches for $c \tau_\phi\sim 1$m.
  
For shorter lifetimes, we employ the ATLAS search for displaced leptons~\cite{ATLAS:2020wjh}. In the auxiliary material provided for this analysis, one can find the case of oppositely charged dimuons final state that can be applied directly to our model, leading to the exclusion region in Figs.~\ref{fig:lepto2} and~\ref{fig:lepto} colored with light green (tagged DL ATLAS).

For even shorter lifetimes, we exploit the CMS \ac{DL} analysis~\cite{CMS:2021kdm}, since as we have already discussed in Sec.~\ref{sec:searches}, the weaker requirement on the transverse impact parameter makes it more sensitive to smaller lifetimes compared to the ATLAS \ac{DL} search. We can see this by comparing in Figs.~\ref{fig:lepto2} and~\ref{fig:lepto} the dark green region (tagged DL CMS) excluded by CMS with the light green region (tagged DL ATLAS) excluded by ATLAS. The former was obtained by using the results for a left-right degenerate slepton decaying to muons and rescaling them with the corresponding cross section, see App.~\ref{app:DL} for more details.

Finally, we add one note about the sensitivity of prompt lepton searches including missing energy to the leptophilic model under study with a short-lived (but still macroscopic) mediator. This sensitivity arises from the fact that in the prompt lepton searches by CMS~\cite{CMS:2020bfa} and ATLAS~\cite{ATLAS:2019lff}, a small experimental error is allowed so that the reconstructed lepton does not have to point exactly to the primary vertex, but within some uncertainty. Ref.~\cite{Bernreuther:2022bdw} has shown that there is indeed no gap in coverage of the $c\tau$ range between prompt and displaced lepton searches, making them very complementary. However, an observation of physics giving rise to a prompt lepton plus missing energy will not yield us any information about the corresponding cosmological history. Hence, we will not include this signature in our further discussion.

In addition to collider constraints, the Lyman-$\alpha$ observations mentioned in Sec.~\ref{sec:BoundTR} impose a conservative bound of $m_X\gtrsim 10$~keV and further limit the viable parameter space for DM. In the left panel of Fig.~\ref{fig:lepto2}, the light gray region is excluded by the conservative Lyman-$\alpha$ constraints. In the right panel of Fig.~\ref{fig:lepto2}, we also show our conservative $m_X\gtrsim 10$~keV Lyman-$\alpha$ bound with a light gray area as a guide for the eye. In fact, a dedicated analysis of FI as non-cold DM arising from an EMD era is still missing in the literature. Finally, in some regions of the parameter space, DM may eventually achieve thermodynamical equilibrium and the FI computations presented in Sec.~\ref{sec:fi_emd} do not hold. We shade the corresponding parameter space regions in dark gray in Figs.~\ref{fig:lepto2} and~\ref{fig:lepto}.

\paragraph{Results for fixed $T_{R}$.} The parameter space under investigation, including also the EMD cosmology, is four-dimensional and it is described by the mediator mass $m_\phi$ and proper decay length $c \tau_\phi$, the DM mass $m_{\rm DM}$ and the reheating temperature $T_R$. In Fig.~\ref{fig:lepto2}, we show slices that identify the cosmological history, or equivalently we fix the value of $T_{R}$.

In the left panel of Fig.~\ref{fig:lepto2}, we consider a standard cosmological scenario with a rather high reheating temperature, $T_{R}= 10^5$~GeV $\gg T_{FI}$. We show in the upper part with different colored lines, each one corresponding to a different value of the DM mass, the relation between the mediator mass and its decay length necessary to account for the observed DM abundance. The colored areas tagged with the labels HSCP, KT, DL ATLAS and CMS are excluded or can be probed by the corresponding collider searches as discussed above. The only relevant searches for this case are those for charged tracks, such as HSCP, because the mediator will typically cross the detector entirely. LLP searches targeting displaced vertices can essentially not test the viable DM space allowed by Lyman-$\alpha$ constraints. Although the detection of a charged track associated to $\Phi_B$ would be a spectacular signal of new physics, it would not let us learn much about the possible connection to DM or the physics of the early universe. A similar conclusion can be reached looking at the bottom part, where the colored curves are associated to different decay lengths in the $(m_\phi,m_{\rm DM})$ plane. 

In the right panel of Fig.~\ref{fig:lepto2}, we illustrate how this situation can be turned into a more promising one when considering \ac{FI} occurring during an \ac{EMD} era with $T_{R}= 20$~GeV. The upper part shows how displaced searches can probe DM with masses $m_{\rm DM}\lesssim 1$~GeV. From the bottom panel, it appears that the decay length of the mediator ranges from the order of a millimeter to hundreds of meters while still accommodating the correct relic abundance when $m_\text{DM}> 10$~keV and $100~\GeV<m_\phi<600~\GeV$. Leptophilic FIMPs can thus be probed at colliders through HSCP, KT and DL signatures.  The HSCP searches probe charged mediator masses up to $m_\phi\approx 400$~GeV and $c\tau_\phi \gtrsim 5$~m. The DL ATLAS search has a very strong reach, excluding mediator masses up to $\sim~500$~GeV for $c\tau_\phi \sim 10$~cm, and covering a large portion of the interesting displacements, i.e.  0.3~cm~$\lesssim c\tau_\phi \lesssim$~80~cm. For even shorter decay lengths, the DL CMS search provide further sensitivity up to $\ctau_\phi \approx 0.1$~cm. For intermediate lifetimes, it is interesting to observe how a KT search (or any search sensitive to such signature) can close the gap between HSCP and DL searches.

\paragraph{Results for fixed $m_{\rm DM}$.}
Measuring the DM mass from a displaced event is quite challenging. However, we can still learn something about the early universe, and in particular extract an upper bound on $T_{R}$, from a displaced event. We illustrate this point, and exemplify the discussion in Sec.~\ref{sec:BoundTR}, with a comparison in Fig.~\ref{fig:lepto} between a 10 keV DM scenario (right panel) and a case with heavier FIMP (left panel). In the upper parts, colored lines identify in the
($c\tau_\phi$, $m_\phi$) plane the values of $T_{R}$ reproducing the observed DM abundance. Likewise, in the lower parts they give the values of $c\tau_\phi$ (again reproducing the DM relic density) in the ($T_R$, $m_\phi$) plane. The colored areas show the regions probed by the collider searches as discussed above.

In the top-left panel of Fig.~\ref{fig:lepto}, the case for $m_\chi = 1$~GeV, all continuous curves illustrate that FI occurs in an \ac{EMD} epoch, i.e.~$T_{R}<T_{FI}$. In order to get a lower mediator lifetime for a given $m_\phi$, we have to reduce $T_{R}$, as also shown in the bottom-left panel. Reducing $c \tau_\phi$ corresponds to effectively enhancing the coupling constant $\lambda_{\chi}$ in Eq.~\eqref{eq:lagr}, which in turn increases the DM production rate and gives a larger FI contribution. Therefore we need the \ac{EMD} era to last longer in order to provide more dilution. The same panel also shows that, for a given value of the mediator lifetime, larger $m_\phi$ imply larger values of $T_{R}$. Indeed, by increasing $m_\phi$, we suppress $Y_X^{\infty}$ and we need a shorter duration of the \ac{EMD} epoch to provide less dilution, see Eq.~\eqref{eq:YXMDfinal}. The strong dependence of the final abundance on $T_R$ in Eq.~\eqref{eq:YXMDfinal} also explains the narrow range of values of $T_R$ that reproduce the correct relic density in the top-left panel of Fig.~\ref{fig:lepto}.

Similar conclusions can be drawn from the bottom-left panel of Fig.~\ref{fig:lepto}, where we plot contours of fixed mediator lifetime accounting for the correct relic abundance in the plane $(m_\phi,T_{R})$. In the upper part of the plot, the contours become vertical since we recover FI in a \ac{RD} era $T_R\gg T_{FI}$. Focusing on the bottom curves, an increase in the reheating temperature implies an increase in the decay length, and then a decrease in $\lambda_{\chi}$. A larger reheating temperature (with $T_{R}< T_{FI}$) means less entropy dilution between $T_{FI}$ and $T_R$, and in order to compensate for this effect there should be less DM produced during FI, i.e.~the decay width should decrease.

The picture changes when we consider the lowest value of the DM mass compatible with the Lyman-$\alpha$ constraints, $m_\chi$ = 10~keV. Since the relic density is proportional to the DM mass, less dilution is needed to give displaced signatures at the LHC. Equivalently, the relic density can be reproduced for larger values of $T_R$ as it is well visible comparing the two panels of Fig.~\ref{fig:lepto}. Furthermore, the colored curves in the top-right panel do not always satisfy $T_{R}< T_{FI}$, i.e.~both cases of FI occurring during an \ac{EMD} era and a RD era are possible. Indeed, the contours for $T_{R} = 10^4~\text{GeV}$ and $10^3~\text{GeV}$ are superposed. In the bottom panels, the contours at fixed values of $c\tau_\phi$ become vertical for $T_{R}\gg m_\phi$.

The contours in the right panel of Fig.~\ref{fig:lepto} are associated to lightest FIMP allowed by Lyman-$\alpha$ constraints, and they inform us on the maximal $T_{R}$ allowed compatible with FI production (in particular, not larger than $\Omega_\chi h^2=0.12$). For instance, suppose that a long-lived charged scalar decaying into a muon and MET is observed and its mass and decay length are measured to be $m_\phi \approx 500$~GeV, $c\tau_\phi \approx 1$~cm. In such a case, even without a measurement of the DM mass, Fig.~\ref{fig:lepto} tells us that this particle can be a FI mediator compatible with the observed DM relic density only if $T_R \lesssim 100$~GeV. Such information would have a profound impact on our understanding of the early universe (for instance inflation, baryogenesis, etc). Given that for heavier DM the upper bound on the reheating temperature only becomes more stringent, the discussion of the following models will focus on the $m_\text{DM}=10$~keV case only.

As a final remark, we comment on the possible impact of future experiments. First, assuming that a LLP search at the LHC remains background free, we can estimate the reach of HL-LHC (with an integrated luminosity of 3000 fb$^{-1}$) by (linearly) rescaling its sensitivity on the mediator production cross section with the luminosity. For instance, by studying the cross section dependence on the mediator mass $m_\phi$, we can estimate that the DL ATLAS analysis will probe masses up to $\sim 900$ GeV. This will considerably extend the explored parameter space of the model also to regions where lower $T_{R}$ upper bounds could be inferred (see Fig.~\ref{fig:lepto} top right).

On the other hand, we observe that future dedicated detectors targeting larger values of $c\tau_{\phi}$ (such as MATHUSLA~\cite{Curtin:2018mvb} or CODEX-b~\cite{Aielli:2019ivi}) will not constitute a very sensitive probe for this specific simplified model. First, the long-lived particle is electromagnetically charged and hence the regions of large $c {\tau}_{\phi}$ will be covered efficiently by HSCP searches at HL-LHC. Such experiments are also typically shielded (to create a zero background environment) so that charged particles do not reach the decay volume. Second, large $c {\tau}_{\phi}$ corresponds to very small couplings and hence to scenarios where dilution is not necessary in order to obtain the correct relic abundance. As a consequence, even in the event of a discovery, it will not be possible to set an absolute upper bound on $T_R$ in the region of large $c {\tau}_{\phi}$, as it is visible from the ``overlapping'' contours in the top-right panel of Fig.~\ref{fig:lepto}.

The above discussion changes in the case of models with non-renormalizable interactions and UV sensitivity, as we will discuss at the end of Section \ref{sec:SingletTriplet}, where future facilities dedicated to LLP signatures would be beneficial in exploring further the parameter space and the connection with $T_R$.

\section{Topphilic scenario}
\label{sec:VLtop}
The second simplified model we study is the one with a real scalar DM particle $\phi$ and a vectorlike fermion mediator $\Psi_B$ with the quantum numbers of right-handed up-type quarks. We consider a ``topphilic'' scenario where BSM particles couple only to the top
\begin{equation}
 {\cal L}_{\mathcal{F}_{t_R\phi}}  \supset \partial_{\mu}\phi
 \ \partial^{\mu}\phi -\frac{m_{\rm DM}^2}{2} \phi^2 + \frac{1}{2} \bar{\Psi}_B \gamma^{\mu} D_{\mu}
 \Psi_B -m_\psi \bar \Psi_B \Psi_B \ - \ \lambda_{\phi}  \bar \Psi_B  t_R \phi\ + \ h.c.\,,
 \label{eq:lagr-top}
 \end{equation}
where $\lambda_\phi$ is a dimensionless Yukawa coupling. As usual, both the DM field and the mediator are odd under an unbroken $\mathbb{Z}_2$ symmetry, and we neglect possible extra contributions to DM production from a DM-Higgs portal coupling~\cite{Lebedev:2019ton,Heeba:2018wtf} so as to focus on the cubic interaction sketched in Fig.~\ref{fig:fi_model_diagram}. Decay and scattering processes can both produce DM in this scenario. Moreover, the higher (color) degrees of freedom and the larger gauge coupling are such that scattering processes are expected to play a more important role than in our previous leptophilic model, see also the discussion in~\cite{Garny:2018ali}. As a consequence, for fixed values of $m_{\rm DM}$ and $T_R$, one will need a longer lifetime in order to account for the relic abundance in this topphilic scenario than in models of the class ${\cal  F}_{\ell\phi}$ (or ${\cal S}_{\ell\chi}$ studied above).
\begin{figure}
	\centering
	\subfloat[]{\includegraphics[width=0.48\textwidth]{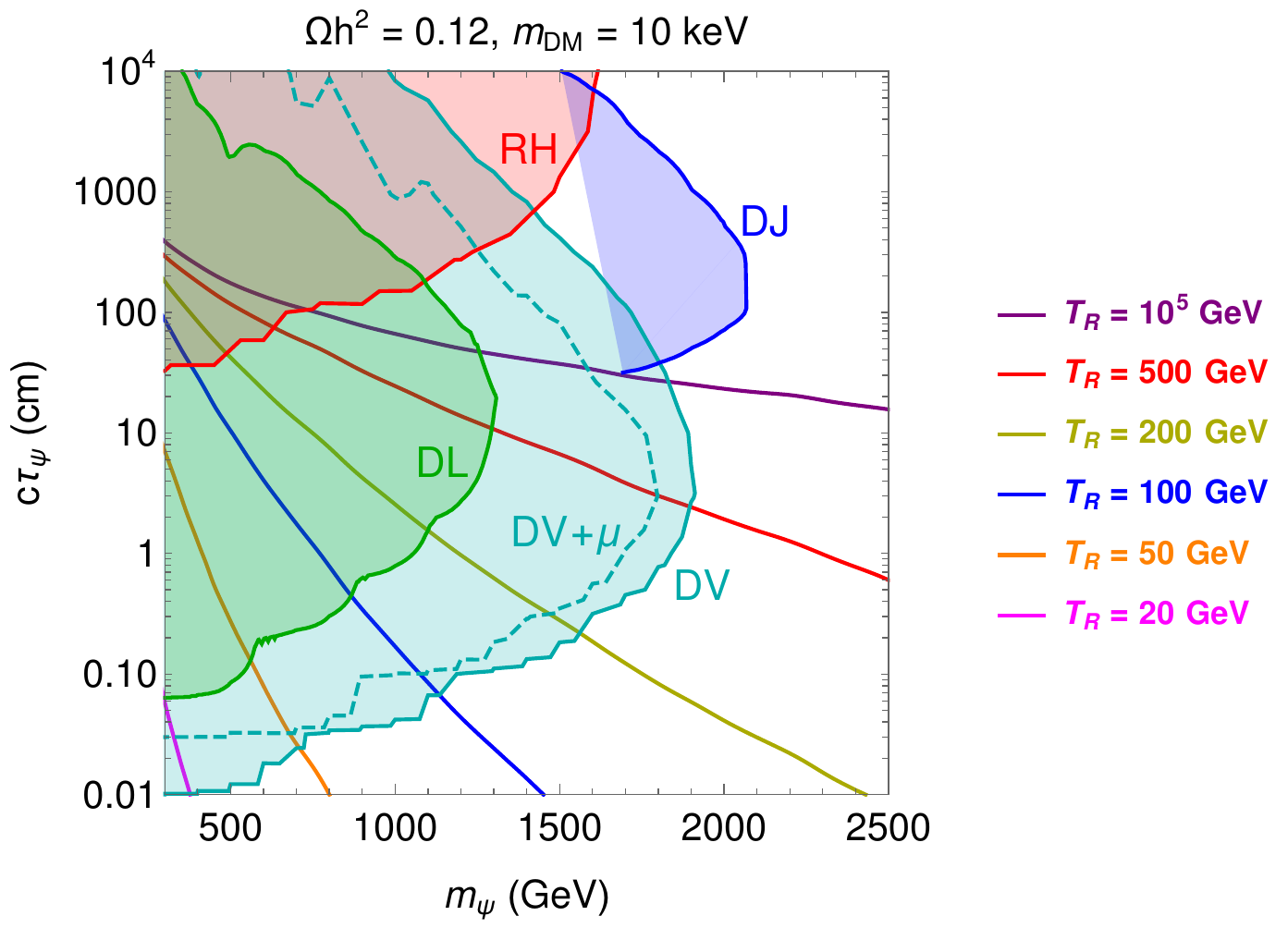}}
	\hspace{0.02\textwidth}
	\subfloat[]{\includegraphics[width=0.48\textwidth]{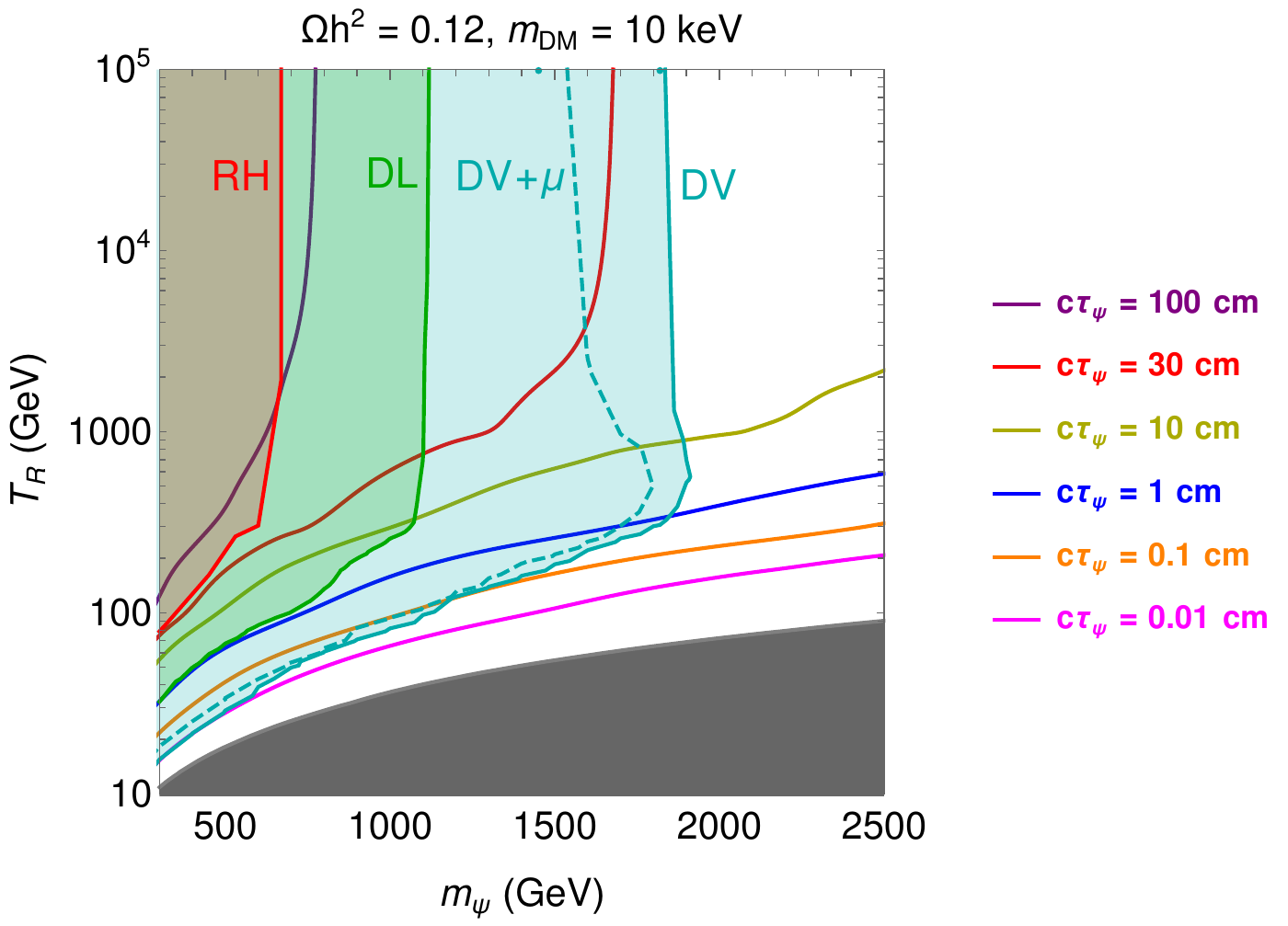}}
	\caption{Tophilic DM with $m_\chi = 10$~keV. Contours for fixed value of $T_R$ account for the DM relic abundance in the $(m_\psi,\,c\tau_\psi)$ plane (left), and contours of fixed $c\tau_\psi$ do the same in the $(m_\psi,\,T_R)$ plane (right). The value of $T_R$ associated to a given $(m_\psi,c\tau_\psi)$ provides an upper bound on $T_R$ for any dark matter mass as $m_\chi \gtrsim 10$~keV due to small scale structures constraints. The blue shaded area is excluded by the DV+MET search, the red area by R-hadron searches, the blue dashed area by DV+$\mu$ search, the dark blue area by the DJ search, and the green area by the DL search.}
	\label{fig:top}
\end{figure}

Since the \ac{CDFO} regime in a topphilic DM model has already been studied in \cite{Garny:2018icg}, we will here focus on the regime where $m_{\rm DM} \ll m_\psi$ where the DM relic abundance is set by FI in an \ac{EMD} universe. In Fig.~\ref{fig:top} we consider the topphilic scenario for a DM mass of 10 keV, a value that saturates the conservative Lyman-$\alpha$ constraints and allows to extract an upper bound on $T_R$, as illustrated at the end of Sec.~\ref{sec:lepto@lhc_emd}. In the left panel, colored lines identify the value of $T_R$ needed to reproduce the relic density in the $(m_\psi,\,c\tau_\psi)$ plane. Likewise, colored lines in the right panel show the values of $c \tau_\psi$ needed in the $(m_\psi, T_R)$ plane. Comparing contours of Fig.~\ref{fig:top} with those of the right panel of Fig.~\ref{fig:lepto}, we can see that longer lifetimes can be reached for fixed value of the mediator mass and $T_R$ in model ${\cal F}_{t_R\phi}$. This can be explained by the difference in the contribution of the scattering processes, as in the topphilic scenario, the scattering cross section scales with the strong coupling constant compared to the electroweak coupling constant in the leptophilic scenario. Except from the LHC searches, this is the main difference between the leptophilic and topphilic scenarios. The dependence on the masses and temperature scaling are the same since for both cases all processes producing DM are renormalizable and thus IR-dominated.

\paragraph{Collider signatures and constraints.} Looking at Tab.~\ref{tab:models-searches}, one can infer that the topphilic model can lead to a rich variety of signatures at the LHC. The mediator decays into a top quark whose decays can in turn give both fully hadronic and semi-leptonic signatures. In order to assess the most relevant searches, we first obtain the production cross section of the colored vectorlike fermion mediator $\Psi_B$ at the LHC by multiplying the LO result of {\tt MadGraph5\_aMC@NLO} by a flat K-factor of 1.6. This K-factor reproduces well the ratio of LO to NNLO cross-section for squark pair production tabulated in~\cite{tR-xsec} and we assume that the same K-factor can be applied to the case of colored vectorlike fermion mediator.
 
A long-lived $\Psi_B$ forms a \acf{RH} before decaying~\cite{Buchkremer:2012dn} which can leave a highly ionized track when it has a long enough lifetime, as discussed in Sec.~\ref{sec:searches}. The region excluded at 95\% \ac{CL} by RH searches according to our recasting is shown as a red area in Fig.~\ref{fig:top}. To obtain these results, we employed the public code {\tt SModelS}~\cite{Ambrogi:2018ujg,Heisig:2018kfq} and the one obtained from the ``LLP recasting repository''~\cite{LLPrepos}, and applied the analysis done for a stop mediator in both CMS~\cite{CMS:2016ybj} and ATLAS~\cite{ATLAS:2019gqq} searches (see App.~\ref{app:HSCP}) to our vectorlike top mediator as the spin of the produced particle is not expected to affect the sensitivity~\cite{Buchkremer:2012dn}.\footnote{We assume that the RH has the same lifetime as the heavy vectorlike fermion $\Psi_B$. This is supported by e.g.~\cite{ParticleDataGroup:2020ssz,Lenz:2014jha} where the lifetime of hadrons containing heavy quarks, in particular $b$, is discussed.}
 
For shorter decay lengths, the decay happens before the R-hadron leaves the detector with a top quark in the final state that itself decays dominantly into a $bW$ pair. This in turn can decay hadronically to form jets, giving rise to \acf{DV} or \acf{DJ} signatures. For details about these searches and our implementations we refer to, respectively, App.~\ref{app:DV} and~\ref{app:DJ}. Concerning the DV analysis, let us emphasize that the algorithm built to reconstruct the events in Ref.~\cite{ATLAS:2017tny} combines displaced vertices that are located within 1~mm from each other. This is of particular relevance when the mediator couples to a top quark which dominantly decays to a $b$-quark and a $W$ bosons. In the latter  case, two displaced vertices possibly arise from the decay of the long-lived mediator since jets originating from a b-quark are displaced themselves (with a proper decay length of $\mathcal{O}(0.5)$~mm). Treating this properly is beyond the scope of this work. Therefore, we conservatively take only the tracks originating from the $W$ into account for the reconstruction of the displaced vertex. The region excluded by the DV+MET search is shown with a light blue region while the region excluded by the DJ+MET is shown in dark blue.  We see that the DJ+MET analysis can probe larger $c\tau_\psi$ than DV+MET, as already noted in Sec.~\ref{sec:searches}. We also display with a dashed blue line the area probed by DV+$\mu$ (cf.~App.~\ref{app:DVmu}) which is similar to the one of DV searches. This captures the topologies where one of the top quark decays leptonically to a muon. The sensitivity is comparable to the DV searches since the suppression of the signal due to the leptonic branching ratio of the $W$ is approximately compensated by the higher integrated luminosity employed in the analysis.

Alternatively, the two tops can both decay leptonically inside the detector leading to a displaced lepton pair. This signature can be covered by the DL searches~\cite{CMS:2021kdm,ATLAS:2020wjh}. We focus on the ATLAS analysis~\cite{ATLAS:2020wjh} which employs the largest set of data at $\sqrt{s}=13$ TeV. In the App.~\ref{app:DL} we discuss the details of our recasting procedure, whose uncertainty is significant for some values of the LLP mass. Nevertheless our analysis provides an indicative estimate of the LHC reach on the leptonic channel, which we display as a uniform green region in Fig.~\ref{fig:top}. Even if the signal yield is limited by the leptonic branching ratio of the $W$ boson, the search has a certain coverage of the parameter space, which is however superseded by the DV + MET search for all ranges of $c\tau_{\psi}$.

LLP searches for the topphilic ${\cal F}_{t_R\phi}$ model probe a larger parameter space than in the leptophilic ${\cal S}_{\ell_R\chi}$ of Sec.~\ref{sec:leptophilic@lhc}. The main reason is a larger LHC production cross section because the mediator is colored, and fermions display a higher production cross section than scalars. We see that RH searches probe the mediator masses up to $m_\psi \approx 1.6$ TeV for $c\tau_\psi\gtrsim$ 30 cm. The DV searches cover a large range of parameter space corresponding to masses up to $m_\psi \approx 1.9$ TeV and $c\tau_\psi\gtrsim$ 0.1 mm, partially overlapping with RH searches. Complementary constraints are obtained by making use of DJ searches that cover 1.5 TeV~$\lesssim m_\psi \lesssim$ 2.1 TeV and 30 cm $\lesssim c\tau_\psi \lesssim 100$~m. As in the case of the leptophilic model considered above, the plots in Fig.~\ref{fig:top} show how a measurement of an LLP mass and lifetime could be employed to bound the reheating temperature.

Also for this model, the impact of HL-LHC searches will considerably extend the mass reach and hence the coverage of regions corresponding to  low $T_R$ bounds. By naively scaling the sensitivity of the DV+MET search on the production cross section of the colored mediator with the luminosity, we estimate that HL-LHC could reach $m_\psi \approx 2.8$ TeV.

\section{Singlet-triplet}
\label{sec:SingletTriplet}

The two examples considered so far feature renormalizable FIMP interactions and therefore DM production is dominated by physics in the IR. We now discuss the $\mathcal{F}_{W\chi}$ model where interactions are non-renormalizable and scattering processes contributing to DM production are UV-dominated. Looking at Tab.~\ref{tab:models-searches}, this model appears to be sensitive to a large set of existing LLP searches. The two BSM fields of the $\mathcal{F}_{W\chi}$ model are an extra singlet fermion DM and an $SU(2)_L$ triplet fermion mediator that we denote with
\begin{equation}
	\chi_S,	\qquad \chi_T= 
	\begin{pmatrix}
	\chi_T^0/\sqrt{2} & \chi_T^+ \\
	\chi_T^- & - \chi_T^0/\sqrt{2}
	\end{pmatrix}
	\,,
\end{equation}
and with masses $m_S$ and $m_T$ respectively. Both new fields are odd under an unbroken $\mathbb{Z}_2$ symmetry and couple to the EW gauge bosons through the Lagrangian:
\begin{align}
  \mathcal{L} \supset& - \frac{m_S}{2} \bar\chi_S \chi_S - \frac{m_T}{2} \text{Tr}\left[\bar\chi_T \chi_T \right] + \frac{1}{2} \text{Tr}\left[ \bar\chi_T i \cancel{D}_\mu \chi_T \right]  \nonumber\\ 
  &+\frac{1}{\Lambda} (W^a_{\mu \nu} \bar \chi_S \sigma^{\mu \nu} \chi_T^a + \text{h.c.}),
\label{eq:unphysL}
\end{align}
where $W^a_{\mu \nu}$ is the $SU(2)_L$ field strength and $\Lambda$ is the scale of new dynamics responsible for generating the higher dimensional operator. More gauge invariant interactions exist at dimension five (for instance $H \chi_T H \chi_S$, see for details App.~\ref{app:13}). However, our goal is to explore interactions of the kind depicted in Fig.~\ref{fig:fi_model_diagram}, and we will assume that other possible dimension-five operators are suppressed, possibly by a higher mass scale, to the extent that they can be neglected for what concerns both DM production and LHC phenomenology. We only consider reheating scenarios with $T_{\textsc{max}} < \Lambda$ in order for the effective description in Eq.~\eqref{eq:unphysL} to be valid during DM production. The singlet-triplet model has been also considered in Ref.~\cite{Filimonova:2018qdc} in the context of freeze-out DM production with compressed spectra.

We work in the $m_T\gg m_S$ regime and we neglect any Higgs portal coupling. As a result, singlet/triplet mixing will have minor influence and the lightest neutral fermion under the $\mathbb{Z}_2$ symmetry, i.e.~the DM candidate, is essentially the singlet fermion $\chi \simeq \chi_S$ with mass $m_\text{DM}\simeq m_S$. The mediator multiplet has a neutral component that is essentially the neutral component of the triplet $\Psi_B^0\simeq \chi^0_T$ with mass $m_T$. The charged components, corresponding to the charged components of the triplet $\Psi_B^{\pm}=\chi_T^{\pm}$, have a slightly higher mass $m_C= m_T+\Delta m$ with a splitting induced by EW quantum corrections $\Delta m \simeq 160$~MeV, see App.~\ref{app:13} for details. As a result, the mass
hierarchy of the new fields in our analysis is always: $m_C\gtrsim m_T\gg m_\text{DM}=m_S$. This mass spectrum allows for multiple decay channels of the charged mediator, as is shown in Fig.~\ref{fig:spectrum}, which will be discussed later.
\paragraph{DM production in the early universe.} The processes reported in Tab.~\ref{tab:processes} of App.~\ref{app:13}, decays and scatterings, contribute to DM production. As discussed in Sec.~\ref{sec:fi_emd_UV}, scattering processes mediated by non-renormalizable operators are UV dominated. The t-channel exchange of a photon in $\Psi_B^0e^-\to \chi e^-$ gives rise to a transition matrix $\propto (g/\Lambda)^2$ (where $g$ is the $SU(2)_L$ gauge coupling) that has a form similar to that shown in Eq.~(\ref{eq:Mt}) given that all particles involved are much lighter than the mediator particle $\Psi_B^0$.\footnote{We regulated the IR divergence of the diagram with a photon thermal mass, $m_\gamma \sim eT$, in the propagator.} 

\begin{figure}[!t]
  \centering
	\subfloat[]{\includegraphics[width=0.41\textwidth]{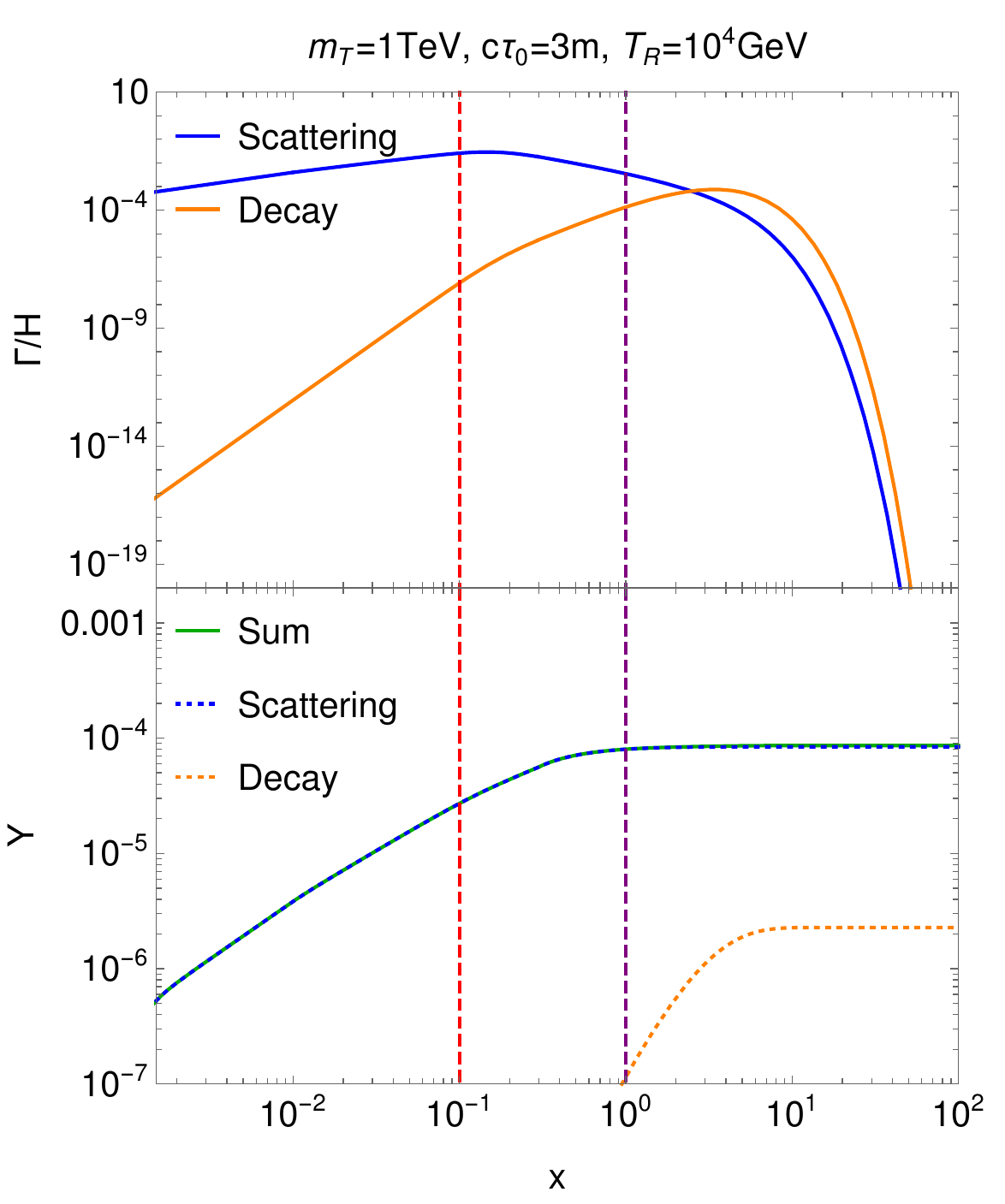}}
	\hspace{0.05\textwidth}
	\subfloat[]{\includegraphics[width=0.41\textwidth]{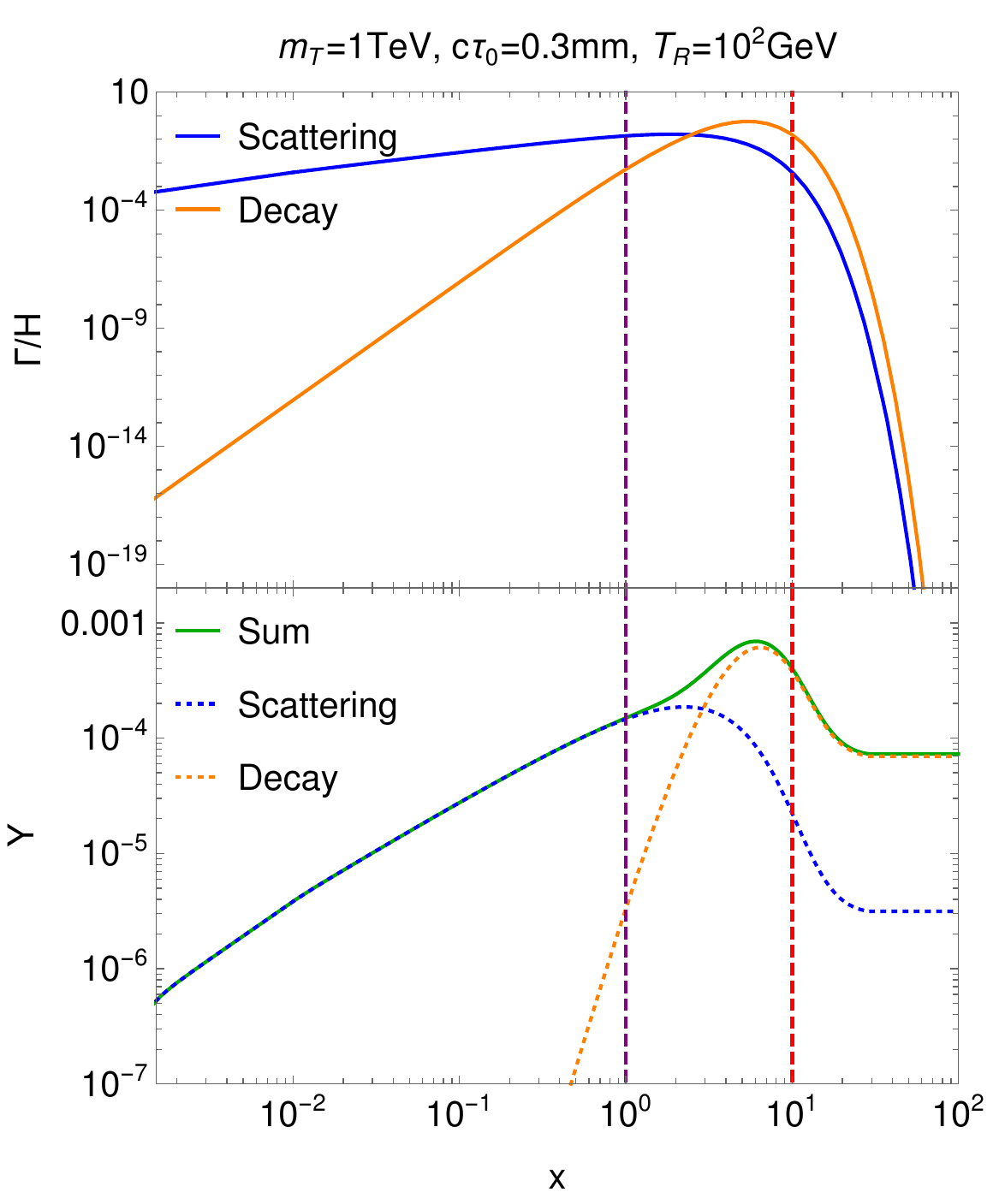}}
	\caption{Singlet-triplet model. Above: production rate/Hubble ratio for decay (blue) and scattering (orange). Below: DM yield. We consider $T_R>T_{FI}$ (left) and $T_{R}<T_{FI}$ (right). The dashed lines denote $T=T_{R}$ (red) and $T=m_T$ (purple) as a proxy for $T_{FI}$. For these parameters, the observed DM abundance is reproduced for $m_{\rm DM} = 10$~keV.}
        \label{fig:rates13}
\end{figure}

Similarly to Fig.~\ref{fig:rates}, we compare in the top panels of Fig.~\ref{fig:rates13} the production rate from scatterings (in blue) to the one of the decays (in orange) normalized by the Hubble rate as a function of $x=m_T/T$. In the bottom panels, we plot the contributions of decay and scatterings (orange and blue dotted, respectively) to the total DM yield (continuous green). We fix $m_T=1$~TeV, and the prefactor $1/\Lambda$ reproduces the measured relic density for $m_\text{DM}=10$~keV; this gives a lifetime for the neutral mediator $\Psi_B^0$ of $c\tau_0=3$~m when $T_{R}\gg T_{FI}$ and $c\tau_0=0.3$~mm when $T_{R}< T_{FI}$. The main contribution to DM production in the left panel comes from scatterings consistently with our discussion in Sec.~\ref{sec:fi_emd_UV}. A larger reheating temperature implies an enhanced DM production, leading ultimately to a smaller DM-mediator coupling (i.e.~a longer mediator lifetime) needed to account for the relic abundance. In contrast, in the right panel of Fig.~\ref{fig:rates13} we recover the yield dilution post freeze-in for $x>x_{FI}$ observed both for renormalizable and non-renormalizable operators in Figs.~\ref{fig:FIearlyMD} and~\ref{fig:rates}. In this second case, scattering processes play a subleading role.

\begin{figure}[t]
	\centering
	\subfloat{\includegraphics[width=0.48\textwidth]{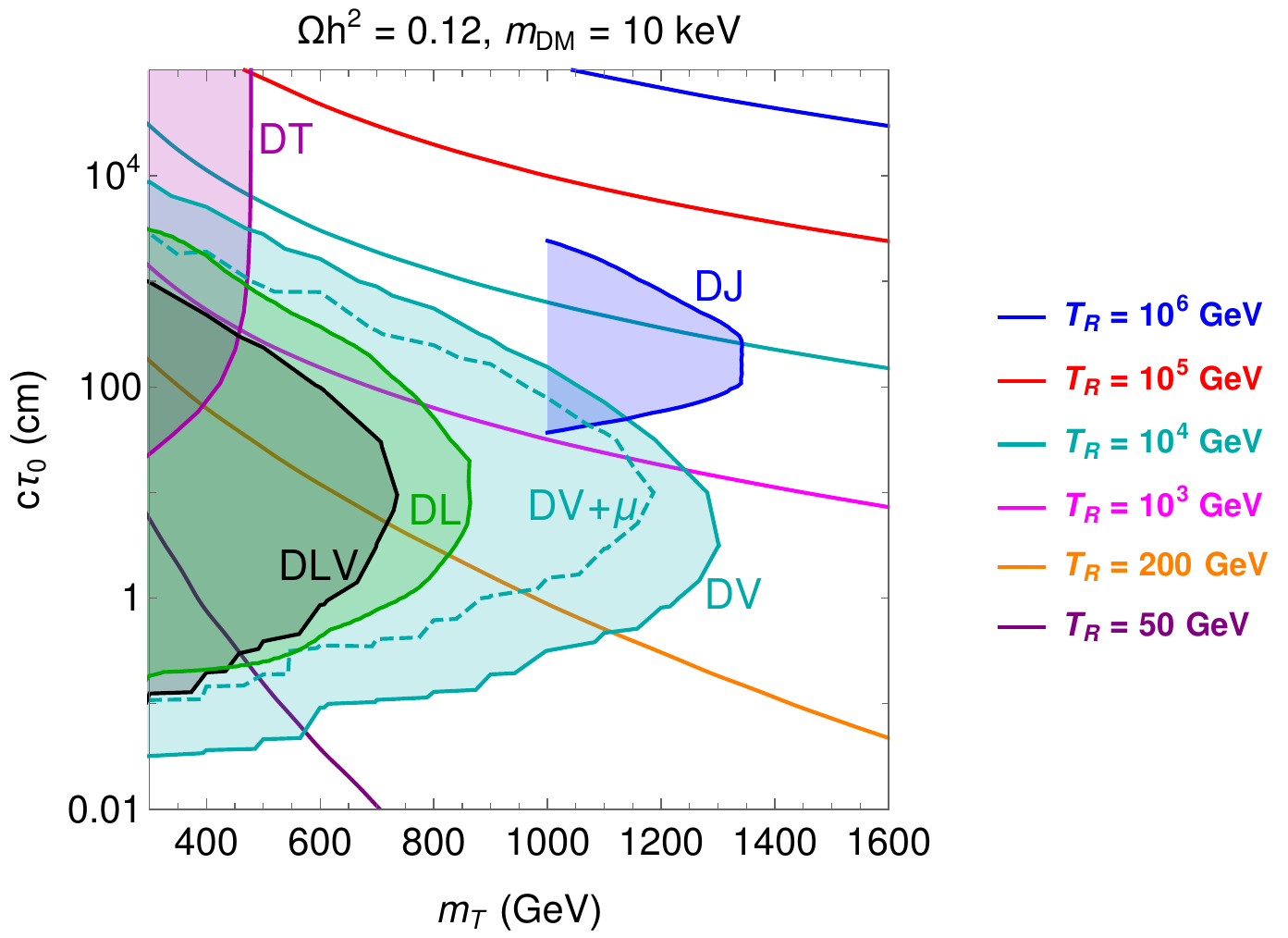}}
	\hspace{0.02\textwidth}	
	\subfloat{\includegraphics[width=0.48\textwidth]{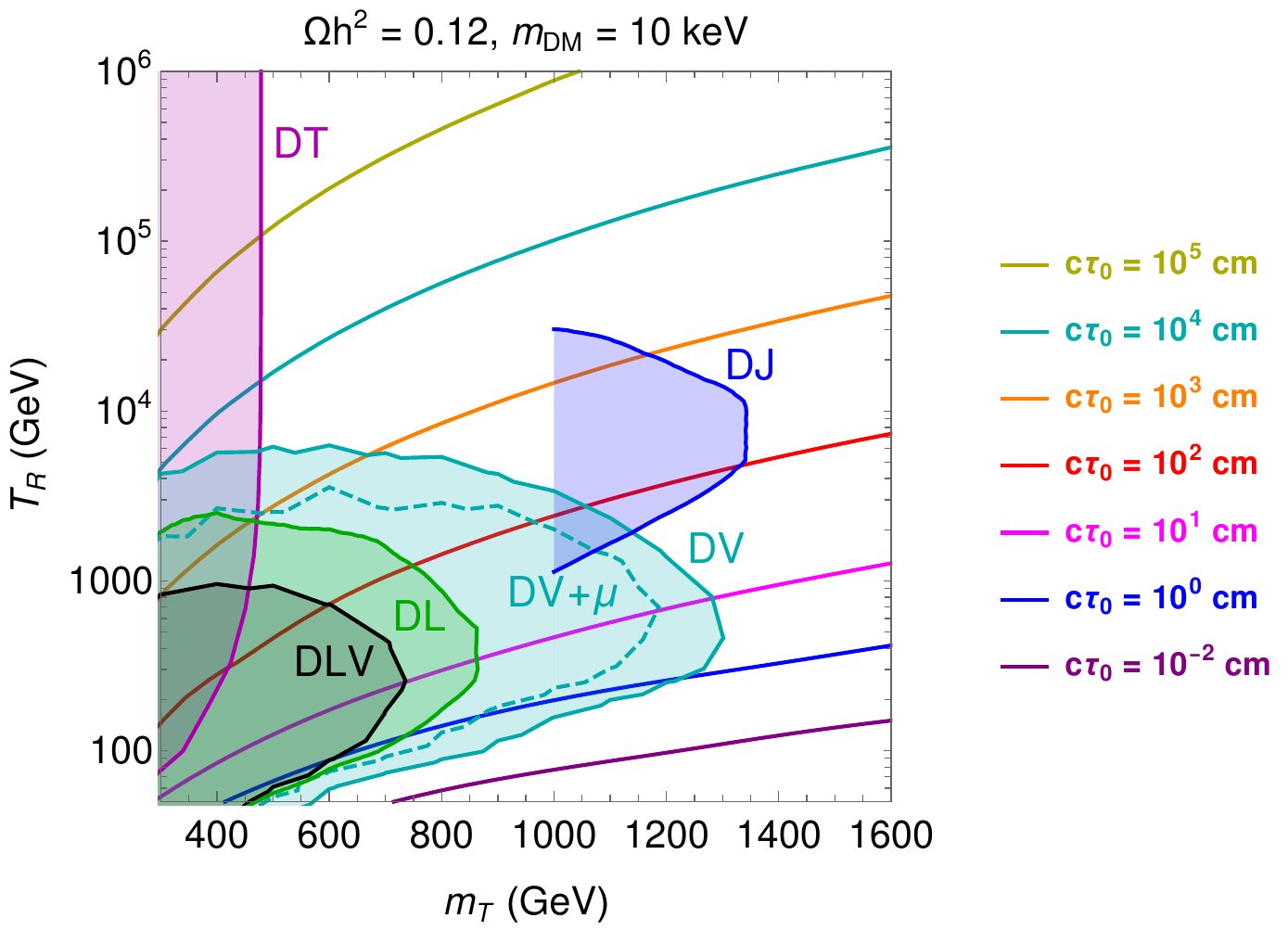}}
	\caption{Contours of the value of $T_R$ accounting for the whole DM relic abundance in the  $(m_T,\,c\tau_0)$ plane (left), contours of the neutral mediator lifetime $c\tau_0$ in the $(m_T,\,T_R)$ plane (right), for the singlet-triplet DM model with $m_{\rm DM}=10$~keV. The value of $T_R$ associated to a given $(m_T,c\tau_0)$ provides an upper bound on $T_R$ for any dark matter mass as $m_{\rm DM} \gtrsim 10$~keV due to small scale structures constraints. The colored areas are excluded by LHC searches for displaced vertices + MET (blue), displaced vertices + $\mu$ (dashed blue), delayed jets (dark blue), displaced lepton vertices (black), displaced leptons (green), and the disappearing tracks (purple). 
              }
	\label{fig:ExclRegion13}
\end{figure}

\begin{figure}[t]
  \centering
	\subfloat[]{\includegraphics[width=0.43\textwidth]{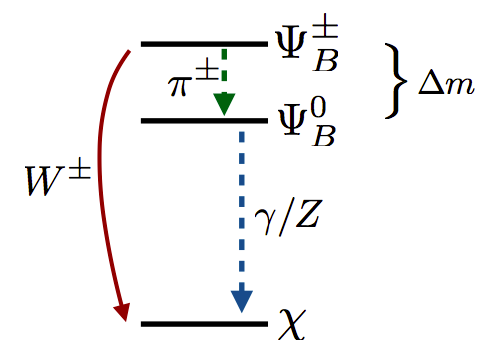}}
	\hspace{0.05\textwidth}
\subfloat[]{\includegraphics[width=0.45\textwidth]{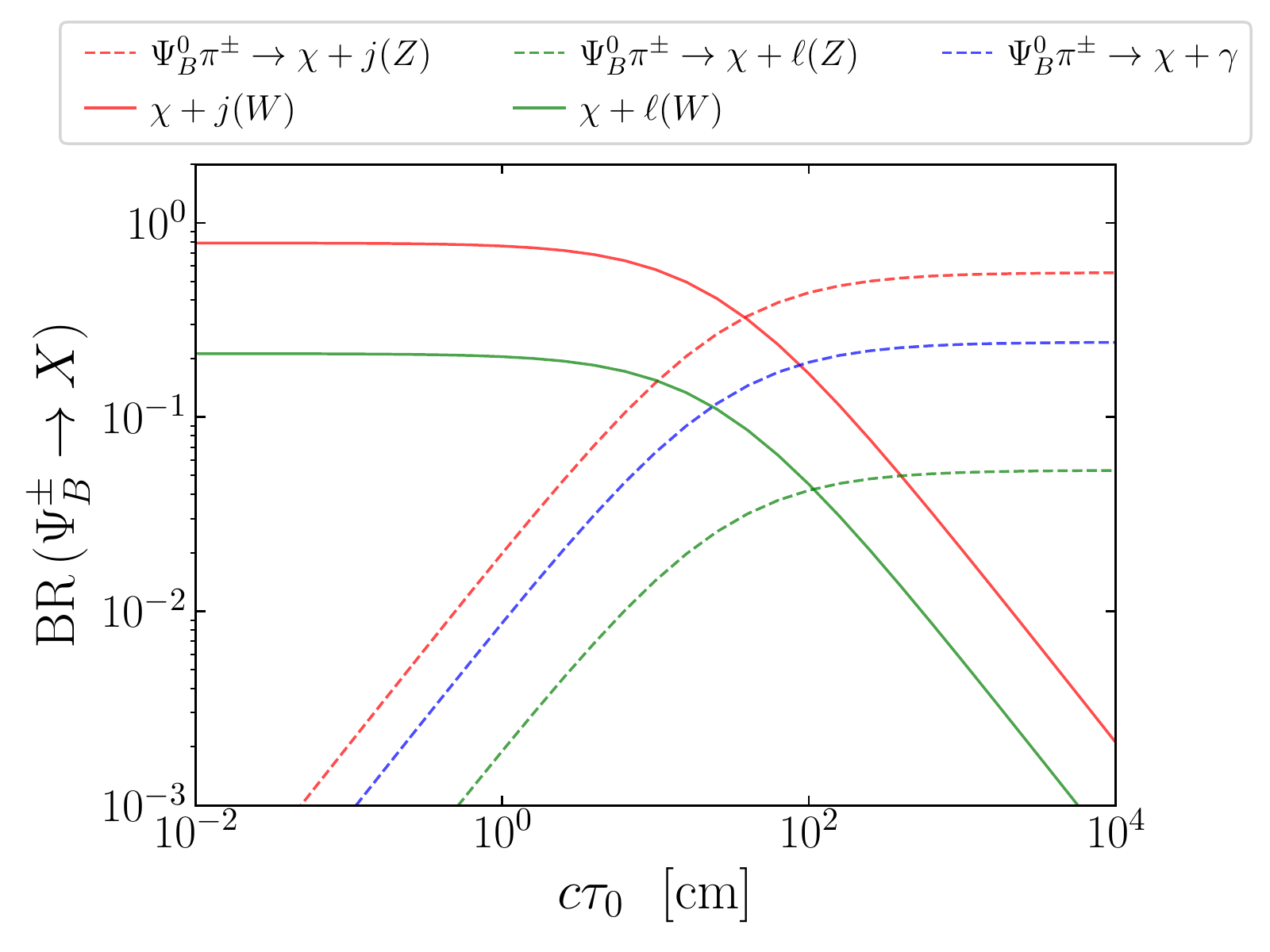}}
	\caption{Typical decay processes of the singlet-triplet model (left). Charged mediator branching ratio to the products listed in the legend as a function of the neutral mediator decay length $c\tau_0$ as a proxy for the scale $\Lambda$ for a mediator mass $m_T$ = 500 GeV (right).}
	\label{fig:spectrum}
\end{figure}

We consider again $m_\text{DM}=10$~keV, and we show in Fig.~\ref{fig:ExclRegion13} contours of the values of $T_{R}$ giving the correct relic density in the  $(m_T,\,c\tau_0)$ plane (left panel) and $c\tau_0$ contours satisfying the same constraints in the $(m_T,\,T_R)$ plane (right panel). In the left panel, we see how higher $T_{R}$ induces larger values of the proper mediator lifetime without reaching the saturation effect at large $T_{R}$ that was observed in the case of renormalizable operators (see the top-right panel of Figs.~\ref{fig:lepto} and~\ref{fig:top}). Furthermore, $c\tau_0$ contours in the right panel never become $T_R$ independent. These are direct implications of the UV-dominated scatterings at work when $T_{R}\gg T_{FI}$ for non-renormalizable operators, requiring smaller couplings (increasing $c \tau_0$) when increasing~$T_{R}$.

\paragraph{Collider constraints.} It is helpful to visualize the mass spectrum and the decay channels of the triplet components illustrated in Fig.~\ref{fig:spectrum}. The neutral heavy fermion $\Psi_B^0$ can decay into DM emitting a $\gamma$ or a $Z$ boson through the dimension-five operator, and the decay width scaling as $\sim {m_T^3}/{\Lambda^2}$ typically assures macroscopic decay lengths in the parameter space relevant to FI. The heavy charged fermion can either decay into the neutral heavy fermion plus a soft pion through the exchange of a $W$ or directly into DM plus a $W$ boson. The first decay mode is purely due to gauge interactions, while the second one is due to the dimension-five operator, hence it is controlled by $\Lambda$. Different $\Psi_B^{\pm}$ decay products give different signatures at the LHC. In the right panel of Fig.~\ref{fig:spectrum}, we illustrate how the different branching ratios (BR) of $\Psi_B^\pm$ decay products depend on the neutral mediator lifetime $c\tau_0$, which we take as a proxy for the scale $\Lambda$ (as $c\tau_0\propto \Lambda^2$). The $\Lambda$ parameter indeed drives the relative importance of the $\Psi^{\pm}_B$ decays induced by the dimension-five operator relative to the gauge-induced decays. The decay length of the $\Psi_B^{\pm}$ is at most few cm, when the decay is dominated by the gauge-induced interactions. Shorter lifetimes can be obtained only when $\Lambda$ is small such that the decay channels into DM dominate.

We distinguish in Fig.~\ref{fig:spectrum} between the hadronic and leptonic subsequent decays of the $Z$ and $W$ bosons. For small $c\tau_0$ (small $\Lambda$), $\Psi_B^{\pm}$ principally decays into $\chi W^{\pm}$. For larger $c \tau_0$ (larger $\Lambda$), $\Psi_B^{\pm}$ mainly decay into a soft $\pi^\pm$ and a heavy neutral fermion $\Psi_B^0$ which in turn decays to $\chi$ + jets, leptons or $\gamma$. Let us also emphasize that, at the LHC, mediator pair production involves the production of charged fermions $\Psi_B^{+} \Psi_B^{-}$ or the associated production of the $\Psi_B^{\pm} \Psi_B^0$ states, in all cases through s-channel electroweak bosons exchange.\footnote{There is no vertex involving a SM gauge boson and a pair of neutral heavy fermions $\Psi_B^0$.}

We denote with colored areas in Figure~\ref{fig:ExclRegion13} the regions excluded by LLP searches at LHC. For large values of the $\Psi_B^0$ lifetime, the strongest bound come from searches for disappearing tracks (DT)~\cite{ATLAS:2017oal,CMS:2018rea} excluding the purple region up to $m_C\simeq m_T \lesssim 480$~GeV. This topology is relevant when producing a charged $\Psi_B^\pm$ decaying to $\pi^\pm\Psi_B^0$ with the pion being too soft to be detected and $\Psi_B^0$ too long-lived to decay inside the detector. If instead $\Psi_B^0$ decays inside the detector, into DM plus a $Z$ or a photon, DT searches might not be sensitive since extra hits could be recorded in outer parts of the detector and associated to the track of the charged mediator. In order to avoid this issue, we conservatively require that the decay of the neutral component $\Psi_B^0$ occurs outside the tracker for the DT search to be applicable (see App.~\ref{app:DT} for details). This is the reason why DT searches appear to be sensitive only for $c\tau_0\gtrsim 1$~m. In the plots of Fig.~\ref{fig:ExclRegion13}, we show the DT exclusion region following from the ATLAS search. Indeed, ATLAS performs better than CMS for small decay lengths (see the discussion in Sec.~\ref{sec:searches}) which seems to be more relevant for the model under consideration.

Other signatures arise when the neutral component of the triplet decays inside the detector, or when the charged component decays directly to the singlet and a $W$ boson.\footnote{The charged component will only decay to the singlet when $\Lambda$ (equivalently, $c\tau_0$) is small enough that $\Gamma_{\Psi_B^\pm \rightarrow \chi W^\pm} > \Gamma_{\Psi_B^\pm \rightarrow \Psi_B^0 \pi^\pm}$. This predominantly happen inside the detector, see the right panel of Fig.~\ref{fig:spectrum}.} If the $Z$ or $W$ boson decays hadronically, a displaced vertex (DV) or delayed jet (DJ) + MET signature could be observed. The heavy neutral $\Psi_B^0$ is in some cases produced in the decay chain of charged $\Psi_B^\pm$, which is itself long-lived (with a decay length of at most few cm), together with a soft pion (see Fig.~\ref{fig:spectrum}). In this case, we assume that $\Psi_B^0$ is emitted in the same direction as $\Psi_B^\pm$, and hence we add the two displacements to define the total displacement of the resulting gauge boson. We consider all combinations of production and decay modes (with the corresponding branching ratios) to obtain the final states leading to displaced jets plus missing energy. Final states always contain two DM particles, consistently with $\mathbb{Z}_2$ conservation, and two gauge bosons which can be either $W$ or $Z/\gamma$ (plus possibly soft pions). The relevant decay chains for these searches are the ones leading to jets (i.e.~the hadronic decays of $W$ or $Z$). The hadronic BR of $\Psi_B^{\pm}$, as shown in Fig.~\ref{fig:spectrum}, depends on the neutral component decay length. The BR of $\Psi_B^0$ into hadronic final states is instead independent on the value of its decay length. In Fig.~\ref{fig:ExclRegion13}, the regions excluded by the DV~\cite{ATLAS:2017tny} and DJ \cite{CMS:2019qjk} searches are shown in light and dark blue, respectively. For moderate values of the lifetime ($c\tau_0 \lesssim 10$~m) they put strong bounds on the mass of the triplet, excluding up to $m_T \lesssim 1.3$~TeV.

The other LLP signatures shown in Tab.~\ref{tab:models-searches} arise in this model. Leptonic decays of the $Z$ and $W$ boson lead to a rich variety of topologies, including displaced vertices + $\mu$~\cite{ATLAS:2020xyo}, displaced lepton vertices~\cite{ATLAS:2019fwx} and displaced leptons~\cite{ATLAS:2020wjh,CMS:2021kdm}. However, because of the comparatively small leptonic BR of the electroweak gauge boson, we expect these searches to be less constraining that the ones targeting hadronic final states plus missing energy. From Fig.~\ref{fig:ExclRegion13}, we observe for instance that the displaced jets+muon search is once again slightly less constraining than the DV search, as already happened in the topphilic model.  In Fig.~\ref{fig:ExclRegion13} we show also some representative sensitivity curves for purely leptonic LHC searches. In uniform green we display the reach of the DL ATLAS analysis~\cite{ATLAS:2020wjh}. Note that this search targets same-flavor ($e^{\pm} e^{\mp}$ and $\mu^{\pm} \mu^{\mp}$) as well as different flavors ($e^{\pm} \mu^{\mp}$) final states,  which are all possible topologies for the singlet-triplet model (from leptonically decaying $W$ and/or $Z$). By combining all these channels we obtain the sensitivity in Fig.~\ref{fig:ExclRegion13}. The reach of this search cannot overcome the search focusing on DV+MET, but it provides nevertheless a  significant coverage of the parameter space of the model. For completeness we also show the reach of the the displaced lepton vertex search~\cite{ATLAS:2019fwx} (black region denoted as DLV in Fig.~\ref{fig:ExclRegion13}), see App.~\ref{app:DLV} for details. This search requires the two leptons to point to the same displaced vertex, hence effectively targets displaced $Z$ for our model. 

We have already discussed for the previous simplified models how the large luminosity of HL-LHC can further increase the mass reach of the LLP searches. Here we would like instead to emphasize the impact on this model of future detectors targeting particles with large ($\mathcal{O}(10-100)$ m) decay lengths, such as MATHUSLA~\cite{Curtin:2018mvb} and CODEX-b~\cite{Aielli:2019ivi}. First, in the singlet-triplet model the long-lived particle is neutral and can naturally reach the far detector, hence we expect that these facilities will significantly probe the large $c\tau$ region (a dedicated investigation as the one performed for the singlet-doublet model in Ref.~\cite{No:2019gvl} is left for future works). Second, the freeze-in DM production is UV sensitive. As a consequence, also for very small couplings (corresponding to large $c{\tau}$), the value of $T_R$ is set by the relic abundance constraint, as it can be seen by the contours in the left panel of Fig.~\ref{fig:ExclRegion13}, and thus one can infer an absolute upper bound on the allowed $T_R$. For instance, as we can see from the figure, observing a signal of a $1$~TeV mediator with lifetime $c{\tau} \simeq 10^4$~cm would imply, in the singlet-triplet model, an upper bound on $T_R$ of approximately $T_R \lesssim 10^5$~GeV. This would have relevant implications for well-motivated cosmological scenarios such as leptogenesis and for models of inflation.

Finally, comparing the results in Fig.~\ref{fig:ExclRegion13} with those for the other models in Figs.~\ref{fig:lepto} and~\ref{fig:top}, one can see how the relative sensitivity of different LPP searches differs in the three scenarios. This would provide an important handle to discriminate among possible models, in the (lucky) case of the observation of a pattern of signals from different LLP searches. The problem of determining the theoretical model underlying the LLP production is quite generic. For significant progress in this direction, see e.g.~\cite{Barron:2020kfo}.

\chapter{MeV-scale inelastic dark matter in vector portal models}
\label{chap:lightDM}

In Chapter~\ref{chap:model_class} and \ref{chap:LHC_cons}, we focused on simplified models constructed on the basis of a cubic interaction involving the DM, mediator and a SM particle. These models are often referred to as t-channel models. This nomenclature stems from the WIMP regime, where the DM annihilation process is achieved by a t-channel annihilation process with the mediator in the t-channel. In this chapter, we will switch gears and focus on s-channel or portal DM models~\cite{Abdallah:2015ter}, where the dark and SM sector are connected by a portal. This can be for instance a (pseudo-)scalar portal or vector portal, or potentially even both~\cite{Duerr:2016tmh}. These portals arise when the \ac{DS} contains a scalar $\Phi$ or vector $V^\mu$ particle. Even when those particles are completely neutral under the SM gauge group, they couple to the SM through mixing with the Higgs or hypercharge boson respectively,
\begin{align}
    \mathcal{L}_{\rm Scalar} &\supset -\lambda |\Phi|^2 |H|^2, \\
    \mathcal{L}_{\rm Vector} &\supset \frac{\epsilon}{2} B_{\mu \nu} F'^{\mu\nu}, \label{eq:vect_portal}
\end{align}
where $H$ is the Higgs boson and $B_{\mu \nu}$ and $F'^{\mu \nu}$ are the field strength tensors of the hypercharge and DS vector boson respectively. The parameters $\lambda$ and $\epsilon$ determine the strength of the mixing. Here, we will focus on the vector portal coupling to a \ac{DS} containing multiple fermions to the SM sector. 

\section{Inelastic Dark Matter}

In the simplest scenario, the DS consists out of a DM particle $\chi$ neutral under the SM gauge symmetry but charged under a new abelian gauge $U(1)_D$ symmetry. Therefore, the DM particle interacts with the gauge boson, referred to as the dark photon $A'_\mu$, that kinetically mixes with the SM hypercharge gauge boson as shown in Eq.~\eqref{eq:vect_portal}. This provides a portal between $\chi$ and the SM fermions. When the DM is a Dirac fermion, the Lagrangian reads
\begin{align}
    \mathcal{L} \supset g_D A'_\mu J^\mu_D + \epsilon e A'_\mu \sum_f Q_f \bar{f} \gamma^\mu f,
\end{align}
where $g_D$ is the dark coupling constant, $J^\mu_D=\bar{\chi}\gamma^\mu\chi$ is the dark current, $e$ is the electromagnetic coupling and $f$ sums over all SM fermions with electric charge $Q_f$. The different production mechanisms in both the WIMP and FIMP regime for this model have been studied in the literature for a massless~\cite{Chu:2011be} and massive dark photon~\cite{Hambye:2019dwd}.\footnote{The dark photon can obtain a mass though a dark Higgs mechanism. When this dark Higgs boson is much heavier than the other DS particles, it will not influence the DM production dynamics.} The case with a pseudo-scalar portal has been studied by Ref.~\cite{Bharucha:2022lty}.

Due to the impressive sensitivity that direct detection experiments reach today, they do not only constrain the thermal WIMP regime, but also the FIMP regime of this simple vector portal model, as shown by Ref.~\cite{Hambye:2018dpi}.\footnote{Ref.~\cite{Belanger:2020npe} has shown that also in the scalar portal model, direct detection experiments start to reach the FIMP regime.} An elegant option to evade direct detection constraints is to consider \emph{inelastic Dark Matter}, where elastic nucleon scattering through the vector portal is absent or parametrically suppressed and inelastic up-scattering into a heavier dark partner is kinematically suppressed~\cite{Tucker-Smith:2001myb}.

In a minimal realization of inelastic Dark Matter, commonly dubbed iDM, the two Weyl fiels $\xi_1$ and $\xi_2$ contained in the Dirac state $\chi$ obtain a Majorana mass term. These mass terms are typically small compared to the Dirac mass as they spontaneously break the $U(1)_D$ symmetry. In the diagonal mass basis, the spectrum is splitted into two nearly degenerate mass states. For instance, when the Majarona mass terms for $\xi_1$ and $\xi_2$ are equal, the mass eigenstates are $\chi_1=i(\xi_1-\xi_2)/\sqrt{2}$ and $\chi_2=(\xi_1+\xi_2)/\sqrt{2}$ with eigenvalues $m_{1,2} = m_D \pm m_M$, where $m_D$ and $m_M$ are respectively the Dirac and Majorana mass. The dark current coupling to the dark photon equals
\begin{align}
    \label{eq:current_idm}
    J^\mu_D = i(\chi_1^\dagger \bar{\sigma}^\mu \chi_2 - \chi_2^\dagger \bar{\sigma}^\mu \chi_1).
\end{align}
This off-diagonal interaction with the dark photon drives the DM freeze-out in the early universe and the relic abundance is set through co-annihilation $\chi_1\chi_2 \to A'^\ast \to f\bar{f}$ into Standard Model (SM) fermions $f$.\footnote{When both Majorana masses are not equal, diagonal currents coupling to the dark photon arise. This coupling is proportional to the mass difference relative to the Dirac mass scale. Since the off-diagonal current couples to the dark photon without this suppression, it still dominates the freeze-out dynamics.} The abundance measured today favors iDM candidates in the MeV-GeV mass range. The phenomenology of this predictive scenario has been investigated in detail~\cite{Izaguirre:2015zva, Izaguirre:2015yja,Bramante:2016rdh,Izaguirre:2017bqb,Berlin:2018pwi,Berlin:2018bsc,Tsai:2019buq,Duerr:2019dmv,Duerr:2020muu,Kang:2021oes,Baryakhtar:2020rwy,Batell:2021ooj,CarrilloGonzalez:2021lxm,Bell:2021xff}, specifically for direct detection~\cite{Bramante:2016rdh,Baryakhtar:2020rwy,CarrilloGonzalez:2021lxm,Bell:2021xff}, at colliders~\cite{Duerr:2019dmv,Duerr:2020muu,Kang:2021oes}, and at fixed-target experiments~\cite{Izaguirre:2017bqb,Berlin:2018pwi,Berlin:2018bsc,Tsai:2019buq,Batell:2021ooj}. Taken together, searches for light dark particles in all three areas have excluded most of the parameter space of iDM. Inelastic dark matter from thermal freeze-out via coannihilation appears to be strongly constrained.

In Ref.~\cite{Filimonova:2022pkj}, we have introduced \emph{inelastic Dirac Dark Matter} (i2DM) as a new model for feebly interacting DM. We promote the two dark fermions in iDM to Dirac fields, one being charged and the other one uncharged under the new $U(1)_D$ gauge symmetry. This symmetry is spontaneously broken by the vacuum expectation value of a dark scalar, which induces mixing between the dark fermions and gives the dark gauge boson a mass. We will discuss the details about the spontaneous symmetry breaking and mixing within this model in Sec.~\ref{sec:i2dm_model}.

A priori, such a dark sector could be realized at any mass scale. However, just as for iDM, the relic density calculations prefer masses in the MeV-GeV range. Throughout the following sections, we therefore focus on dark particles in this mass range. The dark current coupling to the dark photon in i2DM reads
\begin{align}\label{eq:current_i2DM}
    J^\mu_D = \sin^2\theta  \bar{\chi}_1 \gamma^\mu \chi_1 - \sin\theta \cos\theta (\bar{\chi}_1 \gamma^\mu \chi_2 + \text{h.c.}) + \cos^2\theta \bar{\chi}_2 \gamma^\mu \chi_2
\end{align} 
where $\theta$ is the mixing angle between the dark fermions. When taking the kinetic mixing properly into account, the dark current also couples to the $Z$ boson. This coupling is suppressed by a at least one factor of $\epsilon$, see Sec.~\ref{sec:i2dm_model}.

The mass mixing between the dark fermions determines the relative coupling strength of $\chi_1$ and $\chi_2$ to the dark photon. For $\theta \to 0$, the dark fermion $\chi_1$ decouples. Throughout our analysis, we assume that the dark scalar responsible for the mixing is much heavier than the other dark particles and does not affect the observables we consider. In general, the presence of a light dark scalar could lead to interesting effects \cite{Duerr:2020muu,Baek:2020owl} and deserves a dedicated analysis.

Just as in iDM, elastic scattering off atomic nuclei is suppressed in i2DM models for small dark fermion mixing $\theta$. Despite similar predictions for nucleon scattering, i2DM and iDM feature very distinct DM dynamics in the early universe. The difference lies mostly in efficient $A'\chi_2\bar{\chi}_2$ interactions in i2DM, which are suppressed or completely absent in iDM. This moderate modification of iDM leads to a very different cosmology since other processes than the co-annihilation process can set the relic abundance. More specifically partner annihilation $\chi_2 \chi_2 \to f\bar{f}$ or even coscattering $\chi_1 f\to \chi_2 f$ processes can dominate the freeze-out dynamics. We have already discussed the general features of these mechanisms in Chapter~\ref{chap:DM_prod}. In Sec.~\ref{sec:freeze-out}, we will apply them to i2DM and discuss in details the effect of decoupling from kinetic equilibrium. This work is based on Ref.~\cite{Filimonova:2022pkj}.

\section{Formalism for i2DM}
\label{sec:i2dm_model}

The main interactions at leading order of the kinetic mixing parameter $\epsilon$ of the fermionic dark mass eigenstates with the dark photon can be parameterized by the dark current in Eq.~\eqref{eq:current_i2DM}. However, when we take into account the kinetic mixing properly, this dark current also couples to the SM $Z$ boson and obtains higher order corrections in $\epsilon$. The kinetic mixing also changes the predictions of the $Z$ boson mass compared to its SM value. To illustrate this, we will discuss i2DM in full detail, starting from the gauge-invariant interactions of the dark sector with the Standard Model in the unbroken phase of the underlying theory,
\begin{align}\label{eq:lagrangian}
	\mathcal{L} = & \ i\,\bar{\chi}_D \gamma^\mu D_\mu \chi_D -m_D \bar{\chi}_D \chi_D + i\, \bar{\chi}_0 \gamma^\mu \partial_\mu \chi_0 - m_0 \bar{\chi}_0 \chi_0\\\nonumber
	& \qquad + (D^\mu \phi_D)^\dagger (D_\mu \phi_D) + V(\phi_D) - (\lambda \phi_D \bar{\chi}_D \chi_0 + h.c.)\\\nonumber
	& \qquad - \frac{1}{4} \hat{F}_{D\mu\nu} \hat{F}_D^{\mu\nu} - \frac{1}{2} \frac{\epsilon}{c_W} \hat{B}_{\mu\nu} \hat{F}_D^{\mu\nu},
\end{align}
where $D_\mu = \partial_\mu + i g_D Q_D\hat{A}_{D\mu}$ is the covariant derivative for the gauge field $\hat{A}_D$ of the dark $U(1)_D$ symmetry. The field strength tensors for the dark and the hypercharge gauge fields are denoted as $\hat{F}_{D}^{\mu\nu}$ and $\hat{B}^{\mu \nu}$ respectively. The dark fermion field $\chi_D$ is charged under $U(1)_D$, while $\chi_0$ is a singlet under all gauge interactions.

The scalar $\phi_D$ with potential $V(\phi_D)$ and charge $Q_D=1$ facilitates spontaneous $U(1)_D$ breaking, once it develops a vacuum expectation value $v_D$, so that $\phi_D=(v_D+\varphi_D)/\sqrt2$.  Upon symmetry breaking, the dark photon $\hat{A}_D$ acquires a mass $m_{\hat{A}_D} = g_D v_D$, and the dark fermions $\chi_D$ and $\chi_0$ mix through the dark Yukawa coupling $\lambda$, resulting in two mass eigenstates
\begin{equation}
	\begin{pmatrix}
	\chi_1 \\ \chi_2
	\end{pmatrix}
	=
	\begin{pmatrix}
	\cos \theta & -\sin \theta \\
	\sin \theta & \cos \theta
	\end{pmatrix}
	\begin{pmatrix}
	\chi_0 \\ \chi_D
	\end{pmatrix},\qquad \sin(2 \theta) = \frac{\sqrt{2}\lambda v_D}{\Delta m_1}\,.
        \label{eq:rotferm}
\end{equation}
The corresponding masses are
\begin{align}
	m_{1,2} = \frac{1}{2}\big(m_D+m_0 \pm \Delta m_1\big) \quad {\rm with}\quad (\Delta m_1)^2 = (m_D-m_0)^2+2 (\lambda v_D)^2\,,
	\label{eq:mass_split_i2dm}
\end{align}
where $\Delta = (m_2-m_1)/m_1$ is the relative mass difference. In order to obtain canonical kinetic terms in Eq.~\eqref{eq:lagrangian}, we first redefine the $U(1)_Y$ and $U(1)_D$ gauge bosons to absorb the kinetic mixing via
\begin{equation}
	\begin{pmatrix}
	\hat{B}^\mu  \\ \hat{A}^\mu_D
	\end{pmatrix}
	= G(\hat{\epsilon})
	\begin{pmatrix}
	\tilde{B}^\mu \\ \tilde{A}^\mu_D
	\end{pmatrix}
	=
	\begin{pmatrix}
	1  && -\frac{\hat{\epsilon}}{\sqrt{1-\hat{\epsilon}^2}}\\
	0  && \frac{1}{\sqrt{1-\hat{\epsilon}^2}}
	\end{pmatrix}
	\begin{pmatrix}
	\tilde{B}^\mu  \\ \tilde{A}^\mu_D
	\end{pmatrix},
        \label{eq:trsfkin}
\end{equation}
where $\hat{\epsilon}=\epsilon/\cos \theta_W$. Subsequently, we transform the fields to the physical eigenstates of weak interactions, $A^\mu,\,Z^\mu,\,A'^{\mu}$, via two rotations: one rotation, $R_\xi$, mixing the $SU(2)_L$ gauge field $W^3_\mu$ and $\tilde{A}^\mu_D$ with an angle $\xi$; and a second rotation, $R_W$, mixing $W^3_\mu$ and $\tilde{B}_\mu$ with the Weinberg angle $\theta_W$. The overall transformation takes the form
\begin{equation}
\begin{pmatrix}
\tilde B^\mu \\ W^{3\mu} \\ \tilde A^\mu_D
\end{pmatrix}
=
\begin{pmatrix}
c_W && -c_\xi s_W && s_\xi s_W  \\
s_W && c_\xi c_W && -s_\xi c_W \\
0 && s_\xi && c_\xi
\end{pmatrix}\,
\begin{pmatrix}
A^\mu \\ Z^{\mu} \\ A'^\mu
\end{pmatrix},
\label{eq:rotgauge}
\end{equation}
where
\begin{eqnarray}
  &&\tan 2 \xi = \frac{2 \eta s_W}{1-(\eta s_W)^2-\delta} \label{eq:xi}\\\nonumber
  &&{\rm with}\quad \eta=\hat{\epsilon} / \sqrt{1-\hat{\epsilon}^2}, \quad \delta =\hat{\delta}/(1-\hat{\epsilon}^2) ,\quad  \hat{\delta}=m_{\hat{A}_D}^2/m_{\hat{Z}}^2\,,
\end{eqnarray}
and $m_{\hat{Z}}= g v/(2c_W)$, with the $SU(2)_L$ gauge coupling, $g$, and the SM Higgs vacuum expectation value, $v$. The masses of the $Z$ boson and the dark photon are
\begin{eqnarray}
    m^2_Z=m_{\hat{Z}}^2\left((c_\xi+\eta s_\xi s_W)^2+s_\xi^2\delta\right)\quad {\rm and}\quad m^2_{A'}=m_{\hat{Z}}^2\left((- s_\xi + c_\xi\eta s_W)^2+c_\xi^2\delta\right),
  \label{eq:gbmasses}
\end{eqnarray}
while the photon remains massless. In the limit of small kinetic mixing, $\{\epsilon, \eta,\xi \} \ll 1$, one recovers $ m_Z\simeq m_{\hat{Z}}$ and $m_{A'}\simeq m_{\hat{A}_D}$~\cite{Wells:2008xg,Curtin:2014cca,Freitas:2015hsa,Foldenauer:2019vgn}.

In terms of mass eigenstates, the Lagrangian finally reads\footnote{To derive the Feynman rules and transition amplitudes, we implemented the Lagrangian into {\tt FeynRules}~\cite{Alloul:2013bka}. In particular, we have adapted the \emph{Hidden Abelian Higgs Model}~\cite{Wells:2008xg}, which is available from the {\tt FeynRules} model data base to i2DM. Our {\tt FeynRules} model for i2DM is available at~\cite{FeynRulesModel}.}
\begin{align}\label{eq:lag-mass}
	\mathcal{L} \supset & \ -e \big(A_\mu - \epsilon A'_{\mu}\big) J^\mu_{em} + \frac{e}{2 s_W c_W} Z_\mu J^\mu_Z - g_D \left(A'_{\mu} + \epsilon\, t_W Z_\mu\right) J_D^\mu \\ \nonumber
				& - \frac{\lambda}{\sqrt{2}} \varphi_D\big(c_{2\theta} (\bar{\chi}_1 \chi_2 + \bar{\chi}_2 \chi_1) - s_{2\theta} (\bar{\chi}_1 \chi_1 - \bar{\chi}_2 \chi_2)\big)\\ \nonumber
			&+\frac{g_D^2}{2} \big( 2 v_D \varphi_D + \varphi_D^2\big) \big(A'_{ \mu} A'^\mu + \epsilon\,t_W(A'_{ \mu} Z^\mu+Z_\mu A'^\mu) + \epsilon^2 t_W^2 Z_\mu Z^\mu \big)\,,
\end{align}
where $t_W$ refers to the tangent of the Weinberg angle $\theta_W$. The dark current $J^\mu_D$ is defined in Eq.~\eqref{eq:current_i2DM}. The SM fermion currents are given by 
\begin{align}
    J^\mu_Z = \bar{f}(c_V \gamma^\mu - c_A \gamma^\mu \gamma^5)f,\quad J^\mu_{em} = Q_f \bar{f} \gamma^\mu f,
\end{align}
where $f$ are the SM fermions and $c_V=T_f^3 - 2s_W^2 Q_f$, $c_A= T_f^3$ are the electroweak charges, with $T_f^3$ the weak isospin quantum number and $Q_f$ the electric charge in units of $e$. Here we have only included the leading terms in $\epsilon$ and provided the SM fermion couplings in unitary gauge.

From the transformations of Eqs.~(\ref{eq:trsfkin}) and (\ref{eq:rotgauge}) and the Lagrangian in Eq.~\eqref{eq:lag-mass}, it should be clear that the photon $A^\mu$ has no field component from the dark $U(1)_D$ gauge field $\hat{A}^\mu_D$. As a result, the photon does not couple directly to the dark fermions $\chi_1,\,\chi_2$. In particular, the DM candidate $\chi_1$ carries no millicharge and is not subject to otherwise strong constraints~\cite{Hambye:2019dwd}.

The phenomenology of i2DM is described by six independent parameters
\begin{align}
    \big\{m_1,\Delta,m_{A'},\alpha_D,\epsilon,\theta\big\}
\end{align}
with $\alpha_D = g_D^2/(4\pi)$. Throughout this work we focus on the mass hierarchy
\begin{align}\label{eq:mass-hierarchy}
    m_{A'} > 2 m_2\,,
\end{align}
so that decays of the dark photon into dark fermions are kinematically allowed.\footnote{For smaller $m_{A'}$, the phenomenology can change significantly. In particular, for $m_{A'} < m_1$, pair annihilations $\chi_1\chi_1 \to A'A'$ are important to set the relic abundance (see e.g.~\cite{DAgnolo:2015ujb}) and dark photons decay exclusively into SM fermions.} The total decay width of the dark photon is given by
\begin{align}
    \Gamma_{A'} & = \alpha_D \Gamma(A' \to \chi\overline{\chi}) + \epsilon^2 \alpha_{\rm EM} \Gamma(A' \to \rm{SM}),
\end{align}
where $\alpha_{\rm EM}=e^2/(4\pi)$ is the fine structure constant, while $\Gamma(A' \to \chi\overline{\chi})$ and $\Gamma(A' \to \rm{SM})$ denote the (normalized) decay rates into dark fermions and into leptons and hadrons, respectively. Dark photon decays into pairs of dark fermions, $\chi_i\overline{\chi}_j$, or leptons, $\ell_i \bar{\ell}_j$, are described by the kinematic function
\begin{align}
    \Gamma(m_i,m_j) & = \frac{m_{A'}}{2}
    \left[\left(1 - \frac{(m_i + m_j)^2}{m_{A'}^2}\right)\left(1 - \frac{(m_i - m_j)^2}{m_{A'}^2}\right)\right]^{\frac{1}{2}}\\\nonumber
    & \qquad\qquad \times \left(1 - \frac{(m_i - m_j)^2 - 4 m_i m_j}{2m_{A'}^2} - \frac{(m_i^2 - m_j^2)^2}{2 m_{A'}^4}\right),
\end{align}
with $m_i$ and $m_j$ the masses of the decay products. Dark photon decays into hadrons can be computed by rescaling the leptonic decay rate with $e^+e^-$ data~\cite{Ilten:2018crw}. For $\epsilon^2\alpha_{\rm EM} \ll \alpha_D$, decays into SM particles are suppressed and the dark photon mostly decays into dark fermions. The corresponding decay rate is
\begin{align}
\Gamma(A' \to \chi\overline{\chi}) = \sin^4\theta\,\Gamma(m_1,m_1) + \sin^2(2\theta) \,\Gamma(m_1,m_2) + \cos^4\theta\, \Gamma(m_2,m_2)\,.
\end{align}
For $\Delta \ll 1$, the branching ratios are determined to a good approximation by the dark fermion mixing, so that
\begin{align}\label{eq:brs}
    \mathcal{B}(A'\to \chi_1\overline{\chi}_1) &\approx \sin^4\theta\,,\\
    \mathcal{B}(A'\to \chi_1\overline{\chi}_2,\chi_2\overline{\chi}_1) &\approx \sin^2\theta \cos^2\theta\,,\\ 
    \mathcal{B}(A'\to \chi_2\overline{\chi}_2) &\approx \cos^4\theta\,.
\end{align}

The freeze-out dynamics rely on dark fermion annihilation into SM particles. Annihilations into leptons via $\chi_i \chi_j \to A'^\ast \to \ell^+\ell^-$ can be calculated in perturbation theory. Annihilations into hadrons can be predicted by rescaling the cross section for annihilation into muons with the measured ratio~\cite{Ilten:2018crw,Zyla:2020zbs}
\begin{align}\label{eq:hadronic-decays}
    R(s) = \frac{\sigma(e^+e^- \to \text{hadrons})}{\sigma(e^+e^- \to \mu^+\mu^-)}\,.
\end{align}
The total cross section for dark fermion annihilation at a center-of-mass energy $\sqrt{s}$ is then given by
\begin{align}
    \sigma_{\chi_i \chi_j \to \text{SM}}(s) = \sum_{\ell = e,\mu,\tau} \sigma_{\chi_i \chi_j \to \ell^+\ell^-}(s) + \sigma_{\chi_i \chi_j \to \mu^+\mu^-}(s)\,R(s)\,.
\end{align}
In dark fermion interactions with SM fermions, the dark photon acts as a virtual mediator. As a consequence, for $m_{A'} \gg m_{1,2}$, the scattering and decay rates of dark fermions scale as~\cite{Izaguirre:2015yja}
\begin{align}
  y = \epsilon^2\,\alpha_D \left(\frac{m_1}{m_{A'}}\right)^4.
\end{align}
As long as the dark photon is heavy compared to the momentum scale probed in observables, the dark sector interactions are described in terms of the four parameters
\begin{align}
    \{m_1,\Delta,y,\theta\}\,.
\end{align}
The phenomenology of i2DM crucially relies on the properties of the dark partner. For $m_2 \lesssim 1\,$GeV, the dark partner decays to almost 100\% into leptons~\cite{Duerr:2020muu}. The decay rate via a heavy virtual dark photon is given by
\begin{align}\label{eq:decay-rate}
   \Gamma(\chi_2 \to \chi_1 \ell^+ \ell^-) & = \frac{4 \alpha}{15 \pi} \tan^2\theta\cos^4\theta\, y\,m_1 \Delta^5 + \mathcal{O}\left(\frac{m_1^2}{m_{A'}^2}\right),
\end{align}
where we have neglected the lepton mass in the final state. We neglect hadronic decays in our analysis.

\section{Freeze-out at feeble couplings}
\label{sec:freeze-out}
Dark matter relics in the MeV-GeV range must be feebly coupled to the thermal bath in order to account for the observed DM abundance, $\Omega_\chi h^2 = 0.12$. Moreover, viable scenarios of inelastic dark matter typically require a compressed spectrum of dark-sector particles,\footnote{A small mass splitting is only technically natural when the Dirac masses $m_D$ and $m_0$ in Eq.~\eqref{eq:lagrangian} are equal, see Eq.~\ref{eq:mass_split_i2dm}.} to ensure that co-annihilation or mediator annihilation processes can contribute to setting the DM relic abundance. For i2DM, this leads to the parameter region of interest
\begin{equation}
  \{y, \theta, \Delta\} \ll 1\,.
  \label{eq:i2DMframework}
\end{equation}
Within this regime, the relic abundance can be set by several of the various mechanisms discussed in Chapter~\ref{chap:DM_prod}. As we will see, the coscattering process $\chi_1 e^- \to \chi_2 e^-$ will be always efficient at high temperatures in the parameter space we consider. The DM candidate will hence be in kinetic and chemical equilibrium over a period of time before the relic density is completely set. The production mechanism in such scenarios is typically a freeze-out process, and we expect the dark fermions to be non-relativistic around the freeze-out temperature
\begin{align}
    T_{\rm fo} \ll 2 m_2 < m_{A'}.
\end{align}
The freeze-out temperature $T_{\rm fo}$ is determined by the time  $x_{\rm fo} = m_1/T_{\rm fo}$ at which the comoving DM density approaches the DM abundance $Y_0$ observed today,
\begin{equation}
    Y_1(x_{\rm fo}) = Y_0\,.
\label{eq:fo}
\end{equation}
In our numerical analysis, we determine $x_{\rm fo}$ by requiring that the DM density at freeze-out satisfies
\begin{equation}\label{eq:freeze-out-temp}
    \frac{Y'_1(x_{\rm fo})}{Y_1^{(0)}(x_{\rm fo})} =  - 0.1\,.
\end{equation}
In Sec.~\ref{sec:CDFO}, we discussed that for feeble couplings, chemical and/or kinetic equilibrium with the thermal bath can be lost before the freeze-out temperature, leading to alternative production mechanisms such as the coscattering mechanism. In such regime, freeze-out as defined in Eq.~\eqref{eq:fo} does not necessarily coincide with chemical decoupling, as in the case of vanilla WIMP DM~\cite{Bringmann:2006mu}. We therefore define the times $x_1$ and $x_2$, where $\chi_1$ and $\chi_2$ chemically decouple from the bath, corresponding to the decoupling temperatures $T_1$ and $T_2$.

As already briefly mentioned in Sec.~\ref{sec:CDFO} for the coscattering mechanism, assuming kinetic equilibrium without introducing additional degrees of freedom is not trivial. Hence, we need to track the time evolution of the density distribution function $f_i(x,q_i)$ (expressed in terms of the dimensionless quantities $x=m_1/T$ and $q_i=p_i/T$) of a particle species $i$ using the Boltzmann equation
\begin{equation}
 E_i \, H x \, \partial_x f_i(x,q_i)=C[f_i(x,q_i),f_j(x,q_j)]\,.
  \label{eq:UnintBE}
\end{equation}
Here $E_i=E_i(x,q_i)$ is the energy associated with a species of mass $m_i$ and momentum $q_i$. The collision term $C$ describes interactions of species $i$ with all other involved species $j$. We emphasize that Eq.~(\ref{eq:UnintBE}) is only valid as long as the number of relativistic degrees of freedom in the universe is constant.\footnote{For an i2DM candidate within the parameter range we will consider ($10\MeV < m_1 < 1\GeV$), freeze-out happens between the QCD phase transition and neutrino decoupling and Eq.~\eqref{eq:UnintBE} applies.} In App.~\ref{app:BE_beyond_KE}, we provide for a more detailed derivation and expressions for the collision terms of the relevant processes in i2DM.

In what follows, we will refer to Eq.~\eqref{eq:UnintBE} as the \emph{unintegrated Boltzmann equation}. As we have discussed in Sec.~\ref{sec:FO+coann}, the set of $N$ partial integro-differential equations for species $i,j = \{1,\dots N\}$ can be reduced to a set of ordinary differential equations if kinetic equilibrium for all particle species involved can be assumed. The relic abundance can then be calculated from the time evolution of the number densities $n_i(x)$~\cite{Griest:1990kh,Gondolo:1990dk}. The corresponding evolution equations are referred to as \emph{integrated Boltzmann equations}. For the computations in this chapter we have used our own solver to numerically solve the Boltzmann equation. We have verified that our results for the relic abundance agree with the results obtained from micrOMEGAs~\cite{Belanger:2018ccd} in the regime where conversion processes are always efficient, since this is always implicitly assumed in micrOMEGAs.

In the remainder of this section, we will discuss the freeze-out dynamics specifically for i2DM. In Sec.~\ref{sec:relevant-processes}, we analyze all relevant annihilation, scattering and decay processes that can play a role in setting the DM relic abundance. Depending on the relative importance of these processes, we encounter different phases of freeze-out. In Sec.~\ref{sec:phases}, we discuss these phases in detail and explain how to account for deviations from chemical and kinetic equilibrium when computing the relic abundance. Note that this mechanism has only been considered in toy models~\cite{DAgnolo:2017dbv}, while we are the first ones considering it in a more realistic model and taking properly into account deviations from kinetic equilibrium.

\subsection{Dark matter interactions}
\label{sec:relevant-processes}

For i2DM, the relevant interactions of the dark fermions with the thermal bath entering the collision term $C$ in Eq.~\eqref{eq:UnintBE} are
\begin{align}
\text{(co-)annihilation:} & \quad \chi_i\chi_j  \to A'^{\ast} \to f\bar{f}\,,\\\nonumber
\text{(co)scattering:} & \quad \chi_i f \to \chi_j f\,,\\\nonumber
\text{(inverse) decay:} & \quad \chi_i \to \chi_j f\bar{f}\,.
\end{align}
At temperatures below the GeV scale, $f$ denotes all (hadronized) quarks and leptons that are in equilibrium with the thermal bath. To determine the relevance of the various processes for DM freeze-out, we investigate the thermally averaged interaction rates $\langle \Gamma \rangle$ of the dark fermions with the bath. We define the interaction rates for annihilation, scattering and decay of particle $i$ in terms of the reaction density defined in Eq.~\eqref{eq:react_rate_gen}
\begin{equation}\label{eq:interactions}
 \Gamth_{ij} = \frac{\gamma_{ij\to kl}}{n_i^{(0)}}, \qquad \Gamth_{i\to j}^{scat} = \frac{\gamma_{ik\to jl}}{n_i^{(0)}}, \qquad \Gamth_{i\to j}^{dec} = \frac{\gamma_{i\to jkl}}{n_i^{(0)}}\,.
\end{equation}
Here and below, $i,j = \{1,2\}$ denote the dark fermions $\chi_1$ and $\chi_2$, and $k,l$ label the SM fermions $f$. The brackets indicate the thermal average. The reaction densities $\gamma$ can be expressed as
\begin{eqnarray}
\gamma_{ab\to cd} = n_a^{(0)}n_b^{(0)}\,\sigmav_{ab\to cd},\qquad
\gamma_{a\to bcd} = n_a^{(0)}\,\Gamma_{a\to bcd}\, \frac{K_1(m_a/T)}{K_2(m_a/T)} \,,\label{eq:gam-min}
\end{eqnarray}
where $a,b,c,d$ can denote particles from both the visible and the dark sector, $\sigmav_{ab\to cd}$ is the thermally averaged cross section, $\Gamma_{a\to bcd}$ is the decay rate of particle $a$ in its rest frame, and $K_1,\,K_2$ denote the modified Bessel functions of the first and second kind. For simplicity, we label the reaction densities only by the involved dark fermions $i,j$ as
\begin{align}    \label{eq:gamij}
     \gamma_{ij} & \equiv \gamma_{ij \to kl} =\gamma_{kl\to ij}\,,\\\nonumber
     \gamma_{i\to j}^{scat} & \equiv \gamma_{ik\to jl} = \gamma_{jl\to ik}\,,\\\nonumber
     \gamma_{i\to j}^{dec} & \equiv \gamma_{i\to jkl} = \gamma_{jkl\to i}\,.
\end{align}

The relevance of the various interaction rates at a certain temperature can be inferred from their scaling with the time variable $x = m_1/T$ and with the model parameters. For non-relativistic dark sector particles $\chi_1$ and $\chi_2$ and relativistic involved SM fermions around the time of freeze-out, the reaction densities for (co-)annihilations in thermal equilibrium scale as
\begin{align}\label{eq:annihilation}
 \gamma_{11} & = n_1^{(0)} n_1^{(0)}\, \sigmav_{\chi_1 \chi_1 \to f\bar{f}} \propto x^{-3} e^{-2x}  \, y \, \sin^4\theta \,,\\\nonumber
 \gamma_{12} & = n_1^{(0)} n_2^{(0)}\, \sigmav_{\chi_1 \chi_2 \to f\bar{f}} \propto x^{-3}  e^{-2x}  \, y \,  \sin^2(2\theta)  \, e^{-x\Delta}
 \,,\\\nonumber
  \gamma_{22} & = n_2^{(0)} n_2^{(0)}\, \sigmav_{\chi_2 \chi_2 \to f\bar{f}} \propto x^{-3} e^{-2x} \, y \,\cos^4\theta \, e^{-2x\Delta}.
\end{align}
The (co)scattering reaction densities scale as
\begin{align}\label{eq:scattering}
  \gamma_{1\to1}^{scat} & = n_1^{(0)} n_f^{(0)}\,  \sigmav_{\chi_1 f \to \chi_1 f} \propto x^{-9/2} e^{-x} \, y \,\sin^4\theta\,  v_{1f}^3\,,\\\nonumber
    \gamma_{2\to1}^{scat} & = n_2^{(0)} n_f^{(0)}\, \sigmav_{\chi_2 f \to \chi_1 f} \propto x^{-9/2} e^{-x} \, y \,\sin^2(2\theta) \, v_{2f}^3\, e^{-x \Delta}
    \,,\\\nonumber
    \gamma_{2\to2}^{scat} & = n_2^{(0)} n_f^{(0)}\, \sigmav_{\chi_2 f \to \chi_2 f} \propto x^{-9/2} e^{-x} \, y  \, \cos^4 \theta\, v_{2f}^3 \, e^{-x \Delta}\,,
\end{align}
where $v_{a,b}$ is the M{\o}ller velocity defined in Eq.~\eqref{eq:moller_v}. Three-body decays and inverse decays yield the reaction density
\begin{equation}
    \gamma_{2\to1}^{dec} = n_2^{(0)}\,\Gamma_{\chi_2 \to \chi_1 f\bar{f}}\,\frac{K_1(x (1+\Delta))}{K_2(x (1+\Delta))} \propto x^{-3/2} e^{-x}\, y\,\sin^2(2\theta)\, \Delta^5 \,  e^{-x\Delta},
    \label{eq:decay}
\end{equation}
given that $K_1(x)/K_2(x)\to 1$ for $x\gg 1$.\footnote{In Eqs.~\eqref{eq:annihilation}, \eqref{eq:scattering} and \eqref{eq:decay} we have neglected subleading contributions in $m_1^2/m_{A'}^2$, $m_f^2/m_{A'}^2$, $\Delta$ and $v_{ij}$, but include them in our numerical analysis.} For processes with leptons in the final state, we have calculated all rates analytically. To compute annihilations into hadrons, we rescale the cross section as described in Sec.~\ref{sec:i2dm_model}, following Ref.~\cite{Duerr:2020muu}. For scattering processes we only include the dominant scatterings $\chi \ell^\pm \to \chi \ell^\pm$ off leptons $\ell = \{e,\mu\}$, which are still relativistic at sub-GeV temperatures. We neglect scatterings off hadrons, whose number densities are Boltzmann-suppressed for masses above the muon mass.

In the expressions above, the variable $y$ determines the effective strength of the dark force and thereby the overall efficiency of dark sector interactions. As we will discuss in Chapter~\ref{chap:i2dm_pheno}, $y$ is constrained by cosmology and laboratory searches. In addition, the dark fermion mixing $\theta$ must be small to circumvent bounds from direct detection experiments. Therefore DM annihilation $\gamma_{11}$ and scattering $\gamma_{1\to 1}$ must be strongly suppressed. In the absence of further interactions, such a suppression would lead to an overabundance of DM today. As a consequence, feebly interacting i2DM candidates cannot be thermal WIMPs in the classical sense, where the relic abundance is determined by the WIMP pair-annihilation rate at freeze-out.

Indeed, for feebly coupling i2DM, DM pair annihilation and scattering with the thermal bath play no role for the temperature evolution of the DM density. Instead, the evolution of $\chi_1$ is driven by interactions with the dark partner $\chi_2$. The dark partner is kept in chemical and kinetic equilibrium with the bath via efficient annihilation and scattering, driven by the reaction densities $\gamma_{22}$ and $\gamma_{2\to 2}$. However, at freeze-out, all interaction rates of $\chi_2$ are exponentially Boltzmann-suppressed by powers of $e^{-x\Delta}$, compared to the interactions of $\chi_1$. For $\chi_2$ interactions to impact the freeze-out of $\chi_1$, the mass difference $\Delta$ must be small, as in Eq.~\eqref{eq:i2DMframework}. Viable scenarios of i2DM with small DM couplings hence require a compressed spectrum of dark fermions.

Conversions $\chi_1\leftrightarrow \chi_2$ play an essential role in i2DM freeze-out; they keep DM in equilibrium with the thermal bath and impact the evolution of the number density. Both coscatterings and decays contribute to the conversion rate of $\chi_i$,
\begin{align}\label{eq:conversion-rate}
   \Gamth_{i\to j} \equiv \Gamth_{i\to j}^{scat} + \Gamth_{i\to j}^{dec} = \frac{\gamma_{i\to j}^{scat} + \gamma_{i\to j}^{dec}}{n_i^{(0)}}\,.
\end{align}
Due to the respective scaling of the reaction densities with $x$, see Eqs.~\eqref{eq:scattering} and \eqref{eq:decay}, the suppression of the thermal rates at low temperatures is stronger for coscattering ($\gamma_{2\to 1}^{scat} \propto x^{-9/2}$) than for decays ($\gamma_{2\to 1}^{dec} \propto x^{-3/2}$). Depending on their relative amplitude at a given temperature, either process can dominate the conversion rate $\Gamth_{i\to j}$ and thus the evolution of
the number densities.

\begin{figure}
   \centering
  \includegraphics[width=0.4\textwidth]{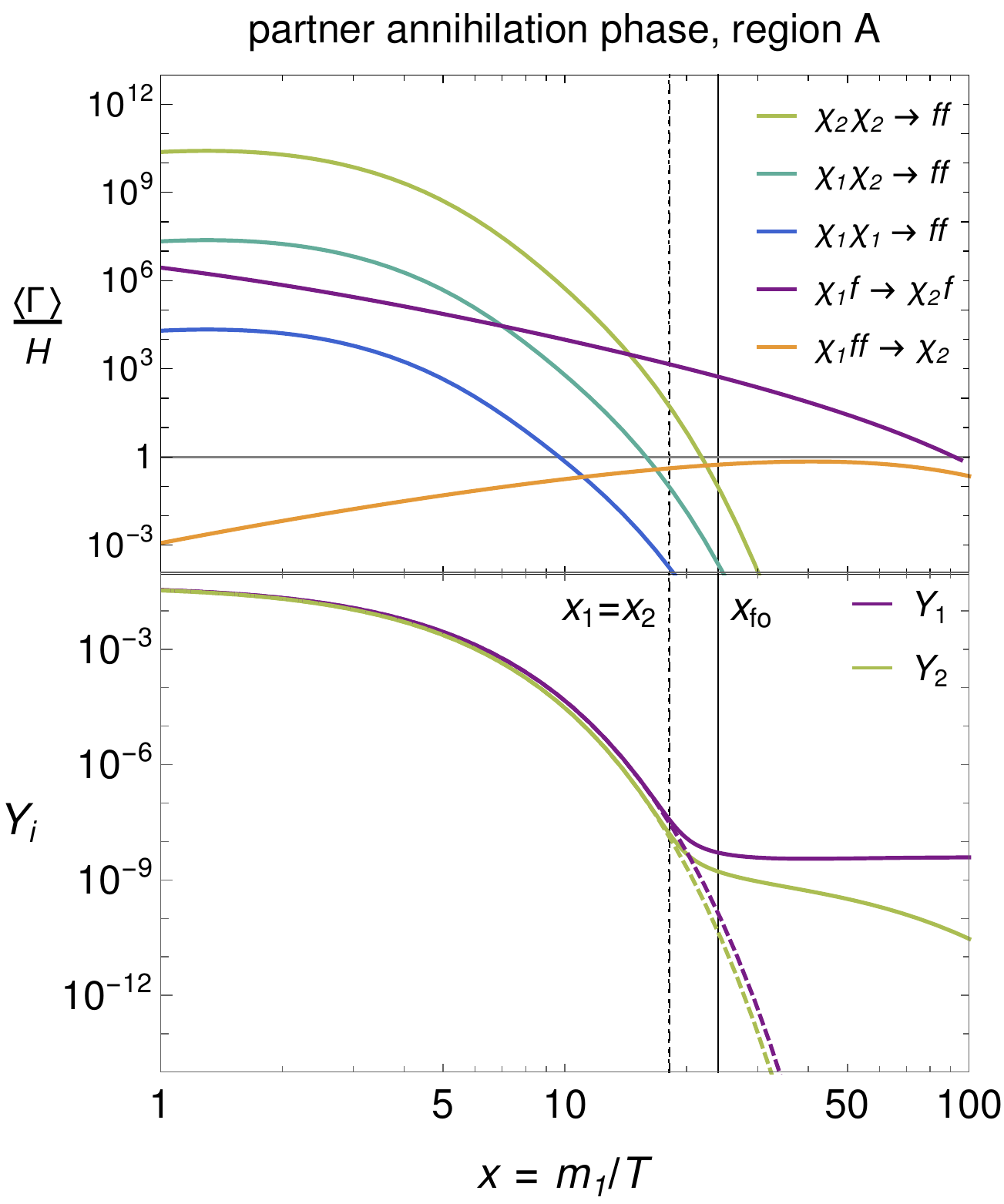}
  \hspace{0.03\textwidth}
  \includegraphics[width=0.4\textwidth]{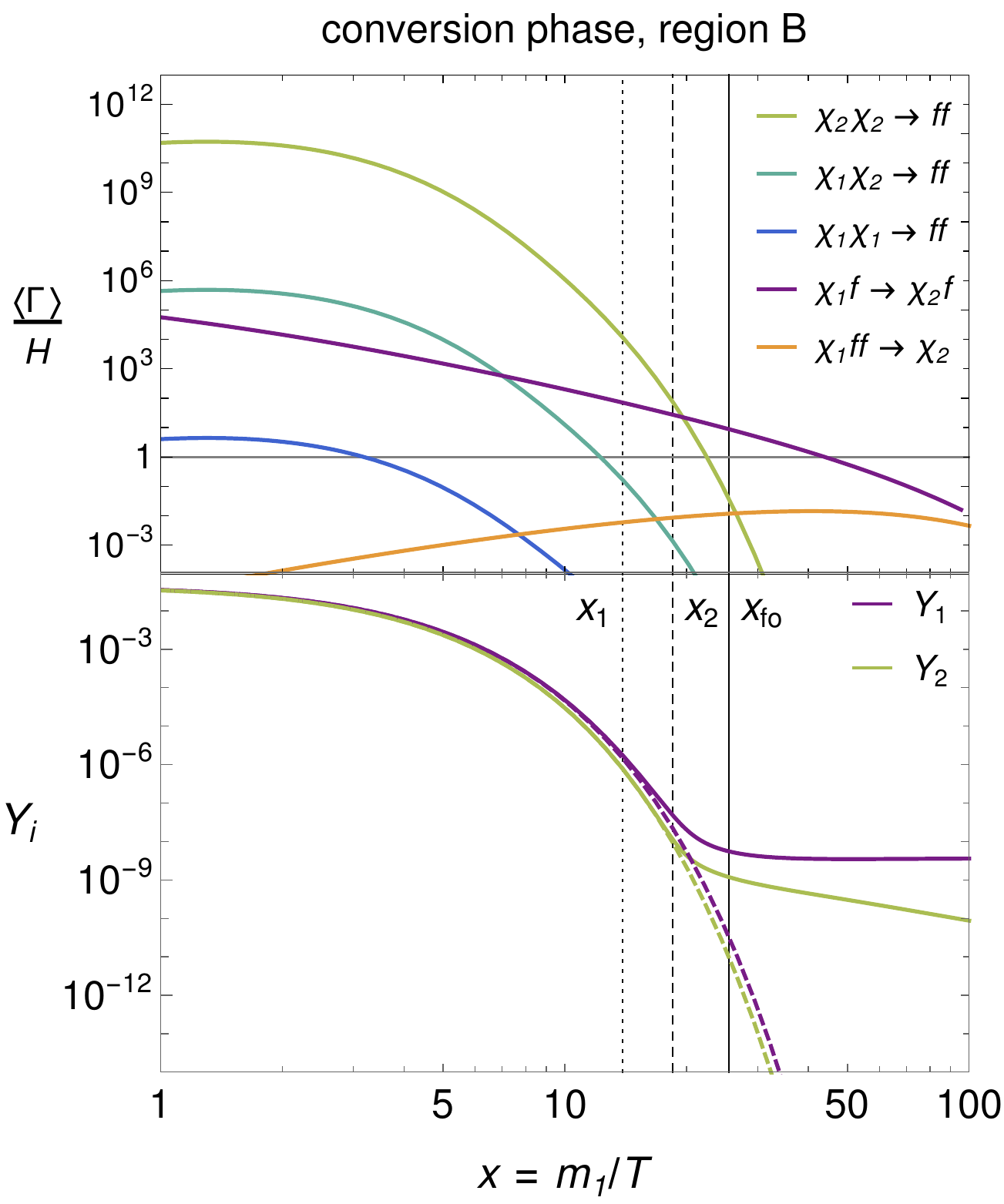}
  \vspace{0.0001\textheight}
  \includegraphics[width=0.4\textwidth]{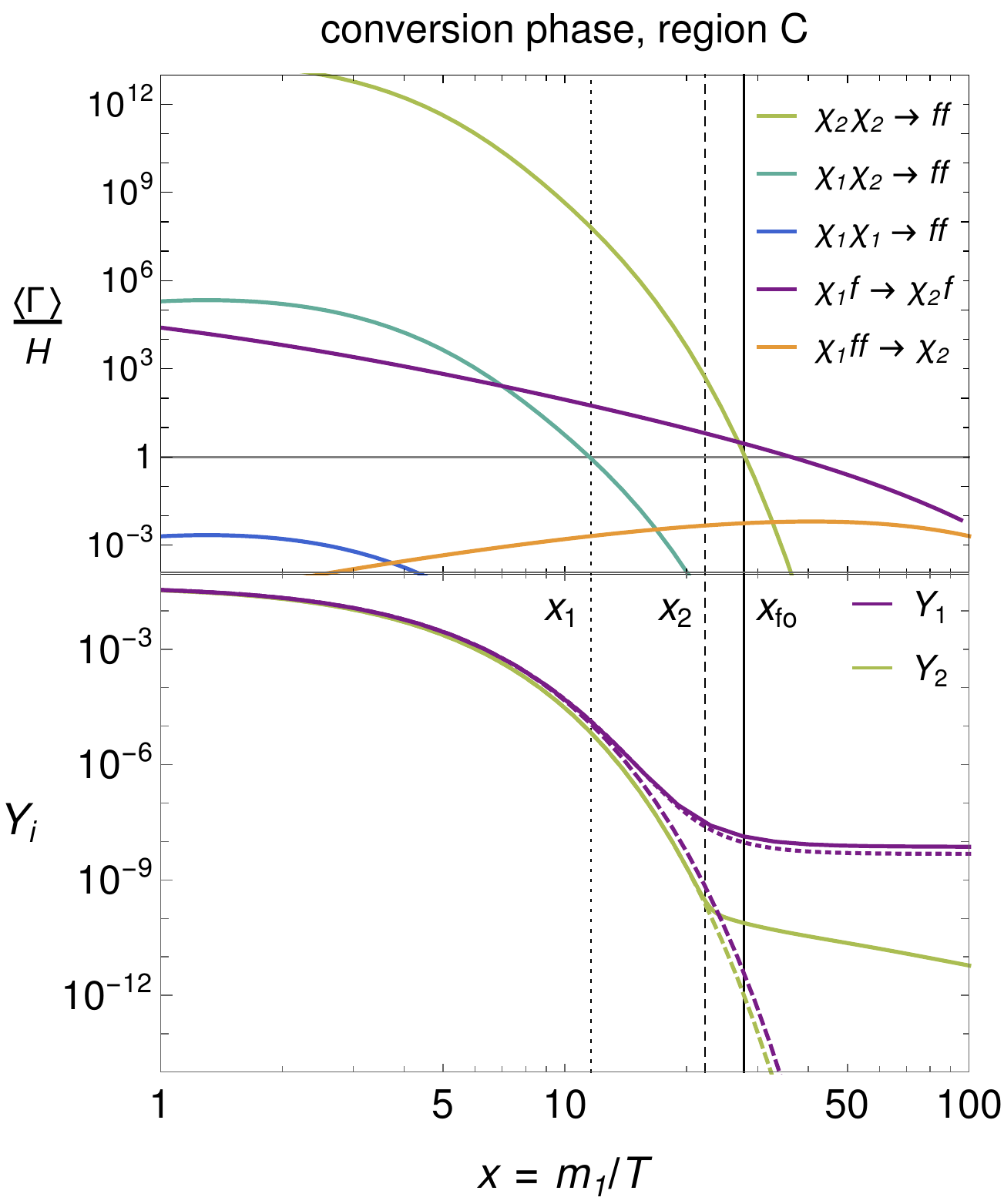}
  \caption{Evolution of the dark fermion interaction rates $\langle \Gamma \rangle\equiv \{\langle \Gamma\rangle_{ij},\langle \Gamma\rangle_{i\to j}\}\cdot n_i^{(0)}/(n_1^{(0)})$, $i,j=\{1,2\}$ (top panels) and comoving number densities $Y_i$ of dark fermions $\chi_i$ (solid curves) and equilibrium number densities $Y_i^{(0)}$ (dashed curves) (lower panels) as a function of $x=m_1/T$ for three i2DM benchmarks with $m_1 = 60\,$MeV, $\Delta = 0.05$ and $\alpha_D = 1/4\pi$. The relic abundance is set by partner annihilation (upper left), conversions in (upper right) and out of kinetic equilibrium (below), happening in regions (A), (B) and (C), respectively. The model parameters $(\tan\theta, y)$ are set to $(0.02,4\cdot 10^{-10})$, $(0.003, 10^{-9})$ and $(10^{-4},4\cdot 10^{-7})$. In the lower plot, the naive $\chi_1$ number density obtained from the integrated Boltzmann equations (dotted purple curve) is shown for comparison with the correct result from unintegrated Boltzmann equations (solid curve). Vertical black lines indicate the times for $\chi_1$ (dotted) and $\chi_2$ chemical decoupling (dashed), and $\chi_1$ freeze-out (solid).}
  \label{fig:ratiogamH}
 \end{figure}

To illustrate the impact of the various processes on i2DM, in Fig.~\ref{fig:ratiogamH} we show the time evolution of the dark sector interaction rates $\langle \Gamma \rangle$ and the yield $Y$ for three i2DM benchmarks. The three benchmarks belong to different phases of freeze-out, which we will discuss in detail in Sec.~\ref{sec:phases}.

\subsection{Phases of freeze-out}
\label{sec:phases}
From the discussion in Sec.~\ref{sec:relevant-processes}, it becomes clear that the freeze-out dynamics should be very sensitive to the parameters $y,\, \theta$ and $\Delta$. When successively decreasing the dark interaction strength $y$, we identify three different \emph{phases of freeze-out}, distinguished by the processes that set the DM relic abundance:
\begin{enumerate}
\item\label{item:coan} {\bf coannihilation phase}: $\Omega_\chi h^2$ set by $ \chi_1 \chi_2 \leftrightarrow f\bar{f}$ and $\chi_2 \chi_2 \leftrightarrow f\bar{f}$,
\item\label{item:partn} {\bf partner annihilation phase}: $\Omega_\chi h^2$ set by $ \chi_2 \chi_2 \leftrightarrow f\bar{f}$,
\item\label{item:conv} {\bf conversion phase}: $\Omega_\chi h^2$ set by $\chi_1 f \leftrightarrow \chi_2 f$ and/or $\chi_2 \leftrightarrow \chi_1 f\bar{f}$.
\end{enumerate}
In addition, the thermal history of the DM candidate depends on whether departures from chemical or kinetic equilibrium with the bath have occurred prior to freeze-out. This critically depends on the efficiency of $\chi_1\leftrightarrow \chi_2$ conversions as this process can maintain a state of kinetic equilibrium for $\chi_1$. Further, efficient conversions also keep $\chi_1$ chemically coupled with $\chi_2$, so that both dark fermions are in chemical equilibrium with the thermal bath until the time $x_2$ at which $\chi_2$ chemically decouples.\footnote{In i2DM, $\chi_2$ is always in kinetic and chemical equilibrium at high temperatures due to efficient elastic scatterings and pair annihilation processes.} We hence distinguish between three regions of parameter space, where the following conditions are satisfied:
  \begin{eqnarray}
        (\text{A}) & &\qquad\  \ \frac{\Gamth_{1\to 2}}{H}(x_{2}) \gtrsim 100
        \label{eq:equ}\\
        (\text{B}) & &\ 10 \lesssim \frac{\Gamth_{1\to 2}}{H}(x_{2}) \lesssim 100 \label{eq:chemdec}\\
        (\text{C}) & &\qquad\ \ \frac{\Gamth_{1\to 2}}{H}(x_{2})  \lesssim  10 \,,\label{eq:kindec}
    \end{eqnarray}
where $\Gamth_{1\to 2}$ denotes the conversion rates from Eq.~\eqref{eq:conversion-rate} and $H(x_2)$ is the Hubble expansion at the time $x_2$. These three regions correspond to the regions where we can safely assume kinetic and chemical equilibrium for $\chi_1$ until freeze-out (A); only assume kinetic equilibrium and have to take into account deviations from chemical equilibrium (B); and a region where we cannot simply assume chemical nor kinetic equilibrium for $\chi_1$ (C). This classification will allow us to systematically study the effects of DM chemical and kinetic decoupling before freeze-out on the relic abundance.

Chemical decoupling of $\chi_2$ plays hence an important role in the evolution of $\chi_1$. This decoupling occurs when the $\chi_2$ (co-)annihilation rate drops below the Hubble rate. If conversions $\chi_1 \leftrightarrow \chi_2$ are efficient around $x_2$, the decoupling time can roughly be estimated using
\begin{align}
   \big(\Gamth_{22}(x_2) + 2\Gamth_{21}(x_2)\big) \frac{r_2}{r} \approx  H(x_2)\,,
   \label{eq:x22dec}
\end{align}
where $r_2$ and $r$ are defined as in Eq.~\eqref{eq:r_i}. Due to the efficient conversions, $\chi_1$ decouples at the same time as $\chi_2$. However, if conversions are absent, $\chi_2$ chemically decouples around
\begin{align}
  \Gamth_{22}(x_2) \approx H(x_2)\,.
\end{align}
Numerically we determine the time $x_i$ where $\chi_i$ chemically decouples by requiring that the density yield deviates from equilibrium by 20\%,
\begin{align}\label{eq:chi-dec}
 x_i = \frac{m_1}{T_i}:\qquad \frac{Y_i(x_i)}{Y_i^{(0)}(x_i)} = 1.2\,,\qquad i=\{1,2\}.
\end{align}

\begin{figure}[!t]
    \centering
    \subfloat[]{\includegraphics[width=0.46\textwidth]{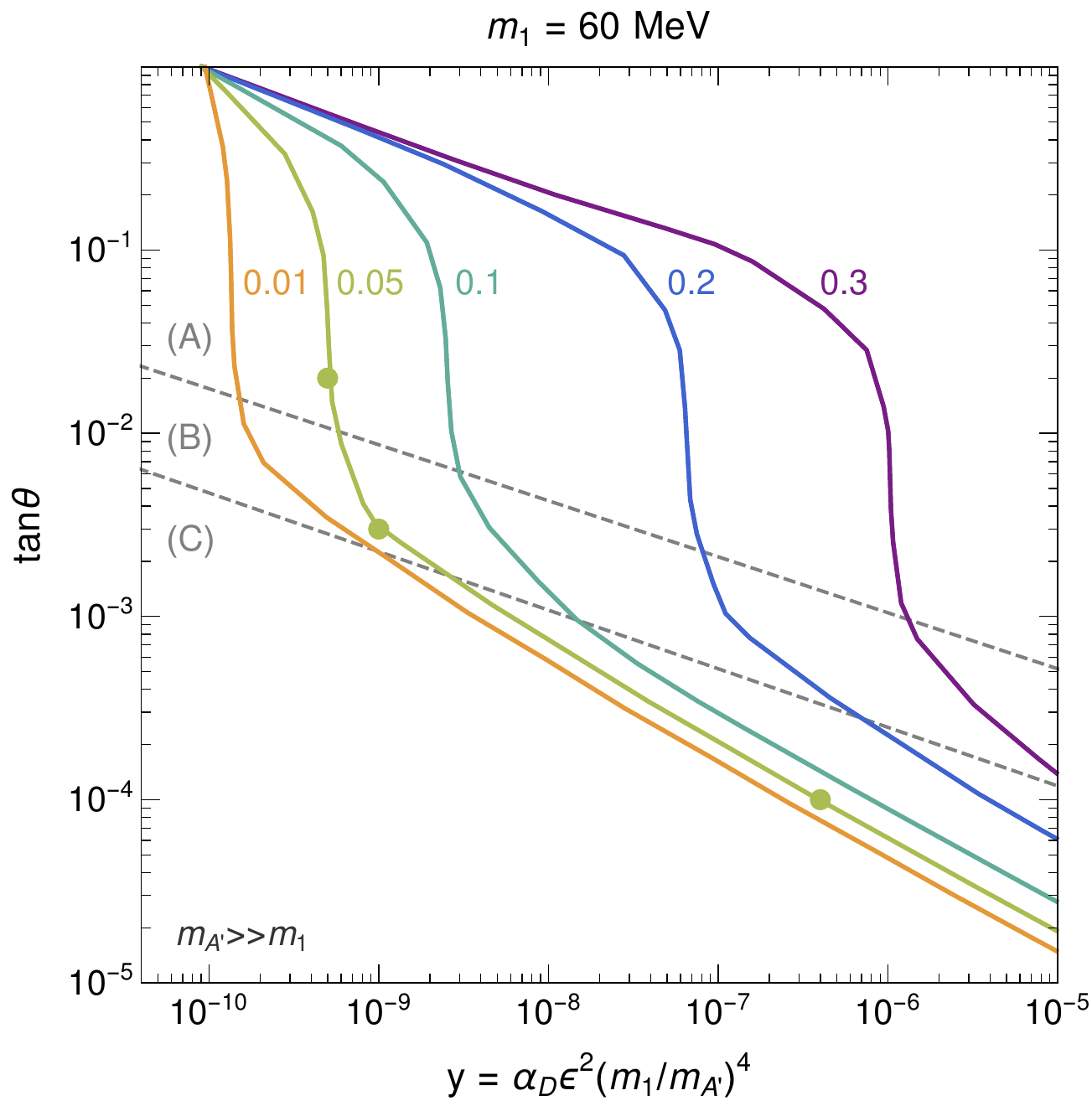}\label{fig:phases_left}}
	\hspace{0.05\textwidth}
    \subfloat[]{\includegraphics[width=0.46\textwidth]{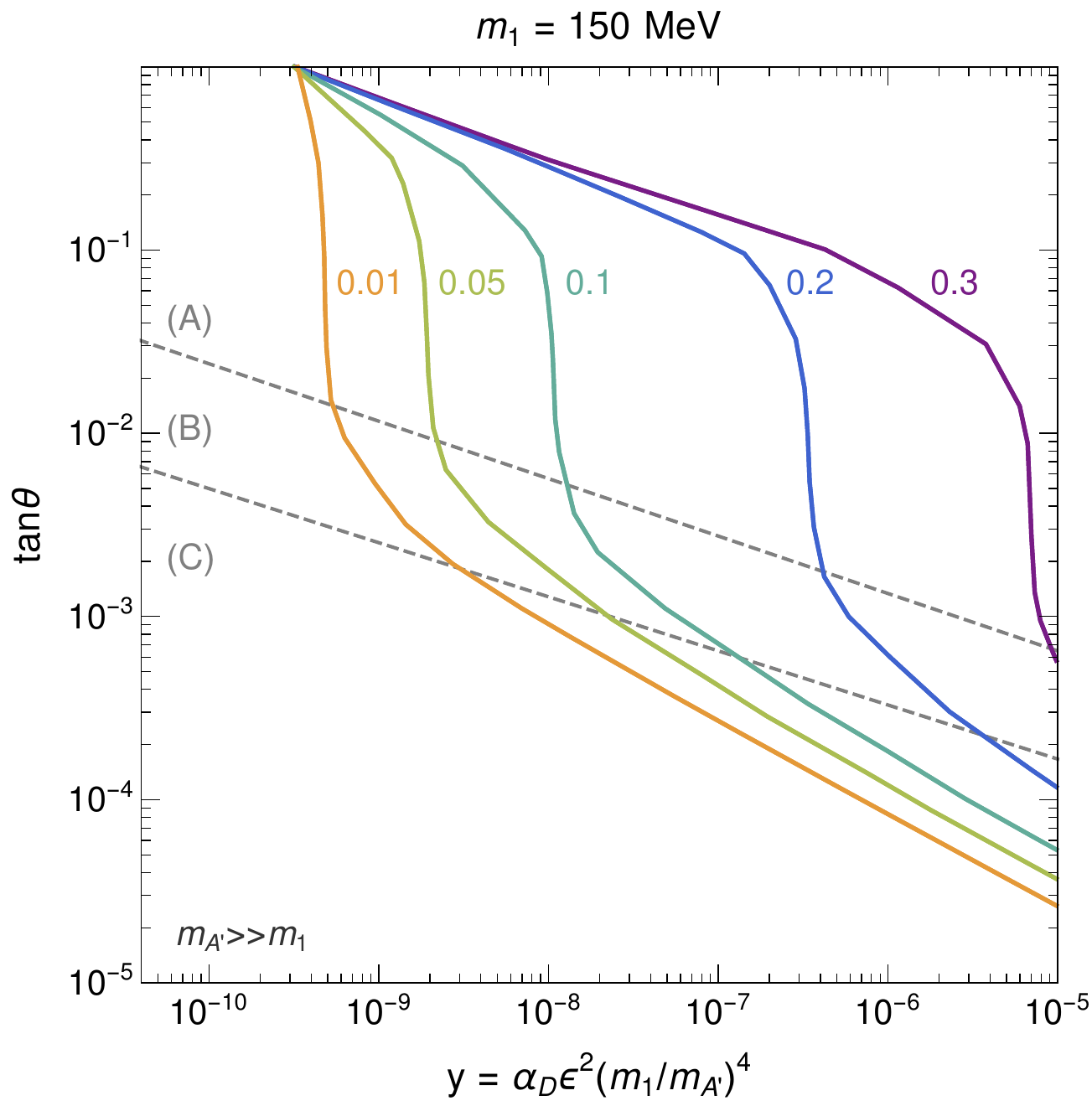}\label{fig:phases_right}}
    \caption{Dark matter relic abundance for i2DM as a function of the dark interaction strength, $y$, and the dark fermion mixing, $\tan\theta$. The dark matter mass is fixed to $m_1 = 60\,$MeV (top) and 150 MeV (bottom). The observed abundance $\Omega_{\chi} h^2=0.12$ is obtained along the colored contours for different dark fermion mass splittings $\Delta=0.01 \dots 0.3$. The decoupling regions (A), (B) and (C) are separated by dashed gray lines, which satisfy Eqs.~(\ref{eq:kindec1}) and~(\ref{eq:kindec2}). In region (A), the dark sector is in chemical and kinetic equilibrium with the bath. In regions (B) and (C), $\chi_1$ successively decouples from chemical and kinetic equilibrium with $\chi_2$ prior to freeze-out. In the top panel, the bullets on the green contour for $\Delta=0.05$ mark the position of the three i2DM benchmarks displayed in Fig.~\ref{fig:ratiogamH}.
    \label{fig:phases}}
\end{figure}

The classification made above allows us to understand the density evolution of the dark fermions shown in Fig.~\ref{fig:ratiogamH}. The three i2DM benchmarks correspond to the freeze-out phases of partner annihilation (upper left plot) and conversion (upper right and lower plots). Partner annihilation is relevant in region (A), while conversion can prevail either in region (B) in kinetic equilibrium (upper right plot), or in region (C) beyond kinetic equilibrium (lower plot). The interplay between the different freeze-out phases and decoupling regions is shown in Fig.~\ref{fig:phases} as a function of the model parameters $y$ and $\tan\theta$ for two scenarios with fixed DM masses. In the upper plot of Fig.~\ref{fig:phases}, the three benchmark i2DM scenarios from Fig.~\ref{fig:ratiogamH} are marked as green bullets. Below we first discuss Fig.~\ref{fig:ratiogamH} in detail and then turn to Fig.~\ref{fig:phases}.

In Fig.~\ref{fig:ratiogamH}, all benchmarks correspond to fixed parameters $m_1 = 60\,$MeV, $\Delta = 0.05$ and $\alpha_D = 1/4\pi$. In each of the plots, the top panel shows the evolution of the various interaction rates $\langle\Gamma\rangle$, normalized to the Hubble rate. The conversion rate $\langle\Gamma\rangle_{1\to 2}$ that distinguishes regions (A), (B), and (C) is driven by $y\tan^2 \theta$, see Eq.~\eqref{eq:scattering}. It decreases when going  from partner annihilation in region (A) (upper left plot) to  conversion beyond kinetic equilibrium in region (C) (lower plot). The relative scaling of the interaction rates is determined by the reaction densities from Eqs.~\eqref{eq:annihilation}, \eqref{eq:scattering} and \eqref{eq:decay} discussed in Sec.~\ref{sec:i2dm_model}. In particular, the reaction densities for (co-)annihilations, $\gamma_{ij}\sim e^{-2\Delta x}$, drop faster at low temperatures than scattering and decays, $\gamma_{i\to j}\sim e^{-\Delta x}$. The relative strength of the (co-)annihilation processes depends exponentially on the mass splitting $\Delta$ and also on the dark fermion mixing $\theta$. Due to the small splitting and mixing in the three benchmarks, partner annihilation dominates (light green), followed by co-annihilation (dark green) and suppressed DM annihilations (blue). As mentioned in Sec.~\ref{sec:i2dm_model}, conversions through scattering (purple) decrease faster with time than inverse decays (orange). Around freeze-out, however, conversions dominate over decays in all three benchmarks.

In the bottom panels of Fig.~\ref{fig:ratiogamH}, we show the time evolution of the dark fermion comoving number densities, $Y_i(x)$, (solid) and the equilibrium yield, $Y_i^{\rm  eq}(x)$, for comparison (dashed). In the lower plot, we also indicate the evolution that is obtained when neglecting the kinetic decoupling of DM (dotted). To guide the eye, we highlight the times for freeze-out and chemical decoupling with black vertical lines, determined by Eqs.~\eqref{eq:freeze-out-temp} and~\eqref{eq:chi-dec}.

We now turn our attention to Fig.~\ref{fig:phases}, which illustrates the different phases of freeze-out for i2DM as a function of the dark interaction strength $y$ and the dark fermion mixing $\tan\theta$ for two fixed DM masses $m_1=60$ MeV (left) and 150 MeV (right). The regions (A), (B) and (C), corresponding to decreasingly efficient conversions as suggested  by Eqs.~\eqref{eq:equ}-\eqref{eq:kindec}, are delineated with dashed gray lines. The exact relations between the conversion rate and the Hubble rate along these lines are given in Eqs.~\eqref{eq:kindec1} and~\eqref{eq:kindec2}. The observed relic abundance $\Omega_{\chi} h^2 = 0.12$ is obtained along the solid colored contours in the $(y,\tan \theta)$ plane for fixed values of the mass splitting $\Delta$. When increasing the DM mass $m_1$, the contours shift to the right, meaning that the observed abundance is obtained for larger values of $y$. This is easily understood, as all (co-)annihilation and conversion rates scale as $\Gamth_{ij}, \Gamth_{i\to j}\propto y/m_1^2$. In the upper part of the plots, all contours converge and the mass splitting $\Delta$ plays no role in setting the relic abundance. Here the abundance is set by pair annihilations $\chi_1 \chi_1 \to f\bar{f}$, see Eq.~(\ref{eq:annihilation}). In Chapter~\ref{chap:i2dm_pheno}, we will see that laboratory searches exclude this region of parameter space. As a result, we focus on DM candidates with small couplings, corresponding to the phases of co-annihilation and partner annihilation in region (A), and on conversions in regions (B) and (C).

In region (A), efficient conversion rates satisfying Eq.~(\ref{eq:equ}) keep $\chi_1$ in chemical and kinetic equilibrium with the thermal bath until freeze-out. In particular, efficient conversions ensure that $\chi_1$ and $\chi_2$ have equal chemical potentials, so that their number densities are related by
\begin{align}\label{eq:chemical-eq}
    \frac{n_1}{n_2} = \frac{n_1^{(0)}}{n_2^{(0)}}\,.
\end{align}
In this case, the freeze-out conditions are similar to the ones of a thermal WIMP and the observed relic DM abundance is obtained for a freeze-out time~\cite{Griest:1990kh,Gondolo:1990dk,Edsjo:1997bg}
\begin{equation}
 x_{\rm WIMP}\simeq 25\,.
\label{eq:xfoWIMP}
\end{equation}
This is illustrated by the density evolution for the benchmark in the upper left plot of Fig.~\ref{fig:ratiogamH}, which corresponds to the upper green bullet in Fig.~\ref{fig:phases_left}. At large $\tan \theta$ of region (A), we note the rapid bending of the curves. This happens when co-annihilations and pair annihilations of the lightest dark state start playing a significant role in the decoupling of $\chi_1$. The relative impact of the mentioned reactions is not important for our discussion as long as condition (A) is satisfied.

When entering in region (B), defined by Eq.~\eqref{eq:chemdec}, conversion processes are less efficient and the rates $\Gamth_{1\to 2}(x_2)$ are about ten to one hundred times larger than the Hubble rate. As a result, the DM density departs from chemical equilibrium prior to freeze-out. The effect is visible in the second benchmark displayed in the upper right plot of Fig.~\ref{fig:ratiogamH}, corresponding to the second green bullet in Fig.~\ref{fig:phases_left}.

Once the conversion rate $\Gamth_{1\to 2}(x_2)$ is further suppressed, DM cannot be expected to be kept in kinetic equilibrium until freeze-out. We enter region (C), defined by Eq.~\eqref{eq:kindec} and illustrated by a benchmark in the lower plot of Fig.~\ref{fig:ratiogamH}, corresponding to the lower green bullet of Fig.~\ref{fig:phases_left}. To quantify the effect of kinetic decoupling, in Fig.~\ref{fig:ratiogamH} we display the physical DM yield obtained by respecting kinetic decoupling (solid purple curve) compared to the same yield obtained when neglecting kinetic decoupling (dotted purple curve).

In what follows, we will discuss the physics of the three freeze-out phases in detail and how the DM relic abundance can be calculated, paying particular attention to non-equilibrium effects in the conversion phase.

\subsubsection{Coannihilation}
\label{sec:coannihilation}
Coannihilation sets the relic abundance when the rate of $\chi_1\chi_2 \leftrightarrow f \bar f$ is larger than the Hubble rate and dominates over DM pair annihilations $\chi_1\chi_1 \leftrightarrow f \bar f$ around the freeze-out time, i.e., when
\begin{align}
    \Gamth_{12}(x_{\rm fo}) > \Gamth_{11}(x_{\rm fo})\,.
    \label{eq:Gamcoa}
\end{align}
According to Eqs.~\eqref{eq:interactions} and \eqref{eq:annihilation}, in i2DM the ratio of these rates scales as
\begin{align}\label{eq:relative-annihilation}
 \frac{\Gamth_{11}(x)}{\Gamth_{12}(x)} \sim \frac{n_1^{(0)}}{n_2^{(0)}}\, \tan^2\theta \sim e^{x\Delta} \tan^2\theta\,.
\end{align}
One might deduce that co-annihilation sets the relic abundance if $e^{x_{\rm fo}\Delta}\tan^2\theta \lesssim 1$. However, this is only a necessary condition, because in this regime the relic abundance could also be driven by partner annihilation $\chi_2\chi_2 \to f\bar{f}$. To determine the relative impact of partner annihilation, one needs to consider the \emph{weighted} ratio of co-annihilation and partner annihilation rates, $r_1\Gamth_{12}(x)/r_2\Gamth_{22}(x)$.\footnote{The effective co-annihilation and partner annihilation rates entering in the computation of the DM relic abundance have to be weighted by the ratio of the dark species equilibrium densities to the total one, when the dark sector species are in chemical equilibrium, see e.g.~\cite{Griest:1990kh}.} In i2DM, this ratio scales as $e^{x\Delta}\tan^2\theta$, exactly as in Eq.~(\ref{eq:relative-annihilation}). As a result, in the co-annihilation phase the comoving number density $Y_1(x)$ freezes once the weighted sum of co-annihilation and partner annihilation drops below the Hubble rate,\footnote{This relation assumes that $\chi_1$ and $\chi_2$ are both in kinetic and chemical equilibrium prior to freeze-out.}
\begin{align}\label{eq:fo-coa}
     2 \frac{r_1}{r} \Gamth_{12}(x_{\rm fo}) + \frac{r_2}{r} \Gamth_{22}(x_{\rm fo}) \approx H(x_{\rm fo})\,,\qquad x_{\rm fo}\simeq x_{\rm WIMP}\,.
\end{align}
The relic DM abundance is set around the freeze-out temperature of a thermal WIMP. Freeze-out through co-annihilation is realized in the upper part of region (A) in Fig.~\ref{fig:phases}. Due to the scaling $\Gamth_{12} \sim y\,\sin^2(2\theta)$, the relic abundance contours tend to larger interaction strength $y$ at smaller mixing $\theta$ in this regime.

In the co-annihilation phase, i2DM resembles iDM, where only the off-diagonal current of Eq.~\eqref{eq:current_idm} couples to the dark photon and co-annihilations are the dominant number-changing interactions of the DM candidate with the bath. However, lowering the dark interaction strength $y$ leads to inefficient co-annihilation around freeze-out. In iDM, this results in an overabundance of DM. In i2DM, suppressed co-annihilations can be compensated by efficient partner annihilations and conversions, thus explaining the observed DM abundance even if the dark sector is feebly coupled.

\subsubsection{Partner annihilation}
\label{sec:partner-annihilation}
If the coannihilation rate is suppressed compared to the Hubble rate around $x=x_{\rm WIMP}$, the relic DM abundance can be set by partner annihilations $\chi_2\chi_2 \to f\bar{f}$. In this phase, the freeze-out condition reads\footnote{This condition assumes that Eq.~(\ref{eq:equ}) holds.}  \begin{align} 
    \frac{r_2}{r}\,\Gamth_{22}(x_{\rm fo}) \approx H(x_{\rm fo})\,, \qquad  x_{\rm fo}\simeq x_{\rm WIMP}\,.
    \label{eq:Gampart}
\end{align}
As in the coannihilation phase, the freeze-out time is fixed to $x_{\rm fo} \simeq x_{\rm WIMP}$, up to a moderate logarithmic dependence on the model parameters~\cite{Griest:1990kh}. In Fig.~\ref{fig:phases}, the phase of partner annihilation lies in region (A) and is characterized by vertical lines. Indeed, due to the scaling $r_2/r\langle \Gamma_{22} \rangle \sim y \cos^4\theta\, e^{- 2 x\Delta}$, the relic DM abundance is essentially independent of the mixing as $\cos^4\theta \approx 1$. To obtain the observed abundance, variations of the dark interaction strength $y$ can be compensated by the mass splitting $\Delta$, as illustrated by the various contours. In the partner annihilation phase, $\chi_1\leftrightarrow \chi_2$ conversions have to be efficient enough to satisfy the decoupling condition from Eq.~(\ref{eq:equ}). As a consequence, $\chi_1$ and $\chi_2$ chemically decouple from the bath around the same time.  This is visible in the upper left plot of Fig.~\ref{fig:ratiogamH}. Notice that $\chi_2 \to \chi_1$ decays happen on a time scale shorter than the period of chemical decoupling and do not affect the freeze-out time $x_{\rm fo} \approx x_{\rm WIMP}$.

\subsubsection{Conversion}
\label{sec:conversion}
As we discussed above, partner annihilation can set the relic DM abundance even if $\chi_1\chi_1\to f\bar{f}$ and $\chi_1 \chi_2\to f\bar{f}$ annihilations are suppressed. If the dark fermion mixing $\theta$ is very small, $\chi_1\leftrightarrow \chi_2$ conversion rates can be comparable to or even fall below the Hubble rate around $\chi_2$ chemical decoupling, as in Eqs.~\eqref{eq:chemdec} and \eqref{eq:kindec}. The DM abundance is now set by conversion processes, i.e., by coscattering and/or (inverse) decays.

In i2DM, conversion-driven freeze-out can occur in regions (B) and (C) in Fig.~\ref{fig:phases}. Here, coscattering dominates over decays in the thermal history around freeze-out. Due to the scaling $\Gamth_{1\to 2} \propto \sin^2(2\theta)$, the contours of constant $\Omega_\chi h^2$ are sensitive to the mixing angle $\theta$.  For a fixed interaction strength $y$ and mass splitting $\Delta$, the relic abundance in this regime is generally larger than what one would expected from partner annihilation. The reason is that $\chi_1\leftrightarrow   \chi_2$ conversions are less efficient and the DM yield $Y_1(x)$ can start to deviate from the equilibrium yield $Y_1^{(0)}(x)$ well before the freeze-out time $x=x_{\rm fo}$ ~\cite{Garny:2017rxs,DAgnolo:2018wcn}. The latter effect is illustrated in the upper right and lower plots of Fig.~\ref{fig:ratiogamH}. The increased yield has to be compensated by a larger interaction strength $y$, causing the contours in Fig.~\ref{fig:phases} to bend towards the lower right corner. Notice that efficient partner annihilation is essential for conversions to explain the observed relic abundance. Suppressed partner annihilations would result in an overabundance of DM. 

To describe deviations of the DM density from chemical and kinetic equilibrium within the conversion phase, it is convenient to distinguish three key moments:
\begin{enumerate}
    \item the time $x_1$ at which $\chi_1$ chemically decouples from the bath;
    \item the time $x_{2}$ at which $\chi_2$ chemically decouples from the bath;
    \item the DM freeze-out time $x_{\rm fo}$ at which the DM yield becomes constant, which can differ from $x_{\rm WIMP} \simeq 25$ in the conversion phase.
\end{enumerate}
Numerically, we determine $x_1$, $x_2$ and $x_{\rm fo}$ using Eqs.~\eqref{eq:chi-dec} and~\eqref{eq:freeze-out-temp}. Thanks to efficient elastic scattering, the dark partner $\chi_2$ is kept in kinetic equilibrium with the bath throughout the DM freeze-out process and in particular for $x > x_2$ after $\chi_2$ chemical decoupling. For $\chi_1$, deviations only from chemical equilibrium prior to freeze-out typically occur when
\begin{align}
    x_1 \lesssim x_2 < x_{\rm fo}\,,
\end{align}
while deviations from chemical \emph{and} kinetic equilibrium can occur for
\begin{align}
    x_1 \ll x_2 < x_{\rm fo}\,.
\end{align}
Below, we discuss how to threat these deviations properly.

\begin{figure}[!t]
    \centering
    \includegraphics[width=0.9\textwidth]{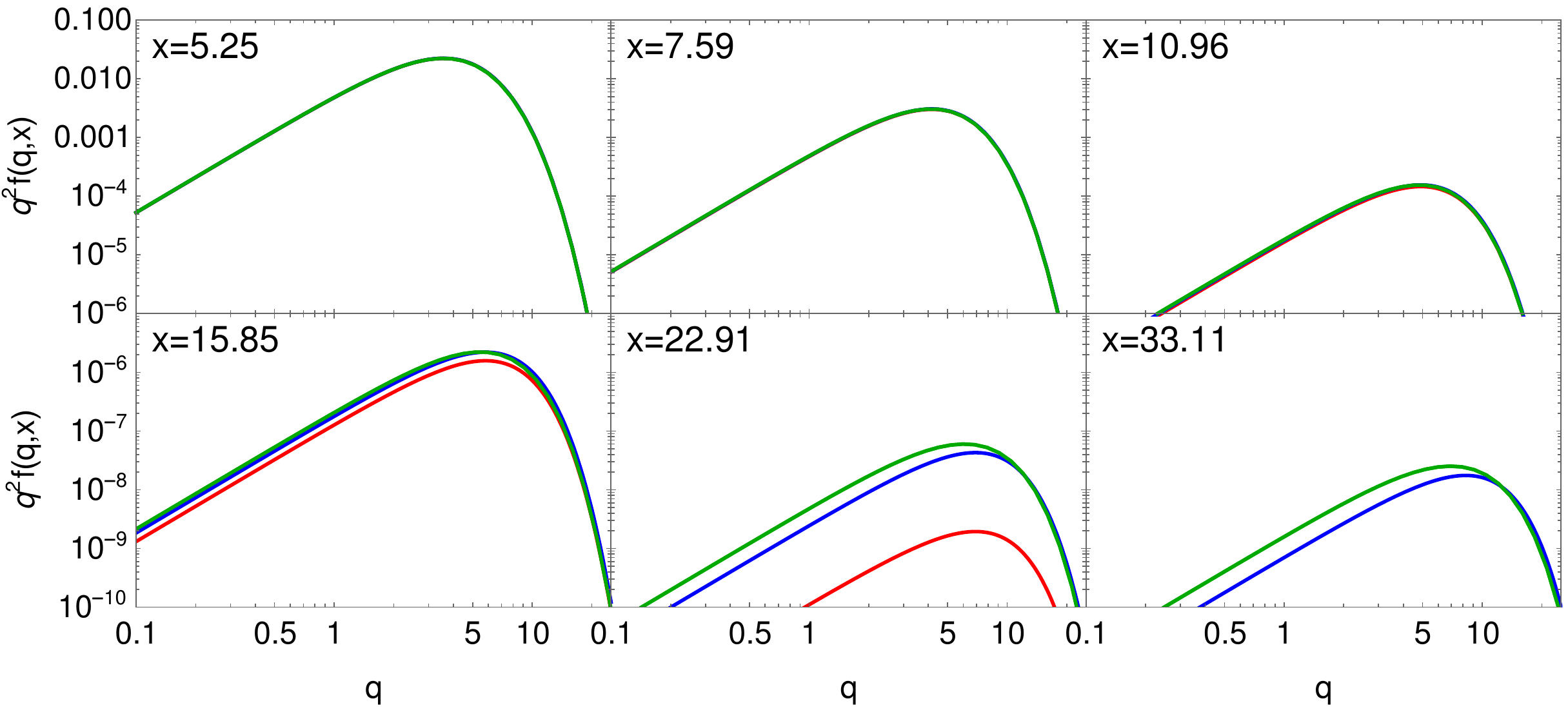}
    \caption{The phase space density functions for the benchmark in right upper panel of Fig.~\ref{fig:ratiogamH} in region B, obtained by solving the unintegrated Boltzmann equation (green) compared to the phase space density assuming kinetic equilibrium (blue) and chemical equilibrium (red). The different panels show different values of $x$.}
    \label{fig:psd_B}
\end{figure}

\paragraph{Deviations from chemical equilibrium} If $x_1 \lesssim x_2$, conversions are barely efficient during $\chi_2$ chemical decoupling, as in Eq.~\eqref{eq:chemdec}. This scenario corresponds to region (B) in Fig.~\ref{fig:phases}. Numerically we find that the boundary between regions (A) and (B) corresponds to
\begin{align}
    \Gamth_{1\to 2}(x_2)= 200\, H(x_2)\,,
    \label{eq:kindec1}
\end{align}
using Eq.~(\ref{eq:chi-dec}) to evaluate $x_2$. Below this boundary, the $\chi_1$ density departs from chemical equilibrium before $\chi_2$. This effect is illustrated for a benchmark in the upper right plot of Fig.~\ref{fig:ratiogamH}, where the freeze-out process for $Y_1(x)$ terminates around $x_{\rm fo} > x_1$. At that time, we expect that conversions are still sufficiently active to keep $\chi_1$ in kinetic equilibrium with the thermal bath via $\chi_2$. In particular, we assume that the DM distribution function $f_1(x,q)$ is well approximated by the Boltzmann distribution $f_1^{(0)}(x,q)$ throughout the entire evolution process. In Fig.~\ref{fig:psd_B}, we specifically check this assumption for the benchmark on the top right of Fig.~\ref{fig:ratiogamH} by solving the unintegrated Boltzmann equation following the procedure described in App.~\ref{app:BE_beyond_KE}. We show in green the resulting phase space distribution and compare it to the cases where we assume kinetic equilibrium in blue, i.e. where the phase space density of $\chi_1$ is given by
\begin{equation}
    f_1(q,x) = f_1^{(0)}(q,x) \frac{n_1(x)}{n_1^{(0)}(x)},
\end{equation}
Further, we also show the phase space density assuming chemical equilibrium, $f_1(q,x) = f_1^{(0)}(q,x)$, in red. One can see that $\chi_1$ starts decoupling from chemical equilibrium for $x\approx11$, which is in correspondence with the top right panel of Fig.~\ref{fig:ratiogamH}, while still maintaining kinetic equilibrium. Small deviations from kinetic equilibrium occur at $x\approx15$, however, the evolution of the phase space density still tightly follows the one when assuming kinetic equilibrium, even until $\chi_1$ freezes out and the phase space distributions stops evolving in time. Hence, kinetic equilibrium can be safely assumed until freeze-out, but deviations from chemical equilibrium occur before and must be taken into account.\footnote{Conversion-driven freeze-out with deviations from chemical equilibrium was studied before in Ref.~\cite{Garny:2017rxs}, where $f_1(x,q)$ was observed to depart from $f_1^{(0)}(x,q)$ prior to DM freeze-out. In i2DM, however, we expect that the distribution function of $\chi_1$ resembles $f_1^{(0)}(x,q)$ before freeze-out, because $\chi_1$ is kept in kinetic equilibrium at early times via $\chi_2 \to \chi_1$ decays and coscatterings (with $\chi_2$ being in kinetic equilibrium with the bath), see Fig.~\ref{fig:ratiogamH}. This was not the case for the DM model studied in Ref.~\cite{Garny:2017rxs}.}

The Boltzmann equations commonly used for (co-)annihilating DM (see Eq.~\eqref{eq:coan_boltz} and Ref.~\cite{Griest:1990kh}) assume chemical equilibrium between $\chi_1$ and $\chi_2$. In order to threat deviations from this behavior, we have to be supplement these equations by explicitly including the conversion rate in the coupled system of evolution equations for $Y_{1}(x)$ and $Y_2(x)$, as already discussed in Sec.~\ref{sec:CDFO}. For i2DM specifically, the coupled system of Boltzmann equations reads\footnote{Similar Boltzmann equations exist for the antiparticles $\bar{\chi}_1$ and $\bar{\chi}_2$. In i2DM these equations are completely equivalent and we  obtain the total abundance for $\chi_1+\bar{\chi}_1$ by simply doubling the yield for $\chi_1$.}
\begin{align}
    \label{eq:BE1}
    \frac{dY_1}{dx}=\frac{-1}{\overline{H} xs} & \left[\gamma_{1 1} \left(\frac{Y^2_{1}}{{Y_{1}^{(0)}}^2}-1 \right) + \gamma_{1 2} \left(\frac{Y_{1} Y_{2}}{Y_{1}^{(0)} Y_{2}^{(0)}}-1 \right) -   \gamma_{2   \to 1 }\left(  \frac{Y_{2}}{Y_{2}^{(0)}}-\frac{Y_{1}}{Y_{1}^{(0)}} \right)  \right],\\
    \label{eq:BE2}
    \frac{dY_{2}}{dx}=\frac{-1}{ \overline{H}xs} & \left[\gamma_{22} \left( \frac{Y^2_{2}}{{Y_{2}^{(0)}}^2}- 1\right) + \gamma_{12} \left(\frac{Y_{1} Y_{2}}{Y_{1}^{(0)} Y_{2}^{(0)}}-1 \right)  +\gamma_{2\to1}\left(  \frac{Y_{2}}{Y_{2}^{(0)}}-\frac{Y_{1}}{Y_{1}^{(0)}} \right) \right],
\end{align}
where $\gamma_{ij}$ are the reaction densities defined in Eq.~\eqref{eq:gamij} and $\gamma_{i\to j}= \gamma_{i\to  j}^{dec}+\gamma_{i\to j}^{scat}$.\footnote{We have checked that we can neglect quantum statistical effects for i2DM in the relevant density distributions $f_i$ and use the Maxwell-Boltzmann equilibrium distributions $f_i^{(0)}$.}

\begin{figure}[!t]
    \centering
    \includegraphics[width=0.9\textwidth]{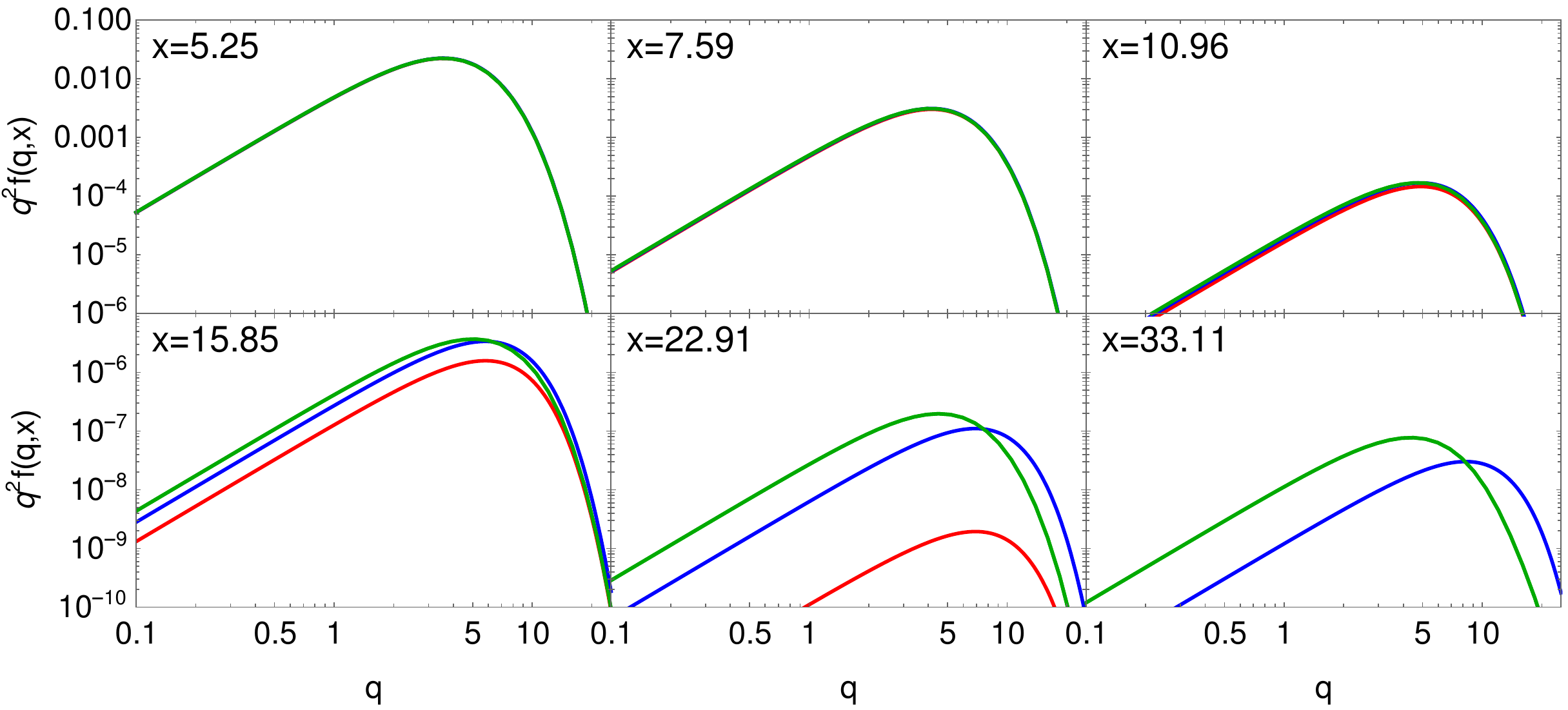}
    \caption{Same as Fig.~\ref{fig:psd_B} for the benchmark in the lower panel of Fig.~\ref{fig:ratiogamH} in region C.}
    \label{fig:psd_C}
\end{figure}
\paragraph{Deviations from chemical and kinetic equilibrium}  For $x_1 \ll x_{2}$, conversions become inefficient before $\chi_2$ chemically decouples, and the condition of Eq.~(\ref{eq:kindec}) is satisfied. This scenario corresponds to region (C) in Fig.~\ref{fig:phases}. Numerically, we find that the boundary between regions (B) and (C) is given by
\begin{align}
    \Gamth_{1\to 2}(x_2)= 20\, H(x_2)\,,
    \label{eq:kindec2}
\end{align}
with $x_2$ evaluated with Eq.~\eqref{eq:chi-dec}. Below this line, the mixing $\theta$ is so small that conversions fail to keep DM in kinetic equilibrium. This effect is visualized in Fig.~\ref{fig:psd_C}, where just as in Fig.~\ref{fig:psd_B}, we compare the phase space density obtain by numerically solving the unintegrated Boltzmann equation (green lines) with the ones where we assume kinetic (blue) and chemical (red) equilibrium. One can see that soon after $\chi_1$ decouples chemically at $x\approx10$, deviations from kinetic equilibrium start to occur. These deviations start to grow fast as time evolves and the phase space density only stops evolving after freeze-out. Since the time between DM chemical decoupling and freeze-out is now stretched over a larger range between $x_1$ and $x_{\rm fo}$, the deviations from kinetic equilibrium become more severe as the conversion rate continues to decrease over time. Hence, we can not simply assume a generic MB phase space density for $\chi_1$ and solving Eq.~\eqref{eq:BE1} will not yield the correct DM relic density. This effect is also demonstrated in the lower panel of Fig.~\ref{fig:ratiogamH} where the DM yield $Y_1(x_{\rm fo})$ including departures from kinetic equilibrium (solid curve) is larger than under the assumption of kinetic equilibrium (dotted curve). We took the departures of kinetic equilibrium into account by solving the unintegrated Boltzmann equations for $\chi_1$ (Eq.~\eqref{eq:UnintBE}) together with the integrated equation for $\chi_2$ (Eq.~\eqref{eq:BE2}), as this species will maintain kinetic equilibrium due to efficient elastic scatterings. More details on how we solve this system of equations can be found in App.~\ref{app:BE_beyond_KE}. 

While the deviations from kinetic equilibrium have a significant effect on the shape of the phase space distribution shown in Fig.~\ref{fig:psd_C}, the deviations in the relic abundance in the lower panel of Fig.~\ref{fig:ratiogamH} are modest, ranging around 35\%. In the remaining discussion about i2DM, and in particular in Figs.~\ref{fig:phases},~\ref{fig:mass-vs-y} and~\ref{fig:mass-vs-coupling}, we use the set of integrated Boltzmann equations from Eqs.~\eqref{eq:BE1}-\eqref{eq:BE2}, unless specified otherwise. This method is computationally much faster than solving the unintegrated Boltzmann equations and reproduces the exact results to a good approximation, which we have checked explicitly for some of the contours in Fig.~\ref{fig:phases}.

\chapter{Probing light dark sectors containing long-lived particles}
\label{chap:i2dm_pheno}

We already established in Chapters~\ref{chap:LHC_cons} that searches for long-lived particles at the LHC can be a very good probe for feebly interacting DM models with mediators in the GeV-TeV range. Mediators much lighter than this mass scale cannot produce high $p_T$ objects, so that they would disappear into the SM background and will not pass the trigger selection. High energetic proton-proton colliders will hence not be able to probe light, i.e. MeV-GeV scale, DM models as we have introduced in Chapter~\ref{chap:lightDM}.

In this Chapter, we will focus on how to probe feebly interacting \ac{DS} in the MeV-GeV mass range. To do this, we will study the inelastic DM models presented in Chapter~\ref{chap:lightDM}, with our main focus on i2DM. Light DS particles can be easily produced in experiments such as electron-positron colliders or beam-dump experiments, so searches for LLPs in such experiments provide a good way of testing these models. The presence of light new particles also affects the thermal history of the universe in two distinctive ways. First, the presence of additional particles in the thermal bath changes the evolution of the Hubble expansion and of the entropy density of the universe. This may significantly affect the QCD phase transition, Big Bang Nucleosynthesis (BBN), the Cosmic Microwave Background (CMB) or supernova cooling. Second, the late-time decay of long-lived particles injects energy into the thermal plasma, inducing excitation, ionisation and heating effects. Such effects can modify BBN and the CMB compared to the predictions of the cosmological standard model.

We will first focus on the cosmological and astrophysical probes in Sec.~\ref{sec:cosmo} before moving on and study how the remaining parameter space can be constrained by laboratory searches, which will be discussed in Sec.~\ref{sec:colliders}. In the latter, we will also discuss near-future experiments that could potentially probe the i2DM parameter space completely. The work presented in this chapter is based on Ref.~\cite{Filimonova:2022pkj}.

\section{Bounds from cosmology and astrophysics}
\label{sec:cosmo}
The presence of MeV scale dark particles can influence the cosmological history and the evolution of astrophysical objects. In this section, we will discuss the main cosmic and astrophysical probes that can constrain inelastic DM. 
\subsection{QCD phase transition}
\label{sec:qcd-pt}
If new particles freeze-out around the GeV scale, the QCD phase transition around $T_{\rm QCD}\sim 200\,$MeV affects the relic abundance~\cite{Steigman:2012nb}. The confinement of quarks and gluons into hadrons reduces the effective number of relativistic degrees of freedom contributing to the entropy density, $h_{\rm eff}$~\cite{Olive:1980dy}. Calculations of $h_{\rm eff}$ around the phase transition are subject to significant uncertainties, leading to variations of about ten percent in the relic abundance~\cite{Hindmarsh:2005ix,Laine:2006cp,Drees:2015exa}. However, the effect of the QCD phase transition on the thermal evolution of light new particles can be much larger than the mentioned uncertainties. In particular, it can affect the relic abundance of dark matter candidates that freeze-out during or shortly after the phase transition.

In Sec.~\ref{sec:phases}, we have investigated the freeze-out dynamics of i2DM separately from effects of the QCD phase transition. If the freeze-out occurs in thermal equilibrium, dark matter candidates with masses
\begin{align}\label{eq:qcd-pt-bound}
    m_1\lesssim 1\,\text{GeV}
\end{align}
decouple from the bath around $T_{\rm WIMP} \approx m_1/x_{\rm WIMP} \ll T_{\rm QCD}$, late enough to neglect effects of the QCD phase transition on dark matter decoupling~\cite{Drees:2015exa}. In i2DM this holds in the phases of coannihilation and partner annihilation. However, in the conversion phase, chemical and kinetic equilibrium are not guaranteed until freeze-out, see Sec.~\ref{sec:conversion}. In this case, the dark fermions decouple from chemical equilibrium at earlier times $x_{1,2} \lesssim x_{\rm WIMP}$, which might be affected by the QCD phase transition if the corresponding decoupling temperature is similar to $ T_{\rm QCD}$. On the other hand, dark matter freeze-out is expected to be unaffected by the phase transition, because the dark states decouple at temperatures $T \lesssim T_{\rm WIMP}$. In any case, bounds from laboratory searches exclude i2DM dark matter candidates with masses near 1\,GeV, see Sec.~\ref{sec:colliders}. Even if non-equilibrium effects can change the decoupling times of the dark states, we do not expect effects from the QCD phase transition to affect the thermal history of dark matter candidates with masses well below the GeV scale.

\subsection{Big Bang Nucleosynthesis}
\label{sec:bbn}
The formation of primordial light nuclei starts around $T_{\rm BBN} \sim 0.1\,$MeV. New physics can affect the nuclei abundances in multiple ways. First of all, new dark particles can affect BBN by modifying the Hubble rate or the entropy density of the universe. Second, efficient annihilation of MeV-GeV-scale dark particles to electrons or photons may alter the rate at which light elements form~\cite{Depta:2019lbe}. Third, dark particles decaying to electrons or photons at later times can destroy the already formed primordial nuclei~\cite{Depta:2020zbh,Depta:2020mhj}. In i2DM, dark matter annihilation and dark partner decays around $T \sim T_{\rm BBN}$ are not strong enough to cause observable effects. Late decays of dark partners with lifetimes $\tau_2 > t_{\rm BBN}$, however, can destroy the newly formed elements through photodisintegration and constrain parts of the i2DM parameter space. Below we will discuss photodisintegration in i2DM in detail and briefly argue why annihilation and decays during BBN are not efficient.

For annihilating dark matter during BBN, a lower mass bound of $m_{\rm DM} \gtrsim 10\,$MeV has been derived in~\cite{Depta:2019lbe}. This bound applies for vanilla WIMP annihilation with $\langle\sigma v\rangle \approx 10^{-26}-10^{-28}$~cm$^3$/s. In the coannihilation and conversion phases of i2DM, however, the cross section for annihilations $\chi_1\chi_1\to \{e^+e^-,\gamma \gamma\}$ is much smaller than for vanilla WIMPs, resulting in weaker bounds on the dark matter mass. We expect these bounds to be superseded by constraints on $\Delta N_{\rm eff}$ from CMB measurements, see Sec.~\ref{sec:DNeff}.

On the other hand, decays $\chi_2 \to \chi_1 \ell^+\ell^-$ and $\chi_2 \to \chi_1\gamma\gamma$ of dark partners with MeV-GeV masses and lifetimes $\tau_2 > 10^3\,$s can produce an electromagnetic cascade of photons with energies above the binding energy of light nuclei, consequently disintegrating them.\footnote{For lifetimes $\tau_2\sim t_{\rm BBN}\sim 10^2$s, we have estimated from the analysis of~\cite{Depta:2020zbh} that no further constraints arise. This expectation is justified by comparing our i2DM predictions with BBN bounds on decaying dark scalars with a certain lifetime, shown in~Fig.~4 in~\cite{Depta:2020zbh}. Compared to~\cite{Depta:2020zbh}, in i2DM we expect a smaller branching ratio to electrons and photons and softer spectra of the 3-body decays products; freeze-out temperatures significantly lower than $10^{-2}\,$GeV for MeV-GeV particles; and an exponentially suppressed abundance of the decaying dark partner $\chi_2$.} Disintegration is efficient only if the emitted photons do not lose their energy too rapidly before reaching the target nucleus. This condition is satisfied if the energy of the photons lies below the di-electron threshold, $E^{\rm th}_{e^+ e^-} \simeq m_e^2/(22\,T)$~\cite{Hufnagel:2018bjp}.\footnote{Qualitatively this condition can be understood from the requirement that the center-of-mass energy of the injected photon and the thermal bath photon scaling as $E_\gamma E_{\gamma_{\rm th}}$ is of the order of $ m_e^2$, where $E_\gamma$ is the energy of the injected photon and $E_{\gamma_{\rm th}} \sim T$ is the average energy of a photon from thermal bath.} On the other hand, for photodisintegration to take place, $E^{\rm th}_{e^+ e^-}$ should lie well above the binding energy of light elements. Photodisintegration is thus efficient for photon energies
\begin{align}
    \mathcal{O}(\text{MeV}) \lesssim E_\gamma \lesssim \frac{m_e^2}{22\,T}\,,
\end{align}
so that photodisintegration only starts at temperatures below a few~keV. We calculate the effects of photodisintegration on the nuclei abundances starting from a continuous electron spectrum originating from $\chi_2 \to \chi_1 e^+e^-$ decays. We neglect loop-suppressed direct photon production, $\chi_2 \to \chi_1 \gamma \gamma$, as well as final-state radiation, resulting in a conservative bound on the electromagnetic flux obtained from partner decays~\cite{Forestell:2018txr}. To investigate photodisintegration for i2DM, we use the public code ACROPOLIS~\cite{Depta:2020mhj}. The code calculates the modified primordial abundances of light elements induced by photodisintegration, accounting for the redistribution of the energy injected by the decaying dark particle in the plasma. In particular, ACROPOLIS includes the exponential suppression in the photon spectrum for energies $E_{\gamma}>E^{th}_{e^+ e^-}$, which was neglected in previous studies,\footnote{Previous studies used the universal photon spectrum, which is an analytic approximation working very well for energies $E_{\gamma}<E^{th}_{e^+ e^-}$, but neglects photon with energies above this threshold.} up to energies $E_{\gamma}<E_0$, where $E_0$ is a model-dependent upper limit on the photon spectrum. We set $E_0$ to the maximum kinematically allowed energy. For the initial abundances of light nuclei, we use the \emph{Standard BBN} prediction extracted from the code AlterBBN~\cite{ARBEY20121822}. The input number density of $\chi_2$ is extracted from our system of Boltzmann equations.

The resulting constraints from photodisintegration on partner decays in i2DM are shown in Fig.~\ref{fig:mass-vs-y}. Bounds on the displayed parameter space are only visible in the lower right corner of the left panel with $\tan \theta=10^{-4}$ and $\Delta=0.05$, far from the correct relic abundance (the green dotted line). At larger $\tan \theta$, photodisintegration is not efficient because the dark partners decay too early. The lifetime of $\chi_2$ also determines the upper edge of the excluded region. At dark matter masses, $m_1 \lesssim 120\,$MeV, photodisintegration becomes inefficient due to the small absolute mass splitting between the dark fermions, $m_2-m_1$, which causes too soft decay products. For larger mass splittings $\Delta=0.1$, in the right panel, photodisintegration is sensitive to smaller dark matter masses. However, the lifetime of $\chi_2$ is generally smaller so that smaller couplings would be necessary for photodisintegration to take place. As a consequence, the excluded region lies below the plotted area in Fig.~\ref{fig:mass-vs-y} and BBN bounds are irrelevant for viable i2DM relics.  In summary, in i2DM photodisintegration excludes dark partners with lifetimes much larger than in cosmologically viable scenarios.

\subsection{Cosmic Microwave Background}
\label{sec:cmb}
Dark sector particles that annihilate or decay around the time of recombination can affect the overall shape of the CMB black-body spectrum, as well as its temperature and the polarization anisotropy spectra. Measurements of the CMB temperature and the polarization anisotropy spectra set strong constraints on the annihilation cross section of WIMP-like dark matter and also on decaying new particles with lifetimes $\tau> 10^{13}\,$s. For particles with shorter lifetimes, extra constraints can be obtained by studying deformations of the black-body spectrum before recombination, usually referred to as spectral distortions. In addition, dark matter with couplings to neutrinos, photons or electrons can shift the effective number of neutrinos around the time of recombination $\Delta N_{\rm eff}(T_{\rm CMB})$, and affect the CMB anisotropies. We will discuss the constraints arising from $\Delta N_{\rm eff}$ in Sec.~\ref{sec:DNeff}.

Charged SM particles that can arise from decays or annihilations of dark particles can inject energy into the plasma in the form of heat. Heat injection at redshift $z\lesssim 2\cdot 10^6$ induce spectral distortions in the CMB. Measurements of spectral distortions are sensitive to dark particles with lifetimes $\tau\gtrsim 10^4\,$s.\footnote{In~\cite{Chluba:2013wsa} it has been shown that spectral distortions cannot set competitive bounds on dark matter annihilation compared to bounds from CMB anisotropies.} However, existing bounds from the COBE-FIRAS experiment~\cite{Fixsen:1996nj} are largely superseded by the BBN bounds discussed in Sec.~\ref{sec:bbn}.\footnote{See also~\cite{Bolliet:2020ofj} for FIRAS/Planck constraints on decays into low-energy photons.}  Future CMB missions similar to PiXie could strengthen the bounds on particles with lifetimes $\tau\gtrsim 10^4\,$s by up to two orders magnitude~\cite{Poulin:2015opa,Lucca:2019rxf}. In contrast, BBN bounds are not expected to improve as much in the future.

Charged particles from decays or annihilations of dark sector particles can also ionize the plasma. Ionization at redshift $z\lesssim 10^3$ modifies the CMB anisotropy spectra.\footnote{Any energy release into the plasma at redshifts earlier than $z\sim 1400$ hardly affects the ionization history and has little impact on the CMB anisotropies~\cite{Chluba:2009uv,Bolliet:2020ofj}. As a result, MeV-GeV dark partners can only affect the CMB through spectral distortions.} This ionized fraction of the energy deposit induces a broadening of the surface of last scattering; for instance, it attenuates the CMB power spectrum on scales smaller than the width of the surface~\cite{Padmanabhan:2005es}. Measurements by the Planck collaboration constrain this effect and set strong upper limits on the cross section for $s$-wave dark matter annihilation~\cite{Planck:2018vyg}. In i2DM, dark matter annihilation $\chi_1 \chi_1 \to f\bar{f}$ is suppressed well below these limits.

Compared to ionization effects on the CMB anisotropies, searches for dark matter annihilation in indirect detection experiments impose much weaker bounds in the MeV-GeV mass range~\cite{Leane:2018kjk,e-ASTROGAM:2017pxr}. Future missions such as e-astrogram~\cite{e-ASTROGAM:2017pxr} can provide competitive bounds on the annihilation cross section, which however are still far above the suppressed annihilation rates in i2DM.

\subsection{Effective number of neutrinos $ N_{\rm eff}$}
\label{sec:DNeff}
As mentioned above, stable dark matter coupled to neutrinos, photons or electrons can change the effective number of neutrinos by $\Delta N_{\rm eff}$. This modification can affect the abundance of light nuclei (set at $T_{\rm BBN } \sim 0.1$ MeV) and the CMB anisotropies (set around $T_{\rm CMB } \sim 0.4$ eV). Both measurements can thus set bounds on $\Delta N_{\rm eff}$.

Efficient dark matter scattering with the electromagnetic or neutrino bath can induce an entropy transfer to these species after neutrino decoupling (at $T_\nu \sim$ MeV), thus modifying the neutrino-to-photon temperature ratio compared to the SM prediction. In practice, this information is encapsulated in $\Delta N_{\rm eff}$. For dark matter that only couples to neutrinos, the entropy transfer reheats the neutrino bath and induces a positive shift $\Delta N_{\rm eff} > 0$. For dark matter coupling to either electrons or photons, the electromagnetic bath gets reheated, which induces a negative shift $\Delta N_{\rm eff} < 0$.

To estimate the effect of light dark particles on $N_{\rm eff}$, we follow the detailed analysis of~\cite{Escudero:2018mvt}, which uses the constraint on $\Delta N_{\rm eff}$ by the Planck collaboration~\cite{Planck:2018vyg},
\begin{align}
    N_{\rm eff}(T_{\rm CMB})=2.99^{+0.34}_{-0.33}\qquad \text{at } 95\%\,\text{CL}.
\end{align}
From this constraint, the authors of~\cite{Escudero:2018mvt} derive a  lower bound on the mass of a Dirac fermion dark matter candidate that couples efficiently to electrons, finding
\begin{equation}
    m_{\rm DM} > 9.2\, {\rm MeV}\qquad \text{at } 95\%\,\text{CL}.
    \label{eq:DNeffbound}
\end{equation}
This result agrees with the estimates in~\cite{Boehm:2013jpa,Depta:2019lbe}, for instance. We emphasize that a modification of $N_{\rm eff}$ can only be observed in CMB data if~\cite{Boehm:2013jpa}
\begin{itemize}
    \item[(i)] the dark matter is in kinetic equilibrium with either electrons, photons or neutrinos at temperatures above and below $T_\nu \sim$~MeV;
    \item[(ii)] the dark matter becomes non-relativistic at temperatures $T \lesssim T_\nu$, typically for masses below a few tens of MeV.
\end{itemize}
As a result, in i2DM the bound from Eq.~(\ref{eq:DNeffbound}) only holds in regions (A) and (B), defined in Eqs.~(\ref{eq:equ}) and (\ref{eq:chemdec}), where the dark fermions are in kinetic equilibrium prior to freeze-out.  For simplicity, we restrict ourselves to dark matter candidates with $m_1 > 10\,$MeV in all three regions.

\subsection{Supernova cooling}
\label{sec:supernovae}
Feebly interacting dark sector particles can be produced in proto-neutron stars and freely escape them. The energy carried by the dark particles speeds up the supernova cooling and therefore can constrain models with light dark sectors. In i2DM, dark fermions are mostly produced via decays of dark photons produced through bremsstrahlung during neutron-proton collisions or directly in collisions of light SM fermions~\cite{Chang:2018rso}. Existing bounds on the cooling of the supernova SN1987A~\cite{Kamiokande-II:1987idp,PhysRevLett.58.1494} constrain the i2DM parameter space only at very small couplings that are irrelevant for our purposes. Based on the results for iDM in~\cite{Chang:2018rso}, we estimate that in i2DM supernova cooling constrains dark couplings up to at most $y \approx 10^{-13}$ for masses $m_1\lesssim 200\,$MeV; a parameter space where the dark matter relics would be overabundant. The bounds could potentially be even weaker if different core-collapse simulations were used~\cite{Bar:2019ifz}. Valid i2DM relics are thus not subject to bounds from supernova cooling.

\section{Laboratory searches}\label{sec:colliders}
In this section we discuss the phenomenology of inelastic dark matter in direct detection experiments, at particle colliders and at fixed-target experiments. Despite the original motivation of inelastic dark matter to evade direct detection, recent investigations reveal sensitivity to certain scenarios of iDM. Colliders and beam-dump experiments set strong bounds on the parameter space of iDM. We show that i2DM can evade some of these bounds, but can be conclusively tested at future experiments. Indirect detection searches are not sensitive to i2DM, because the pair-annihilation rate of dark matter is suppressed by small kinetic mixing and small dark fermion mixing, well below the reach of current and projected future experiments.

\subsection{Direct detection}
\label{sec:direct-detection}
In scenarios of feebly interacting inelastic dark matter, elastic scattering $\chi_1 X\to \chi_1 X$ off nucleons or electrons, $X=n$ or $e$, is suppressed below the sensitivity of current direct detection experiments. In the MeV-GeV mass range, the recoil energy in nucleon scattering is typically too small to be observed, but electron recoils are a promising road to detection~\cite{Ema:2018bih}. The strongest current bound on dark matter-electron scattering by Xenon1T lies around $\sigma(\chi_1 e\to \chi_1 e) \sim 10^{-40}$ cm$^2$~\cite{XENON:2019gfn} for $m_1\approx 100\,$MeV. For comparison, in i2DM with $m_1= 100\,$MeV, $m_{A'}= 3  m_1$ and large mixings $\sin \theta = 0.1$ and $\epsilon = 0.02$, the predicted cross section is $\sigma(\chi_1 e\to \chi_1 e)=6\times 10^{-41}\,{\rm cm}^2$. Future direct detection experiments might reach a higher sensitivity to dark matter-electron scattering, see Ref.~\cite{Essig:2022dfa} for an overview. Whether they can be competitive with other probes for i2DM will depend on the progress in collider searches (see Sec.~\ref{sec:idm-colliders}), which can already probe kinetic mixing down to $\epsilon \approx 10^{-3}$ and potentially suppress the target cross section as $\sigma\sim \epsilon^2$.

In inelastic DM models, the off-diagonal coupling of the dark sector fermions to the dark photon can give rise to sizable inelastic scattering processes. It as been shown that for a mass splitting $\Delta$ between the dark states smaller than a few 100~keV, inelastic up-scattering $\chi_1 n \to \chi_2 n$ can produce an observable nuclear recoil signal, provided that the threshold of the experiment is low enough to detect the recoil energy, see Refs.~\cite{Tucker-Smith:2001myb,Bramante:2016rdh}. In i2DM, the inelastic scattering cross section is only suppressed by $\sin^2\theta$ compared tot $\sin^4\theta$ for elastic scattering. However, for larger values of the mass splitting (comparable to the ones considered in this work), the process becomes kinematically suppressed and extra energy is needed to overcome the mass splitting and still exceed the minimal recoil energy required to give an observable signal at current direct detection experiments.

Even when the nuclear recoil threshold is high, direct detection experiments have some sensitivity to MeV-GeV scale DM elastic and inelastic (up-)scattering processes of nuclei. If the DM particle is accelerated through interactions with cosmic rays in the atmosphere around the earth, it gains energy and can overcome the minimal energy requirement to be observable through nuclear recoils~\cite{Bringmann:2018cvk}. It is hence inevitable that at least a sub-dominant population of boosted DM arises due to scattering processes of cosmic rays, via $\chi_1 p \to \chi_1 p$, that have enough energy to give observable nuclear recoil signals. Since in inelastic DM models (both iDM and i2DM), this process is either absent or parametrically suppressed, the population of boosted DM will be small. However, a similar mechanism can create a sub-dominant population of $\chi_2$ via the up-scattering process $\chi_1 p \to \chi_2 p$\cite{Bell:2021xff}.\footnote{Up-scattering can also occur in the Sun or the Earth~\cite{Baryakhtar:2020rwy,Emken:2021vmf}, but is not efficient enough to be probed in current direct detection experiments for the dark matter scenarios considered in this work.} These dark partners can induce nuclear recoils via down-scattering $\chi_2 n\to \chi_1 n$, provided that their lifetime is long enough to reach the experiment~\cite{Graham:2010ca,Bell:2021xff,CarrilloGonzalez:2021lxm}. Also down-scattering off electrons $\chi_2 e\to \chi_1 e$ is an interesting alternative, see e.g. Ref.~\cite{Aboubrahim:2020iwb}. For inelastic dark matter, efficient down-scattering only occurs if the lifetime is longer than several years~\cite{Bell:2021xff}, corresponding to mass splittings much smaller than those considered in this work.

For i2DM, we expect that the rates for up-scattering and down-scattering are generally suppressed compared to iDM, due to the smaller $A'\chi_1\chi_2$ coupling proportional to $\tan\theta$. On the other hand, in i2DM scenarios with small mass splitting, dark partners produced from cosmic-ray up-scattering are long-lived enough so that a substantial fraction of them could reach the detector before decaying. In this case, elastic scattering via $\chi_2 n \to \chi_2 n$ or $\chi_2 e \to \chi_2 e$ should dominate and leave an interesting characteristic signature of i2DM. In such scenarios, also down-scattering is expected, however parametrically suppressed compared to the elastic $\chi_2$ scattering. If the dark states are heavy and compressed enough to induce an observable nuclear recoil, up-scattering of i2DM off cosmic rays followed by elastic scattering off nuclei in the detector could be directly probed at experiments with a low energy threshold, see e.g. Ref.~\cite{CarrilloGonzalez:2021lxm}. This seems to be the most promising avenue to detect the i2DM dark matter candidate in direct detection experiments, but a dedicated analysis goes beyond the scope of this work. We will anyway show in the remainder of Sec.~\ref{sec:colliders} that existing and near-future accelerator experiments will already be able to fully probe the viable i2DM parameter space.

\subsection{Electroweak precision observables}
\label{sec:ewpo}
A general bound on kinetic mixing of a dark photon is obtained from electroweak precision tests. In electroweak observables measured at LEP, Tevatron and the LHC, kinetic mixing modifies the $Z$ boson mass and couplings to SM fermions at $\mathcal{O}(\epsilon^2)$. For dark photons with masses well below the $Z$ resonance, a global fit to electroweak precision data yields a 95\% CL upper bound of~\cite{Curtin:2014cca}
\begin{align}\label{eq:ewp-bound}
    \epsilon \lesssim 0.02 \qquad \text{for}\qquad  m_{A'} \lesssim 10\,\text{GeV}.
\end{align}
Stronger bounds apply for dark photons with masses closer to the $Z$ pole.
For $m_{A'} \lesssim 10\,$GeV, slightly stronger bounds have also been obtained from $e^\pm$ scattering off protons at HERA~\cite{Kribs:2020vyk}.

\subsection{Collider searches}
\label{sec:idm-colliders}
At $e^+e^-$ colliders, dark fermions coupling via a dark photon can be produced via three main processes:
\begin{align}
    e^+e^- & \to A' \gamma \to \chi_1 \chi_1 \gamma\\\nonumber
    e^+e^- & \to A' \gamma \to \chi_1 \chi_2 \gamma \to \chi_1 \chi_1 \ell^+\ell^- \gamma\qquad \qquad \text{ (iDM)}\\\nonumber
    e^+e^- & \to A' \gamma \to \chi_2 \chi_2 \gamma \to \chi_1 \ell^+\ell^- \chi_1 \ell^+\ell^- \gamma \qquad \text{(i2DM)}.
\end{align}
In models of inelastic dark matter, the first process is suppressed by construction. The second process dominates in iDM scenarios, which typically rely on the coupling of the dark photon to $\chi_1$ and $\chi_2$. The third process is characteristic of i2DM, since the coupling $A'\chi_1\chi_2$ is suppressed by $\tan\theta$ and the dark photon mostly decays via $A' \to \chi_2\chi_2$, see Eq.~\eqref{eq:brs}. Dark photon decays into SM fermions are suppressed as $\epsilon^2 \alpha_{\rm EM}/\alpha_D$, so that resonance searches at BaBar~\cite{BaBar:2014zli} and LHCb~\cite{LHCb:2017trq,LHCb:2020ysn} are not sensitive to the scenarios investigated in this work.

Collider signals of inelastic dark matter depend on whether the dark partners decay within or outside the detector. Below the hadronic threshold, the lifetime of $\chi_2$ is mostly determined by $\chi_2\to \chi_1 \ell^+\ell^-$ decays, which strongly depend on  $\epsilon\tan\theta$ and the mass splitting $\Delta$, see Eq.~\eqref{eq:decay-rate}. If one or two dark partners decay within the detector, the signature consists of one or two prompt or displaced vertices of charged leptons, in association with a photon and missing energy. In iDM, the phenomenology of this signature has been investigated in detail for the Belle II experiment~\cite{Duerr:2019dmv,Mohlabeng:2019vrz,Kang:2021oes}. For sufficiently large mass splitting $\Delta$, Belle II will be able to probe scenarios of iDM in the GeV range.

For a smaller mass splitting, the decay products of $\chi_2$ are too soft to be detected, leading to a signal with a photon and missing energy. The same signature is expected if $\chi_2$ decays outside the detector. A search for mono-photon signals at BaBar has set an upper bound on the kinetic mixing of \emph{invisible} dark photons~\cite{BaBar:2017tiz},
\begin{align}\label{eq:mono-photon-bound}
    \epsilon \lesssim 10^{-3}\qquad \text{for}\qquad  m_{A'} \lesssim 5\,\text{GeV}.
\end{align}
In iDM, this bound excludes most of the parameter space for dark matter candidates below the GeV scale. A similar search at Belle II can probe even smaller dark sector couplings and thereby improve the sensitivity to iDM~\cite{Duerr:2019dmv}.

\begin{figure}[t!]
	\centering
	\includegraphics[width=0.44\textwidth]{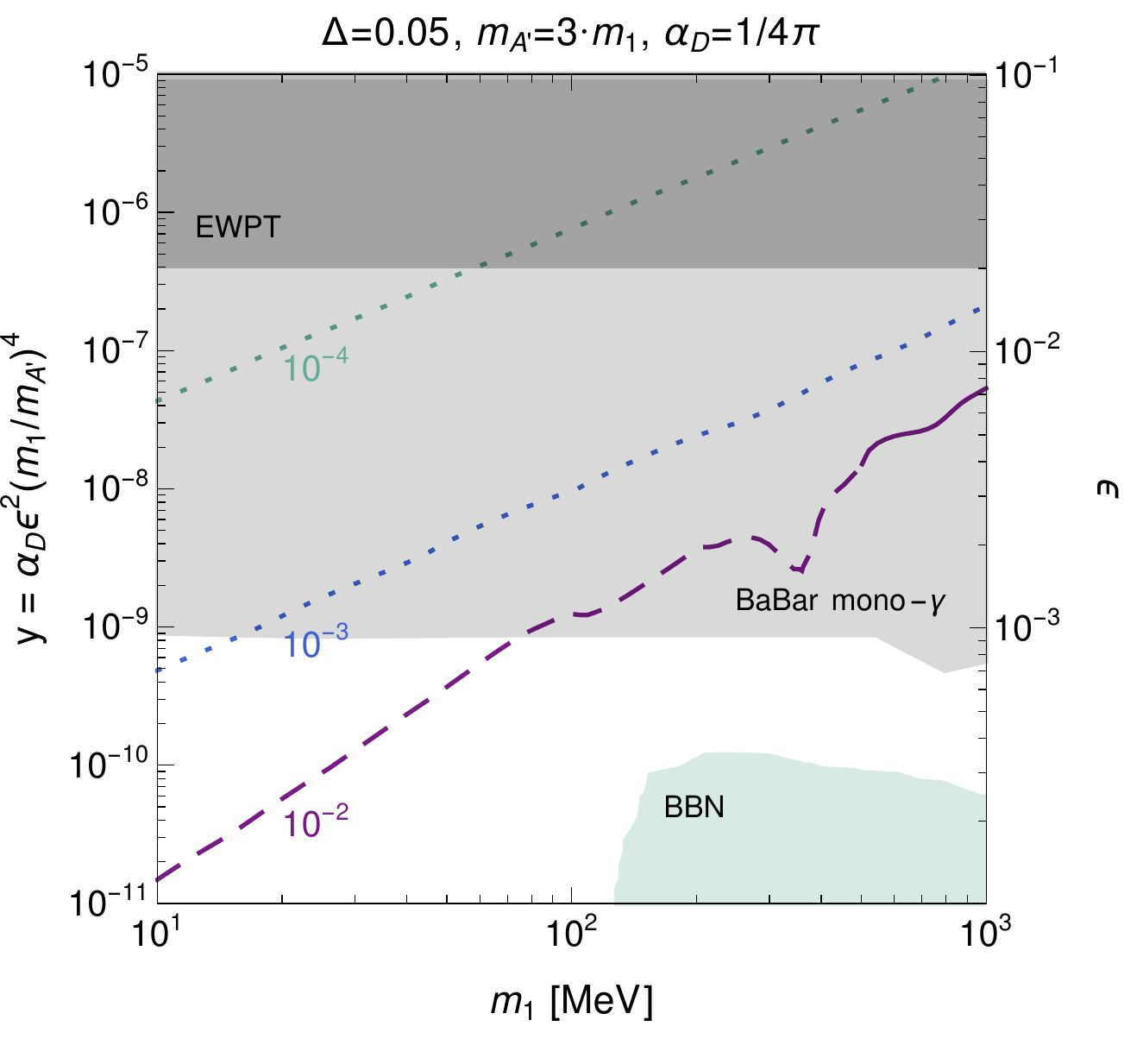}
	\hspace{0.05\textwidth}
	\includegraphics[width=0.44\textwidth]{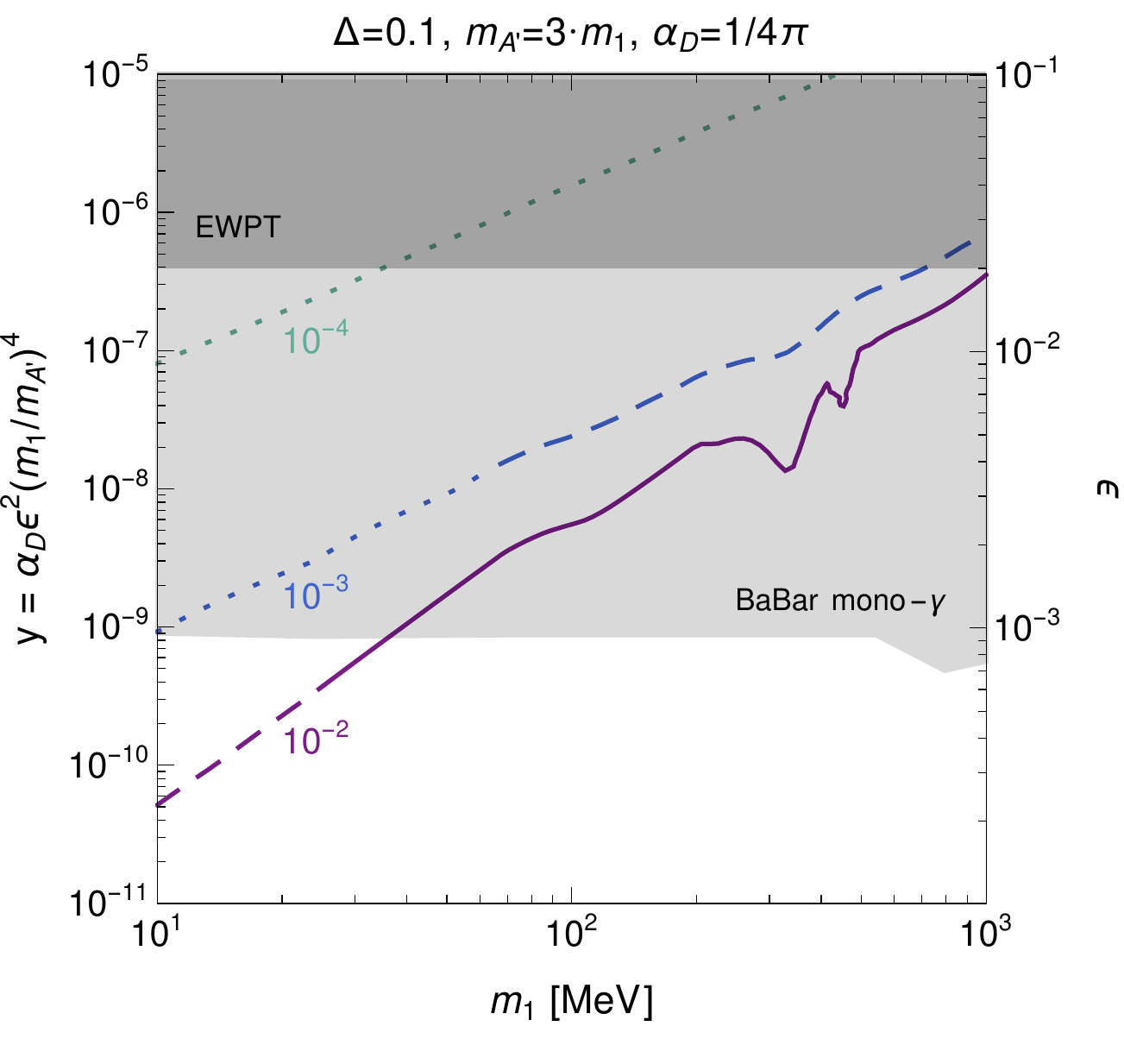}
	\caption{Collider bounds in the $(m_1,y)$ plane of the i2DM. The dark gray area is excluded by electroweak precision observables bound of Eq.~(\ref{eq:ewp-bound}) while the light gray area is excluded by BaBar's mono-photon search, see Sec.~\ref{sec:idm-colliders}. The green area (BBN) is excluded for $\tan \theta=10^{-4}$ due to late photodisintegration of light nuclei, see Sec.~\ref{sec:bbn}. Colored curves correspond to contours of $\Omega_\chi h^2=0.12$ for fixed values of $\tan\theta=10^{-2}$ (purple), $10^{-3}$ (blue) and $10^{-4}$ (green) for fixed relative mass splitting $\Delta=0.05$ (left) and 0.1 (right). On the curves, the three regions introduced in Sec.~\ref{sec:phases} are indicated as plain (A), dashed (B), and dotted (C).
\label{fig:mass-vs-y}}
\end{figure}

In i2DM, the dark photon decays close to its production point due to efficient $A' \to \chi_2\chi_2$ decays. However, the dark partners have larger decay lengths than in iDM, because $\chi_2\to \chi_1 f\bar{f}$ decays are suppressed by $\tan^2\theta$, see Eq.~\eqref{eq:decay-rate}. Therefore the dark photon does not leave any trace in the detector and the bound of Eq.~\eqref{eq:mono-photon-bound} from BaBar's mono-photon search applies. In Fig.~\ref{fig:mass-vs-y}, we display the bounds on kinetic mixing from collider searches in the parameter space of i2DM for $\Delta=0.05$ (left) and 0.1 (right). The observed relic abundance is obtained along the contours for fixed values of the dark fermion mixing $\tan\theta=10^{-2}$, $10^{-3}$ and $10^{-4}$. Small values of $\tan\theta$ need to be compensated by a larger effective interaction strength $y$ to avoid overabundance in the coscattering regime, see Fig.~\ref{fig:phases}. BaBar's mono-photon search translates into a strong upper bound on $y$. In the remaining parameter space, the dark matter abundance is set by partner annihilation in region (A) (plain) and, for small masses, by coscattering in regions (B) (dashed) or (C) (dotted). Departure from kinetic equilibrium, occurring in region (C), mainly happen in regions of parameter space excluded by the mono-photon search. Most viable i2DM candidates are therefore in kinetic equilibrium before freeze-out.
%

\subsection{Bounds from beam-dump experiments}
\label{sec:fixed-target}

\begin{figure}[!h]
  \centering
  \includegraphics[width=0.45\textwidth]{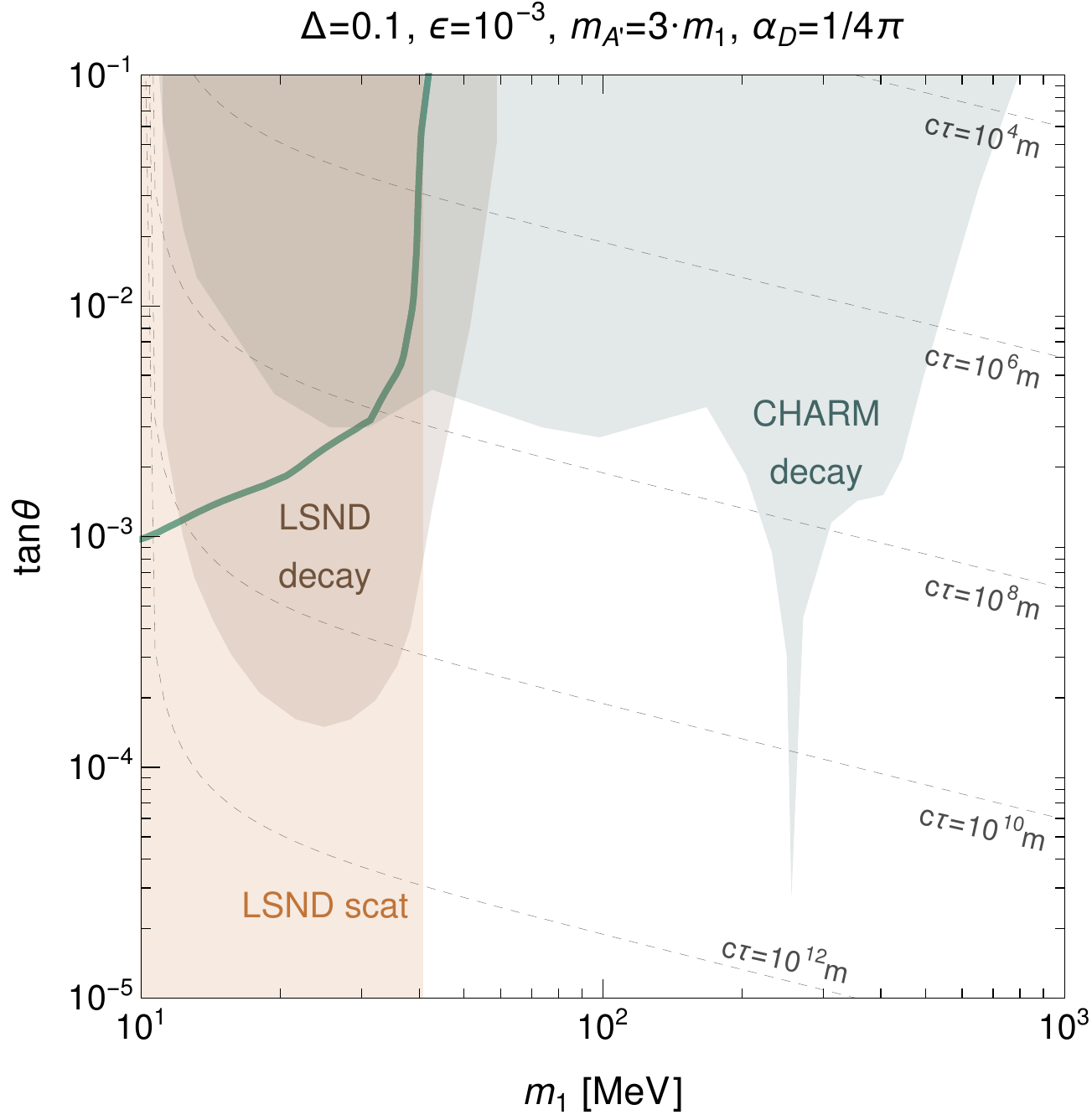}\hspace*{0.02\textwidth}
  \includegraphics[width=0.45\textwidth]{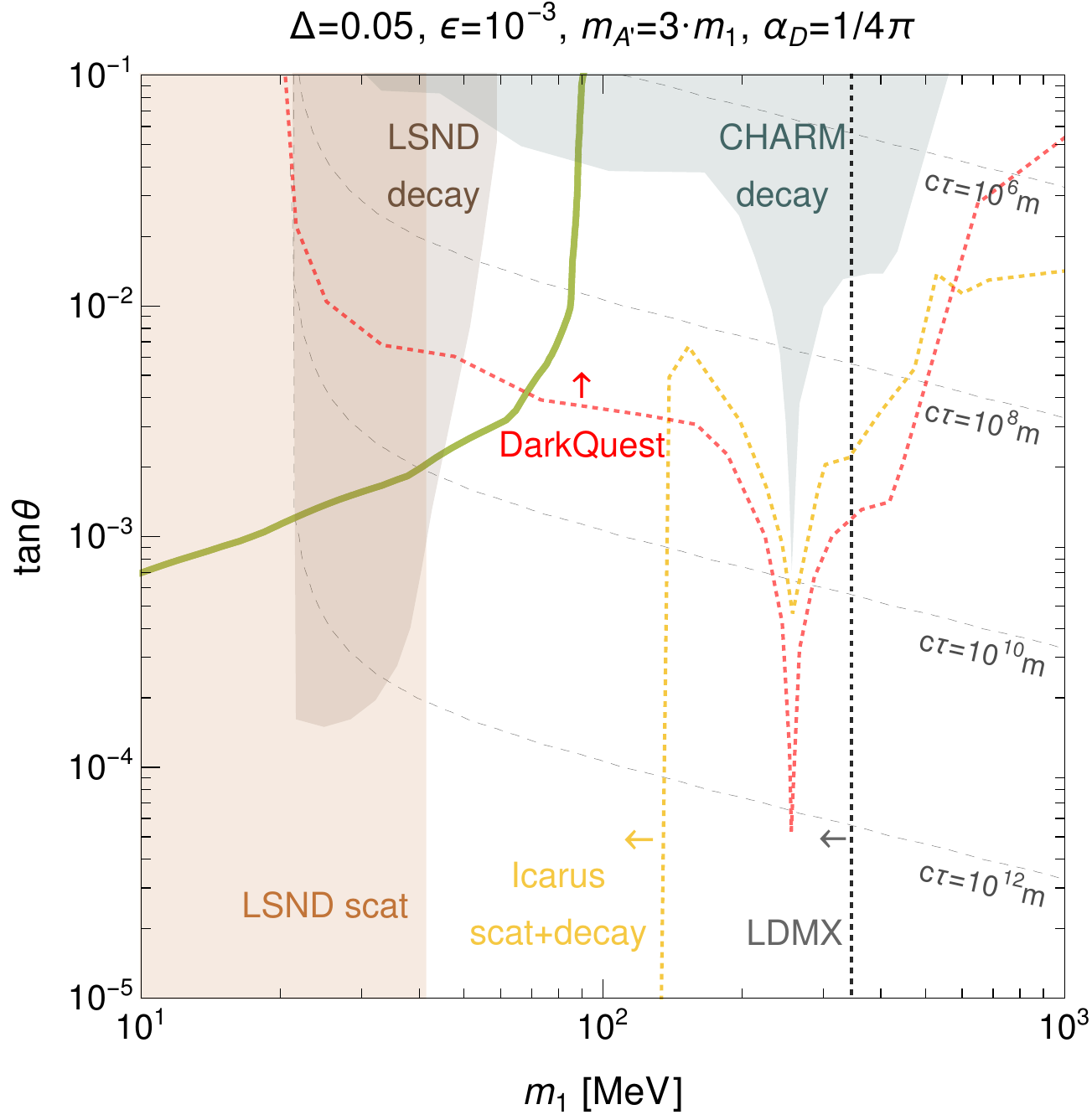}\\
  \vspace*{0.005\textheight}
  \includegraphics[width=0.45\textwidth]{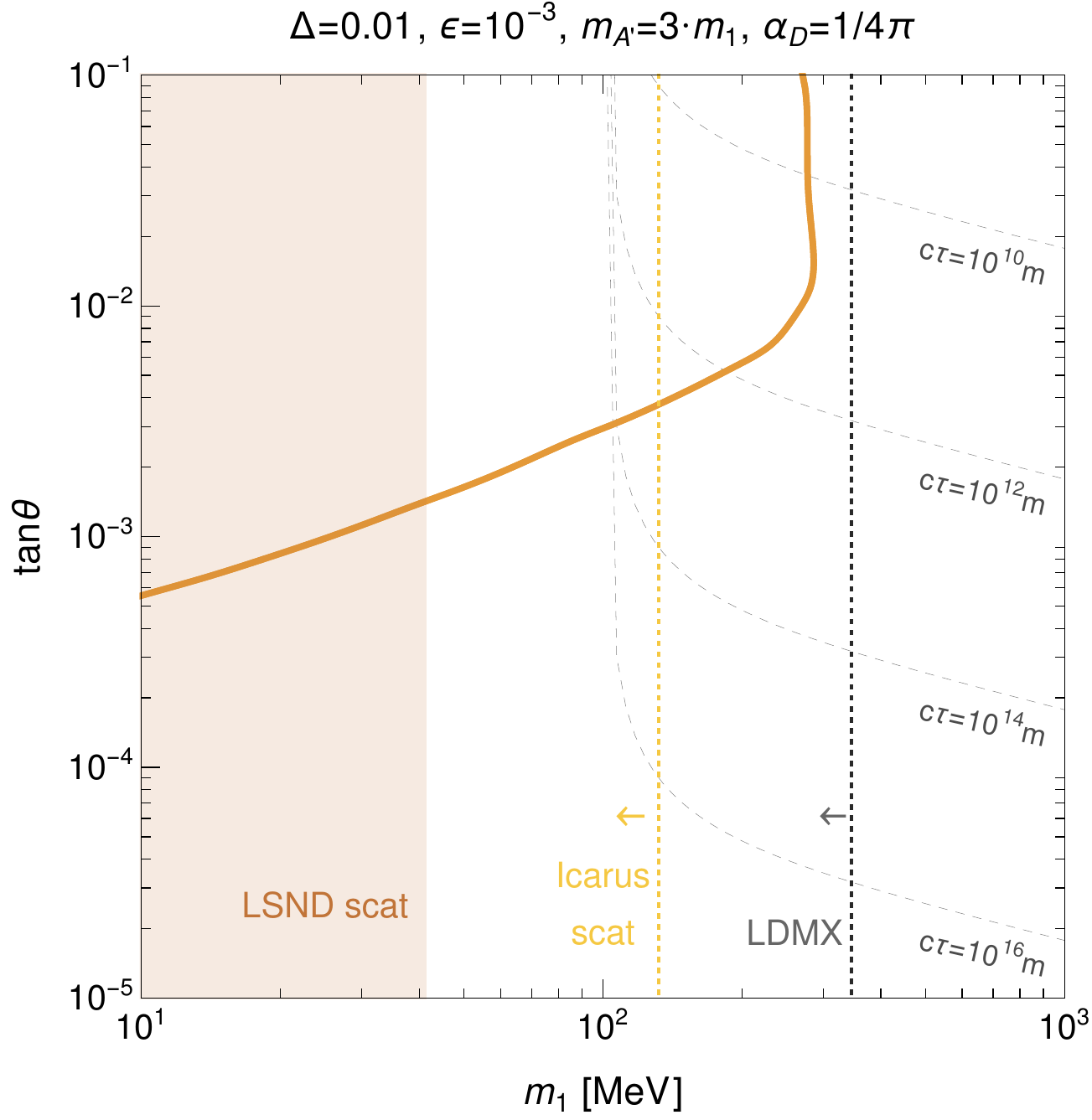}
  \caption{Bounds on i2DM from beam-dump experiments in the ($m_1$,$\tan \theta$) plane for three scenarios with a relative mass splitting $\Delta = 0.1$, $0.05$, $0.01$ and fixed $\epsilon= 10^{-3}$, $m_{A'}= 3 m_1$ and $\alpha_D=1/(4\pi)$. 
  The observed relic abundance is obtained along the colored lines. The gray dashed contours indicate the proper decay length of the dark partner (for $m_1 < 2m_e/\Delta$, the dark partner can only decay into photons or neutrinos, resulting in a very long lifetime). 
  We show existing bounds from CHARM (grey areas) and LSND (red areas). The projected sensitivity of future experiments ICARUS (yellow), DarkQuest (red) and LDMX (grey) is illustrated by dotted lines; the arrow indicates the direction in parameter space that will be probed. See text for details.
\label{fig:mass-vs-coupling}}
\end{figure}

Beam-dump experiments with a large separation of the particle source and the detector are particularly sensitive to particles with a long decay length. Searches for long-lived particles with sub-GeV masses have been performed at various beam-dump and fixed-target experiments and have been reinterpreted for iDM scenarios, for instance in Refs.~\cite{Gninenko:2012eq,Izaguirre:2017bqb,Tsai:2019buq}. In general, a beam of  particles is dumped on a target material, producing light  mesons such as pions or kaons. In iDM and i2DM, dark photons can be produced either in meson decays, through the Primakoff process or from bremsstrahlung. Subsequently the dark photons decay into pairs of dark fermions $\chi_1$ and/or $\chi_2$.

Depending on the model parameters, inelastic dark matter can be detected via three signatures: (displaced) decays of dark partners, $\chi_2 \to \chi_1 \ell^+\ell^-$; (up-/down-)scattering of dark fermions off the detector material, $\chi_i N \to \chi_j N$; or missing energy from dark fermions that are stable at the scales of the experiment. The relative sensitivity to each signature depends on the lifetime of the dark partner, as well as on the experimental setup: short-lived dark partners are mostly observed through decays, while long-lived partners scatter inside the detector material or decay after passing through the detector. In what follows, we discuss the relevant signatures for i2DM and derive bounds from searches at beam-dump experiments.

\paragraph{Partner decays} For partner decays, the expected event rate in the far detector is given by\footnote{Here we neglect dark partners produced via upscattering $\chi_1 \to \chi_2$.}
\begin{align}
    N_{\rm dec} \approx N_{A'}\Big(\mathcal{B}(A'\to \chi_1\chi_2) + 2 \mathcal{B}(A'\to \chi_2\chi_2)\Big) \frac{1}{N_2}\sum_{k=1}^{N_2}\,P_{\rm dec}(d_k)\,,
\end{align}
where $N_{A'} \propto \epsilon^2$ is the total number of  dark photons produced in a given experiment and $N_2$ is the total number of $\chi_2$ states resulting from $N_{A'}$ dark photon decays. The branching ratios are $\mathcal{B}(A'\to \chi_1\chi_2)\approx 1$ for iDM and $\mathcal{B}(A'\to \chi_2\chi_2)\approx 1$ for i2DM when $\tan\theta \ll 1$.\footnote{In our numerical analysis of i2DM, we neglect $A'\to \chi_1\chi_2$ decays.} Finally, $P_{\rm dec}(d_k)$ is the probability to detect the decay products of particle $k$ with decay length $d_k$ inside the detector. The decay length $d_k = (\beta\gamma)_k c\tau_2$ depends on the boost, $(\beta\gamma)_k$, and on the lifetime, $\tau_2$, of the dark partner. For proper lifetimes longer than the baseline, the decay probability scales as $P_{\rm dec}(d) \propto 1/d$. In this regime, the expected event rates for iDM and i2DM depend on the model parameters as
\begin{align}\label{eq:partner-decays}
    N_{\rm dec}^{\rm iDM} \propto \epsilon^2 y \propto \epsilon^4 \alpha_D\,,\qquad N_{\rm dec}^{\rm i2DM} \approx 2 \tan^2\theta \cos^4\theta\, N_{\rm dec}^{\rm iDM}\,.
\end{align}
In i2DM, the lifetime of the dark partner scales as $\tau_2 \sim 1/\tan^{2}\theta$. For small $\tan\theta$, the dark partner tends to decay after passing the detector, which reduces the event rate. The factor of 2 accounts for the two dark partners produced in $A'\to \chi_2\chi_2$ decays. For fixed parameters $\alpha_D,\,m_1,\,m_{A'},\,\Delta$, a bound on $\epsilon$ obtained from searches for dark partner decays in iDM translates into a bound on $\epsilon \cdot (2\tan^2\theta\cos^4\theta)^{1/4}$ in i2DM. This relation can be used to map the bounds obtained for iDM in Refs.~\cite{Gninenko:2012eq,Izaguirre:2017bqb,Tsai:2019buq} on the parameter space of i2DM, see Fig.~\ref{fig:mass-vs-coupling}.

\paragraph{Decays at CHARM} Strong bounds on long-lived dark particles decaying into electrons have been set at the neutrino experiment CHARM. At CHARM, dark fermions can be efficiently produced from $\pi^0$ or $\eta$ meson decays, $\pi^0(\eta) \to \gamma A' \to \gamma \chi_i\chi_j$. Null results of a beam-dump search for heavy neutrinos decaying into electron pairs~\cite{CHARM:1983ayi} have been reinterpreted for $\chi_2\to \chi_1 e^+ e^-$ decays in iDM~\cite{Tsai:2019buq}. In Fig.~\ref{fig:mass-vs-coupling}, we show the resulting bounds in the parameter space of i2DM, using Eq.~\eqref{eq:partner-decays} to rescale the predicted event rates. The three scenarios are distinguished by the mass splitting $\Delta$, while we have fixed $\epsilon = 10^{-3}$ to evade the collider bounds from Sec.~\ref{sec:idm-colliders}.

The observed relic abundance is obtained along the colored contours. The lifetime of $\chi_2$ varies strongly with the mass splitting, $\tau_2 \propto 1/\Delta^{5}$, see Eq.~\eqref{eq:decay-rate}. For $\Delta = 0.1$, most of the dark partners decay within the considered decay volume and the search is sensitive to even small mixing $\tan\theta$. For $\Delta = 0.05$, the sensitivity decreases due to the longer lifetime and softer $e^+e^-$ momenta, which are less likely to pass the analysis cuts. For $\Delta = 0.01$, the dark partner is essentially stable compared to the length of the decay volume and the search becomes insensitive to i2DM.

\paragraph{Decays at LSND} For light dark sectors, even stronger bounds have been obtained from the neutrino experiment LSND. At LSND, dark fermions can be abundantly produced from pion decays via $\pi^0 \to \gamma A' \to \gamma \chi_i\chi_j$. In Ref.~\cite{Izaguirre:2017bqb}, LSND data has been interpreted in terms of $\chi_2 \to \chi_1 e^+ e^-$ decays in iDM, under the conservative assumption that the $e^+e^-$ pair is not resolved in the calorimeter and can mimic elastic neutrino-electron scattering. In Fig.~\ref{fig:mass-vs-coupling}, we show the resulting bounds rescaled for i2DM (labelled `LSND decay'). LSND is very sensitive to dark partners with $m_2 < m_{\pi^0}/2$. The lower cutoff is determined by the kinematic threshold for $\chi_2\to \chi_1 e^+e^-$ decays. The sensitivity could be improved with a dedicated analysis of three-body decays, rather than a re-interpretation of neutrino-electron scattering.

\paragraph{Dark fermion scattering} In addition to decays, long-lived dark fermions can be detected through up-scattering $\chi_1 \to \chi_2$, down-scattering $\chi_2\to \chi_1$, or elastic scattering $\chi_2 \to\chi_2$ inside the detector. We neglect up-scattering, which is typically sub-dominant due to the kinematic suppression. The expected scattering rate is then given by
\begin{align}
    N_{\rm scat} \approx N_{A'}\Big(\mathcal{B}(A'\to \chi_1\chi_2) + 2 \mathcal{B}(A'\to \chi_2\chi_2)\Big) \frac{1}{N_2}\sum_{k=1}^{N_2}\,P_{\rm scat}(d_k)\,,
\end{align}
where $P_{\rm scat}(d_k)$ is the probability for particle $k$ to scatter off the material inside the detector,\footnote{$P_{\rm scat}(d_k)$ is proportional to scattering cross section and the probability for $\chi_2$ to decay after traversing the detector. The latter is roughly one for proper lifetimes much longer than the baseline of the experiment.} with $d_k \to \infty$ for $\chi = \chi_1$, and $N_2$ is the number of produced dark fermions. Neutrino experiments are particularly sensitive to dark fermion scattering, which mimics neutrino-electron scattering. Among various experiments, LSND sets the currently strongest bounds on inelastic dark matter. For iDM and sub-GeV masses, the relevant process is down-scattering $\chi_2 N \to \chi_1 N$~\cite{Izaguirre:2017bqb}. For i2DM, the dark photon mainly produces a $\chi_2$ pair for $\tan\theta \ll 1$ so that elastic scattering $\chi_2 \to \chi_2$ dominates. The respective event rates scale as
\begin{align}
    N_{\rm scat}^{\rm iDM} \propto \epsilon^4\alpha_D\,,\qquad N_{\rm scat}^{\rm i2DM} \approx 2 \cos^4\theta\, N_{\rm scat}^{\rm iDM}
    \label{eq:Nscat}
\end{align}
for decay lengths $d_k$ larger than the distance between the target and the detector. Similar as for decays, we can use Eq.~\eqref{eq:Nscat} to map bounds obtained for iDM on the parameter space of i2DM.

\paragraph{Scattering at LSND} The LSND bounds on scattering are derived from the same analysis as for partner decays. Again, we translate the results for iDM from Ref.~\cite{Izaguirre:2017bqb} to i2DM, shown in Fig.~\ref{fig:mass-vs-coupling} as `LSND scat'. For $\tan\theta \lesssim 0.3$, the bounds are insensitive to dark fermion mixing and exclude small dark matter masses. The bounds disappear for small kinetic mixing $\epsilon \lesssim 10^{-4}$, where LSND loses its sensitivity due to the low dark photon production rate.\\

It is interesting to compare these results for i2DM with iDM. For $\tan\theta = 1$, the $A'\chi_1\chi_2$ interaction strength is similar in iDM and i2DM. For the benchmark scenarios shown in Fig.~\ref{fig:mass-vs-coupling}, iDM is excluded by CHARM and LSND, unless partner decays into electrons are kinematically forbidden. Viable scenarios of sub-GeV iDM require a stronger interaction strength $y$ for efficient coannihilation, while keeping the kinetic mixing $\epsilon$ small to evade bounds from mono-photon searches~\cite{Batell:2021ooj}, see Sec.~\ref{sec:idm-colliders}. In turn, i2DM scenarios with small $\tan\theta$ can evade the CHARM bounds due to the longer lifetime of the dark partners. In this regime, the relic abundance is set by partner annihilation and/or coscattering. In summary, null searches at current fixed-target experiments are a severe challenge for iDM, while i2DM is a viable option due to the impact of partners on dark matter freeze-out.

\subsection{Prospects of future beam-dump experiments}
\label{sec:fixed-target-future}
As we discussed in Sec.~\ref{sec:fixed-target}, existing beam-dump/fixed-target experiments are very sensitive to sub-GeV inelastic dark matter. However, i2DM scenarios with a compressed dark sector currently still escape detection in certain regions of parameter space. In order to fully explore the parameter space of i2DM, we study the discovery potential of proposed searches at beam-dump/fixed-target experiments, which could be realized in the near future. While a number of experiments can be sensitive to i2DM, here we focus on a few promising proposals.

\paragraph{SBN}
The Short-Baseline Neutrino (SBN) program at Fermilab~\cite{Machado:2019oxb} is a planned facility to probe neutrinos and light dark sectors. The facility uses the Booster 8 GeV proton beam hitting a beryllium target. Three detectors are placed downstream of the target at varying distances. The SBND detector is located at around 110\,m, while the experiments MicroBooNE and ICARUS are placed further away, at 470\,m and 600\,m respectively.

At SBN, similarly to other beam-dump experiments discussed in Sec.~\ref{sec:fixed-target}, dark partners can be produced from decaying dark photons, which are created in the interaction of the proton beam with the target. The SBN detectors can be used to search for $\chi_2 \to \chi_1 \ell^+\ell^-$ decays or scattering $\chi_i \rightarrow \chi_j$ off the detector material.

Dark sector searches at neutrino experiments inevitably feature a large neutrino background. At SBN, there are two proposals to reduce this background. One option is to deflect the proton beam around the target into an iron absorber placed 50\,m downstream, which has previously been done at MiniBooNE to study light dark sectors \cite{MiniBooNE:2017nqe,MiniBooNEDM:2018cxm}. This mode is referred to as ``off-target''. The second option is to use the NuMI 120\,GeV proton beam, impacting on a graphite target. ICARUS and MicroBooNE are placed at angles of 6 and 8 degrees against the NuMI beam direction. This option is known as ``off-axis''. Both off-target and off-axis options have been studied for iDM~\cite{Batell:2021ooj}, where SBND has the best sensitivity in the off-target mode, while ICARUS can set the strongest bound in the off-axis mode. In Fig.~\ref{fig:mass-vs-coupling}, we show our reinterpretation of these predictions for i2DM, following the procedure described in Sec.~\ref{sec:fixed-target}. We present the results for ICARUS, assuming that the off-axis mode can be realized with much less technological effort than the off-target option. Compared to existing fixed-target experiments, ICARUS is substantially more sensitive to i2DM, provided that the dark partners are sufficiently long-lived to induce enough signal in the detector.

\paragraph{DarkQuest}
Originally developed to study the sea quark content of the proton with a 120 GeV proton beam and various targets, the Fermilab experiment SeaQuest has a good potential to probe dark sectors \cite{Gardner:2015wea,Berlin:2018pwi,Apyan:2022tsd}. It is already equipped with a displaced muon trigger to study exotic long-lived particles decaying to muons, and could be supplemented by an electromagnetic calorimeter to also probe electron signals. Such an update would improve the sensitivity of the SeaQuest experiment to long-lived decaying dark states drastically. Due to its sensitivity for light dark sectors, this extension of the experiment is referred to as DarkQuest~\cite{Apyan:2022tsd}. In Ref.~\cite{Berlin:2018pwi}, the sensitivity of DarkQuest to dark partner decays in iDM has been studied for three different decay volumes. In Fig.~\ref{fig:mass-vs-coupling}, we show the corresponding predictions for i2DM for the largest possible decay volume. Compared with CHARM and ICARUS, DarkQuest can improve the sensitivity to i2DM for dark matter with masses near the $\eta$ resonance. As discussed in Ref.~\cite{Berlin:2018pwi}, the reach of DarkQuest could be further enhanced by running the experiment without the magnet, which however would require a dedicated analysis of the experimental setup.

\paragraph{LDMX}
The proposed Light Dark Matter eXperiment (LDMX)~\cite{LDMX:2018cma} is an electron beam-dump experiment designed primarily for probing light dark matter models. Its search strategy relies on dark sector particles being produced in the beam-dump that escape the detector, which extends up to about $1\,$m downstream from the target. This gives rise to a signature of missing energy. All charged particles in an event are vetoed, except for the soft remnant of the incoming electron. A signal of inelastic dark matter is detected if a substantial fraction of dark partners decay \emph{after} passing through the detector. The LDMX collaboration has investigated the projected sensitivity to many MeV-GeV dark sector models, including iDM~\cite{Berlin:2018bsc}. In Fig.~\ref{fig:mass-vs-coupling}, we show our interpretation for i2DM for the most conservative design option. LDMX is well suited to probe i2DM scenarios with very small mass splitting $\Delta$, which are difficult to detect in experiments that probe the decay products of the dark partner.\\

We summarize our projections for future beam-dump and fixed-target experiments for the three i2DM benchmarks in Fig.~\ref{fig:mass-vs-coupling}. For $\Delta=0.1$, the dark matter target is already probed by existing experiments. We therefore do not show projections for future experiments, but note that they could be sensitive to other regions of the parameter space. For $\Delta=0.05$, DarkQuest alone can improve the sensitivity to dark partner decays, but cannot fully probe the dark matter target. ICARUS can complement DarkQuest by also detecting dark fermion scattering, which is predominant for small dark matter masses. LDMX, searching for missing energy, can extend the reach of ICARUS at larger masses and long lifetimes. Either ICARUS or LDMX could conclusively probe this scenario. For even smaller mass splitting $\Delta=0.01$, dark partner decays cannot be observed due to the long lifetime and the softness of the SM decay products. The sensitivity of ICARUS and LDMX through scattering and missing energy, however, is kept and allows to conclusively probe the dark matter scenario. Larger dark photon masses $m_{A'} > 3 m_1$ or smaller kinetic mixing $\epsilon$ reduces the rate of produced dark photons and thus the sensitivity of any  experiment. For smaller dark couplings $\alpha_D$, the lifetime of the dark partners is enhanced and dark partner decays close behind the target are less abundant. In this case, scattering and missing energy are more promising signals, especially for small dark matter masses. All in all, beam-dump and fixed-target experiments have a high potential to conclusively test i2DM in the near future, provided that they are built and successfully run.

\chapter{Conclusion and outlook}

Even today, the exact nature of DM remains one of the main open questions in physics, despite the strong experimental effort of the recent years to answer this question. The \acf{WIMP} paradigm, where the DM is a new BSM particle interacting weakly with the SM particle content, can nicely explain the observed effects that DM leaves on our universe today. However, direct and indirect DM experiments tried to detect these \acp{WIMP}, without any success, placing stringent bounds on the \ac{WIMP} paradigm. It is therefore timely to study the theoretical possibilities of alternative DM candidates, and look for potential avenues of detecting them. 

In this thesis, we focused on \acfp{FIMP}, particles that interact way more feeble with the SM compared to \acp{WIMP}. While \acp{WIMP} are generally produced through the freeze-out mechanism in the early universe, \acp{FIMP} are not always able to reach or stay in chemical and kinetic equilibrium until the final DM abundance is set. We started out in Chapter~\ref{chap:DM_prod} by reviewing in detail the freeze-out dynamics setting the relic abundance of \acp{WIMP}, before studying potential production mechanisms for \acp{FIMP}, such as freeze-in, conversion driven freeze-out or the superWIMP mechanism. After discussing the main features of those mechanisms, we showed how they can be realized by considering a simplified leptophilic model. Depending on the value of the interaction coupling connecting the \acf{DS} with the SM, we showed that different mechanisms set the relic DM abundance, and we discussed the details of the conversion driven freeze-out and freeze-in regime as these have not been studied yet in the literature within this model. In Chapter~\ref{chap:alt_cosmo}, we also showed how DM production can occur in a non-radiation dominated universe. Such an alternative cosmological history can have an important impact on the relic abundance calculations. We mainly focused on the possibility of freeze-in happening during inflationary reheating, i.e. when the reheating temperature is smaller than the temperature at which the main freeze-in dynamics happen. In such a scenario, the universe is dominated by the decaying inflaton, a process which injects entropy in the thermal bath. This entropy injection causes the DM yield to be diluted, and only remains constant after reheating has finished and we have entered the radiation dominated era in the evolution of our universe.

After studying several production mechanisms for \acp{FIMP} and even considering DM production during reheating, we turned our attention to potential detection avenues for \acp{FIMP}. Since the interactions with the SM are very suppressed, direct and indirect detection experiments generally do not have great sensitivity to \ac{FIMP} models. In order to identify viable detection strategies, we first focused on what is referred to as t-channel DM models. In those models, the DM interacts with the SM through a cubic interaction involving another BSM particle called the mediator. Since the DM has to be neutral under the SM gauge group, the mediator inherits the quantum gauge charges of the SM particle involved in the cubic interaction. Therefore, this mediator can be copiously produced at collider experiments, and due to the small coupling dictating the decay of the mediator, this particle can be long-lived. Since LEP already put stringent constraints on charged BSM particles with masses below 100~GeV, we focus on heavier mediators so that the searches for displaced signatures at the LHC can have great discovery potential for the FIMP regime of t-channel DM models.

In order to structure the discussion of probing \ac{FIMP} models at colliders, we proposed in Chapter~\ref{chap:model_class} a classification scheme based on the cubic interaction described above, focusing on scalar and fermionic DM and mediator particles. Starting from this classification, we listed possible observable displaced signatures at the LHC, either based on the track of the long-lived mediator or on its displaced decay products. Most of those signatures have already been searched for by the CMS and ATLAS collaborations, however, we were able to identify one novel signature, dubbed \acf{KT}. In Chapter~\ref{chap:LHC_cons}, we reinterpreted already performed searches for displaced new physics and applied them to a selection of models taken from our classification, namely a leptophilic scenario with a fermionic DM candidate (where we discussed both a compressed and light FIMP regime), a topphilic scenario involving a scalar DM candidate and finally the singlet-triplet model. In all three scenario, we stressed the interplay between the reheating temperature and the DM freeze-in production. We showed that in the case of a discovery of new displaced physics, we can use the measured kinematic variables of this decay to infer information about the evolution of the early universe, in particular the reheating temperature. This is however dependent on the model under consideration. Our proposed classification might play an important role in this procedure, as we identified in Chapter~\ref{chap:LHC_cons} which models show which signatures. It is hence of great importance to keep probing a large variety of signatures to keep sensitivity to a large variety of models, but also to identify the model once a discovery has occurred.

As we have demonstrated in Chapters~\ref{chap:model_class} and~\ref{chap:LHC_cons}, the experimental effort in probing long-lived particles at the LHC has already lead to a variety of searches performed at the ATLAS and CMS experiments. However, since these detectors are located very close to the proton-proton collision, they are not optimized for probing lifetimes much longer than the detectors length. Since in the freeze-in regime, the mediators in our DM models presented in Chapter~\ref{chap:LHC_cons} can easily reach such long lifetimes for large values of the reheating temperature compared to the mediator mass, potential secondary experiments placed further away from the interaction point are proposed. Some examples of such experiments are MATHUSLA~\cite{MATHUSLA:2018bqv,MATHUSLA:2020uve}, CODEX-B~\cite{Aielli:2019ivi} and ANUBIS~\cite{Bauer:2019vqk}. These experiments are placed in the transverse direction compared to the beamline and are typically shielded to create an almost complete background free decay detector. Due to this shielding, these experiments will only be sensitive to models including neutral \acp{LLP}, while most of the models proposed in our classification from Chapter~\ref{chap:model_class} contain charged or colored mediators. The long lifetime regime for these models can hence only be probed by the HSCP/R-hadron searches performed by ATLAS and CMS. Only models containing $SU(2)_L$ multiplets, such as the singlet-triplet model discussed in Sec.~\ref{sec:SingletTriplet}, can give rise to neutral \acp{LLP}. These models cannot be probed by the HSCP/R-hadron searches, and far-away detectors can hence become an essential probe in exploring the parameter space of these models, as has been shown in Ref.~\cite{No:2019gvl}.

In Chapter~\ref{chap:lightDM}, we switched gears and started discussing s-channel or portal DM models. In particular, our main focus was on inelastic DM models, where the DS consists of two dark fermions interacting with a dark photon, originating from a dark $U(1)$-symmetry. Due to kinetic mixing with the SM hypercharge boson, the dark photon provides a vector portal to the SM. In a well studied version dubbed iDM, the dark fermions mainly interact off-diagonally with the dark photon and the DM is produced through co-annihilations. We introduced a new realization of inelastic dark matter, named inelastic Dirac Dark Matter or i2DM. Within this realization, the DM can also be produced through mediator driven freeze-out and the coscattering mechanism. These extra production regimes open up the viable DM parameter space. We discussed the details of the different production regimes while carefully taking into account chemical and kinetic decoupling of the dark fermions. We noticed that the relic abundance requirements prefer MeV-GeV scale dark fermions, which is important when discussing the phenomenology in Chapter~\ref{chap:i2dm_pheno}. The presence of such light particles can influence the cosmological evolution as well as the evolution of astrophysical objects. While such probes can constrain iDM, the smaller couplings and longer lifetimes obtained in i2DM evade those constraints. For the same reason, also direct detection constraints are easier to evade. We identified electron-positron colliders and beam-dump/fixed-target experiments as the most promising way of probing inelastic dark matter, as we have seen that i2DM is already majorly constrained by BaBar, CHARM and LSND, while near future beam-dump or fixed-target experiments such as ICARUS, DarkQuest and LDMX are able to entirely probe the viable i2DM parameter space. Beam-dump and fixed-target experiments have a very strong sensitivity to MeV-GeV scale DM models including long-lived mediators, and are hence very complementary to the displaced physics searches performed at CMS and ATLAS. However, this does not mean that MeV-GeV scale inelastic DM models cannot be probed at the LHC. The main reason why CMS and ATLAS have low sensitivity to such light scale models is that the soft signal one expects from such models will be overwhelmed by SM background. As discussed before, low background experiments can be achieved by placing detectors further away from the interaction point and shield them from charged radiation so that they form an optimal environment to probe neutral \acp{LLP}. Detectors as MATHUSLA, CODEX-b and ANUBIS can fulfill this role in the transverse direction. However, particles in the MeV-GeV range are also produced very efficiently in the forward direction. By making use of the available space in underground tunnels more than 100m away from the interaction point, experiments such as FASER~\cite{FASER:2019aik,FASER:2018eoc} and SND~\cite{SNDLHC:2022ihg,Boyarsky:2021moj} are already taking data during run-3 of the LHC, while for the HL-LHC, a forward physics facility~\cite{Anchordoqui:2021ghd,Feng:2022inv} dedicated to host multiple experiments taking data in the forward direction of the proton-proton collisions at the LHC has been proposed. Al these dedicated \ac{LLP} experiments, both in the forward and transverse direction, can have very good sensitivity to inelastic DM models as shown in Refs.~\cite{Berlin:2018jbm,Bertuzzo:2022ozu}.

A final avenue in exploring \acp{LLP} we have not discussed in this thesis is looking for displaced physics at electron-positron colliders. While they are limited in energy compared to proton-proton colliders, they provide a very clean environment to reconstruct displaced decays so that the efficiency is mainly determined by the detector acceptance. This has been shown for instance for the Belle II experiment at the superKEKB collider in Japan, see e.g.~\cite{Duerr:2019dmv,Duerr:2020muu,Filimonova:2019tuy,Ferber:2022rsf}. Also the new GAZELLE detector has been proposed to enlarge the sensitivity for longer lifetimes~\cite{Dreyer:2021aqd}. The main drawback of current electron-positron colliders is that they are limited in energy so that they only probe BSM physics at the GeV-scale and below. Future colliders such as the ILC~\cite{ILC:2007oiw}, CLIC~\cite{CLIC:2012}, CEPC~\cite{CEPCStudyGroup:2018ghi} and FCC-ee~\cite{FCC:2018evy} can serve as precision machines up to the electroweak energy scale. Therefore, sensitivity studies have already been performed to see how such future lepton colliders can perform in probing \acp{LLP}, see e.g.~\cite{Schafer:2022shi,Blondel:2022qqo,Chrzaszcz:2020emg,Wang:2019xvx,Tian:2022rsi}.

\appendix

\chapter{Phase space distribution in kinetic and chemical equilibrium}
\label{app:equilibrium}

The evolution of a particles phase space density $f(x^\mu,p^\mu)$ is governed by the unintegrated Boltzmann equation,
\begin{align}
    L[f]=C[f],
\end{align}
where $L$ is the Liouville operator describing the influence of the cosmological model on the phase space density, while the collision operator $C$ describes the effects from particle physics interactions. The former can be written as
\begin{align}
    L[f] = E \left( \frac{\partial f}{\partial t} - H \frac{p^2}{E} \frac{\partial f}{\partial E} \right).
    \label{eq:unint_BE_app}
\end{align}
in a Friedman-Robertson-Walker (FRW) universe, described by the Hubble rate $H$. The isotropy and homogeneity imply that the phase space density only depends on the magnitude of the three momentum ($p$), or equivalent the energy ($E$), and time ($t$), i.e. $f(x^\mu,p^\mu)=f(E,t)$. Further, the collision term for a general two-to-two process $a,b \to i,j$ reads
\begin{align}
    C[f_a] =& \int d\Pi_b d\Pi_i d\Pi_j
     (2\pi)^4 \delta^4(p_a+p_b-p_i-p_j) \nonumber \\
     &\times|\mathcal{M}|^2 \left[ f_i f_j (1\pm f_a) (1\pm f_b) - f_a f_b (1\pm f_i) (1\pm f_j) \right],
\end{align}
where $f_a,f_b,f_i,f_j$ are the phase space densities of the particles $a,b,i,j$ involved in the reaction, the plus or minus sign depends on the nature of the particle ($+$ for bosons undergoing bose-enhancement, $-$ for fermions undergoing Pauli-blocking) and 
\begin{align}
    d\Pi_k = \frac{g_k}{(2\pi)^3} \frac{d^3p_k}{2E_k},
\end{align}
for each species $k$ involved. We have assumed that the process $a,b \to i,j$ conserves charge and parity transformations so that the Feynman amplitude $|\mathcal{M}|$ of this process is the same in both directions, i.e.
\begin{align}
    |\mathcal{M}_{a,b \to i,j}| = |\mathcal{M}_{i,j \to a,b}| = |\mathcal{M}|.
\end{align}

A particle species is said to be in equilibrium if its phase space density is purely dictated by the evolution of the universe, which means that it cannot be influenced by any particle physics process. This is true if the collision term in Eq.~\eqref{eq:unint_BE_app} equals zero. One obvious way to make it zero is by invoking the requirement of ``detailed balance'',
\begin{align}
    f_i^{eq} f_j^{eq} (1\pm f_a^{eq}) (1\pm f_b^{eq}) = f_a^{eq} f_b^{eq} (1\pm f_i^{eq}) (1\pm f_j^{eq}),
\end{align}
or equivalently
\begin{align}
    \log \left( \frac{f_i^{eq}}{1\pm f_i^{eq}} \right) + \log \left( \frac{f_j^{eq}}{1\pm f_j^{eq}} \right) = \log \left( \frac{f_a^{eq}}{1\pm f_a^{eq}} \right) + \log \left( \frac{f_b^{eq}}{1\pm f_b^{eq}} \right). 
    \label{eq:detailed_balance}
\end{align}
As an example, we will consider two-to-two self-interaction of a particle species $a$, where two particle with incoming energies $E_1$ and $E_2$ collide and form two particles with energies $E'_1$ and $E'_2$. This process ensures kinetic equilibrium when efficient. If this is the case, we know from Eq.~\eqref{eq:detailed_balance} that the logarithm of $f^{\rm KE}(E)/(1 \pm f^{\rm KE}(E))$ is conserved. Since also energy must be conserved during this collision, this logarithm must be a linear combination of the energy,
\begin{align}
    \log \left( \frac{f^{\rm KE}_a(E)}{1\pm f^{\rm KE}_a(E)} \right) = -\beta (E - \mu).
    \label{eq:log_psd_ke}
\end{align}
From this, we obtain that the phase space density of a particle equals
\begin{align}
    f^{\rm KE}_a(E) = \frac{1}{\exp(\beta (E-\mu) ) \mp 1},
    \label{eq:psd_ke}
\end{align}
where $\beta$ can be identified as the inverse temperature and $\mu$ is referred to as the chemical potential so that when self scattering processes are efficient, the particle species is in kinetic equilibrium and its phase space density has the form of a \acf{BE} or \acf{FD} distribution. During this derivation, we omitted the time dependence of the phase space density. When in kinetic equilibrium, this time dependence is captured by the time dependence of the chemical potential.

In many models of DM, the DM particles $\chi$ do not necessarily self-interact efficiently. However, in the early universe, kinetic equilibrium can be initiated by elastic scattering processes $\chi(E_1) a(E_2) \to \chi(E'_1) a(E'_2)$ with bath particles $a$ who are themselves in kinetic equilibrium and hence having a phase space distribution of the form depicted in Eq.~\eqref{eq:psd_ke}. Indeed, assuming the elastic scattering collision term has to be zero, the requirement of detailed balance dictates
\begin{align}
    \log \left( \frac{f_\chi^{eq}(E_1)}{1\pm f_\chi^{eq}(E_1)} \right) - \log \left( \frac{f_\chi^{eq}(E'_1)}{1\pm f_\chi^{eq}(E'_1)} \right) &= \log \left( \frac{f_a^{eq}(E'_2)}{1\pm f_a^{eq}(E'_1)} \right) - \log \left( \frac{f_a^{eq}(E_2)}{1\pm f_a^{eq}(E_2)} \right) \nonumber \\ 
    &= -\beta (E'_2-E_2) \nonumber \\
    &= -\beta (E_1-E'_1),
\end{align}
where, in the last term, we used conservation of energy. The logarithms on the left hand side must hence again be a linear combination of the energy as in Eq.~\ref{eq:log_psd_ke} so that the phase space distribution of DM equals a BE/FD distribution. 

DM models sometimes do not exhibit efficient elastic scattering processes, but only efficient inelastic processes $\chi_1(E_1) a(E_2) \to \chi_2(E'_1) a(E'_2)$ of the DM candidate $\chi_1$ with its partner $\chi_2$. The detailed balance requirement now dictates
\begin{align}
    \log \left( \frac{f_{\chi_1}^{eq}(E_1)}{1\pm f_{\chi_1}^{eq}(E_1)} \right) &= \log \left( \frac{f_{\chi_2}^{eq}(E'_1)}{1\pm f_{\chi_2}^{eq}(E'_1)} \right) + \log \left( \frac{f_a^{eq}(E'_2)}{1\pm f_a^{eq}(E'_2)} \right) - \log \left( \frac{f_a^{eq}(E_2)}{1\pm f_a^{eq}(E_2)} \right) \nonumber \\ 
    &= -\beta (E'_1-\mu_{\chi_2}+E'_2-E_2) \nonumber \\
    &= -\beta (E_1-\mu_{\chi_2}),
\end{align}
so that the DM particle $\chi_1$ again obtains a BE/FD distribution with chemical potential equal to the one of its partner $\chi_2$.

Finally, chemical equilibrium between two particle species, e.g. $\chi$ and $a$, is usual ensure by efficient annihilation processes $\chi(E_1) \Bar{\chi}(E'_1) \to a(E_2) \Bar{a}(E'_2)$. If all species involved are in kinetic equilibrium, their chemical potentials are related by the detailed balance relation,
\begin{align}
    \mu_\chi + \mu_{\Bar{\chi}} = \mu_a + \mu_{\Bar{a}}.
\end{align}

\chapter{Boltzmann equations beyond thermal equilibrium}
\label{app:BE_beyond_KE}
On general grounds, in an isotropic and homogeneous universe, the evolution of a particle species $i$ is described in terms of a distribution function $f_i(t,|\vec p_i|)$, expressed in terms of the physical time $t$ and of the norm of the physical 3-momentum $\vec p_i$, denoted as $|\vec p_i| \equiv p_i$.\footnote{In the text, $p_i$ usually refers to the norm of the 3-momentum, except when it appears in a 4-dimensional delta function which enforces both 3-momentum {\it and} energy conservation.} If a species $i$ interacts with other species $j$, the time evolution of species $i$ is described by the Boltzmann equation
\begin{equation}
  \frac{d f_i(t,p_i)}{dt}=\frac{1}{E_i}\, C[f_i(t,p_i),f_j(t,p_j)]\,,
  \label{eq:UnintBE0}
\end{equation}
where $C[f_i(t,p_i),f_j(t,p_j)]$ is the collision term involving all decay and scattering processes with the other species $j$, and $E_i=\sqrt{ p_i^2+m_i^2}$ is the energy of particles of species $i$ with mass $m_i$. In this appendix, we give more details about the collision terms for coscattering, $C_{\rm coscat}$, and inverse decays, $C_{\rm decay}$.

The total time derivative in Eq.~(\ref{eq:UnintBE0}) can be re-expressed in terms of partial derivatives with respect to time and momentum as
\begin{eqnarray}
    \frac{d f_i(t,p_i)}{dt}&=&\left( \partial_t - H p_i \partial_{p_i}
    \right) f_i(t,p_i) =\overline{H} \left(x\partial_x+ \frac{x}{ 3
      h_{\rm eff}}\frac{d h_{\rm eff}}{dx} \, q_i\,\partial_{q_i} \right) f_i(x,q_i),
  \label{eq:UnintBE1}
\end{eqnarray}
with
\begin{equation}
    \overline{H} = H\left(1-\frac{x}{3 h_{\rm eff}}\frac{dh_{\rm eff}}{dx}\right)^{-1}.
\label{eq:barH}
\end{equation}
In Eq.~(\ref{eq:UnintBE1}) we have used the rescaled time $x$ and momentum variable $q_i=p_i/T$.  In this appendix, we will make use of the variables $(t,p_i)$ or $(x,q_i)$ whenever convenient. Also, $H=H(x)$ is the Hubble expansion rate and $h_{\rm eff}= h_{\rm eff}(x)$ is the effective number of relativistic degrees of freedom contributing to the entropy density $s=h_{\rm eff}\, {2\pi^2}/45 \, T^3$.\footnote{The insertions of $d h_{\rm eff}/dx$ in Eqs.~(\ref{eq:UnintBE1}) and~(\ref{eq:barH}) are due to the usual choice of rescaled momentum $q= p/T$, instead of the time-independent comoving momentum $q=p a$, where $a$ is the scale factor. Another convenient choice of time-independent rescaled momentum would be $q=p/s^{1/3}$ see e.g.~\cite{Belanger:2020npe,Decant:2021mhj}.} When we consider dark matter production in a radiation dominated era, the Hubble rate reduces to
\begin{eqnarray}
 H(x)&=&\frac{m_1^2}{x^{2}M_0}  \quad {\rm with}\quad  M_0=M_P \sqrt{\frac{45}{4 \pi^3 g_{\rm eff}}}\,,
  \label{eq:H}
\end{eqnarray}
where $M_P$ is the Planck mass and $g_{\rm eff}=g_{\rm eff}(x)$ denotes the number of relativistic degrees of freedom  in the  thermal bath at time $x$ contributing to the radiation energy density $\rho_R= g_{\rm eff}\, \pi^2/30\,T^4$. For our numerical analysis, we use tables available in the public code {\tt micrOMEGAs}~\cite{Belanger:2018ccd} to evaluate $h_{\rm eff}(x)$ and  $g_{\rm eff}(x)$. When the number of relativistic degrees of freedom, $h_{\rm eff}$ and $g_{\rm eff}$, can be considered constant around freeze-out, the unintegrated Boltzmann equations of Eq.~(\ref{eq:UnintBE1}) simplify to
\begin{equation}
    x H \, \partial_x f_i(x,q_i)=\frac{1}{E_i} C[f_i(x,q_i),f_j(x,q_j)]\,.
      \label{eq:UnintBE3}
\end{equation}
To compute the dark matter freeze-out beyond kinetic and chemical equilibrium (as for instance in the CDFO or coscattering regime, see Sec.~\ref{sec:CDFO}, one has to use the latter evolution equation.

To see how we can solve Eq.~\eqref{eq:UnintBE3}, we will consider the same scenario as in Sec.~\ref{sec:CDFO}, i.e. we have two DS particles $\chi_1$ and $\chi_2$, where the former is the lightest and our DM candidate. While we cannot simply assume $\chi_1$ to be in kinetic equilibrium, $\chi_2$ will hold KE long after FO as it undergoes efficient elastic scattering. Therefore, we will always assume $f_2(x,q_2)=f^{(0)}_2(x,q_2) Y_2(x)/Y^{(0)}_2(x)$, with $Y_2$ the yield of $\chi_2$. Further, we apply two levels of simplification to Eq.~(\ref{eq:UnintBE3}). First, we neglect the time dependence in $h_{\rm eff}$ for the dark matter density evolution, setting $h_{\rm eff}=h_{\rm eff}(x_{\rm fo})$ when integrating over $f_1(x,q_1)$. This is a good approximation as long as FO does not happen during the QCD phase transition, as is the case for dark fermions in the MeV-GeV mass range or above 100~GeV. Second, we follow Refs.~\cite{DAgnolo:2017dbv, Garny:2017rxs} and only take into account the dominant interaction processes driving the dark matter distribution $f_1(x,q_1)$ towards kinetic equilibrium. For the CDFO/coscattering mechanism, this means that we include conversion processes but neglect co-annihilation.

With these simplifications, the unintegrated Boltzmann equation for $f_1(x,q_1)$ from Eq.~(\ref{eq:UnintBE3}) can be rewritten as
\begin{eqnarray}
  H x\partial_x f_1(x,q)&=& \tilde C_{1\to 2}(x,q) \left(f_1^{(0)}(x,q) \frac{Y_2(x)}{Y_2^{(0)}(x)}-f_1 (x,q)\right),
  \label{eq:UBcoscat}
\end{eqnarray}
where $q=q_1 = p_1/T$ to simplify the notation. The contributions to the collision operator $\tilde C_{1\to 2}(x,q)=\tilde C_{\rm coscat}(x,q)+\tilde C_{\rm 2-decay}(x,q)+\tilde C_{\rm 3-decay}(x,q) $ from coscattering, two-body and three-body decays are spelled out in Secs.~\ref{sec:coscattering},~\ref{sec:two-body-decay} and~\ref{sec:three-body-decay}. The above description assumes that $\chi_2$ and the light SM fermions involved in the conversion processes are in kinetic equilibrium.  We also neglect all spin statistics effect.

The differential equation in Eq.~(\ref{eq:UBcoscat}) can be solved iteratively with $Y_2(x)$ as an input. The latter is obtained from the integrated Boltzmann equation for $Y_2$ in Eq.~(\ref{eq:boltz_cdfo}), which in turn involves $Y_1(x)$, or equivalently the zeroth moment of $f_1(x,q)$ in $q$, obtained by integrating over Eq.~(\ref{eq:UBcoscat}). More details on the integration of Eq.~(\ref{eq:UBcoscat}) will be discussed in Sec.~\ref{sec:simpl-unint-boltzm}.

\section{Coscattering}
\label{sec:coscattering}
%
Here we provide details on the collision term due to coscattering. First we show that the collision term for coscatterings $\chi_1 f \rightarrow \chi_2 f'$ can indeed be written as in Eq.~\eqref{eq:UBcoscat}. Neglecting spin-statistic effects, the collision term takes the form
\begin{align}
  C_{\text{coscat}} = \frac{1}{2}\int d\Pi_f d\Pi_2 d\Pi_{f'} (2\pi)^4 \delta^{(4)}(p_1+p_f-p_2-p'_{f}) |{\cal M}_{\chi_1
f \rightarrow \chi_2 f'} |^2  \left( f_2 f_{f'} - f_1 f_{f} \right).
\end{align}
Assuming that the SM fermions $f$ and $f'$ are in chemical and kinetic equilibrium with the thermal bath and that $\chi_2$ is in kinetic equilibrium throughout the $\chi_1$ freeze-out, we have $f_{x}(t,p_{k})=f^{(0)}_{k} (t,p_{k})$ for $k=\{f,f'\}$ and $f_2(t,p_2) = f_2^{(0)}(t,p_2) \, Y_2(t)/Y_2^{(0)}(t)$, so that
\begin{equation}
    f_2 f_{f'} - f_1 f_{f} = \left( f_2^{(0)} \frac{Y_2}{Y_2^{(0)}} f_{f'}^{(0)} - f_1 f_{f}^{(0)} \right) = f_f^{(0)} \left( f_1^{(0)} \frac{Y_2}{Y_2^{(0)}}  - f_1 \right).\label{eq:distrib-rel1}
\end{equation}
In the second equation we have used the relation of detailed balance, $f_2^{(0)} f_{f'}^{(0)} = f_1^{(0)} f_f^{(0)}$. With  Eq.~(\ref{eq:distrib-rel1}), we can write the collision term as
\begin{equation}
\label{eq:CollTermCoscat}
    \frac{1}{E_1}\, C_{\text{coscat}} = \tilde C_{\text{coscat}} \left( f_1^{(0)} \frac{Y_2}{Y_2^{(0)}}  - f_1 \right),
\end{equation}
with the coscattering collision operator
\begin{eqnarray}
    \tilde C_{\text{coscat}} &=&\frac{g_f T}{ 16 \pi^2 p_1E_1}\int ds\, \sigma_{\text{coscat}}(s)\, (s-m_1^2) \left(e^{-E_{f}^-/T}- e^{-E_{f}^+/T}\right).
\label{eq:Ctildcoscatfin}
\end{eqnarray}
Here the cross section for coscattering process is defined as~\cite{DAgnolo:2018wcn,Garny:2018icg}
\begin{equation}
     \sigma_{\text{coscat}}(s) = \frac{1}{4 \hat{p}_1 \sqrt{s}} \int d\Pi_2 d\Pi_{f'} \ (2\pi)^4 \delta^4(p_1+p_f-p_2-p'_{f}) \ |{\cal M}_{\chi_1
f \rightarrow \chi_2 f'} |^2 \,,
\end{equation}
with the modulus of the dark matter 3-momentum in the center-of-mass frame, $\hat{p}_1$, the squared center-of-mass energy, $s$, and the energy variables
\begin{equation}
     E_f^\pm = \frac{s-m_1^2}{2m_1^2} (E_1 \pm p_1)\,.
\end{equation}

\section{Two-body decays}
\label{sec:two-body-decay}
A second conversion process occurring in Eq.~\eqref{eq:UBcoscat} is the decay process. While we will discuss the collision term for a three-body decay in Sec.~\ref{sec:three-body-decay}, the collision term for a two-body inverse decay $\chi_1 f \to \chi_2$ reads
\begin{eqnarray}
\label{eq:CollTermDecay2}
  C_{\text{2-decay}} = \frac{1}{2}\int\! d\Pi_2 d\Pi_f (2\pi)^4 \delta^{(4)}(p_2-p_1-p_f) |{\cal M}_{\chi_2 \to \chi_1 f}|^2  \left( f_2 - f_1 f_{f} \right).
\end{eqnarray}
Assuming again that the involved SM fermions $f$ are in chemical and kinetic equilibrium with the thermal bath and that $\chi_2$ is in kinetic equilibrium with the bath, the collision term reduces to
\begin{equation}
\label{eq:CollTermDecay}
    \frac{1}{ E_1}C_{\text{2-decay}} = \tilde C_{\text{2-decay}} \left( f_1^{(0)} \frac{Y_2}{Y_2^{(0)}}  - f_1 \right),
\end{equation}
with the collision operator for 2-body decays,
\begin{eqnarray}
    \tilde C_{\text{2-decay}}
    &=& \frac{1}{2 E_1}\int d\Pi_2 d\Pi_f (2\pi)^4 \delta^4(p_2-p_1-p_f) |{\cal M}_{\chi_2 \to \chi_1 f}|^2  f_{f}^{(0)}\,.
\end{eqnarray}
Due to the presence of the 4-dimensional Dirac delta function, we can easily integrate $C_{\text{2-decay}}$ over $d\Pi_2$ and obtain
\begin{eqnarray}
    \tilde C_{\text{2-decay}}
    &=& \frac{g_2}{2 E_1}\int d\Pi_f \frac{2\pi}{E_2} \delta(E_2-E_1-E_f) |{\cal M}_{\chi_2 \to \chi_1 f}|^2  f_{f}^{(0)}\,.
\end{eqnarray}
Applying the methodology from~\cite{DEramo:2020gpr} for 2-body decays and neglecting the the mass of $f$, the collision operator can be rewritten as
\begin{eqnarray}
     \tilde C_{\text{2-decay}} = \frac{g_f g_2}{32\pi E_1 p_1} \int_{E^-_f}^{E^+_f} dE_f |{\cal M}_{\chi_2 \to \chi_1 f}|^2 f_f(x,p_f)\, ,
\end{eqnarray}
with
\begin{eqnarray}
    E^\pm_f = \frac{m_2^2-m_1^2}{2m_1^2} (E_1 \pm p_1).
\end{eqnarray}
\section{Three-body decays}
\label{sec:three-body-decay}
%
The collision for an inverse decay process $ \chi_1 f f'\to \chi_2 $ reads
\begin{align}
  C_{\text{3-decay}} = \frac{1}{2}\int\! d\Pi_2 d\Pi_f d\Pi_{f'} &(2\pi)^4 \delta^{(4)}(p_2-p_1-p_f-p'_{f}) \nonumber \\ 
  & \times |{\cal M}_{\chi_2 \to \chi_1 f f'}|^2  \left( f_2 - f_1 f_{f} f_{f'} \right).
\end{align}
We can again write this in the from of Eq.~\eqref{eq:UBcoscat} if we assume that all involved SM fermions $f,f'$ are in chemical and kinetic equilibrium with the thermal bath and that we can a assume kinetic equilibrium for $\chi_2$,
\begin{equation}
\label{eq:CollTermDecay3}
    \frac{1}{ E_1}C_{\text{3-decay}} = \tilde C_{\text{3-decay}} \left( f_1^{(0)} \frac{Y_2}{Y_2^{(0)}}  - f_1 \right),
\end{equation}
with the collision operator for decays,
\begin{align}
    \tilde C_{\text{3-decay}}
    = \frac{g_{f'}}{2 E_1}\int d\Pi_2 d\Pi_f & \frac{2\pi}{2E'_{f}} \delta(E_2-E_1-E_f-E'_{f}) \nonumber \\ 
    &\times |{\cal M}_{\chi_2 \to \chi_1 f f'}|^2  f_{f}^{(0)} f_{f'}^{(0)}
    \label{eq:CtildeInterediate1}
\end{align}
after integrating over $d\Pi_{f'}$. Following the methodology depicted in Ref.~\cite{DEramo:2020gpr} for 3-body decays and assuming we can neglect the SM fermion masses, the collision operator takes the form,
\begin{align}\label{eq:Ctilddecfin}
    \tilde C_{\text{3-decay}} = \frac{g_f g_{f'} g_2}{256 \pi^3 p_1 E_1} & \int_0^{(\Delta m_1)^2}\!\!\!\! \frac{dm_{ff'}^2}{\sqrt{\lambda(m_{ff'},m_2,m_1)}} \nonumber \\
    & \times \int_{E_2^-}^{E_2^+} dE_2 \,f_{f}^{(0)} f_{f'}^{(0)}   \int_{(m_{1f}^2)^-}^{(m_{1f}^2)^+} dm_{1f}^2 |{\cal M}_{\chi_2 \to \chi_1 f f'}|^2,
\end{align}
where
\begin{eqnarray}
 \lambda(x,y,z) & = & (x^2-(y+z)^2)(x^2-(y-z)^2),\\\nonumber
    m_{1f}^2 &=& (p_1+p_f)^2,\\\nonumber
   m_{ff'}^2 &=& (p_2-p_1)^2.
\end{eqnarray}
The integration boundaries are
\begin{eqnarray}
    &E_2^\pm& = \sqrt{m_2^2+(p_2^\pm)^2},\quad {\rm with} \\\nonumber
    &p_2^\pm& = \frac{p_1(m_1^2+m_2^2-m_{ff'}^2)\pm \sqrt{(m_{ff'}^2+m_1^2)\lambda(m_{ff'},m_2,m_1)}}{2m_1^2},
\end{eqnarray}
and $(m_{1f}^2)^\pm = m_1^2 + m_{ff'} (\hat{E}_1\pm\hat{p}_1)$, where the hatted quantities are evaluated in the reference frame where $\Vec{p}_f = \Vec{p}_{f'}$.

\section{Simplified unintegrated Boltzmann equation}
\label{sec:simpl-unint-boltzm}
%
The collision terms for inverse two- and three-body decays and coscattering have the same form as the right-hand side of the simplified Boltzmann equation in Eq.~(\ref{eq:UBcoscat}), as can be seen in Eqs.~\eqref{eq:CollTermCoscat},~\eqref{eq:CollTermDecay2} and~\eqref{eq:CollTermDecay3}. We can now solve the Boltzmann equation for $\tilde C_{1\to2} = \tilde C_{\text{coscat}} + \tilde C_{\text{decay}}$ (where $\tilde C_{\text{decay}}$ can be the collision term for either 2-body decay, 3-body decay or the sum of both). We first rewrite it to simplify the notation,
\begin{eqnarray}
\label{eq:UnintBEsimp}
    \frac{\partial f_1(x,q)}{\partial x} + g(x,q)f_1(x,q) = g(x,q) h(x,q),
\end{eqnarray}
where
\begin{eqnarray}
    g(x,q) &=& \frac{\tilde C_{2\to1}(x,q)}{xH(x)},\\\nonumber
    h(x,q) &=& f_1^{(0)}(x,q)\frac{Y_2(x)}{Y_2^{(0)}(x)}.
\end{eqnarray}
Multiplying both sides of Eq.~(\ref{eq:UnintBEsimp}) by $u(x,q)=\exp \left[ \int dx \, g(x,q) \right]$ and using $\frac{\partial u(x,q)}{\partial x} = g(x,q)u(x,q)$, we can simplify this equation to
\begin{eqnarray}
    \frac{\partial (u(x,q)f_1(x,q))}{\partial x} = \frac{\partial u(x,q)}{\partial x} h(x,q).
\end{eqnarray}
Integrated by parts, we obtain
\begin{eqnarray}
    u(x,q)f_1(x,q)=u(x_0,q)f_1(x_0,q)+\left[h(x,q) u(x,q) \right]^x_{x_0}+\int_{x_0}^x dz \frac{\partial h(z,q)}{\partial z} u(z,q)\,,
\end{eqnarray}
where we set the initial time to $x_0=1$.

At early times, we can use the boundary conditions that $\chi_2$ is in chemical equilibrium, $Y_2(x_0)=Y_2^{(0)}(x_0)$, such that $h(x_0,q)=f^{(0)}_1(x_0,q)$. The dark matter phase-space distribution then reduces to
\begin{align}\label{eq:f1-simp}
    f_1(x,q) & = f_1^{(0)}(x,q)\frac{Y_2(x)}{Y_2^{(0)}(x)} \nonumber \\
    & \quad - \int_{x_0}^x dz \frac{\partial }{\partial z}\left( f_1^{(0)}(z,q)\frac{Y_2(z)}{Y_2^{(0)}(z)} \right) \exp\left(\int_z^x dy \frac{\tilde C_{2\to1}(y,q)}{yH(y)}\right).
\end{align}
In order to solve this equation, we have to specify the comoving number density of $\chi_2$, which remains in kinetic equilibrium throughout the whole evolution of $\chi_1$. Hence, we can use the integrated Boltzmann equation for $\chi_2$ from Eq.~(\ref{eq:boltz_cdfo}).

Solving both equations for $\chi_1$ and $\chi_2$ together is numerically difficult, so we choose to solve the two equations iteratively. As an initial seed, we can solve the system of integrated Boltzmann equations from Eq.~(\ref{eq:boltz_cdfo}) as if $\chi_1$ is in kinetic equilibrium. This gives us an initial value for $Y_2(x)$. We then solve the unintegrated Boltzmann Eq.~(\ref{eq:UnintBEsimp}) for $\chi_1$ and feed again the integrated Boltzmann equation for $\chi_2$ to obtain the next iteration for $Y_2(x)$. We stop this iteration once the difference in the relic dark matter abundance between the last two iterations is less than one percent.

\chapter{LHC searches for LLPs}
\label{app:recast}
We discuss the details of the recasted searches we used in Chapter \ref{chap:LHC_cons}.

\section{Heavy stable charged particle (HSCP) and R-hadrons (RH)}
\label{app:HSCP}

\paragraph{Description.}
When a charged mediator has a long enough lifetime to cross the detector completely ($c\tau_B > \mathcal{O}(10)$~m), it can leave a highly ionized track in the detector. This type of signature is usually referred to as Heavy Stable Charged Particle (HSCP). When the charged particle also carries QCD color, the long-lived particle is expected to hadronize, hence one refers to ``R-hadron'' (RH) searches. Searches of this type have been performed by both CMS~\cite{CMS:2016ybj} and ATLAS~\cite{ATLAS:2019gqq}. Both collaborations have publicly provided efficiency tables for the cases of slepton-like HSCP and gluino- or squark-like R-hadrons. The efficiency tables of the CMS search are gathered in the public code {\tt SModelS}~\cite{Kraml:2013mwa,Ambrogi:2018ujg}, which we use for recasting this search, see~\cite{Junius:2019dci} for details. In order to reinterpret the ATLAS search, a public code is provided in the ``LLP recasting repository''~\cite{LLPrepos}, a repository on GitHub holding various example codes to recast existing LLP searches. Notice that we use both codes to constrain R-hadrons originating from vectorlike quarks instead of squarks. In the latter case we assume that the differences in efficiency due to the difference in spin are minor.

As can be seen in Tab.~\ref{tab:searches}, one would expect the ATLAS search to be slightly more constraining since the employed luminosity is about three times higher. However, there is an important difference between ATLAS and CMS searches, which unfolds itself when the search strategy is applied to look for charged particles/R-hadrons with a mass of the order of 100-200 $\rm GeV$. The main difference between the two searches relies on how the \acfp{SR} are defined. Both make use of the reconstructed mass to define the \acp{SR}, however the CMS \acp{SR} are defined such that any heavy LLP/R-hadron will fit in at least one of the \acp{SR}, while the ATLAS search only probes heavy LLPs/R-hadrons with a reconstructed mass larger than a certain threshold.\footnote{This limit depends on the type of particle/R-hadron considered.} Hence, the CMS search is more sensitive in the lower mass region. Details on our recasting of these searches are provided in the following.

\paragraph{Recasting.}
Searches for highly ionized tracks originating from heavy stable charged particles (HSCP) and R-Hadrons (RH) have been both performed by CMS~\cite{CMS:2016ybj} and ATLAS~\cite{ATLAS:2019gqq}. As discussed above, the main differences between these two searches are the integrated luminosity and the sensitivity to the low mass range of the HSCPs/RHs.

In order to reinterpret the search performed by CMS, we employed a publicly available code named {\tt SModelS}~\cite{Kraml:2013mwa,Ambrogi:2018ujg}. This code makes use of the available efficiency tables in order to find the upper limit on the cross section in any given model. However, the efficiency tables included in {\tt SModelS} are built by assuming that the HSCPs/RHs are detector stable (i.e.~decay way outside the detector). For HSCPs/RHs with an intermediate lifetime, it can happen that only a fraction of the produced particles traverse the whole detector. {\tt SModelS} takes this effect into account by multiplying the efficiency of a stable HSCP/RH by the probability that the HSCP/RH with a specified lifetime traverses the detector completely. The latter are computed making use of the approximation:
\begin{equation}
  {\cal F}_{long}=\exp\left(-\frac{1}{c\tau} \left\langle\frac{l_{out}}{\gamma \beta}\right\rangle_{eff}\right)\,,
\label{eq:flong}
\end{equation}
where $\beta$ is the velocity of the LLP, $\gamma = (1 -\beta^2)^{-1/2}$, $l_{out}$ is the travel length through the CMS detector (ATLAS analysis are not yet included) and $c\tau$ is the LLP proper decay length. Here we use $\gamma \beta=2.0$ for our reinterpretation, which we obtained by making use of our own Madgraph simulations. The resulting $\gamma\beta$ distribution tends to values larger than 1.43. In particular for a mediator mass between 100 and 350~GeV, we obtain a mean $\gamma \beta$ varying between 34.9 and 2.17 for $\sqrt{s}= 13$~TeV. As a result, here we use the conservative value of $\gamma \beta=2.0$ for our analysis. In addition, we have to provide {\tt SModelS} with the production cross-sections of our mediator. The production cross section for our leptophilic model is equivalent to the one of a right handed slepton pair in a SUSY model.  We took the NLO+NLL cross sections tabulated by the ``LHC SUSY Cross Section Working Group" which have been derived using Resummino~\cite{Fuks:2013lya}. Notice that the latter just simply correspond to the LO cross-sections (that can be obtained with Madgraph) with a $K$-factor correction of roughly $1.3$. For the topphillic model, we simulated the NLO cross section using Madgraph, multiplying it with a flat K-factor of 1.6, corresponding to the K-factor of a stop pair produced from a proton-proton collision.

In order the reinterpret the ATLAS search, we again made use of a publicly available code in the ``LLP recasting repository'' on GitHub~\cite{LLPrepos}. This code does not rely on previously released efficiency tables, but rather makes use of {\tt Pythia 8}~\cite{Sjostrand:2014zea} to perform an event-by-event based analysis such that the dependence on the lifetime can be estimated with greater accuracy. The recasting strategy is based on the information published by the ATLAS collaboration, see~\cite{AtlasRHrecast}. For the leptophilic scenario described in Sec.~\ref{sec:leptophilic@lhc}, all necessary ingredients to extract the constraints are present in both codes. There is however one caveat for the reinterpretation of the RH searches for the topphilic scenario (Sec.~\ref{sec:VLtop}). Both of these codes are able to reinterpret the existing searches only for squark- or gluino-like RHs.\footnote{In {\tt SModelS}, only efficiency tables for squark- and   gluino-like RHs are present. Also {\tt Pythia 8} is only able to consider the formation RHs originating from squarks or gluinos.} Since we wanted to use these codes to constrain an RH originating from a vectorlike top quark, we made some small modification such that they would treat a vectorlike top RH as a stop-like RH. As the only difference between these two cases is the spin of the particle the RH is originating from, we do not expect that full implementation of a vectorlike top would have a large impact on our results.

\section{Disappearing tracks (DT) and kinked tracks (KT)}
\label{app:DT}
\paragraph{Description.}
\acfp{DT} can arise in models with a charged mediator that has a proper decay length $c\tau_B \sim \mathcal{O}(10)$~cm, i.e.~smaller than the size of the inner tracking system of ATLAS and CMS, and its decay products cannot be reconstructed, either because they are neutral or they carry low momentum. Searches for disappearing tracks have been performed at both ATLAS~\cite{ATLAS:2017oal} and CMS~\cite{CMS:2018rea,CMS:2020atg} with 13 TeV data. 

In order to recast DT searches, we make use of the publicly available efficiency table provided by the CMS and ATLAS collaborations, see~\cite{Junius:2019dci} for details. One important difference between the CMS and ATLAS search is the range of lifetimes they probe. The innermost tracking layers in the ATLAS detector have been upgraded and can focus more on very short tracks (tracklets). In practice~ATLAS can probe shorter lifetimes $\tau \sim \mathcal{O}$(1)~ns while CMS is more sensitive to larger lifetimes with $\tau \sim \mathcal{O}$(10)~ns. Below, we reproduce the existing limits on the simplified model with a wino LSP as presented by CMS~\cite{CMS:2020atg} and ATLAS~\cite{ATLAS:2017oal}, in order to validate our re-interpretations.

Note that these searches could also have some sensitivity for models exhibiting kinked track (KT) topologies. Indeed, let's assume that there is a displaced decay producing a stable SM charged particle plus DM. If the track of the charged decay product is not reconstructed, this signature would resemble a disappearing track. Some effort has been already done to verify the sensitivity of the disappearing track searches to a kinked track signature~\cite{Jung:2015boa,Evans:2016zau,Belyaev:2020wok,Belanger:2018sti}. We remind here that DT searches impose a lepton veto and thus they cannot be employed straightforwardly for DM models involving a direct DM coupling to leptons and leading to displaced hard leptons in the final states. We discuss how we handle this issue below to provide an estimate of the DT search reach on the KT signature which is present in that model.

\paragraph{Recasting.}
Here we reinterpret the disappearing track (DT) searches by ATLAS~\cite{ATLAS:2017oal} and CMS~\cite{CMS:2020atg}, performed for a supersymmetric model with long-lived winos. We first focus on the ATLAS search, where the efficiency maps are provided on the HEPData page~\cite{hepdata.78375}. In order to use these efficiency maps to asses the reach of disappearing track searches on our DM model, we need to know how to interpret them.

Besides the event acceptance $E_A$ and efficiency $E_E$, the tracklet\footnote{A tracklet is a track in the detector between 12 and 30 cm.} too has to pass the reconstruction selection requirements. The probability to pass the generator-level requirements is the tracklet acceptance $T_A$. The probability to pass the full pixel tracklet selection at reconstruction level is the tracklet efficiency $T_E$. Both should be applied for every tracklet. Finally, the probability for a tracklet to have a $p_T > 100~$GeV is denoted independently by $P$ and is taken to have a constant value of 0.57 for the charginos.

There are three different processes that can leave a disappearing track in the detector for the wino-like chargino/neutralino model,
\begin{align}
 & p p \rightarrow \chi^+_1 \chi^-_1,  \label{poc:2tracks}\\
 & p p \rightarrow \chi^+_1 \chi^0_1, \label{poc:1tracks1}\\
 & p p \rightarrow \chi^-_1 \chi^0_1, \label{poc:1tracks2}
\end{align}
with each of the processes having approximately the same production cross section, i.e. one third of the total cross section. The efficiency map provided in the HEPData\footnote{We thank the ATLAS exotics conveners for information about the efficiency maps in the HEPData.} denotes the total model dependent efficiency (i.e. taking into account the fact that some of the above processes can leave two tracks in the detector), without taking into account the probability $P$ (it will be reintroduced later). Since in our model, we have only processes that can leave two tracks, we need to obtain the efficiency for two tracklet processes.

In general, the probability $\mathcal{E}_N$ of reconstructing at least one tracklet coming from a process leaving $N$ tracks in the detector has an efficiency of,
\begin{align}
   \mathcal{E}_N &= E_A \times E_E \times (1-(1-T_A \times T_E )^N) \nonumber \\
   &\approx E_A \times E_E \times (1-(1-N \, T_A \times T_E)) \nonumber \\
  &= N E_A \times E_E \times T_A \times T_E \nonumber \\
  &= N \mathcal{E}_1.
  \label{eq:effNtracks}
\end{align}
For the wino-like chargino/neutralino analysis, only process (\ref{poc:2tracks}) can leave two tracks in the detector while the other two can only leave one. Therefore, the model dependent efficiency for the pure wino chargino/neutralino analysis is
\begin{align}
  \mathcal{E}_{full} = \frac{2}{3} \mathcal{E}_1 + \frac{1}{3} \mathcal{E}_2 \approx \frac{4}{3} \mathcal{E}_1 \approx  \frac{2}{3} \mathcal{E}_2,
  \label{eq:FullEff}
\end{align}
where the $2/3$ represents the contribution from the processes with one tracklet (i.e. \eqref{poc:1tracks1} and \eqref{poc:1tracks2}) and the $1/3$ represents the contribution from the processes with two tracklets \eqref{poc:2tracks}, all assumed to have the same production cross section. As mentioned, the full efficiency $ \mathcal{E}_{full}$ is the one reported in the HEPData and we use Eqs.~(\ref{eq:effNtracks}) and (\ref{eq:FullEff}) to derive the efficiency for a two tracklet process, that we use in the analysis of our leptophilic DM model.

In the HEPData page of the CMS search~\cite{hepdata.84707}, a 95\% CL upper limit $\sigma_{\rm Wino}^{\rm UL} $ on the cross section for direct production of charginos as a function of chargino mass and lifetime is provided. In order to accommodate this limit in our analysis, we reinterpret this into a limit for a model always producing two charged tracks by making use of the same approximations as for the ATLAS search, namely that the efficiency used for the search can be approximated by Eq.~(\ref{eq:FullEff}). Together with this approximation, we can obtain a 95\% CL exclusion limit on the charged scalar $\phi$ in the leptophilic model studied in Sec.~\ref{sec:leptophilic@lhc} from the data given on the HEPData page of the experiment as follows
\begin{equation}
	\sigma_{\phi}^{\rm UL} = \frac{\sigma_{\rm vis}^{\rm UL}}{\mathcal{E}_2} \approx \frac{2}{3} \frac{\sigma_{\rm vis}^{\rm UL}}{\mathcal{E}_{\rm full}} = \frac{2}{3} \sigma_{\rm Wino}^{\rm UL}.
\end{equation}

\begin{figure}
	\centering
	\subfloat{\includegraphics[width=0.45\textwidth]{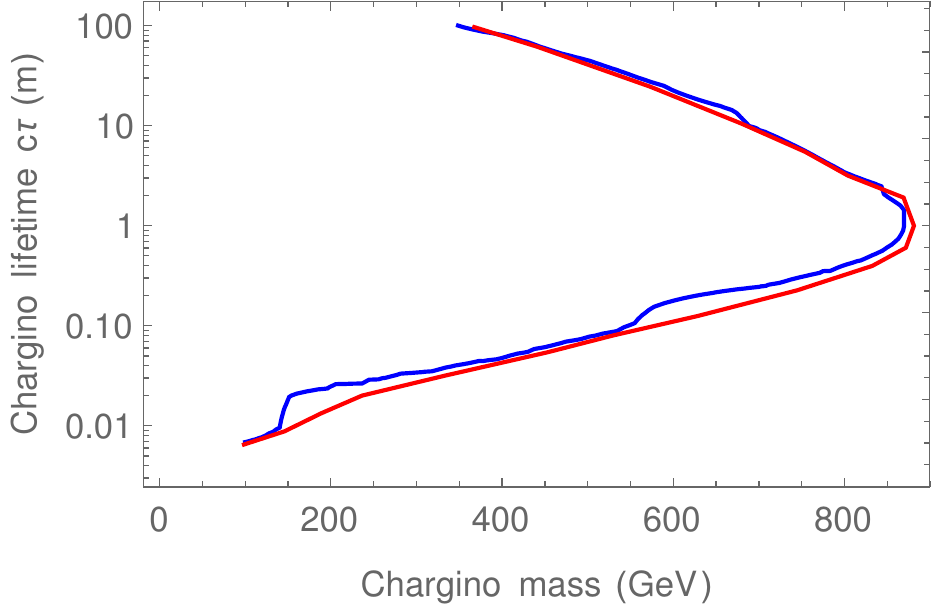}}
	\hspace{0.05\textwidth}
	\subfloat{\includegraphics[width=0.46\textwidth]{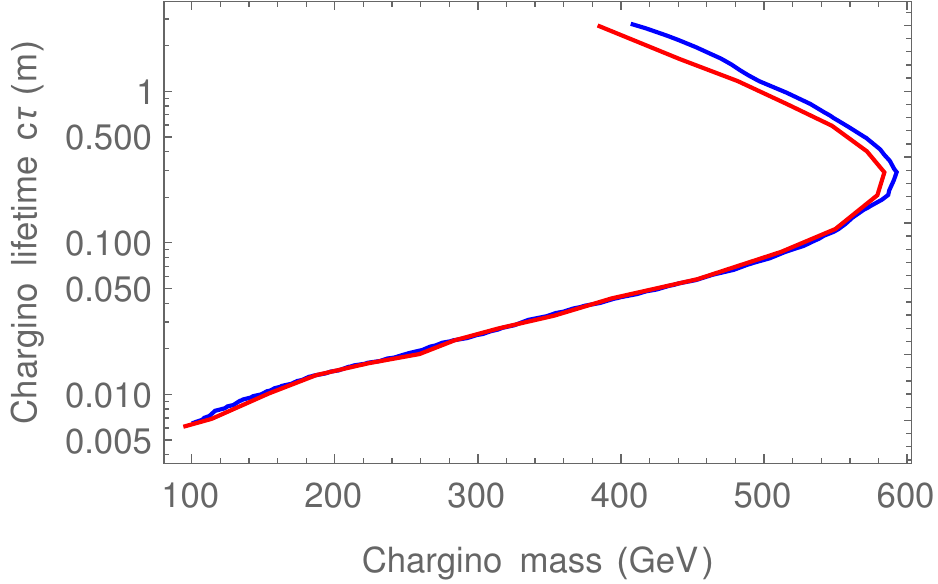}}
	\caption{Comparison between the experimental result (red) and our reinterpreted result (blue) for the latest DT searches performed by CMS~\cite{CMS:2020atg} (left) and ATLAS~\cite{ATLAS:2017oal} (right).}
	\label{fig:ValidationDT}
\end{figure}

We validated our technique for the DT searches used here~\cite{ATLAS:2017oal,CMS:2020atg} by reanalyzing the case of the wino within our framework and comparing our results to the ones presented in the CMS and ATLAS papers. Fig.~\ref{fig:ValidationDT} illustrates that we find a very good agreement.

The DT signature is a smoking-gun signature for the singlet-triplet model studied in Sec.~\ref{sec:SingletTriplet}, which is closely related to the supersymmetric wino model, for which all DT searches are originally performed. The DT originates from the charged component of the triplet decaying into the neutral one in association with a soft pion that evades detection: $\Psi_B^\pm\to \Psi_B^0 +\pi^\pm$. For the charged track to fulfill the disappearance requirement stated in the CMS and ATLAS papers, no hits in the outer layers of the tracker have to be associated to the track. If the lifetime of the neutral component $\Psi_B^0$ is short enough for it to decay inside the detector, hits in the outer layers of the tracker might occur. Therefore, we conservatively require in our analysis $\Psi_B^0$ to decay outside the tracker. In order to take this into account, we multiply the efficiency by the probability for $\Psi_B^0$ to decay inside the tracker by taking $\langle l_\text{outer}/\gamma\beta \rangle_\text{eff}$=0.36 in Eq.~\eqref{eq:flong}, where $l_{outer}$ is now the outer radius of the inner tracker. Notice that the CMS search~\cite{CMS:2020atg} considered here has been performed on a larger data set than the ATLAS search ($\mathcal{L}=140~\text{fb}^{-1}$ compared to $36.1~\text{fb}^{-1}$). Naively, one would then expect CMS to set the strongest constraints on our models. As mentioned above though, the ATLAS detector has been optimized to study shorter disappearing tracks than the those studied by CMS, hence the CMS and ATLAS searches are complementary as one can see from our reinterpretation of Fig.~\ref{fig:ValidationDT}.

We have also estimated the efficiency of the CMS disappearing track search in constraining a kinked-track signature for our leptophilic model (see Sec.~\ref{sec:lepto@lhc_emd}). For the DT search to be sensitive to a kinked track, we require that the angle between the mother and daughter particle tracks be large enough so as to avoid them to be identified as colinear (and hence being disentangled from one single track) and so that the muon is not reconstructed as it does not point back to the collision point. We took the conservative limit on the angle $\Delta R \equiv \sqrt{(\Delta \phi)^2+ (\Delta \eta )^2} > 0.1$.\footnote{We thank Steven Lowette for discussions on this point.} Additionally, the CMS search imposes a cut on the amount of energy deposited in the calorimeters within a cone of $\Delta R < 0.5$ around the disappearing track, which has to be less than 10~GeV. When the charged daughter particle is a muon, it will not leave a large energy deposit in the calorimeter, and hence, we do not need any extra requirements. In contrast, when the charged daughter particle is an electron or some hadronic final state, such an extra cut on the energy deposit in the calorimeter has to be imposed in order to estimate the sensitivity to the kinked track signature.

\section{Displaced leptons (DL)}
\label{app:DL}
\paragraph{Description.}
In cases where the long-lived mediator decays to a lepton within the tracker, one would get displaced lepton signatures. Searches have been performed at $\sqrt{s}=13$~TeV for events containing leptons with large impact parameters, i.e.~displaced leptons, by CMS~\cite{CMS:2021kdm} and ATLAS~\cite{ATLAS:2020wjh}. The CMS search is maximally sensitive to particles with $c\tau_B \sim \mathcal{O}$(few)~cm while the ATLAS search is sensitive to slightly larger values of $c\tau_B \sim \mathcal{O}(10)$~cm.

Three signal regions are defined in both analyses, depending on the lepton flavours, one where a displaced $e^\pm \mu^\mp$ pair is observed and two others where same flavor leptons are observed ($e^\pm e^\mp$ or $\mu^\pm \mu^\mp$). Hence, it is straightforwardly applicable also to cases where the DM couples to one lepton flavor only. More details on the extracted sensitivities are provided below.

\paragraph{Recasting.}

As mentioned above, the DL search performed by ATLAS~\cite{ATLAS:2020wjh} defines three different signal regions (SR-$ee$, SR-$e\mu$ and SR-$\mu\mu$) and hence uses three different trigger strategies, namely single-photon, diphoton, and muon trigger. The photon triggers select events with an energy deposit in the \ac{ECAL} greater than 140 GeV (single-photon) and 50 GeV (diphoton) while the muon trigger select events with a signature in the muon spectrometer with $p_T>$60 GeV and $|\eta|<$1.05. For each event, two signal leptons are defined as the leptons with the highest transverse momentum. These leptons must further have an impact parameter $|d_0|$ between 3 and 300 mm, a transverse momenta $p_T > 65$~GeV and $|\eta|<2.5$. Apart from these cuts, there are two extra event requirements. First, there must be a clear separation between both leptons, $\Delta R>0.2$, and none of the muons can be cosmic tagged. The latter consists in requiring that the two signal muons cannot be produced back to back, a case in which they are considered of cosmic origin and are hence removed. For events passing these selection cuts, the ATLAS collaboration also provides the model-independent reconstruction efficiencies for displaced electrons and muons~\cite{1831504} to be further applied. However, this information is only given for a benchmark with LLP mass of $400$~GeV and proper lifetime of $1$ ns. Nevertheless, we use for concreteness the same efficiency maps also for benchmarks with different LLP mass and lifetime, so we expect some discrepancy with the experimental results in certain regions of the parameter space.

\begin{figure}[t!]
	\centering
	\subfloat{\includegraphics[width=0.45\textwidth]{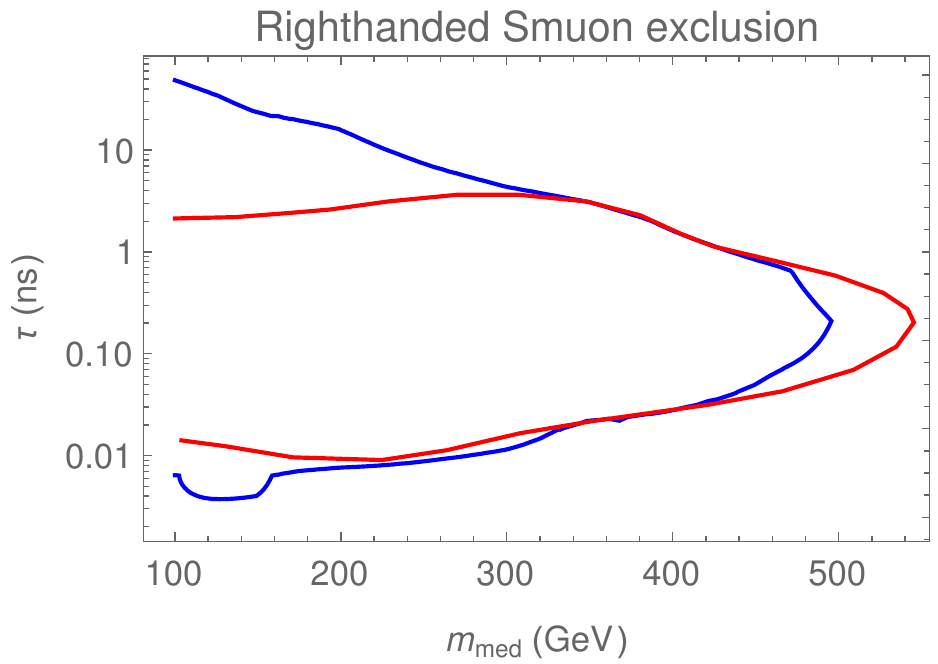}}
	\hspace{0.05\textwidth}
	\subfloat{\includegraphics[width=0.45\textwidth]{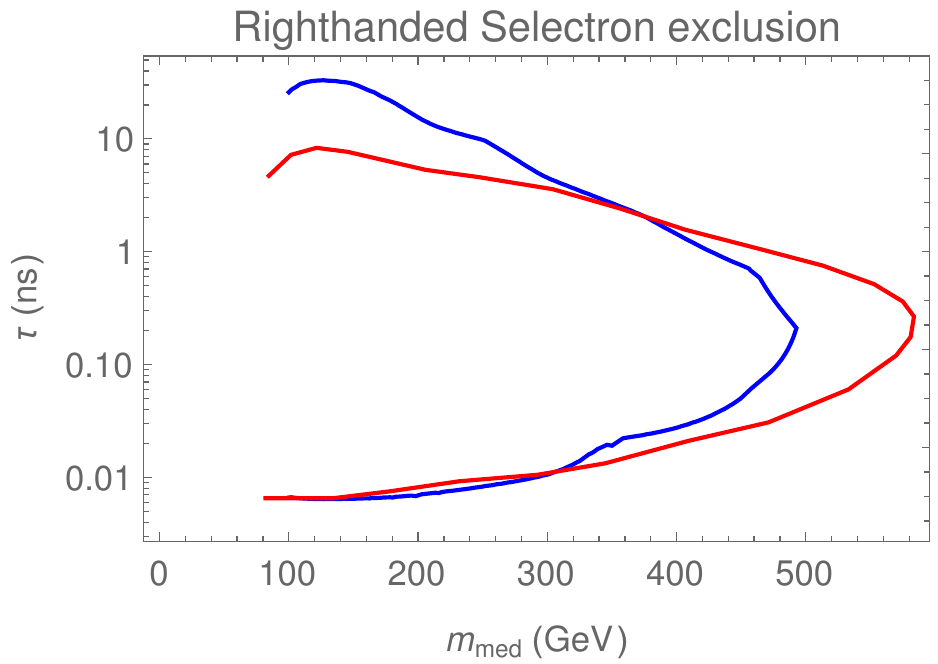}}
	\caption{Validation of the DL performed by ATLAS~\cite{ATLAS:2020wjh}. Left: right-handed smuon production. Right: right-handed selectron production.}
	\label{fig:DL_validation}
\end{figure}

In Fig.~\ref{fig:DL_validation} we show the validation of our recasting procedure for two representative models for which the full exclusion curves are reported in the auxiliary material of the ATLAS paper, that is, production of right-handed smuon and right-handed selectron.
We see that our recasting procedure reproduces very well the experimental exclusion curves around the benchmark point for which detailed information on the electron and muon efficiency are provided by the ATLAS collaboration, as explained above. For masses significantly smaller or larger than $400$~GeV the recasting is instead less precise. We nevertheless observe that, even if not quantitatively precise, our validation shows a qualitative agreement with the experimental results and hence we employ it in the main text to estimate the expected sensitivity for the topphilic and the singlet-triplet model. Since the ATLAS collaboration provides exclusion limits for a right-handed smuon in the lifetime vs. mass plane, we can immediately apply these limits to our leptophilic scenario. The CMS search however only provides limits for a degenerate left-/right-handed smuon model. We rescale these limits with the corresponding cross section to apply them to our leptophilic model. We do not further apply the CMS search to the topphilic and singlet-triplet model as it has a smaller dataset compared to the ATLAS search and there are other searches more constraining in the parameter range were this search would be effective, see Secs.~\ref{sec:VLtop} and~\ref{sec:SingletTriplet}.

\section{Displaced vertices + MET (DV+MET)}
\label{app:DV}
\paragraph{Description.}
We call displaced vertices + MET the case where the long-lived mediator decays to DM and a color charged object before reaching the end of the tracker. A jet will be produced but is not reconstructed using standard jet-clustering algorithms. The events will be rather analyzed by looking at the individual tracks of the jet originating from a displaced vertex, see~\cite{ATLAS:2017tny} for details.  In order to recast this search, we made use of the information provided by the ATLAS collaboration on the HEPdata page of the search~\cite{1630632}. We validated our recasting techniques by applying them on the model studied by the ATLAS collaboration and obtained very similar results, as discussed below.

\begin{figure}[t!]
	\centering
	\subfloat{\includegraphics[width=0.45\textwidth]{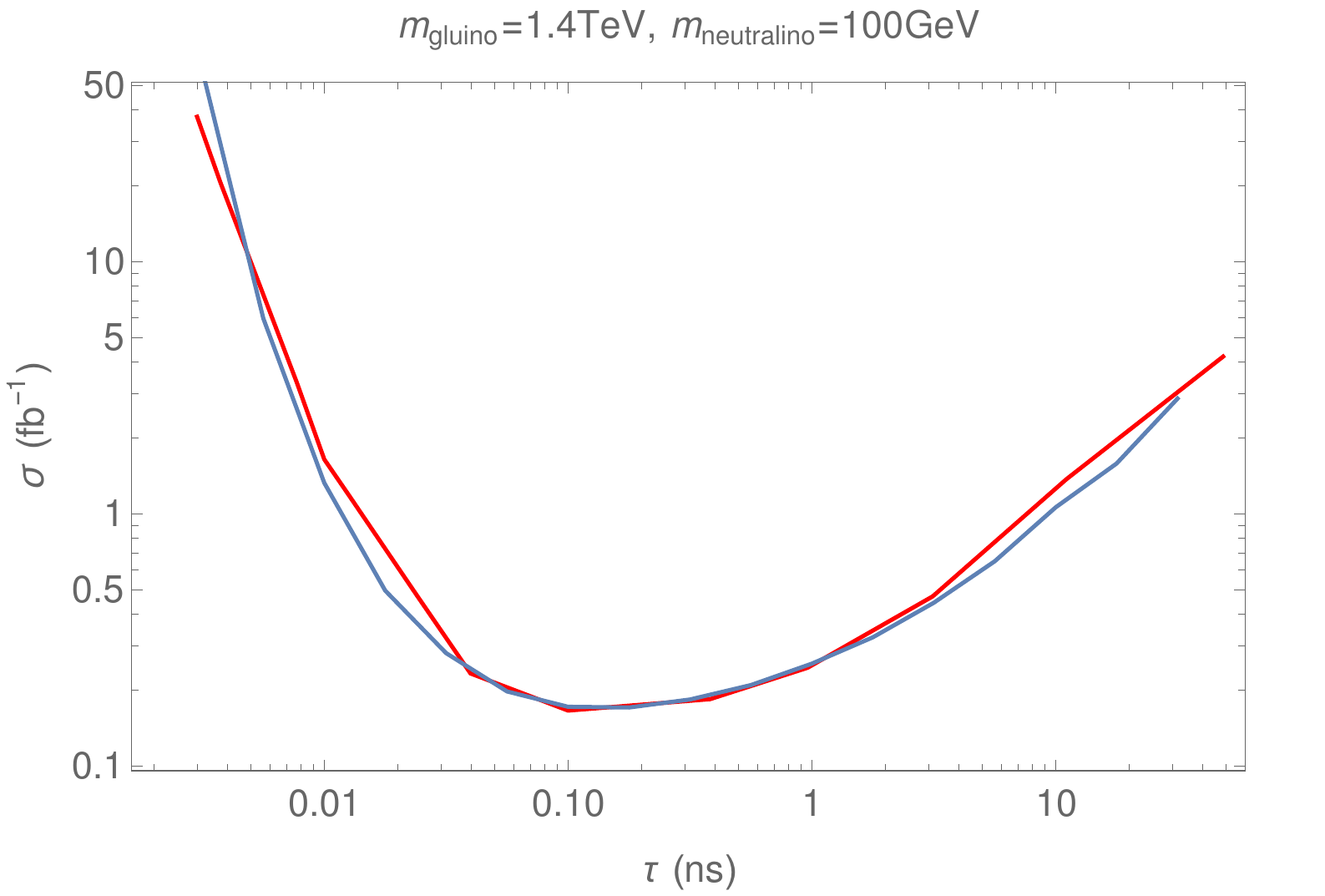}}
	\hspace{0.05\textwidth}
	\subfloat{\includegraphics[width=0.45\textwidth]{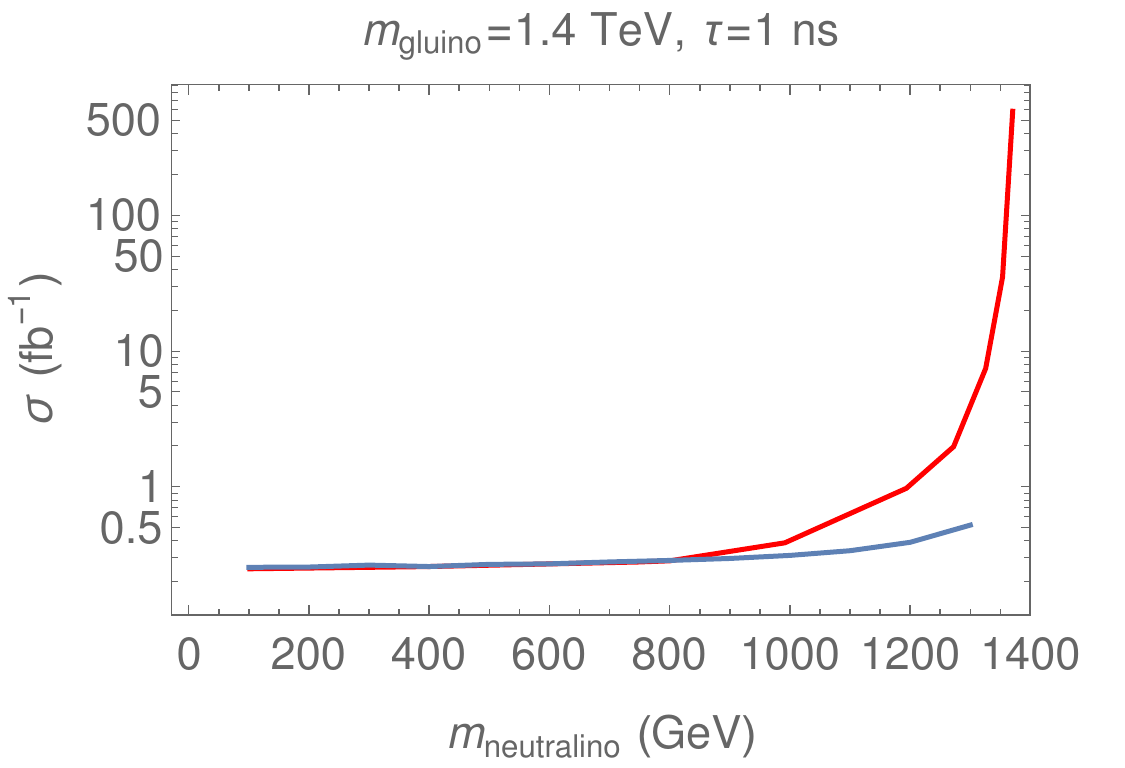}}
	\caption{Validation of the DV + MET search performed by ATLAS~\cite{ATLAS:2017tny} by comparing the ATLAS results (red) with the results from our recasting (blue). Left: Upper limit on the production cross Sec. vs.~the gluino lifetime for a fixed gluino mass (1.4 TeV) and neutralino mass (100 GeV). Right: Upper limit on the production cross Sec. vs~ the mass of the neutralino for a fixed gluino mass (1.4 TeV) and gluino lifetime (1 ns). Notice the discrepancy between the ATLAS limit and our result for a small gluino-neutralino mass splitting ($\Delta m/m \lesssim 0.4$) which is due to the omission of one of the selection criteria (see the text for details).}
	\label{fig:DV_validation}
\end{figure}

\paragraph{Recasting.}
The ATLAS collaboration has released a search for events with displaced vertices in combination with missing transverse momentum in~\cite{ATLAS:2017tny}. They look for events with at least one displaced vertex containing 5 or more tracks and missing transverse energy (MET) larger than or equal to 200 GeV. The efficiency tables are publicly available on HEPData together with a procedure for recasting the search to other models~\cite{1630632}. Following this procedure, we reproduced the results of the ATLAS collaboration for long-lived gluinos decaying into quarks and a neutralino by doing an event-by-event analysis. Our results can be seen in Fig.~\ref{fig:DV_validation}. A deviation between the ATLAS limit and our recasting appears for small values of the mass splitting between the gluino and the neutralino, $\Delta m/m \lesssim 0.4$. This is because we have omitted a cut, placed on the jets transverse momentum for 75\% of the events, that only has a significant impact on the selection efficiency if the mass splitting is small. In our study, we do not explore compressed spectra and hence we can neglect this condition safely. We applied this recasting strategy to two of our models, the  topphilic and  singlet-triplet  scenarios of Secs.~\ref{sec:VLtop} and~\ref{sec:SingletTriplet}. Some special treatments were required for both of these models, as discussed in the following.

In the topphilic scenario, a displaced jet can arise from a long-lived R-hadron decaying into a top quark and the DM scalar. The top quark will predominantly decay to a bottom together with two other quarks (through an intermediate $W$ boson). The latter will promptly form a jet. The bottom quark on the other hand eventually decays to a slightly-displaced jet, with a displacement $\sim\mathcal{O}$(mm) due to the non-negligible $b$ lifetime. For the DV+MET search, reconstructed vertices that are within a distance of 1 mm of each other are seen as one single vertex. Hence the tracks originating from the bottom and from the two other quarks will not necessarily be re-combined into one single vertex. To be conservative, we discard any track originating from decay of the bottom quark in the recasting of the DV+ MET search (we also applied this conservative approach in the DV+$\mu$ search discussed in App.~\ref{app:DVmu}).

In the singlet-triplet model, a displaced jet can originate from the decay of the charged and the neutral component of the triplet. The neutral components can be produced in two ways, through its gauge interactions or from the decay of the charged component. The decay of the charged component to the neutral one (together with soft pions) is slightly displaced, $\mathcal{O}$(cm), and this displacement has to be taken into account in order to correctly interpret the efficiency tables for the singlet-triplet model. Due to the small mass splitting between the charged and the neutral component of the triplet, they will be more or less colinear. Hence, to estimate the total displacement of the displaced jet, we simply add the displacements of the charged and neutral components of the triplet.

\section{Delayed jets + MET (DJ+MET)}
\label{app:DJ}
\paragraph{Description.}
A delayed jet has been defined as a jet that is observed in the calorimeter of the detector at a later time than one would expect from a jet that is produced at the primary vertex. Such a time delay can be due to a heavy, long-lived mediator that slowly crosses the detector before decaying into a jet. This search relies on the timing capabilities of the calorimeter and, as a result, there is no need to reconstruct the displaced vertex. Here we use the search for delayed jets + MET that has been performed by CMS in~\cite{CMS:2019qjk}.

Notice that the delayed jet + MET search is able to probe longer lifetimes ($c\tau_B \sim 1-10$~m) than the displaced vertices + MET search discussed above as the calorimeter lies further from the center than the tracker. The DV+MET search instead is more sensitive to smaller lifetimes ($c\tau_B \sim {\cal O} (1-10)$~cm), and actually, the time delay will not be enough to distinguish between a displaced or a prompt jet. Hence, it is useful to consider both DV+MET and DJ+MET searches together given their complementarity.

\paragraph{Recasting.}
In order to probe larger values of the lifetime of long-lived particles decaying into jets, the CMS collaboration made use of the timing capabilities of the calorimeter to look for delayed jets~\cite{CMS:2019qjk}. They studied explicitly the case of long-lived gluinos decaying into a gluon and a gravitino. The efficiency tables for such a scenario are publicly available, starting from gluino masses of 1 TeV. Since these efficiency maps are blind to any jet activity arising from the primary vertex, we can use them directly  for the scenarios studied here.

The gluino model studied in~\cite{CMS:2019qjk} always gives rise to two delayed jets while in the models listed in Tab.~\ref{tab:classification}, it can happen that only one delayed jet arises.\footnote{Often, this is due to the fact that the jets originate from the hadronic decay of $W$ or $Z$ bosons, and these can also decay leptonically.} In order to address this difference, we assume that the efficiency for an event with one delayed jet is small enough ($\epsilon_{1jet}\ll1$) so as to approximate the efficiency for an event with $N$ delayed jets with $\epsilon_{Njets}=1-(1-\epsilon_{1jet})^N \approx N \epsilon_{1jet}$. With this approximation we can derive the efficiency for an event with one delayed jet, making use of the publicly available efficiency tables for long-lived gluinos (involving events with two delayed jets), and obtain the efficiencies for events with an arbitrary number of delayed jets. By weighting the derived efficiencies with the correct combination of branching ratios, we can obtain model-dependent efficiency tables for the models studied in this paper.

In the singlet-triplet model of Sec.~\ref{sec:SingletTriplet}, we have made extra assumptions in order to extract the corresponding efficiencies. As already mention above, e.g.~in the discussion of DV+MET searches, the neutral component of the triplet can get an extra displacement from the decay of the charged component with $c\tau_C\sim \cal O$(cm). Since the CMS delayed jet search probes relatively long lifetimes of the neutral component ($c\tau_0\sim\mathcal{O}$(m) compared to $c\tau_C\sim \cal O$(cm) ), we assume that we can ignore the extra displacement arising from the decay of the charged triplet component. We have also assumed the charged component decays into the neutral one and soft pions with a 100\% branching ratio, i.e.~BR$(\Psi_B^{\pm}\to \Psi_B^0\pi^\pm)=1$. Indeed, even though $\Psi_B^{\pm}$ can directly decay to DM, this decay width depends on the new physics scale $\Lambda$, while the decay to $\Psi_B^{0}$ is driven by gauge interactions, i.e.~independent of $\Lambda$. Since we probe large lifetimes with the DJ + MET search, we are always in a regime where $\Lambda$ is large such that the direct decay of the charged component to DM is suppressed compared to the decay to the heavy neutral triplet component.

\section{Displaced vertices + muon}
\label{app:DVmu}
\paragraph{Description.}
Another type of displaced vertex search is performed by ATLAS looking for events containing  displaced vertices together with a displaced muon track~\cite{ATLAS:2020xyo,1788448}. The search defines two orthogonal signal regions, one containing events triggered by missing energy, the other one containing events triggered by a high $p_T$ muon. In the context of this thesis, where we focus on simplified DM models, the MET signal region is expected to be the most efficient. The displaced vertex is reconstructed in a very similar way as for the DV+MET search discussed in App.~\ref{app:DV}, so we expect these two searches to constrain similar ranges of lifetimes. The main differences between the two searches are the requirement of a muon and the higher luminosity used in the DV+$\mu$ search. 

\paragraph{Recasting.}
We use the search for displaced vertices and a displaced muon performed by ATLAS in~\cite{ATLAS:2020xyo} (with $\sqrt{s} = 13$ TeV and $\mathcal{L}=132~\text{fb}^{-1}$). It is particularly relevant for our simplified models as the long-lived particles are always produced in pairs and $Z$ and $W$ bosons potentially arising in their decays can partially decay to leptons.

\begin{figure}[t!]
	\centering
	\includegraphics[width=0.55\textwidth]{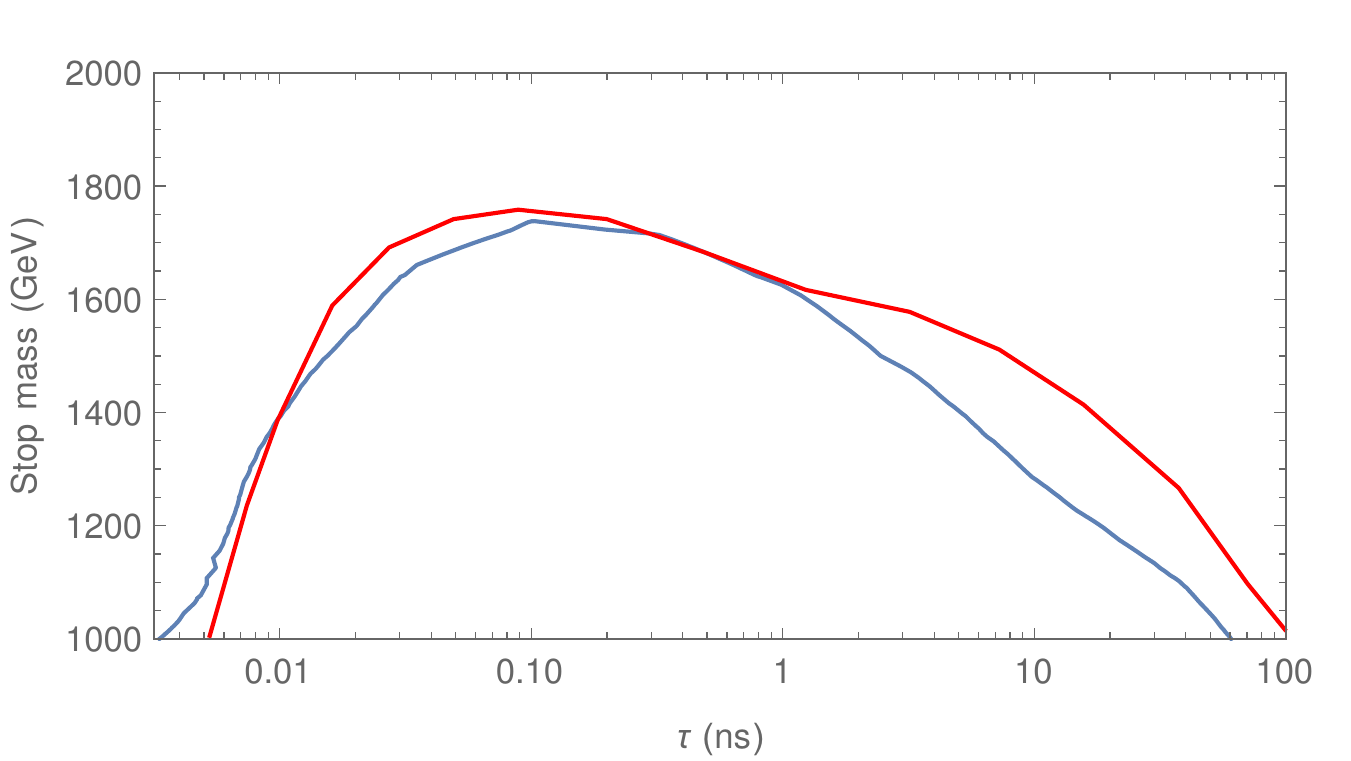}
	\caption{Validation of the DV + $\mu$ search by ATLAS. The red curve is the ATLAS result, the blue curve is our reinterpretation using the available efficiency tables.}
	\label{fig:DVmuon_validation}
\end{figure}

The ATLAS search defines two \acp{SR}: in the first \ac{SR} (SR1) the event is triggered by a large amount of missing energy ($E_T^{miss}>180$~GeV) while in SR2 the event is triggered by a track in the muon spectrometer (with $p_T>62$~GeV, $|\eta|<1.05$). In addition, in SR2, a cut of $E_T^{miss}<180$~GeV is imposed so as to define two orthogonal SRs. The experimental collaboration has provided trigger efficiencies, as well as muon and displaced vertex reconstruction efficiencies. We also make use of the cuts listed in the Tables~1 and 2 of~\cite{ATLAS:2020xyo} to reinterpret this search by doing an event-by-event analysis. We validated our approach by applying our recasting to the model studied in the ATLAS paper. The
results of this validation can be found in Fig.~\ref{fig:DVmuon_validation}.

\section{Displaced dilepton vertex (DLV)} 
\label{app:DLV}
\paragraph{Description.}
This kind of searches targets long-lived neutral particles decaying into a lepton pair ($\mu^+\mu^-$, $e^+e^-$, or $\mu^\pm e^\mp$). In this work, we focus on simplified DM models where the connections between dark and visible sector is governed by a three-particle interaction (see Tab.~\ref{tab:classification}). This interaction also governs the two body decay of the LLP in collider experiments. Since one of the two daughter particles of the LLP is always the DM, there will be only one SM particle involved in this decay. Therefore, a displaced lepton vertex will occur in the considered models only when the SM particle is a $Z$ boson which decays to two leptons. The leptonic branching ratio of the $Z$ boson is only about 10\%, while the hadronic branching ratio is about 70\%. Hence, we expect that the displaced lepton vertex search performed by ATLAS~\cite{ATLAS:2019fwx} will be generically less constraining than the DV+MET or DV+$\mu$ searches discussed in App.~\ref{app:DV} and~\ref{app:DVmu} above. Nevertheless, since the ATLAS collaboration provides useful information to reinterpret this search in a model independent manner~\cite{1745920}, we recast this analysis as discussed in the following and we apply it when relevant for the considered models.

\paragraph{Recasting.}
Here we address the ATLAS search for a pair of oppositely charged leptons originating from the same displaced vertex~\cite{ATLAS:2019fwx}. The ATLAS collaboration has provided a document detailing how to reinterpret this search for any model containing long-lived particles decaying to oppositely charged leptons~\cite{1745920}. In order to calculate the event acceptance, one has to apply some simple kinematic cuts, among which the most relevant ones are that the invariant mass of the lepton pair has to be above 12 GeV and their displacement must be larger than 2 mm. The overall acceptance can be obtained by doing an event-by-event analysis.

\begin{figure}[t!]
	\centering
	\subfloat{\includegraphics[width=0.45\textwidth]{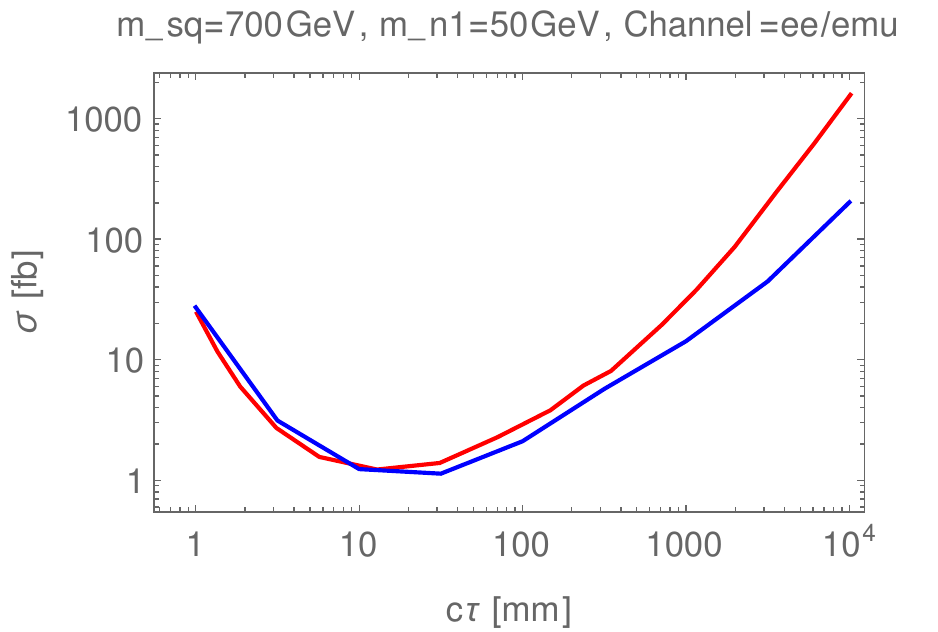}}
	\hspace{0.05\textwidth}
	\subfloat{\includegraphics[width=0.45\textwidth]{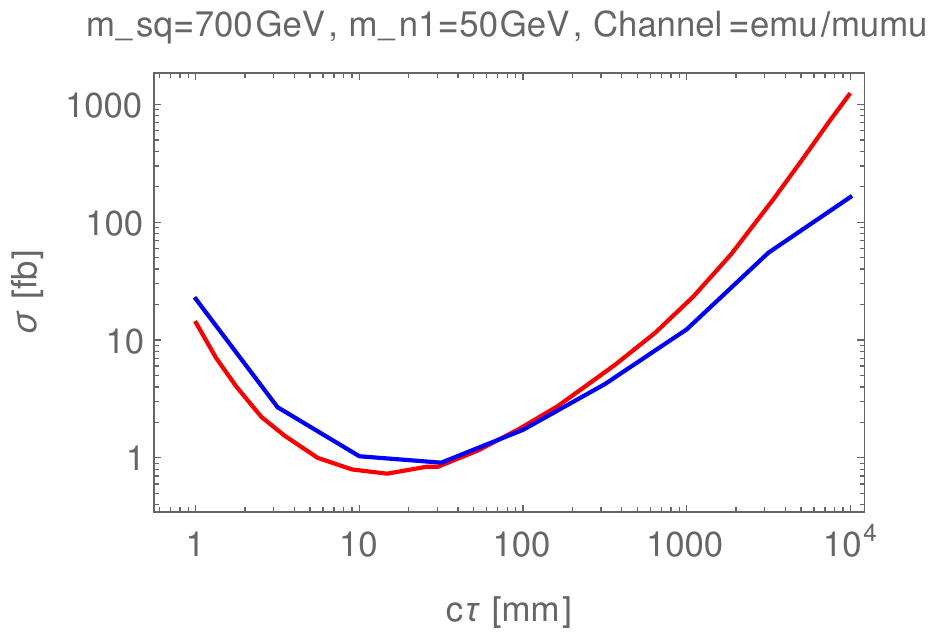}}
	\caption{Validation of the displaced lepton vertex search performed by ATLAS by comparing the upper limit on the cross Sec. obtained by ATLAS (red) with the one from our recasting (blue) for the RPV SUSY model studied in~\cite{ATLAS:2019fwx}. }
	\label{fig:ValidationDLV}
\end{figure}

To obtain the detection efficiency of a displaced lepton pair, the experimental collaboration provides two parameterisations. One for the R-parity violation (RPV) SUSY model and one for a $Z'$ toy model. There are two important differences between these models. First, in the RPV SUSY model, the LLP is a bino-like neutralino produced from the decay of a heavy squark. Due to the large mass splitting between the heavy squarks and the LLP, the LLP will have on average a much higher $p_T$ in the RPV model than in the $Z'$ model. As a result, the physical displacement in the detector will be much larger (due to the larger boost) for the same proper lifetime and the leptons arising in the decay will be also much more collimated in the RPV case than in the $Z'$ case.  Another peculiarity of the RPV SUSY model is that the LLP decays to a lepton pair and a neutralino, such that the displacement vector pointing from the primary vertex to the secondary vertex is not parallel to the momentum vector of the lepton pair. This is also the case for the models studied in this work since the LLP will always decay to a DM + SM (including leptons). As a result, here we use the efficiency parametrisation provided for the RPV SUSY model. We validated the procedure discussed in~\cite{1745920} for the RPV SUSY model in Fig.~\ref{fig:ValidationDLV}. Two different cases are illustrated, one where the LLP decays to either an electron pair or an electron-muon pair (left) and one where it decays to a muon pair or an electron-muon pair (right).

\chapter{Details about the singlet-triplet model}
\label{app:13}
In Sec.~\ref{sec:SingletTriplet}, we have studied the singlet-triplet model, denoted by ${\cal F}_{W\chi}$ in Tab.~\ref{tab:classification}, which is an extension of the SM featuring a singlet and a triplet fermion, $\chi_S$ and $\chi_T$. The most general Lagrangian one can write including interaction terms of dimension $\leq 5$ reads 
\begin{align}
\mathcal{L}_{\rm ST} =&  \frac{1}{2} \bar{\chi}_S i \cancel{\partial} \chi_S + \frac{1}{2} \text{Tr} \left[ \bar{\chi}_T i \cancel{D} \chi_T \right] - \frac{m_S}{2} \bar{\chi}_S \chi_S - \frac{m_T}{2} \text{Tr}\left[\bar{\chi}_T \chi_T \right] \nonumber \\
& + \frac{\kappa_{ST}}{\Lambda} \left( (H^{\dagger} \bar{\chi}_T H) \chi_S + \text{h.c.} \right) - \frac{\kappa_S}{\Lambda} H^\dagger H \bar{\chi}_S \chi_S - \frac{\kappa_T}{\Lambda} H^\dagger H \text{Tr} \left[\bar{\chi}_T\chi_T\right]  \nonumber \\
& + \frac{\kappa'_T}{\Lambda} \left(H^\dagger \bar{\chi}_T\chi_T H + \text{Tr} \left[ \bar{\chi}_T H^\dagger H \chi_T \right]\right) + \frac{\kappa}{\Lambda} (W^a_{\mu \nu} \bar{\chi}_S \sigma^{\mu \nu} \chi_T^a + \text{h.c.}),
\label{eq:LagrST}
\end{align}
where $\kappa,\kappa_{ST},\kappa_S,\kappa_T$ and $\kappa'_T$ are
dimensionless coefficients and $\Lambda$ is a common UV physics scale. 
In Sec.~\ref{sec:SingletTriplet}, we have effectively
assumed that $\kappa\gg\kappa_{ST},\kappa_S,\kappa_T,\kappa'_T$ so as
to focus on the cubic interaction of Fig.~\ref{fig:fi_model_diagram}. 
A complementary analysis focusing on the
Higgs portal to DM for a large range of portal couplings can be found in~\cite{Filimonova:2018qdc}.

\renewcommand{\arraystretch}{1.2}
\begin{table}[t]
	\centering
	\begin{tabular}{ | c | c \mycol c | c | }
		\hline
		\multicolumn{2}{|c\mycol}{Initial state} & \multicolumn{2}{|c|}{Final state}  \\ 
		\hline \hline
		\multicolumn{2}{|c\mycol}{$\Psi_B^\pm$} & $\chi$ & $W^\pm$ \\
		\hline
		\multicolumn{2}{|c\mycol}{$\Psi_B^0$} & $\chi$ & $Z,\gamma$ \\
		\hline
		\hline
		\multirow{5}{*}{$\Psi_B^\pm$} & $Z,\gamma,H$ & \multirow{5}{*}{$\chi$} & $W^\pm$  \\
		\cline{2-2}\cline{4-4}
		& $W^\pm$ & & $Z,\gamma,H$ \\
		\cline{2-2}\cline{4-4}
		& $l^\mp$ & & $\nu_l$ \\
		\cline{2-2}\cline{4-4}
		& $\bar{\nu_l}$ & & $l^\pm$ \\
		\cline{2-2}\cline{4-4}
		& $q$ & & $q'$ \\
		\hline
		\multirow{6}{*}{$\Psi^0_B$} & $W^\pm$ & \multirow{6}{*}{$\chi$} & $W^\pm$  \\
		\cline{2-2}\cline{4-4}
		& $Z$ & & $H$ \\
		\cline{2-2}\cline{4-4}
		& $H$ & & $Z$ \\
		\cline{2-2}\cline{4-4}
		& $l^\pm$ & & $l^\pm$ \\
		\cline{2-2}\cline{4-4}
		& $\nu_l$ & & $\nu_l$ \\
		\cline{2-2}\cline{4-4}
		& $q$ & & $q$ \\
		\hline
		$\Psi_B^+$ & $\Psi_B^-$ & \multirow{2}{*}{$\chi$} & $\Psi_B^0$ \\
		\cline{1-2}\cline{4-4}
		$\Psi_B^\pm$ & $\Psi^0_B$ & & $\Psi_B^\pm$ \\
		\hline
	\end{tabular}
	\caption{Decay and scattering processes, Initial state $\to$ Final state, contributing to the freeze-in production of DM for the singlet-triplet model discussed in Sec.~\ref{sec:SingletTriplet}.}
	\label{tab:processes}
\end{table}

If we neglect the terms that involve the Higgs field, i.e.~setting $\kappa_{ST}=\kappa_S=\kappa_T=\kappa'_T=0$, 
no mixing occurs between $\chi_S$ and $\chi_T^0$ and we can assume that the singlet field is the DM particle while
the triplet plays the role of the freeze-in mediator, that is $\chi\equiv \chi_S$, $\Psi_B^{0,\pm}\equiv\chi_T^{0,\pm}$, as we did in Sec.~\ref{sec:SingletTriplet}. A mass splitting between $\Psi_B^0$ and $ \Psi_B^\pm$ always arises due to quantum corrections. As a
result, the mass of the charged component can be written $m_C = m_T +
\Delta m$, with~\cite{Cirelli:2009uv}
\begin{align}
\Delta m = \frac{\alpha_2 m_T}{4 \pi} \left[(s_w^2-1) f\left(\frac{m_Z}{m_T}\right)+f\left(\frac{m_W}{m_T}\right) \right],
\end{align}
where
\begin{align}
f(r) = r\left[2 r^3 \ln r - 2r + \sqrt{r^2-4}(r^2+2) \ln \left(\frac{r^2-2-r\sqrt{r^2-4}}{2}\right)\right].
\end{align}
Hence the mass splitting between the neutral and the charged components of the triplet
depends on the leading order mass $m_T$. However, such a 
dependence is soft and for $m_T$ in the 100 GeV$-$1~TeV range, $\Delta m
\approx 160$~MeV. As a consequence of this small mass splitting, the charged
component of the triplet can possibly decay into the neutral component
and a soft pion, with a decay width of~\cite{Cirelli:2009uv}
\begin{equation}
\Gamma(\Psi_B^\pm \rightarrow \Psi_B^0 \pi^\pm) = \frac{2 G_F^2 f_\pi^2 \Delta m^3}{\pi} \sqrt{1- \frac{m_\pi^2}{\Delta m^2}}.
\end{equation}

The gauge interactions of the triplet and the cubic-interaction term proportional to $\kappa$\footnote{Notice that in the discussion of Sec.~\ref{sec:SingletTriplet}, we take $\kappa=1$, that is, 
we reabsorb the coupling in the definition of the scale $\Lambda$.} 
in Eq.~(\ref{eq:LagrST}) can be expanded as follows:
\begin{align}
\mathcal{L}_{int} =& \frac{e}{4 s_w} [ \bar\Psi_B^0 \cancel{W}^+\Psi_B^- +  \bar{\Psi}_B^- \cancel{W}^-\Psi_B^0 - \bar{\Psi}_B^+ \cancel{W}^+ \Psi_B^0 - \bar\Psi_B^0  \cancel{W}^- \Psi_B^+ \nonumber \\ 
& \qquad +  \bar{\Psi}^+_B\cancel{W}^0 \Psi_B^+ - \bar{\Psi}_B^- \cancel{W}^0 \Psi_B^-  ] \nonumber \\
& + \frac{\kappa}{\Lambda} [W_{\mu\nu}^- \bar\chi \sigma^{\mu\nu} \Psi_B^+ + W_{\mu\nu}^+ \bar \chi \sigma^{\mu\nu} \Psi_B^- +  W_{\mu\nu}^0 \bar \chi \sigma^{\mu\nu} \Psi_B^0 + \text{h.c.}
],
\end{align}
where $W_{\mu\nu}^\pm = \partial_\nu W^\pm_\mu - \partial_\mu
W^\pm_\nu + g (W^0_\mu W_\nu^\pm - W_\mu^\pm W_\nu^0)$ and
$W_{\mu\nu}^0 = \partial_\nu W^0_\mu - \partial_\mu W^0_\nu + g
(W^+_\mu W_\nu^- - W_\mu^- W_\nu^+)$. This Lagrangian captures all
possible interactions between the DM, the mediator and other SM
particles which give rise to decay and scattering processes that could
lead to freeze-in production of DM. The associated processes are listed in Table~\ref{tab:processes}.

\bibliographystyle{JHEP}
\bibliography{refs.bib}
\end{document}